\documentclass[twocolumn,aps,prd,amssymb,eqsecnum,nofootinbib,superscriptaddress,showpacs]{revtex4}
\usepackage{ifpdf}
\ifpdf
\usepackage{hyperref}
\else
\usepackage[dvipdfm]{hyperref}
\fi

\usepackage{amsmath}
\usepackage{amssymb}
\usepackage{epsfig}
\usepackage{bm}
\usepackage{graphicx}
\usepackage{psfrag}
\usepackage{rotating}
\usepackage{multirow}
\usepackage{wasysym}
\usepackage{overpic}

\usepackage[usenames,dvipsnames,svgnames,table]{xcolor}

\newcommand{\Caltech}{\affiliation{Theoretical Astrophysics 350-17,
    California Institute of Technology, Pasadena, California 91125, USA}}
\newcommand{\Cornell}{\affiliation{Center for Radiophysics and Space
    Research, Cornell University, Ithaca, New York 14853, USA}}
\newcommand{\Oberlin}{\affiliation{Department of Physics and Astronomy, Oberlin College, Oberlin, Ohio 44074, USA}}
\newcommand{\GWPAC}{\affiliation{Gravitational Wave Physics and Astronomy Center, California State University Fullerton, Fullerton, California 92831, USA}}

\begin{document}

\title{Visualizing Spacetime Curvature via Frame-Drag Vortexes and Tidal Tendexes \\ III. Quasinormal Pulsations of Schwarzschild and Kerr Black Holes}

\author{David A.\ Nichols} \Caltech
\author{Aaron Zimmerman} \Caltech
\author{Yanbei Chen} \Caltech
\author{Geoffrey Lovelace} \Cornell \GWPAC
\author{Keith D.\ Matthews} \Caltech
\author{Robert Owen} \Cornell \Oberlin
\author{Fan Zhang} \Caltech
\author{Kip S.\ Thorne} \Caltech
\date{printed \today}

\begin{abstract}
In recent papers, we and colleagues have introduced a way to visualize
the full vacuum Riemann curvature tensor using \emph{frame-drag vortex
lines and their vorticities}, and \emph{tidal tendex lines and their
tendicities}.  We have also introduced the concepts of
\emph{horizon vortexes and tendexes} and \emph{3-D vortexes and tendexes}
(regions on or outside the horizon where vorticities or tendicities are large).
In this paper, using these concepts, we discover a number of previously
unknown features of quasinormal modes of Schwarzschild and Kerr black holes.
These modes can be classified by a radial quantum number $n$, spheroidal harmonic orders $(l,m)$, and
parity, which can be \emph{electric} [$(-1)^l$] or \emph{magnetic}
[$(-1)^{l+1}$]. Among our discoveries are these: (i) There is a near duality
between modes of the same $(n,l,m)$: a duality in which the tendex and vortex 
structures of electric-parity modes are interchanged with the vortex and
tendex structures (respectively) of magnetic-parity modes. (ii) This near
duality is perfect for the modes' complex eigenfrequencies (which are well 
known to be identical) and perfect on the horizon; it is slightly broken in 
the equatorial plane of a non-spinning hole, and the breaking becomes greater 
out of the equatorial plane, and greater as the hole is spun up; but even 
out of the plane for fast-spinning holes, the duality is surprisingly 
good.  (iii) Electric-parity
modes can be regarded as generated by 
3-D tendexes that stick radially out of the horizon.  As these 
``longitudinal,'' near-zone tendexes rotate or oscillate, they 
generate longitudinal-transverse near-zone vortexes and tendexes, 
and outgoing and ingoing gravitational waves. 
The ingoing waves act back on the longitudinal tendexes, driving them to 
slide off the horizon, which results in decay of the mode's strength.
(iv) By duality, magnetic-parity modes are driven in this same manner by longitudinal, near-zone vortexes that stick out of the horizon.  
(v) When visualized,
the 3-D vortexes and tendexes of a $(l,m)=(2,2)$ mode, and also a
$(2,1)$ mode, spiral outward and backward
like water from a whirling sprinkler, becoming outgoing gravitational waves.  
By contrast, a $(2,2)$ mode superposed
on a $(2,-2)$ mode, has oscillating horizon vortexes or tendexes that
eject 3-dimensional vortexes and tendexes, 
which propagate outward becoming gravitational waves; and
so does a $(2,0)$ mode.  
(vi) For magnetic-parity modes of a Schwarzschild black hole, the 
perturbative frame-drag field, and hence also 
the perturbative vortexes and vortex lines,
are strictly gauge invariant (unaffected by infinitesimal magnetic-parity 
changes of time slicing and spatial coordinates).  
(vii) We have computed the vortex and tendex structures
of electric-parity modes of Schwarzschild in two very different gauges and 
find essentially no discernible differences in their pictorial visualizations.
(viii) We have compared the vortex lines, from a numerical-relativity simulation
of a black hole binary in its final ringdown stage, with the vortex lines
of a (2,2) electric-parity mode of a Kerr black hole with the same spin 
($a/M=0.945$)
and find remarkably good agreement.
\end{abstract}

\pacs{04.25.dg, 04.25.Nx, 04.30.-w}

\maketitle

\section{Motivations, Foundations and Overview}
\label{sec:intro}

\subsection{Motivations}

This is the third in a series of papers that introduce a new set of tools
for visualizing the Weyl curvature tensor (which, in vacuum, is the same as 
the Riemann tensor), and that 
develop, explore, and exploit these tools.  

We gave a brief overview
of these new tools and their applications in an initial Physical
Review Letter~\cite{OwenEtAl:2011}.  
Our principal motivation for these tools was described in 
that Letter, and in greater detail in  
Sec.\ I of our first long, pedagogical paper \cite{Nichols:2011pu} (Paper I).
In brief: We are motivated by the quest to understand \emph{the nonlinear
dynamics of curved spacetime} (what John Wheeler has called 
\emph{geometrodynamics}).  

The most promising venue, today, for 
probing geometrodynamics is numerical simulations of the collisions and
mergers of binary black holes~\cite{Centrella:2010}. 
Our new tools provide powerful ways to 
visualize the results of those simulations. As a byproduct, our 
visualizations may motivate new ways
to compute the gravitational waveforms emitted in black-hole 
mergers---waveforms that are needed as templates in LIGO's 
searches for and interpretation of those waves.

We will apply our tools to black-hole binaries in Paper IV of this series.
But first, in Papers I--III, we are applying our tools to 
analytically understood spacetimes, with two goals: (i) 
to gain intuition into the 
relationships between our tools' visual pictures of the vacuum Riemann tensor
and the analytics, and (ii) to gain substantial new insights into phenomena
that were long thought to be well understood.  Specifically, in Paper I
\cite{Nichols:2011pu}, after introducing our tools, we applied them to
weak-gravity situations (``linearized theory''); in Paper II \cite{ZhangPRD2}, 
we applied them to stationary (Schwarzschild and Kerr) black holes; and
here in Paper III we will apply them to weak perturbations (quasinormal
modes) of stationary black holes.  

\subsection{Our new tools, in brief}
\label{sec:NewTools}

In this section, we briefly summarize our new tools.  For details, see 
Secs.\ II, III, and IV of Paper I~\cite{Nichols:2011pu}, and Secs.\ II and III
of Paper II \cite{ZhangPRD2}. 

When spacetime is foliated by a family of spacelike hypersurfaces (surfaces
on which some time function $t$ is constant), the electromagnetic field
tensor $F_{\mu\nu}$ splits up into an electric field $E_{\hat i} = 
F_{\hat i \hat 0}$ and a magnetic field $B_{\hat i} = \frac12 
\epsilon_{\hat i \hat j \hat k} F_{\hat j \hat k}$, which are 3-vector fields
living in
the spacelike hypersurfaces.  Here the indices are components in 
proper reference frames (orthonormal tetrads) of observers who move
orthogonally to the hypersurfaces, and $\epsilon_{\hat i \hat j \hat k}$
is the Levi-Civita tensor in those hypersurfaces.  

Similarly, the Weyl (and vacuum Riemann)
tensor $C_{\mu\nu\lambda\rho}$
splits up into:  (i) a \emph{tidal field} $\mathcal E_{\hat i \hat j} =
C_{\hat i \hat 0 \hat j \hat 0}$, which produces the tidal gravitational
accelerations that appear, e.g., in the equation of geodesic deviation,
$\Delta a_{\hat j} = - \mathcal E_{\hat j \hat k} \Delta x^{\hat k}$
[Eq.\ (3.3) of Paper I];
and (ii) a
\emph{frame-drag field} $\mathcal B_{\hat i \hat j} = \frac12
\epsilon_{\hat i \hat p
\hat q} C_{\hat j \hat 0 \hat p \hat q}$, which produces differential
frame-dragging (differential precession of gyroscopes),
$\Delta \Omega_{\hat j} = \mathcal B_{\hat j \hat k} \Delta x^{\hat k}$
[Eq.\ (3.11) of Paper I].  

We visualize the tidal field $\boldsymbol{\mathcal E}$
by the integral curves of its three eigenvector fields,
which we call \emph{tendex lines}, and also by the 
eigenvalue of each tendex line, which we call the \emph{tendicity} of the
line and we 
depict using colors.  
Similarly, we visualize the frame-drag field $\boldsymbol{\mathcal B}$ by    
frame-drag \emph{vortex lines} (integral curves of its three eigenvector fields)
and their \emph{vorticities} (eigenvalues, color coded).  
See Figs.\ \ref{fig:22MagVortex} and \ref{fig:22SmagEquator} below
for examples.  Tendex and vortex lines are analogs of electric and
magnetic field lines. Whereas through each point in space there pass
just one electric and one magnetic field line, through each point pass 
three orthogonal tendex lines and three orthogonal vortex lines, which
identify the three principal axes of $\boldsymbol{\mathcal E}$ and 
$\boldsymbol{\mathcal B}$.

A person whose body is oriented along a tendex line gets stretched
or squeezed with a relative head-to-foot gravitational acceleration that
is equal to the person's height times the line's tendicity (depicted blue
[dark gray] in our figures for squeezing [positive tendicity] and red [light 
gray] for stretching [negative tendicity]).  
Similarly, if the person's body
is oriented along a vortex line, a gyroscope at her feet precesses around
her body axis, relative to inertial frames at her head, with an angular velocity
equal to her height times the line's vorticity (depicted blue [dark gray] for 
clockwise precession [positive vorticity] and red [light gray] for 
counterclockwise [negative vorticity]).  

We color code the horizon of a black hole by the normal-normal component
of the tidal field, $\mathcal E_{NN}$, to which we give the name
\emph{horizon tendicity}, 
and also by the normal-normal component of
the frame-drag field, $\mathcal B_{NN}$, the \emph{horizon vorticity};
see, e.g., Fig.\ \ref{fig:EvenPertHorizons} below. 
These quantities are boost-invariant along the normal
direction $\bf N$ to the horizon in the foliation's hypersurfaces.

A person hanging radially above the horizon 
or falling into it experiences head-to
foot squeezing (relative acceleration) equal to the horizon tendicity times
the person's height, and a differential head-to-foot precession of 
gyroscopes around the person's body axis with an angular velocity equal to
the horizon vorticity times the person's height.  

For any black hole, static or dynamic, 
the horizon tendicity $\mathcal E_{NN}$ and vorticity
$\mathcal B_{NN}$ are related to the horizon's
Newman-Penrose Weyl scalar $\Psi_2$, and its  
scalar intrinsic curvature
$\mathcal R$ and scalar extrinsic curvature $\mathcal X$ by
\begin{equation}
\mathcal E_{NN} + i \mathcal B_{NN} = 2 \Psi_2 
= - \frac12 (\mathcal R + i \mathcal X) + 2( \mu\rho - \lambda\sigma)\;;
\label{eq:HorizonCurvatures}
\end{equation}
\cite{Penrose1992}, and Sec.\ III of \cite{ZhangPRD2}. 
Here $\rho$, $\sigma$, $\mu$, $\lambda$ are spin coefficients related to the
expansion and shear of the null vectors $\vec l$ and $\vec n$ used in the 
Newman-Penrose formalism
[with $(\vec l + \vec n)/\sqrt2 = \vec u$ the normal to the foliation's
hypersurfaces, $(\vec l - \vec n)/\sqrt2 = \vec N$ the normal to the horizon
in the foliation's hypersurfaces, and $\vec e_2 = (\vec m + \vec m^*)/\sqrt 2$
and $\vec e_3 = (\vec m - \vec m^*)/(i\sqrt 2)$ tangent to the instantaneous
horizon in the foliation's hypersurfaces].
For stationary black holes, $\rho$ and $\sigma$ vanish, and 
$\mathcal E_{NN} = -\frac12 \mathcal R$ and $\mathcal B_{NN} = -\frac12 \mathcal X$.

For perturbations of Schwarzschild black holes, it is possible to 
adjust the slicing at first order in the perturbation, and adjust the associated
null tetrad, so as to make the spin coefficient terms in Eq.\ (\ref{eq:HorizonCurvatures}) 
vanish at first order in the perturbation; whence
$\mathcal E_{NN} = - \frac12  \mathcal R$ and 
$\mathcal B_{NN} = -\frac12  \mathcal X$.  
For perturbations of the Kerr spacetime, however, this is not possible.
See App.~\ref{sec:HorizonDetails} for details.
Following a calculation by Hartle \cite{Hartle:1974}, we show in this appendix 
that for Kerr one can achieve
$\mathcal R + i \mathcal X = - 4(\Psi_2
+ \lambda^{(0)} \sigma^{(1)})$
on the horizon, accurate through first order. 
Here, and throughout this paper, the superscripts $^{(i)}$ (or subscripts 
$_{(i)}$) indicate orders in the perturbation.

For the dynamical black holes described in \cite{OwenEtAl:2011} and for
the weakly perturbed holes in this paper, we found that
the spin terms in Eq.\ (\ref{eq:HorizonCurvatures}) are numerically small 
compared to the other terms, so 
$\mathcal E_{NN} \simeq -\frac12  \mathcal R$ and 
$\mathcal B_{NN} \simeq -\frac12  \mathcal X$. In addition,
in a recent study of the tendexes and vortexes of approximate black hole initial data, Dennison and Baumgarte \cite{Dennison2012} found that these spin terms vanish to a high order in the small velocities of their black holes, giving further evidence that these terms are 
typically negligible.

Because $\mathcal X$ is the 2-dimensional curl of a 2-dimensional vector 
(the H\'aj\'i\v{c}ek field) \cite{Damour1982}, its integral over the black 
hole's 2D horizon vanishes; and by virtue of the Gauss-Bonnet theorem, the 
horizon integral of $\mathcal R$ is equal to $8 \pi$.  
Correspondingly, for fully
dynamical black holes as well as weakly perturbed black holes, the
horizon integrals of $\mathcal E_{NN}$ and $B_{NN}$ have the approximate values
\cite{OwenEtAl:2011}
\begin{equation}
\int_{\mathcal H}\mathcal B_{NN} \simeq 0\; , \quad 
\int_{\mathcal H} \mathcal E_{NN} \simeq - 4\pi\;.
\label{eq:HorizonIntegrals}
\end{equation}

\subsection{Overview of this paper's results}

\subsubsection{Slicing, coordinates and gauges}

Throughout this paper, we use slices of constant Kerr-Schild time $\tilde t$
(which penetrate smoothly through the horizon) 
to decompose the Weyl tensor into its tidal and frame-drag fields; and
we express our quasinormal perturbations, on the slices of constant
$\tilde t$, in Kerr-Schild spatial coordinates (Secs. \ref{sec:BHPandGauge} and
\ref{sec:Methods}, and
also Paper II \cite{ZhangPRD2}). 
In the zero-spin (Schwarzschild)
limit, the Kerr-Schild slices become slices of constant ingoing 
Eddington-Finkelstein time $\tilde t$ and the spatial coordinates become
those of Schwarzschild.  Our choice of Kerr-Schild is dictated by these
coordinates' resemblance to the coordinates that are typically used
in numerical-relativity simulations of binary black holes, at late times, when
the merged hole is settling down into its final Kerr-black-hole state;
see, e.g., Fig.\ \ref{fig:NRComparison} below.  

For a perturbed black hole, the slices and coordinates get modified 
at perturbative order
in ways that depend on the {\it gauge} 
used to describe the perturbations
(i.e., the slicing and spatial coordinates at perturbative order); see
Sec.\ \ref{sec:GaugePerts}.  

For spinning black holes, we perform all our computations
in ingoing radiation gauge (Sec.\ \ref{sec:BHPandGauge} and App.\ \ref{sec:CCKProc}). 
For non-spinning (Schwarzschild) black
holes, we explore gauge dependence by working with two gauges that
appear to be quite different:
ingoing radiation gauge (App.\ \ref{sec:CCKProc}), and Regge-Wheeler gauge 
(App.\ \ref{sec:RWApp}).  Remarkably, for each mode we have explored, 
the field-line visualizations that we have carried out in these two gauges
look nearly the same to the human eye; visually we see little gauge dependence.  
We discuss this and the differences in the gauges, in considerable detail,
in Sec.\ \ref{sec:GaugePerts} and App.~\ref{sec:GaugeCompare}.

For a Schwarzschild black hole, 
we have explored somewhat generally the influence of perturbative 
slicing changes and perturbative coordinate changes on the tidal and
frame-drag fields, and on their tendex and vortex lines, and tendicities
and vorticities (Sec.\ \ref{sec:GaugePerts}). 
We find that the tendicities and vorticities are less affected by perturbative slicing changes, than
the shapes of the tendex and vortex lines. We also find that while coordinate changes affect the shapes of the tendex and vortex lines, the tendicity and vorticity along a line is unchanged, and that in the wave zone a perturbative change in coordinates affects the tendicity and vorticity at a higher order than the effect of gravitational radiation.

For this reason, in this paper we pay considerable
attention to vorticity and tendicity contours, as well as to the shapes
of vortex and tendex lines. 

\subsubsection{Classification of quasinormal modes}

As is well known, the quasinormal-mode, complex eigenfrequencies 
of Schwarzschild and Kerr black
holes can be characterized by three integers: a poloidal quantum number
$l =2, 3, ...$, an azimuthal quantum number $m= -l, -l+1, ... + l$,
and a radial quantum number $n$.  For each $\{n,l,m\}$ and its eigenfrequency
$\omega_{nlm}$, there are actually two different quasinormal modes (a
two-fold degeneracy).
Of course, any linear combination of these two
modes is also a mode. We focus on those linear combinations of modes that have \emph{definite parity} (App.\ \ref{sec:CCKProc}).  

We define a tensor
field to have positive parity if it is unchanged under reflections through
the origin, and negative parity if it changes sign.  A quasinormal mode
of order $(n,l,m)$ is said to have \emph{electric parity} 
[or \emph{magnetic parity}] if the parity of its metric perturbation is 
$(-1)^l$ [or $(-1)^{l+1}$].  The parity of the tidal-field perturbation 
is the same as that of the metric perturbation, but that of the
frame-drag field is opposite.   In much of the literature the phrase 
``even parity'' is used in place of ``electric parity'', and ``odd parity''
in place of ``magnetic parity''; we avoid those phrases because of 
possible confusion with positive parity and negative parity.

In this paper, we focus primarily on the most slowly damped $(n=0)$ quadrupolar
($l=2$) modes, for various azimuthal quantum numbers $m$ and for electric- 
and magnetic-parity. Since we discuss exclusively the 
$n=0$ modes, we will suppress the $n$ index and abbreviate mode numbers as $(l,m)$.	 

\subsubsection{The duality of magnetic-parity and electric-parity 
modes} 
\label{sec:modeDuality}

In vacuum, the exact Bianchi identities for the Riemann tensor become, under
a slicing-induced split of spacetime into space plus time, a set of Maxwell-like
equations for the exact tidal field and frame-drag field [Eqs.\ (2.15) of
Paper I \cite{Nichols:2011pu} in a local Lorentz frame; Eqs.\ (2.13) and (2.4) of
Paper I in general].
These Maxwell-like equations exhibit an exact duality: If one takes
any solution to them and transforms $\boldsymbol{\mathcal E} \rightarrow
\boldsymbol{\mathcal B}$, $\;\boldsymbol{\mathcal B} \rightarrow
- \boldsymbol{\mathcal E}$, they continue to be satisfied (Sec.\ II\,B\,1 of 
Paper I~\cite{Nichols:2011pu}).  

This duality, however, is broken by 
the spacetime geometry of a stationary black hole.  A Schwarzschild black hole
has a monopolar tidal field $\boldsymbol{\mathcal E}$ and vanishing
frame-drag field $\boldsymbol{\mathcal B}$; and a Kerr black hole has
a monopolar component to its tidal field (as defined by a spherical-harmonic
analysis at large radii or at the horizon), but only dipolar and higher-order
components to its frame-drag field.

When a Schwarzschild or Kerr black hole is perturbed, there is a near duality
between its electric-parity mode and its magnetic-parity mode of the same
$(l,m)$; but the duality is not
exact.  The unperturbed hole's duality breaking induces (surprisingly weak) 
duality-breaking imprints in the quasinormal modes.  We explore this duality
breaking in considerable detail in this paper (Secs.\ \ref{sec:SlicingDetails}, 
\ref{sec:HorizonQuant}, \ref{sec:TendVortSchwKerr}, and \ref{sec:ApproximateDual}, 
and Apps.\ \ref{sec:CCKProc} and \ref{App:HorizonFromPsi0}).  

If one tries to see the duality between electric-parity and
magnetic-parity modes, visually, in pictures of the perturbed hole's
tendex and vortex lines, the duality is hidden by the dominant background 
tidal field
and (for a spinning hole) the background frame-drag field.  To see the duality
clearly, we must draw pictures of tendex and vortex lines for the
perturbative parts $\delta \boldsymbol{\mathcal E}$ and 
$\delta \boldsymbol{\mathcal B}$ of the tidal and frame-drag fields,
with the unperturbed fields subtracted off.  We draw many such pictures in
this paper.

We have made extensive comparisons of the least damped ($n=0$) 
electric-parity and magnetic-parity modes with $(l=2,\, m=2)$.  
These two $(2,2)$ modes (for any
chosen black-hole mass $M$ and spin parameter $a$) have identically the same
complex eigenfrequency, i.e., they are degenerate (as has long been known
and as we discussed above).  This frequency degeneracy is an unbroken duality.  

Pictures of the perturbative vortex and tendex lines and their 
color-coded vorticities and tendicities show a strong but not perfect duality:
For a non-spinning hole, the perturbative vortex lines and their vorticities
for the magnetic-parity mode (e.g., Fig.~\ref{fig:22MagVortex})
look almost the same as the perturbative tendex 
lines and their tendicities for the electric-parity mode (Fig.~\ref{fig:TendexVortexEq}); and 
similarly
for the other pair of lines and eigenvalues.  As the hole's spin is increased,
the duality becomes weaker (the corresponding field lines and eigenvalues
begin to differ noticeably); but even for very high spins, the duality is
strikingly strong; see bottom row of Fig.\ \ref{fig:TendexVortexEq} below.
The duality remains perfect on the horizon in ingoing radiation gauge
for any spin, no matter how fast (Sec. \ref{sec:HorizonQuant} and App.\ \ref{sec:HorizonDetails}), 
and there is a sense in which it also remains perfect on
the horizon of Schwarzschild in Regge-Wheeler gauge (last paragraph of 
App.\ \ref{sec:RWZHorizon}).

\subsubsection{Digression: Electromagnetic perturbations of a Schwarzschild 
black hole}
\label{sec:11EM}

As a prelude to discussing the physical character of the gravitational
modes of a black hole, we shall discuss electromagnetic (EM) modes, i.e.,
quasinormal modes of the EM field around a black hole.  The properties of
EM modes that we shall describe can be derived from Maxwell's equations in
the Schwarzschild and Kerr spacetimes, but we shall not give the derivations. 

\begin{figure}
\begin{overpic}[width=0.4\columnwidth]{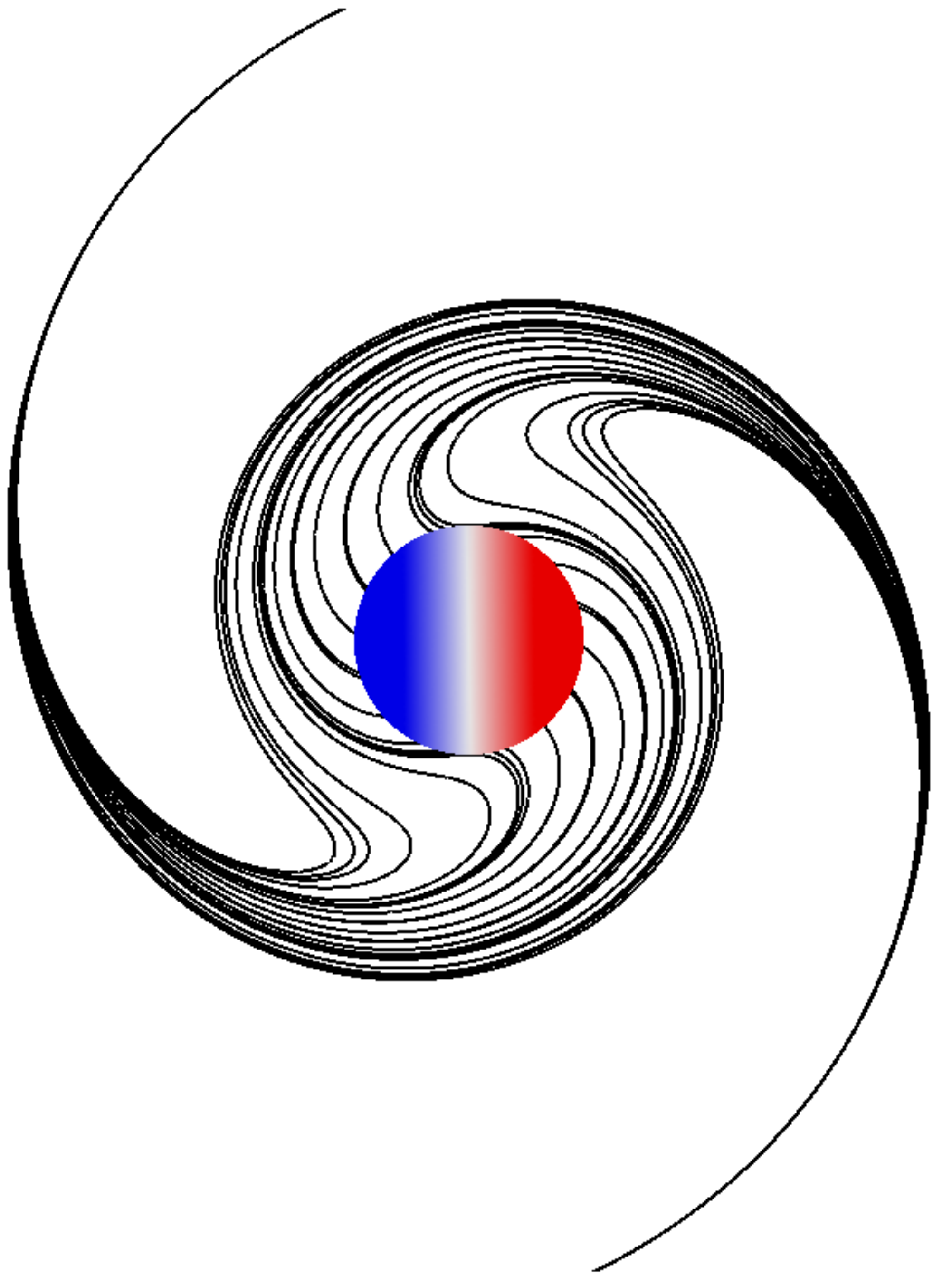}
\put(25,5){(a)}
\end{overpic}
\begin{overpic}[width=0.5\columnwidth]{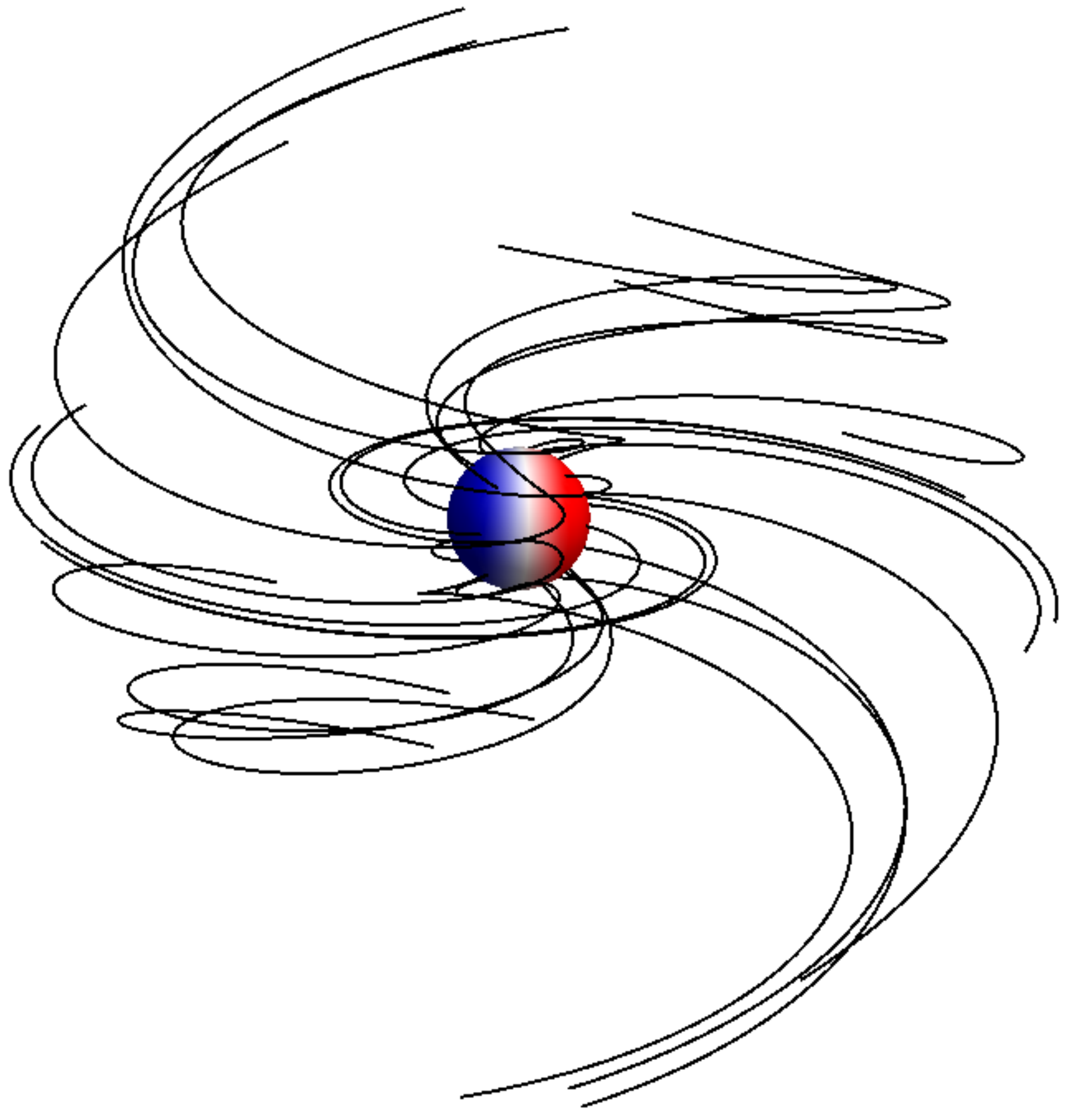}
\put(25,5){(b)}
\end{overpic}
\caption{(color online). 
(a) Some magnetic field lines in the equatorial plane for the
$(1,1)$ quasinormal mode of the electromagnetic field around
a Schwarzschild black hole, with Eddington-Finklestein slicing.  
The horizon is color coded by the sign of the
normal component of the magnetic field.  The configuration rotates 
counterclockwise in time. (b) Some magnetic field lines
for this same quasinormal mode, in 3 dimensions.}
\label{fig:11Mag}
\end{figure}

Because the unperturbed hole has no EM field and the vacuum
Maxwell equations exhibit a perfect duality (they are unchanged when
$\bm E \rightarrow \bm B$ and $\bm B \rightarrow - \bm E$), the 
EM modes exhibit perfect duality. For any magnetic-parity EM mode, the magnetic field pierces the horizon,
so its normal component $B_N$ is nonzero, while
$E_N$ vanishes. By duality, an electric-parity EM mode must have $E_N \ne 0$
and $B_N = 0$. For a magnetic-parity mode, the near-zone magnetic fields that stick out of the
horizon can be thought of as the source of the mode's emitted EM
waves.  We make this claim more precise by focusing on the fundamental
($n=0$), magnetic-parity, $l=1$, $m=1$ mode: 

Figure \ref{fig:11Mag} shows magnetic field lines for
this (1,1) mode, on the left (a) in the hole's equatorial plane, 
and on the right (b) in
3 dimensions with the equatorial plane horizontal.  On the left, we see
a bundle of magnetic field lines that
thread through the horizon and rotate counterclockwise.  As they
rotate, the field lines spiral outward and backward, like water streams from a
whirling sprinkler, becoming the magnetic-field component of an 
outgoing electromagnetic wave.  The electric field lines for this mode (not shown)
are closed circles that represent
the electric part of electromagnetic waves traveling outward at radii
$r \gg 2M$ and inward at radii $r \simeq 2M$. This mode's waves, we claim, are generated
by the near-zone, rotating magnetic field lines that thread the hole 
(Fig.\ \ref{fig:11Mag}a). An analogy will make this clear.  

Consider a rotating (angular velocity $\sigma$), perfectly conducting sphere in 
which is anchored a magnetic
field with the same dipolar normal component $B_N \propto \Re [Y^{11}(\theta,
\phi) e^{-i\sigma t}]$ as the horizon's
$B_N$ for the (1,1) quasinormal mode (the red [light gray] and blue [dark gray]
coloring on the horizon in Fig.\ \ref{fig:11Mag}).  
At some initial moment of time,
lay down outside the conducting sphere, a magnetic-field configuration that
(i) has this $B_N$ at the sphere, (ii) satisfies the constraint equation 
$\boldsymbol{\nabla}\cdot \bm B = 0$, (iii) resembles the field
of Fig.\ \ref{fig:11Mag} in the near zone, i.e., at
$r \lesssim \lambdabar = c/\sigma$ and at larger radii
has some arbitrary form that is
unimportant; and (iv) (for simplicity) specify a vanishing
initial electric field.  Evolve these initial fields
forward in time using the dynamical Maxwell equations.  It should be
obvious that the near-zone, rotating magnetic field will not change much. 
However,
as it rotates, via Maxwell's dynamical equations it will
generate an electric field, and those two fields, interacting, will give rise
to the outgoing electromagnetic waves of a $l=1$, $m=1$ magnetic dipole.
Clearly, the ultimate source of the waves is the rotating, near-zone
magnetic field that is anchored in the sphere.  (Alternatively, one can
regard the ultimate source as the electric currents in the sphere, that
maintain the near-zone magnetic field.)

Now return to the magnetized black hole of Fig.\ \ref{fig:11Mag}, and
pose a similar evolutionary scenario:  At some initial moment of time,
lay down a magnetic-field configuration that (i) has the same normal
component at the horizon as the (1,1) mode, (ii) satisfies the constraint
equation $\boldsymbol{\nabla}\cdot \bm B = 0$, 
and (iii) resembles the field of Fig.\ \ref{fig:11Mag} in the near zone.  
In this case, the field is not firmly anchored in the central
body (the black hole), so we must also specify its time derivative to make
sure it is rotating at the same rate as the (1,1) quasinormal mode.  This
means (by a dynamical Maxwell equation) that we will also be giving a 
nonvanishing electric field that resembles, in the near zone, that of the
(1,1) mode and in particular does not thread the horizon.  Now evolve this
configuration forward in time.  It will settle down, rather quickly,
into the (1,1) mode, with outgoing waves in the wave zone, and ingoing
waves at the horizon.  This is because the (1,1) mode is the most slowly damped 
quasinormal mode that has significant overlap with the initial data.

As for the electrically conducting, magnetized sphere, so also here, the
emitted waves are produced by the rotation of the near-zone 
magnetic field.  But here, by contrast with there, the emitted waves
act back on the near-zone magnetic field, causing the field lines to 
gradually slide off the horizon, resulting in a decay of the
field strength at a rate given by the imaginary part of the mode's complex
frequency. 

This backaction can be understood in greater depth by splitting the
electric and magnetic fields, \emph{near the horizon}, 
into their \emph{longitudinal}
(radial) and \emph{transverse} pieces.  The longitudinal magnetic field
is $B_N$ and it extends radially outward for a short distance; the 
tangential magnetic field is a 2-vector $\bm B^{\rm T}$
parallel to the horizon; 
and similarly for the electric field, which has $E_N = 0$
and so is purely transverse.  The tangential fields actually only look
like ingoing waves to observers who, like the horizon, move outward at
(almost) the speed of light: the observers of a Schwarzschild
time slicing.  As one learns in the Membrane Paradigm for black holes
(Secs.\ III.B.4 and III.C.2 of \cite{Thorne-Price-MacDonald:Kipversion}),
such observers can map all the physics of the event horizon onto a 
\emph{stretched horizon}---a spacelike 2-surface of constant lapse function
$\alpha = \sqrt{1-2M/r} \ll 1$ very close in spacetime to the event horizon.
On the stretched horizon, these observers
see $\bm E^{\rm T} = \bm N \times \bm B^{\rm T}$ (ingoing-wave
condition), and the tangential magnetic field acts back on the 
longitudinal field via
\begin{equation}
{\partial B_N \over \partial t} +  ^{(2)}\boldsymbol{\nabla}
\cdot ( \alpha \bm B^{\rm T}) =0 \;.
\label{eq:EMbackaction}
\end{equation}
Here $^{(2)}\boldsymbol{\nabla} \cdot (\alpha \bm B^{\rm T})$ is the 
2-dimensional divergence in the stretched horizon, and
the lapse function in this equation compensates for the
fact that the Schwarzschild observers see a tangential field that diverges
as $1/\alpha$ near the horizon, due to their approach to the speed of light.

Equation (\ref{eq:EMbackaction}) is a conservation law for magnetic field 
lines on the stretched horizon.  The density (number per unit area) of field
lines crossing the stretched horizon is $B_N$, up to a multiplicative constant;
the flux of field lines (number moving through unit length of some line in
the stretched horizon per unit time) is $\bm B^{\rm T}$, up to the same
multiplicative constant; and Eq.\ (\ref{eq:EMbackaction}) says that the 
time derivative of the density plus the divergence of the flux vanishes: the
standard form for a conservation law.

Return to Fig.~ \ref{fig:11Mag}; the dynamics embodied in this scenario are summarized as follows: 
\emph{The longitudinal magnetic field $B_N(\theta,\phi)$ 
is laid down as an initial condition (satisfying
the magnetic constraint condition).  As it rotates, it generates the
ingoing-wave near-horizon transverse fields 
embodied in $\bm E^{\rm T}$ and $\bm B^{\rm T}$ (and also the outgoing
electromagnetic waves far from the hole); and the divergence of 
$\alpha \bm B^{\rm T}$, via Eq.\ (\ref{eq:EMbackaction}), then acts back on
the longitudinal field that produced it, pushing the field lines 
away from the centers of the blue (dark gray) and red (light gray) spots on the
stretched horizon toward the white ring. 
Upon reaching the white ring, each field line in
the red region attaches onto a field line from the blue region and slips
out of the horizon.  Presumably, the field line then travels outward away
from the black hole and soon becomes part of the outgoing gravitational
waves.  The gradual loss of field lines in this way is responsible for
the mode's exponential decay.}

\subsubsection{The physical character of magnetic-parity and electric-parity 
modes}
\label{sec:ModeCharacter}

For a Schwarzschild black hole, the physical character of the gravitational
modes is very similar to that of the electromagnetic modes:

Just as a magnetic-parity EM mode has nonzero $B_N$ and vanishing $E_N$,
so similarly:
\textit{for a Schwarzschild black hole, 
the \textbf{magnetic-parity} modes of any $(l,m)$
have nonzero (solely perturbative) horizon 
vorticity $\delta \mathcal B_{NN} = \mathcal B_{NN}$, and vanishing 
perturbative horizon tendicity $\delta E_{NN}=0$; and
correspondingly, from the horizon there emerge 
nearly normal vortex lines that
are fully perturbative and no
nearly normal, perturbative tendex lines.} 

Just as in the EM case, the near-zone magnetic fields that emerge from
the horizon are the source of the emitted electromagntic waves, so
also in the gravitational case, 
\emph{for a magnetic-parity mode 
the emerging, near-zone, vortex lines and their vorticities
can be
thought of as the source of the emitted magnetic-parity gravitational
waves} (see the next subsection).  
In this sense, magnetic-parity modes can be thought of as
fundamentally frame-drag in their physical origin. Figure 
\ref{fig:22MagVortex} below depicts a $(2,2)$ example.
We will discuss this example in Sec.\  \ref{sec:Intro22}. 

For a Schwarzschild black hole, 
\textit{the \textbf{electric-parity} modes of any $(l,m)$ 
have nonzero perturbative horizon tendicity $\delta \mathcal E_{NN} \ne 0$, and 
vanishing horizon vorticity $\delta B_{NN} = B_{NN} = 0$; and 
correspondingly, from the horizon there emerge 
nearly normal perturbative tendex lines and no 
nearly normal vortex lines.}  
The emerging, near-zone, perturbative tendex lines can be
thought of as the source of the mode's emitted electric-parity gravitational 
waves.  In this sense, electric-parity modes can be thought of as fundamentally
tidal in their physical origin.

There is a close analogy, here, to the tidal and frame-drag fields of dynamical
multipoles in linearized theory (Paper I~\cite{Nichols:2011pu}):  
Electric-parity (mass) 
multipoles have a tidal field that rises more rapidly, as one approaches
the origin, than the frame-drag field, so these electric-parity 
multipoles are fundamentally tidal in physical origin. 
By contrast, for magnetic-parity (current) multipoles it is the frame-drag 
field that grows most rapidly as one approaches the origin, so they are 
fundamentally frame-drag in physical origin.

When a black hole is spun up, the horizon vorticities of its 
electric-parity modes 
become nonzero, and the horizon tendicities of its magnetic-parity modes 
acquire nonzero perturbations.  However, these spin-induced effects
leave the modes still predominantly tidal near the horizon for 
electric-parity modes, and predominantly frame-drag near the horizon for
magnetic-parity modes (Sec.\ \ref{sec:22Perts}).

\subsubsection{ The $(2,2)$ magnetic-parity mode of a Schwarzschild hole}
\label{sec:Intro22}

In this and the next several subsections, we summarize much of what
we have learned about specific $n=0$, $l=2$ modes (the least-damped
quadrupolar modes), for various $m$.  
We shall focus primarily on magnetic-parity modes,
since at the level of this discussion the properties of electric-parity
modes are the same, after a duality transformation $\delta \boldsymbol{\mathcal E}
\rightarrow \delta \boldsymbol{\mathcal B}$, $\; \delta\boldsymbol{\mathcal B}
\rightarrow - \delta \boldsymbol{\mathcal E}$. 

\begin{figure}
\includegraphics[width=0.95\columnwidth]{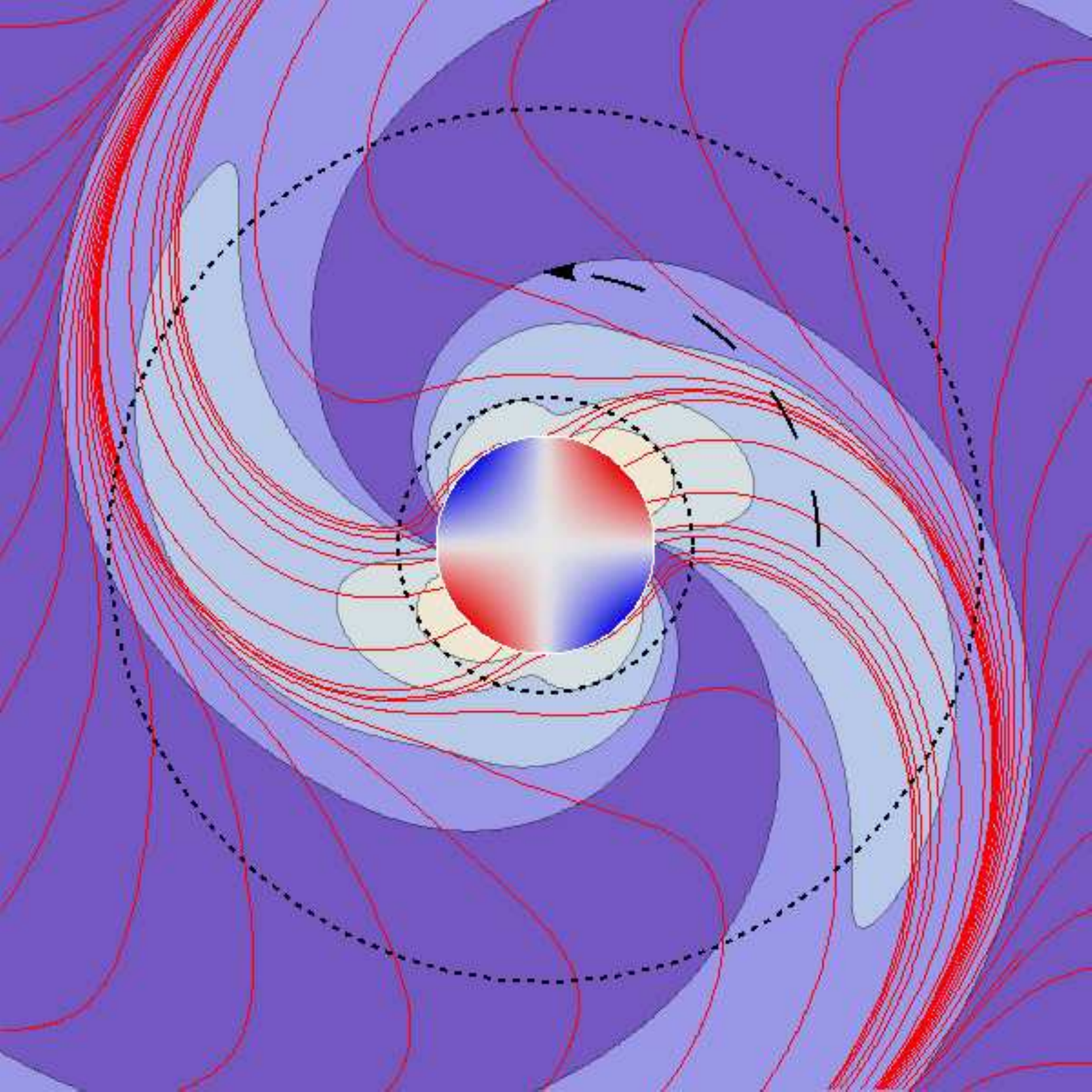}
\caption{(color online).
Some vortex lines (solid, red lines) and contours of vorticity
(shaded regions) in the equatorial plane for the 
$(2,2)$ \emph{magnetic-parity} quasinormal mode of a 
non-rotating, Schwarzschild black hole, with complex eigenfrequency
$\omega = (0.37367-0.08896i)/M$ where $M$ is the hole's mass.  
The horizon (central circle)
is color coded by the horizon vorticity $\mathcal B_{NN}$ as seen by someone
looking down on the black hole; this vorticity is entirely perturbative.  
The thick, solid red curves are one set of vortex lines in the equatorial 
plane---the set with negative vorticity.  
These lines include some that emerge from the horizon in the negative-vorticity
(red) regions, and some that never reach the horizon.  
The other, positive-vorticity, 
equatorial vortex lines are orthogonal to the ones shown, and are identical
to those shown but rotated through 90 degrees around the hole so some
of them emerge from the horizon in the positive-vorticity (blue) regions.  
The contours represent the vorticity of the red (negative-vorticity) vortex 
lines, with largest magnitude of vorticity white and smallest purple 
(dark gray); the contours mark where the vorticity has fallen to 
$50\%$, $25\%$, $10\%$, and $5\%$ of the maximum value attained at the center 
of the horizon vortex. The two dotted circles are drawn at Schwarzschild radii $r = \lambdabar$
and $r= \pi \lambdabar = \lambda/2$.  They mark the approximate 
outer edge of the near zone and the approximate inner edge of the wave zone.
The arrow marks the direction of rotation of the perturbation.}
\label{fig:22MagVortex}
\end{figure}

\begin{figure}
\includegraphics[width=0.95\columnwidth]{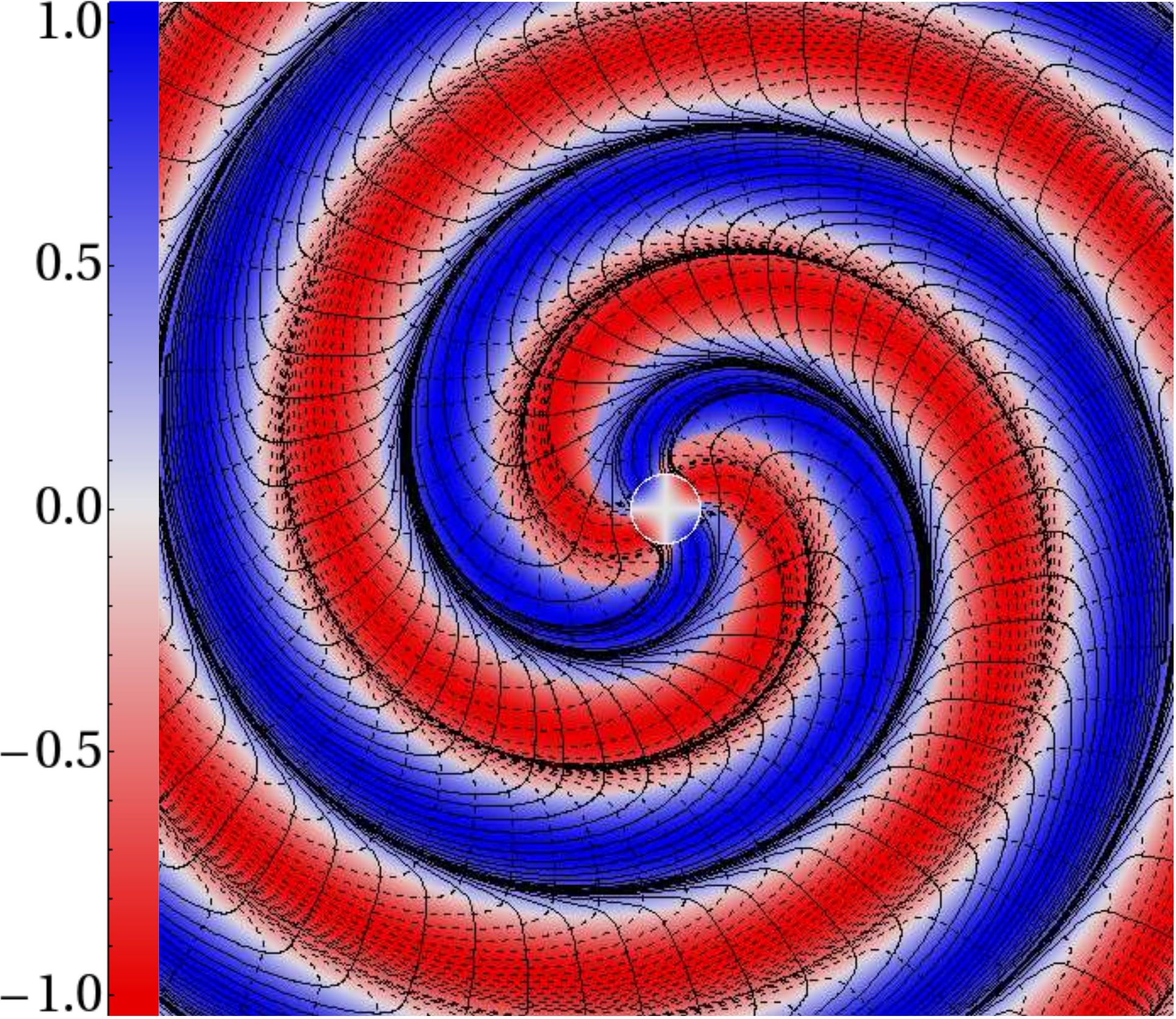}
\caption{(color online).
The vortex lines (solid black for clockwise; dashed black for 
counterclockwise) and color-coded vorticities in the equatorial plane for the 
same magnetic-parity $(2,2)$ mode as in Fig.\ \ref{fig:22MagVortex}.  
This figure differs from Fig.\ \ref{fig:22MagVortex} in ways designed to give 
information about the emitted gravitational waves:  (i) It extends rather far
out into the wave zone.
(ii) It shows the angular structure of the vorticity for the dominant vortex 
lines in each region of the equatorial plane.  More specifically: the color at 
each point represents the vorticity of the equatorial vortex line there which 
has the largest magnitude of vorticity, with radial variations of vorticity 
normalized away (so the linear color code on the left indicates vorticity 
relative to the maximum at any given radius).
The regions of large positive vorticity (blue [dark gray]) are 
\emph{clockwise vortexes}; those of large negative vorticity (red [light gray])
are \emph{counterclockwise vortexes}.}
\label{fig:22SmagEquator}
\end{figure}

\begin{figure}
\includegraphics[width=0.95\columnwidth]{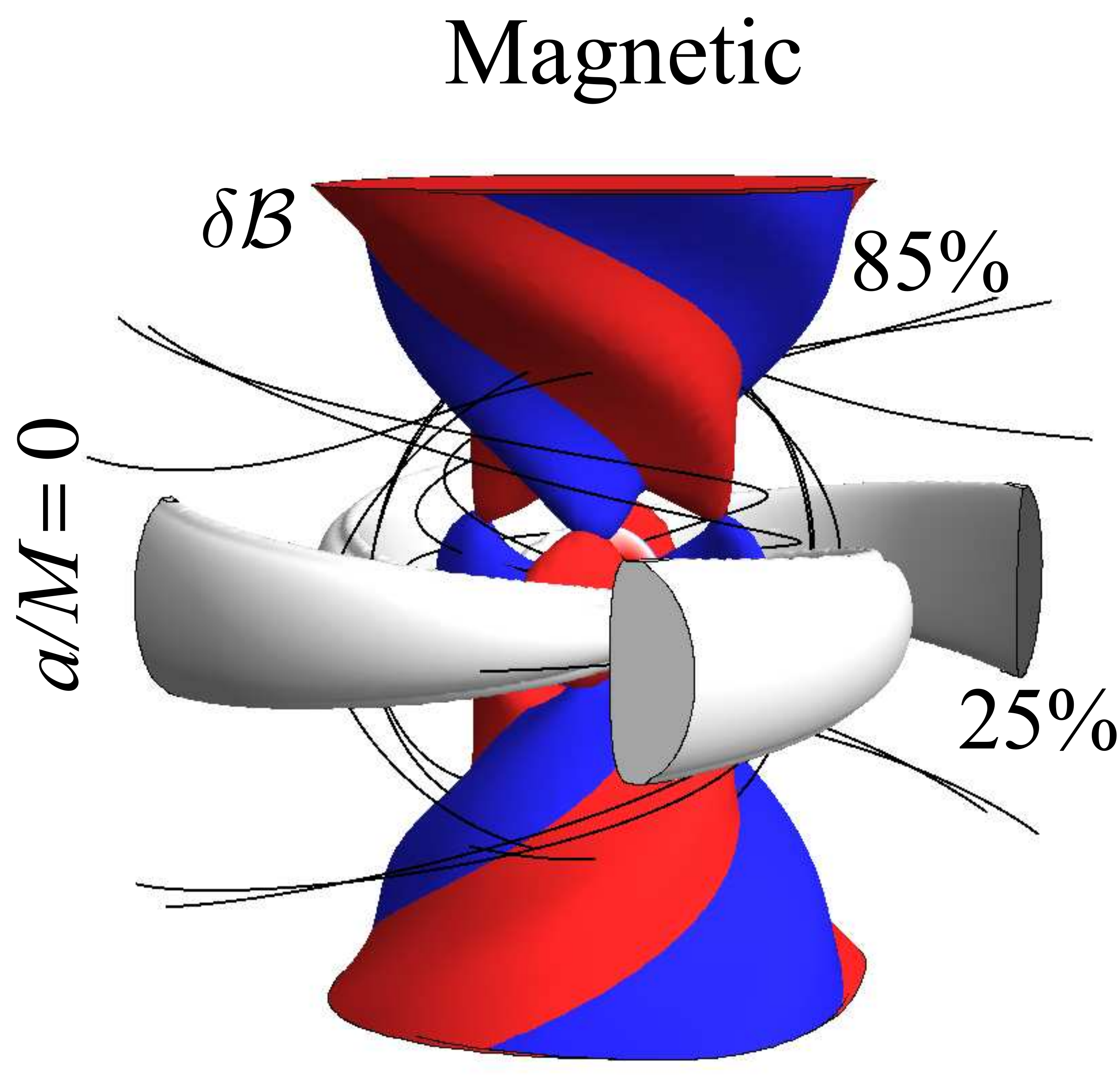}
\caption{(color online).
Some 3-dimensional clockwise vortex lines (shown black) and 
regions of large vorticity (vortexes, shown blue [dark gray] and red [light 
gray]) and small vorticity (shown off-white), for the same magnetic-parity mode
of a Schwarzschild black hole as in Figs.\ \ref{fig:22MagVortex} and 
\ref{fig:22SmagEquator}.  More specifically: the inner sphere is the horizon,
color coded by its vorticity.  The blue region is a clockwise
vortex in which one vortex line has vorticity at 
least 85\% of the maximum value at that radius, and similarly for the counterclockwise
red region.  The four off-white regions are locations where no vortex line has 
magnitude of vorticity in excess of 25\% of the maximum at that radius.}
\label{fig:22Smag3D}
\end{figure}

It is a remarkable fact that, for a magnetic-parity mode of a
Schwarzschild black hole, all gauges share the same slicing, and the mode's
frame-drag field is unaffected by perturbative changes of spatial coordinates;
therefore, the frame-drag field is fully gauge invariant. 
See Sec.\ \ref{sec:GaugePerts}.  This means that 
Figs.\ \ref{fig:22MagVortex}--\ref{fig:22-2Smag} are fully gauge invariant.

We begin with the $(2,2)$ magnetic-parity mode of a Schwarzschild black hole. 
Figure \ref{fig:22MagVortex} depicts the negative-vorticity vortex lines
(red) and contours of their vorticity (white and purple [dark gray]), 
in the hole's equatorial plane.
Orthogonal to the red (solid) vortex lines (but not shown) are
positive-vorticity, vortex lines that also lie in the equatorial plane.
Vortex lines of the third family pass orthogonally through the equatorial plane.  
The entire configuration rotates counterclockwise, as indicated by the thick dashed arrow.  The dotted
lines, at radii $r=\lambdabar$ and $r=\pi \lambdabar$ (where $\lambdabar$
is the emitted waves' reduced wavelength), mark the approximate outer 
edge of the near zone, and the approximate inner edge of the wave zone.  

Just as the near-zone electromagnetic (1,1) perturbations are dominated 
by radial field lines that thread the black hole and have a dipolar
distribution of field strength, so here the \emph{near-zone gravitational
perturbations} are dominated by (i) the radial vortex lines that thread the
hole and have a quadrupolar distribution of their horizon vorticity
$\mathcal B_{NN} = \delta \mathcal B_{NN}$, 
and also by (ii) a transverse, isotropic frame-drag field
$\mathcal B_{\hat\theta \hat\theta} = \mathcal B_{\hat\phi\hat\phi}
= - \frac12 \delta \mathcal B_{NN}$ that is tied to $\mathcal B_{NN}$
in such a way as to guarantee that this dominant part of 
$\boldsymbol{\mathcal B}$ is traceless.

This full structure, the normal-normal field and its accompanying isotropic
transverse field, makes up the \emph{longitudinal,
nonradiative frame-drag field} ${\boldsymbol{\mathcal B}}^{\rm L}$
near the horizon.  (As we shall discuss below, this longitudinal 
structure is responsible for generating the mode's gravitational waves,
and all of the rest of its fields.)
Somewhat smaller
are (i) the longitudinal-transverse components of $\boldsymbol{\mathcal B}$
($\mathcal B_{\hat r \hat\theta}$ and $\mathcal B_{\hat r \hat\phi}$), 
which together make up the \emph{longitudinal-transverse}
part of the frame-drag field,
a 2-vector ${\boldsymbol{\mathcal B}}^{\rm LT}$ parallel to the horizon, and 
give the horizon-piercing vortex lines small non-normal components; and
(ii)
transverse-traceless components $\mathcal B_{\hat\theta\hat\theta}
= - \mathcal B_{\hat\phi\hat\phi}$, which make up the \emph{transverse-traceless}
part of the frame-drag field, a 2-tensor ${\boldsymbol{\mathcal B}}^{\rm TT}$
parallel to the horizon, and are ingoing
gravitational waves as seen by Schwarzschild observers.  
(This decomposition
into L, LT, and TT parts is useful only near the horizon and in the wave 
zone, where there are preferred longitudinal directions associated with 
wave propagation.)

As the near-zone, longitudinal frame-drag field 
$\boldsymbol{\mathcal B}^{\rm L}$ rotates, it generates
a near-zone longitudinal-transverse (LT) perturbative 
frame-drag field $\boldsymbol{\mathcal B}^{\rm LT}$
via $\boldsymbol{\mathcal B}$'s propagation equation (the
wave equation for the Riemann tensor), and it generates a LT 
tidal field
$\delta \boldsymbol{\mathcal E}^{\rm LT}$ via 
the Maxwell-like Bianchi
identity which says, in a local Lorentz frame (for simplicity),
$\partial \boldsymbol{\mathcal E}/\partial t = (\boldsymbol\nabla \times
\boldsymbol{\mathcal B})^{\rm S}$, where the superscript S means ``symmetrize"
[Eq.\ (2.15) of Paper I]. These three fields, $\boldsymbol{\mathcal B}^{\rm L}$,
$\boldsymbol{\mathcal B}^{\rm LT}$, and 
$\delta \boldsymbol{\mathcal E}^{\rm LT}$
together maintain each other
during the rotation via this Maxwell-like Bianchi identity and its
(local-Lorentz-frame) dual $\partial \boldsymbol{\mathcal B}/\partial t 
= -(\boldsymbol\nabla \times \boldsymbol{\mathcal E})^{\rm S}$. They
also generate the transverse-traceless parts of both fields, 
$\boldsymbol{\mathcal B}^{\rm TT}$,
and $\delta \boldsymbol{\mathcal E}^{\rm TT}$, which  become the outgoing
gravitational waves in the wave zone and ingoing gravitational waves at the
horizon.

In the equatorial plane, this outgoing-wave generation process, 
described in terms 
of vortex and tendex structures, is quite pretty, and is 
analogous to the (1,1) magnetic-field mode of Fig.\ \ref{fig:11Mag} and 
Sec.\ \ref{sec:11EM}: 
As one moves outward into 
the induction zone and then the wave zone, the
equatorial vortex lines bend backward into outgoing spirals 
(Fig.\ \ref{fig:22MagVortex}) and gradually
acquire accompanying tendex lines.
The result, locally, in the wave zone, is the 
standard pattern of transverse, orthogonal red and blue vortex lines; 
and (turned by 45 degrees to them) transverse, orthogonal red and blue 
tendex lines,
that together represent plane gravitational waves (Fig.\ 7 of Paper I).

It is instructive to focus attention on regions of space with large
magnitude of vorticity. We call these regions \emph{vortexes}.  Figure
\ref{fig:22SmagEquator} shows that the equatorial frame-drag field consists
of four outspiraling vortexes, two red ([light gray] counterclockwise) and two 
blue ([dark gray] clockwise).

The solid black lines in the figure are clockwise vortex lines.  
In the clockwise vortexes of the wave zone, they have the large magnitude of 
vorticity that is depicted as blue (dark gray), and they are nearly
transverse to the radial wave-propagation direction; so they represent
crests of outgoing waves.  
In the counterclockwise vortexes (red [light gray] regions), these clockwise
vortex lines have very small magnitude of vorticity and
are traveling roughly radially, leaping through a red vortex (a wave trough) 
from one blue vortex (wave crest) to the next.  These clockwise vortex
lines accumulate at the outer edges of the clockwise (blue) vortexes.

The dashed black lines are counterclockwise vortex lines, 
which are related to the red (light gray), counterclockwise
vortexes in the same way as the solid clockwise vortex lines are
related to the blue (dark gray), clockwise vortexes. 

Outside the equatorial plane, this mode also represents outgoing gravitational
waves, once one gets into the wave zone.  We depict the strengths of the vortexes which become those waves in Fig.\ \ref{fig:22Smag3D}. The blue (dark gray) regions are 
locations where one vortex line has vorticity at least 85\% of the maximum at 
that radius; in this sense, they are clockwise vortexes.  
In the near zone, two (blue) clockwise vortexes emerge radially from the 
horizon parallel to the plane of the picture, and two
(red) counterclockwise vortexes emerge radially toward and away from us.
These are 3-dimensional versions of the four vortexes emerging from the
horizon in the equatorial plane of Fig.\ \ref{fig:22SmagEquator}.  In the
wave zone, the ``85\%'' vortexes are concentrated in the polar
regions, because this mode emits its gravitational waves predominantly
along the poles.  The waves are somewhat weaker in the equatorial plane, 
so although there are spiraling vortexes in and near that plane (Fig.\ 
\ref{fig:22SmagEquator}), they do not show
up at the 85\% level of Fig.\ \ref{fig:22Smag3D}.
The off-white, spiral-arm structures in the equatorial
plane represent the four regions where the wave strength is passing through
a minimum.

Turn attention from the wave zone to the horizon. There the ingoing waves,
embodied in ${\boldsymbol{\mathcal B}}^{\rm TT}$ and 
$\delta {\boldsymbol{\mathcal E}}^{\rm TT}$ (which were generated in the
near and transition zones by rotation of 
$\boldsymbol{\mathcal B}^{\rm L}$),
act back on $\boldsymbol{\mathcal B}^{\rm L}$,
causing its vortex lines to gradually slide off the horizon and thereby
producing the mode's exponential decay.  

Just as this process in the electromagnetic case is associated with
the differential conservation law 
(\ref{eq:EMbackaction}) for magnetic field lines threading the horizon, 
$\partial B_N/\partial t
+ {^{(2)}\boldsymbol{\nabla}\cdot (\alpha \bm B^{\rm T})}$, so also here
it is associated with an analogous (approximate) conservation law and 
an accompanying driving equation, given in terms of two Newman Penrose 
equations\ (\ref{eq:DPsiEvoln}) of Appendix \ref{sec:HorizonDetails} and the 
perturbative parts of the Weyl scalars $\Psi_0$, $\Psi_1$, and $\Psi^{(0)}_2$ : 
\begin{subequations}
\begin{align}
& {\bf D} \Psi_2^{(1)}  = ({\bm \delta}^* + 2\pi - 2\alpha) \Psi_1  \, , 
\label{eq:DPsiCons}\\
& ({\bf D} -2\epsilon) \Psi_1  = ({\bm \delta}^* + \pi - 4\alpha) \Psi_0 \, . 
\label{eq:DPsiTTDrive}
\end{align}
\label{eq:DPsiEvolnBody}
\end{subequations}
(Note that only $\Psi_2$ is nonzero for the background spacetime with our 
tetrad choice.)
Here the notation is that of Newman and Penrose: $\bf D$ is a time derivative
on the horizon, $\Psi_2^{(1)}$ is
the mode's
$\delta {\boldsymbol{\mathcal E}}^{\rm L} 
+ i {\boldsymbol{\mathcal B}}^{\rm L}$ (equivalently $\delta \mathcal E_{NN} 
+ i \mathcal B_{NN}$ in disguise), with 
$\delta {\boldsymbol{\mathcal E}}^{\rm L}$ and $\delta \mathcal E_{NN}$
vanishing for our mode; 
$\Psi_1$ is the LT field $\delta {\boldsymbol{\mathcal E}}^{\rm LT} + i \delta {\boldsymbol{\mathcal B}}^{\rm LT}$ (as measured by Schwarzschild observers)
in disguise; 
$\Psi_0$ is the ingoing-wave 
$\delta {\boldsymbol{\mathcal E}}^{\rm TT} + i \delta {\boldsymbol{\mathcal B}}^{\rm TT}$
(as measured by Schwarzschild observers) 
in disguise; ${\bm \delta}^*$ is a divergence in disguise; and
$\epsilon$, $\pi$ and $\alpha$ are NP spin coefficients. 
Equation (\ref{eq:DPsiTTDrive}) says that the ingoing waves embodied in
$\delta {\boldsymbol{\mathcal E}}^{\rm TT} + i  {\boldsymbol{\mathcal B}}^{\rm TT}$ drive the evolution of the quantity $\Psi_1$, and 
Equation (\ref{eq:DPsiCons}) is an approximate differential 
conservation law in which 
this $\Psi_1$
plays the role of the flux of longitudinal vortex lines (number crossing a
unit length per unit time) and $\Psi_2$ (i.e., $\mathcal B_{NN}$) is the
density of longitudinal vortex lines.
This differential conservation law says that the time 
derivative of the vortex-line density plus the divergence of the vortex-line flux
is equal to some spin-coefficient terms that, we believe, are
generally small.  (By integrating this approximate conservation law over
the horizon $\mathcal H$, we see that $\int_{\mathcal H} B_{NN} dA$ must
be nearly conserved, in accord with Eq.\ (\ref{eq:HorizonIntegrals})
above, which tells us that the
horizon integral is nearly zero. In both cases, the 
integral conservation law  (\ref{eq:HorizonIntegrals}) 
and the differential conservation law 
(\ref{eq:DPsiCons}), it is numerically small spin coefficients that slightly
spoil the conservation for vacuum perturbations of black holes. 
In Eq.\ (\ref{eq:LineConservationSuperposed}), 
for a magnetic-parity mode of Schwarzschild and Eddington-Finkelstein slicing, 
we make this conservation law completely concrete and find that
in this case it is precise; there are no small spin coefficients to spoil it.
We plan to investigate this conservation law in numerical simulations as well,
in which there may be additional subtleties related to the formation of 
caustics.

Returning to the evolution of the (2,2) magnetic-parity mode: 
The ingoing waves, via Eqs.\ (\ref{eq:DPsiEvolnBody}), 
push the longitudinal vortex lines away from the centers of the horizon 
vortexes toward their edges (toward the white horizon regions in Figs.\ 
\ref{fig:22MagVortex} and \ref{fig:22SmagEquator}.  At the edges, clockwise
vortex lines from the blue (dark gray) horizon vortex and counterclockwise from
the red (light gray) horizon vortex meet and annihilate each other, leading to 
decay of the longitudinal part of the field and thence the entire mode.

We expect to explore this evolutionary process in greater detail and with
greater precision in future work.

Turn, next, to spinning black holes.  In this case, 
the $(2,2)$ magnetic-parity mode has
qualitatively the same character as for a non-spinning black hole.  The
principal change is due to the spin raising the mode's eigenfrequency,
and the near zone thereby essentially disappearing, so the perturbed
vortex lines that emerge from the horizon have a significant back-spiral-induced
tilt to them already at the horizon.  See Fig.\ \ref{fig:TendexVortexEq} below.  

\subsubsection{The (2,1) magnetic-parity mode of a Schwarzschild hole}
\label{sec:21MagSch}

\begin{figure}
\includegraphics[width=0.6\columnwidth]{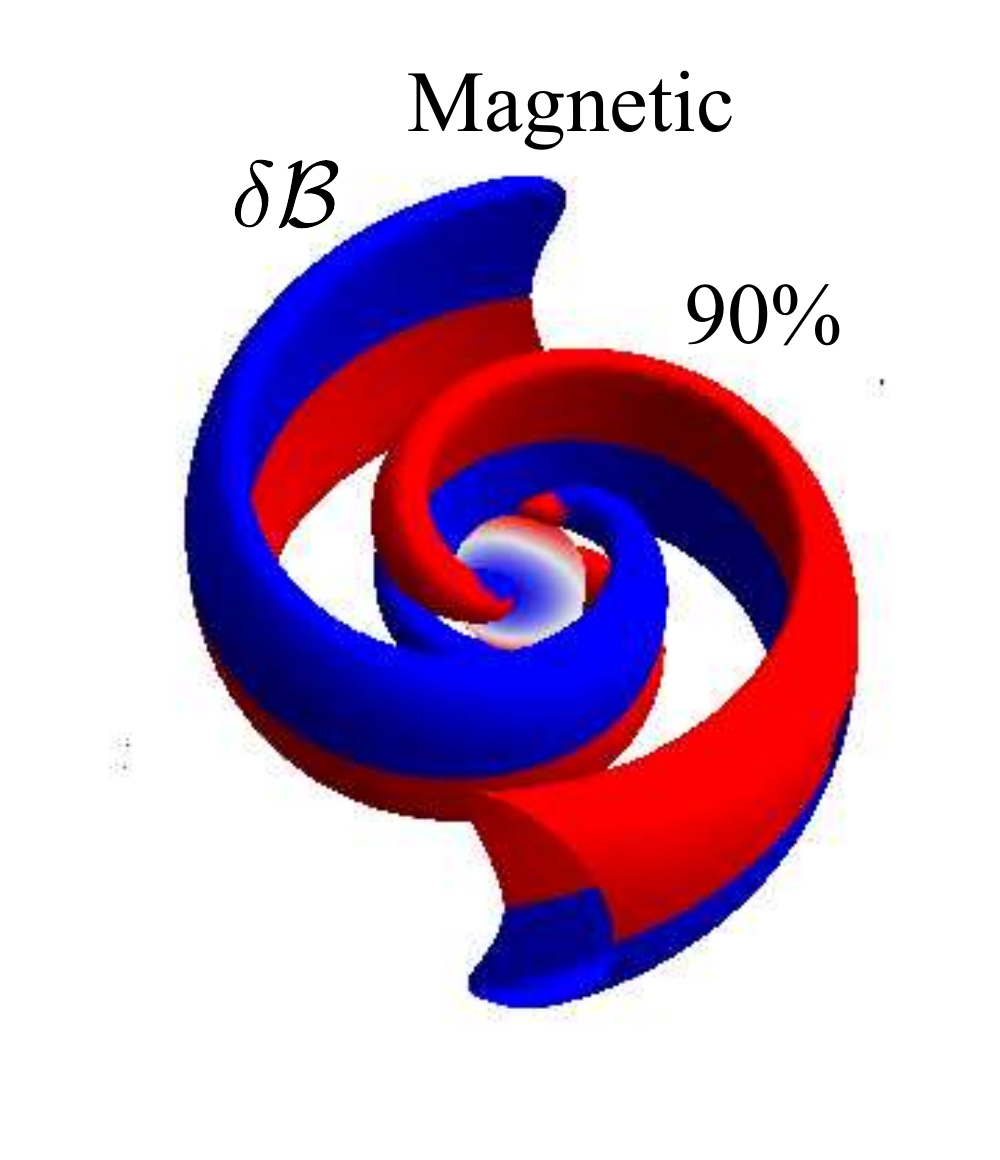}
\caption{(color online).
Three-dimensional vortexes for the magnetic-parity, (2,1) mode
of a Schwarzschild black hole.  The colored surfaces enclose the region
where, for each radius, the vorticity of at least one vortex line
exceeds 90\% of the maximum for that radius.  In the blue and red (dark and 
light gray) regions, the clockwise and counterclockwise vortex lines, 
respectively, have the larger vorticity.}
\label{fig:21_3DVorts}
\end{figure}

For the (2,1) magnetic-parity mode of a Schwarzschild black hole, there are
two horizon vortexes in the hole's northern hemisphere
(one counterlockwise, the other clockwise),
and two in the southern hemisphere.
From these emerge the longitudinal part of the frame-drag field, in the
form of four 3D vortexes (Fig.\ \ref{fig:21_3DVorts}).  

These four vortexes actually form two spiral arms, each
of which contains vortex lines of both signs (clockwise and 
counterclockwise).  The surface of each arm
is color coded by the sign of the vorticity that is largest in magnitude
in that region of the arm.  This dominant vorticity
flips sign when one passes through
the equatorial plane---from positive (i.e., blue [dark gray]; clockwise) on one
side of the equator to negative (i.e., red [light gray]; clockwise) on the 
other side.  The reason for this switch is that
for $m=1$ the $e^{i m \phi}$ angular dependence means reflection 
antisymmetry through the polar axis, which combined with the positive
parity of the $l=2$ frame-drag field implies reflection antisymmetry 
through the equatorial plane. The (2,2) mode of the previous section, by
contrast, was reflection symmetric through both the polar axis and the
equatorial plane.

By contrast with the (2,2) mode, whose region of largest vorticity
switched from equatorial in the near zone to polar in the wave zone
(Fig.\ \ref{fig:22SmagEquator}), for this (2,1) mode, the region of largest vorticity
remains equatorial in the wave zone.  In other words, this mode's 
gravitational waves are stronger in near-equator directions than in
near-polar directions.  (Recall that in the wave zone, the vortexes
are accompanied by tendexes with tendicities equal in magnitude to
the vorticities at each event, so we can discuss the gravitational-wave 
strengths without examining the tidal field.)

Close scrutiny of the near-horizon region of Fig.\ \ref{fig:21_3DVorts}
reveals a surprising feature: Within the 90\% vortexes (colored surfaces), 
the sign of the
largest vorticity switches as one moves from the near zone into the
transition zone---which occurs not very far from the horizon; see 
the inner dashed circle
in Fig.\ \ref{fig:22MagVortex} above).  This appears to be due to the
following:  The near-zone vortexes are dominated by the longitudinal
part of the frame-drag field $\delta {\mathcal B}^{\rm L}$, 
which generates all the other fields including $\delta {\boldsymbol{\mathcal B}}^{\rm LT}$
via its rotation [see discussion of the (2,2) mode above].  The
longitudinal-transverse field $\delta {\boldsymbol{\mathcal B}}^{\rm LT}$ is 
strong throughout the near zone and comes to dominate over 
$\delta {\boldsymbol {\mathcal B}}^{\rm L}$ as one moves into the transition zone.  Its
largest vorticity has opposite sign from that of 
$\delta {\boldsymbol{\mathcal B}}^{\rm L}$, causing the flip of the dominant vorticity 
and thence the color switch as one moves into the transition zone.  (Note that a similar switch in the sign of the strongest vorticity occurs for the magnetic-parity (2,2) mode vortexes illustrated in Fig.~\ref{fig:22Smag3D}, although there the transition occurs farther out, at the edge of the wave zone.) 

In Secs.\ \ref{sec:21Mag} and \ref{sec:21Elec}, 
we explore in considerable detail this magnetic-parity (2,1) mode and 
also its near dual, the electric-parity
(2,1) mode, focusing especially on the shapes of their vortexes.

\subsubsection{The (2,0) magnetic-parity mode of a Schwarzschild hole}
\label{sec:20Mag}

The (2,0) magnetic-parity mode has very 
different dynamical behavior from that of the
(2,1) and (2,2) modes.  Because of its axisymmetry, this mode
cannot be generated by longitudinal, near-zone vortexes 
that rotate around the polar axis, and its waves cannot consist of
outspiraling, intertwined vortex and tendex lines.

\begin{figure}
\includegraphics[width=0.95\columnwidth]{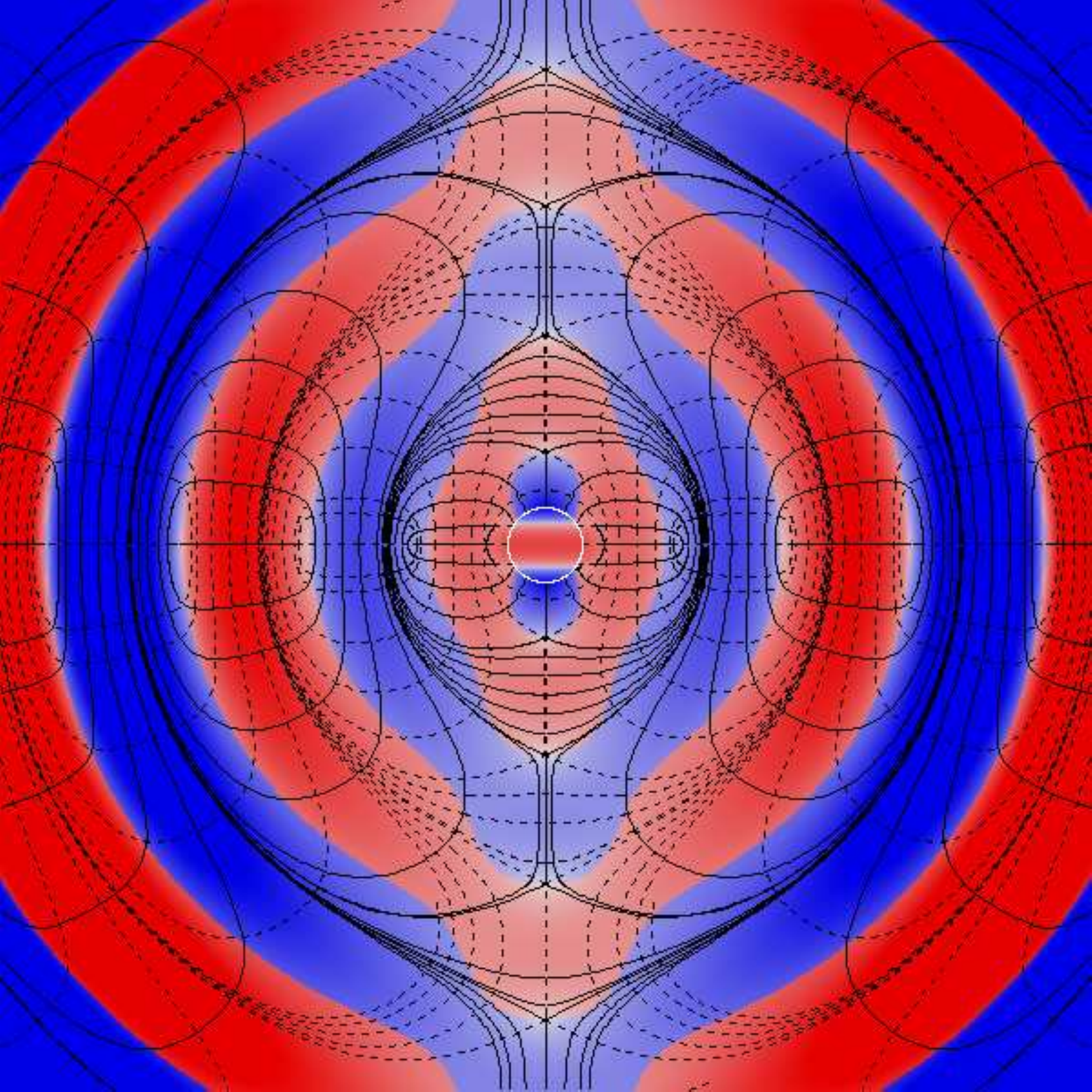}
\caption{(color online).
Vortex lines and vorticities for magnetic-parity (2,0) mode
of Schwarzschild in a surface $\mathcal S_\phi$ of constant $\phi$.
The line and coloring conventions are the same as in Fig.\ 
\ref{fig:22SmagEquator} (solid lines for clockwise, dashed for 
counterclockwise; color shows vorticity of the vortex line with largest
magnitude of vorticity, with radial variations removed and intensity of
color as in the key on right edge of Fig.\ \ref{fig:22SmagEquator}).
The central circle is the horizon, color coded by the horizon vorticity.}
\label{fig:20Sphi}
\end{figure}

Instead, this mode is generated by longitudinal, near-zone vortexes
that oscillate, and its waves are made up of intertwined vortex lines and
tendex lines that wrap around deformed
tori.  These gravitational-wave tori resemble smoke rings and 
travel outward at the speed of light.  More specifically:

Because of axisymmetry, the (2,0) magnetic-parity mode has one family 
of vortex lines that are
azimuthal circles of constant $r$ and $\theta$, and two families that
lie in  surfaces $\mathcal S_\phi$ of constant $\phi$. 
Figure \ref{fig:20Sphi} is a plot in one of these $\mathcal
S_\phi$ surfaces. (The plot for any other $\phi$ will be identical
to this, by axisymmetry.) This plot shows the vortex lines that lie in 
$\mathcal S_\phi$,
and by color coding at each point, the vorticity of the strongest of
those lines. 

Notice that, at this phase of oscillation, there are clockwise (solid)
vortex lines sticking nearly radially out of the horizon's polar regions and
counterclockwise (dashed) vortex lines sticking nearly radially out of the
horizon's equatorial region.  
A half cycle later the poles will be red (light gray) and equator blue 
(dark gray).  These near-zone vortex lines are predominantly the 
longitudinal part of the frame-drag field 
$\delta {\boldsymbol{\mathcal B}}^{\rm L}$, which we
can regard as working hand in hand with the near-zone, longitudinal-transverse
tidal field $\delta {\boldsymbol{\mathcal E}}^{\rm LT}$ to generate 
the other fields.

As  we shall see in Sec.\ \ref{sec:20MagVortexes} (and in more convincing 
detail for a different oscillatory mode in 
Sec.\ \ref{sec:MagSuperposedDynamics}), the dynamics of
the oscillations are these:  Near-zone energy
\footnote{We use the term {\it energy} in a generalized and descriptive sense here and elsewhere in this paper.  We note, however, that with a suitable (nonunique) definition of local energy, we can make these notions more precise. For example, the totally symmetric, traceless Bel-Robinson tensor serves as one possible basis for this. In vacuum it is $T_{\mu \nu \rho \sigma}=1/2(C_{\mu \alpha \nu \beta}C_\rho{}^\alpha{}_\sigma{}^\beta + C^*_{\mu \alpha \nu \beta}C^*_\rho{}^\alpha{}_\sigma{}^\beta)$ with $*$ denoting the Hodge dual, and it is completely symmetric and obeys the \emph{differential} conservation law $\nabla_{\mu} T^\mu{}_{\nu \rho \sigma} = 0$. Given a unit timelike slicing vector $\vec{u}$ we conveniently have $W(\vec{u}) = T_{\mu \nu \rho \sigma} u^\mu u^\nu u^\rho u^\sigma = 1/2(E_{ij}E^{ij} + B_{ij} B^{ij}) \geq 0$ as a positive-definite {\it superenergy} built from the squares of the tidal and frame-drag fields in a given slice (see the reprint of Bel's excellent paper~\cite{Bel2000} for motivation and definition, e.g., Penrose and Rindler~\cite{Penrose1992} for the spinor representation of the Bel-Robinson tensor, and e.g., \cite{Brown1999} for its relation to notions of quasilocal energy). As another example,
magnetic-parity modes of Schwarzschild are describable by the Regge-Wheeler
function $Q(r_*,t)$ which satisfies the Sturm-Liouville equation
$Q_{,{r_*}{r_*}} - Q_{,tt} - \mathcal V(r_*)Q = 0$ [Eq.\ (\ref{Qdeq})
but with the $e^{-i\omega t}$ time dependence absorbed into $Q$]. The
\emph{integral} conservation law associated with this Sturm-Liouville equation is
$\partial/\partial t \int_a^b \left( Q_{,{r_*}}^2 + Q_{,t}^2 + \mathcal V Q^2
\right) dr_* = 2 Q_{,{r_*}} Q_{,t} {\big|}_a^b$. The quantity inside the
integral can be regarded as an energy density, and the quantity on the right
hand side an energy flux.  For the $(2,m)$ magnetic-parity mode, Eqs.\ (\ref{Bij22}) and
(\ref{eq:RWB1m}) express $Q$ in terms of the time derivative of the
longitudinal part of $\delta \boldsymbol{\mathcal B}$ with its angular dependence $Y^{2m}$ removed:
$Q=(r^3/12) \partial \delta \mathcal B_{\hat r \hat r}/ \partial t$. Others of
Eqs.\ (\ref{eq:Bij22RWZ}) and (\ref{eq:magneticSuperposedTidal})
relate $Q_{,{r_*}}$ to the LT parts of
$\delta \boldsymbol{\mathcal B}$ and $\delta \boldsymbol{\mathcal E}$.  
This could be the foundation for a second
way to make more precise the notion of energy fed back and forth between
the various parts of $\delta {\boldsymbol{\mathcal B}}$ and
$\delta {\boldsymbol{\mathcal E}}$.}
oscillates back and forth
between the near-zone $\delta {\boldsymbol{\mathcal B}}^{\rm L}$,
and the near-zone 
$\delta {\boldsymbol{\mathcal B}}^{\rm LT}$ and
$\delta {\boldsymbol{\mathcal E}}^{\rm LT}$.
As $\delta {\boldsymbol{\mathcal B}}^{\rm L}$
decays, its vortex lines slide off the hole and (we presume) form closed loops,
lying in $\mathcal S_\phi$, which encircle outgoing deformed tori of perturbed tendex
lines that become the transverse-traceless gravitational waves.  
Only part of the energy in $\delta {\boldsymbol{\mathcal B}}^{\rm L}$
goes into the outgoing waves.  Some goes into the TT ingoing waves, and
the rest (a substantial fraction of the total energy) goes into 
$\delta {\boldsymbol{\mathcal B}}^{\rm LT}$ and
$\delta {\boldsymbol{\mathcal E}}^{\rm LT}$, which then use it to regenerate 
$\delta {\boldsymbol{\mathcal B}}^{\rm L}$, with its horizon-penetrating vortex
lines switched in sign (color), leading to the next half cycle of oscillation.  

The vortex lines that encircle the gravitational-wave tori are
clearly visible in Fig.\ \ref{fig:20Sphi}.  Each solid (clockwise) line is 
tangential (it points nearly in the $\theta$ direction) when it is 
near the crest (the maximum-vorticity surface) of a blue (dark gray), 
lens-shaped gravitational-wave vortex.  As it
nears the north or south pole, it swings radially outward becoming very
weak (low vorticity) and travels across the red trough of the wave,
until it nears the next blue crest.  There it swings into the transverse,
$\theta$ direction and travels toward the other pole, near which it
swings back through the red trough and joins onto itself in the
original blue crest. 

Each dashed (counterclockwise) closed vortex line behaves in this same manner,
but with its transverse portions lying near red (light gray) troughs (surfaces 
of most negative vorticity). Near the red troughs, there are blue azimuthal vortex lines (not shown)
that encircle the hole in the $\phi$ direction, and near the blue
crests, there are red azimuthal lines.

\begin{figure}
\includegraphics[width=0.95\columnwidth]{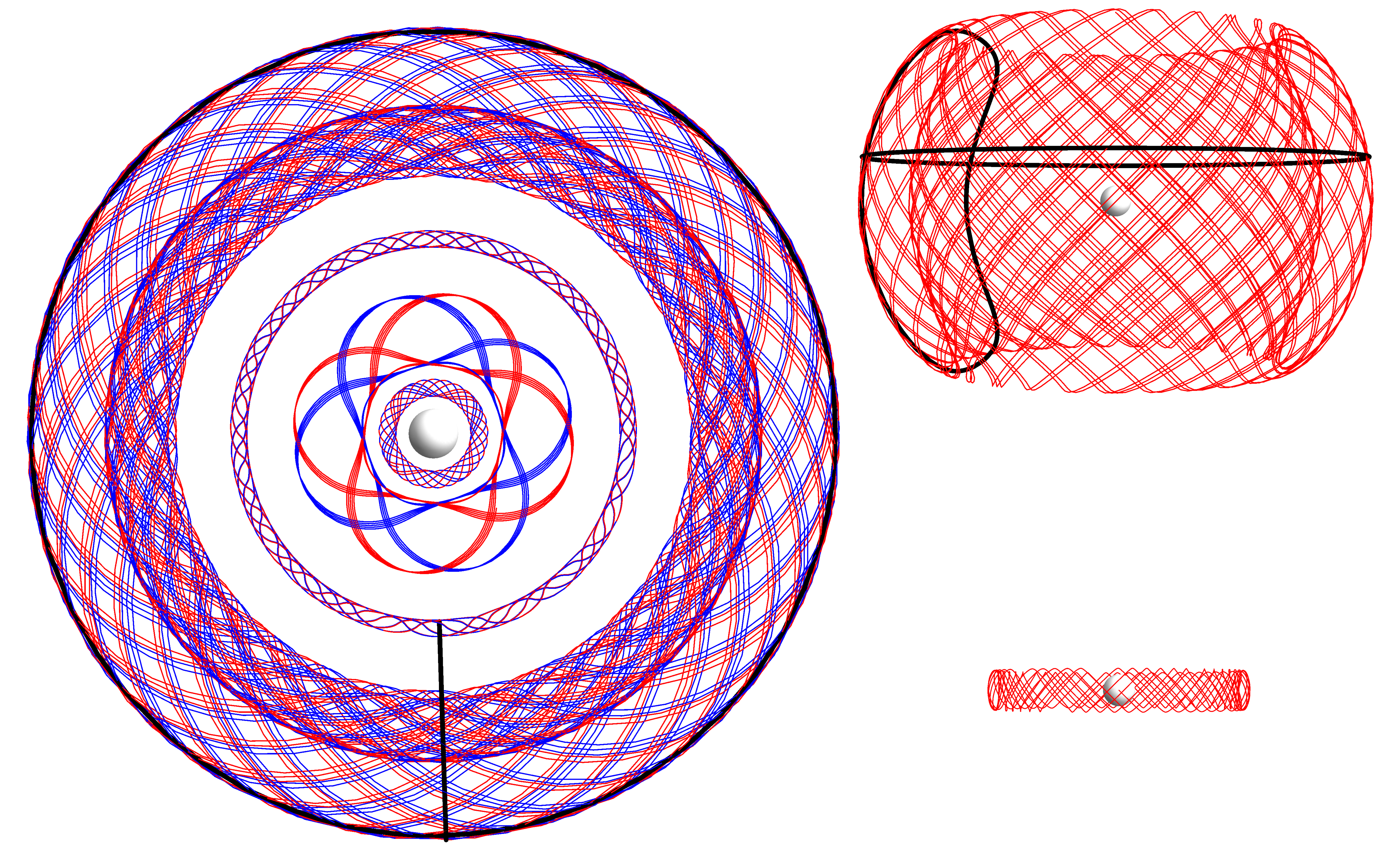}
\caption{(color online).
Positive-tendicity (blue) and negative-tendicity (red) perturbative
tendex lines of a $(2,0)$ magnetic-parity perturbation of a Schwarzschild black
hole. These lines spiral around deformed tori of progressively larger diameter. 
The viewpoint is looking down onto 
the equatorial plane from the positive symmetry axis. 
{\it Upper right inset}: The negative tendex line spiraling around the 
outermost torus, viewed in cross section from the equatorial plane. 
{\it Lower right inset}: The negative tendex line spiraling around the small 
torus third from the center, viewed in cross section from the 
equatorial plane.  
Also shown, in black in the main drawing and the large inset, are two 
of this mode's vortex lines, one from Fig.\ \ref{fig:20Sphi} 
wrapping around the outermost torus in a $\mathcal S_\phi$ plane; 
the other an azimuthal circle wrapping around that torus 
in the $\phi$ direction.
This figure was actually drawn depicting vortex lines
of the electric-parity mode discussed in Sec.\ \ref{sec:20ElectricVortexes}; 
but by duality
(which is excellent in the wave zone), it also represents the tendexes
of the magnetic-parity mode discussed in this section.}
\label{fig:Schw20EvenWaveZone}
\end{figure}

Figure \ref{fig:Schw20EvenWaveZone} sheds further light on these 
gravitational-wave tori.  It shows in three dimensions some of the perturbative tendex lines for
the (2,0) magnetic-parity mode that we are discussing.  (For this mode,
two families of perturbative tendex lines, one red [counterclockwise] and the 
other blue [clockwise], have nonzero tendicity and the third family has 
vanishing tendicity.)  As is required by the
structure of a gravitational wave (transverse tendex lines rotated by
45 degrees relative to transverse vortex lines), these perturbative
tendex lines wind around tori with pitch angles of 45 degrees; one family
winds clockwise and the other counterclockwise, and at each point the
two lines have the same magnitude of vorticity. 

A close examination of Fig.\ \ref{fig:Schw20EvenWaveZone} reveals that
the tori around which the perturbative tendex lines wrap are half as thick
as the tori around which the vortex lines wrap.  Each tendex-line torus
in Fig.\ \ref{fig:Schw20EvenWaveZone} is centered on a
single node of the gravitational-wave field; the thick red torus in the
upper right panel reaches roughly
from one crest of the wave to an adjacent trough.  By contrast, each 
vortex-line torus (Fig.\ \ref{fig:20Sphi} and black poloidal curves in 
Fig.\ \ref{fig:Schw20EvenWaveZone}) reach from crest to crest or trough to
trough and thus encompass two gravitational-wave nodes.

Each node in the wave zone has a family of nested tendex-line
tori centered on it. The four tendex-line tori shown in 
Fig.\ \ref{fig:Schw20EvenWaveZone} are taken from
four successive families, centered on four successive nodes. 
The second thin torus is from near the center of one nested family; it tightly hugs a node and therefore has
near vanishing tendicity.  The two thick tori are from the outer reaches of their nested families.

For further details of the (2,0) modes, see Secs.\ \ref{sec:20MagVortexes}
and \ref{sec:20ElectricVortexes} below.

\subsubsection{The superposed $(2,2)$ and $(2,-2)$ 
magnetic-parity mode of a Schwarzschild hole}
\label{sec:magneticSuperposed}

As we have seen, the magnetic-parity, $(2,2)$ mode of a Schwarzschild
black hole represents vortexes that rotate counterclockwise around
the hole, spiraling outward and backward (Figs.\ \ref{fig:22MagVortex},
\ref{fig:22SmagEquator} and \ref{fig:22Smag3D} above). 
If we change the sign of the azimuthal quantum number to $m=-2$, the
vortexes rotate in the opposite direction, and spiral in the opposite
direction.  If we superpose these two modes (which, for Schwarzschild,
have the same eigenfrequency), then, naturally, we get a non-rotating, 
oscillatory mode---whose dynamics are similar to those of the (2,0) mode
of the last subsection. See Sec.\ \ref{sec:SuperposedPerts} for details.  

\begin{figure}
\includegraphics[width=.95\columnwidth]{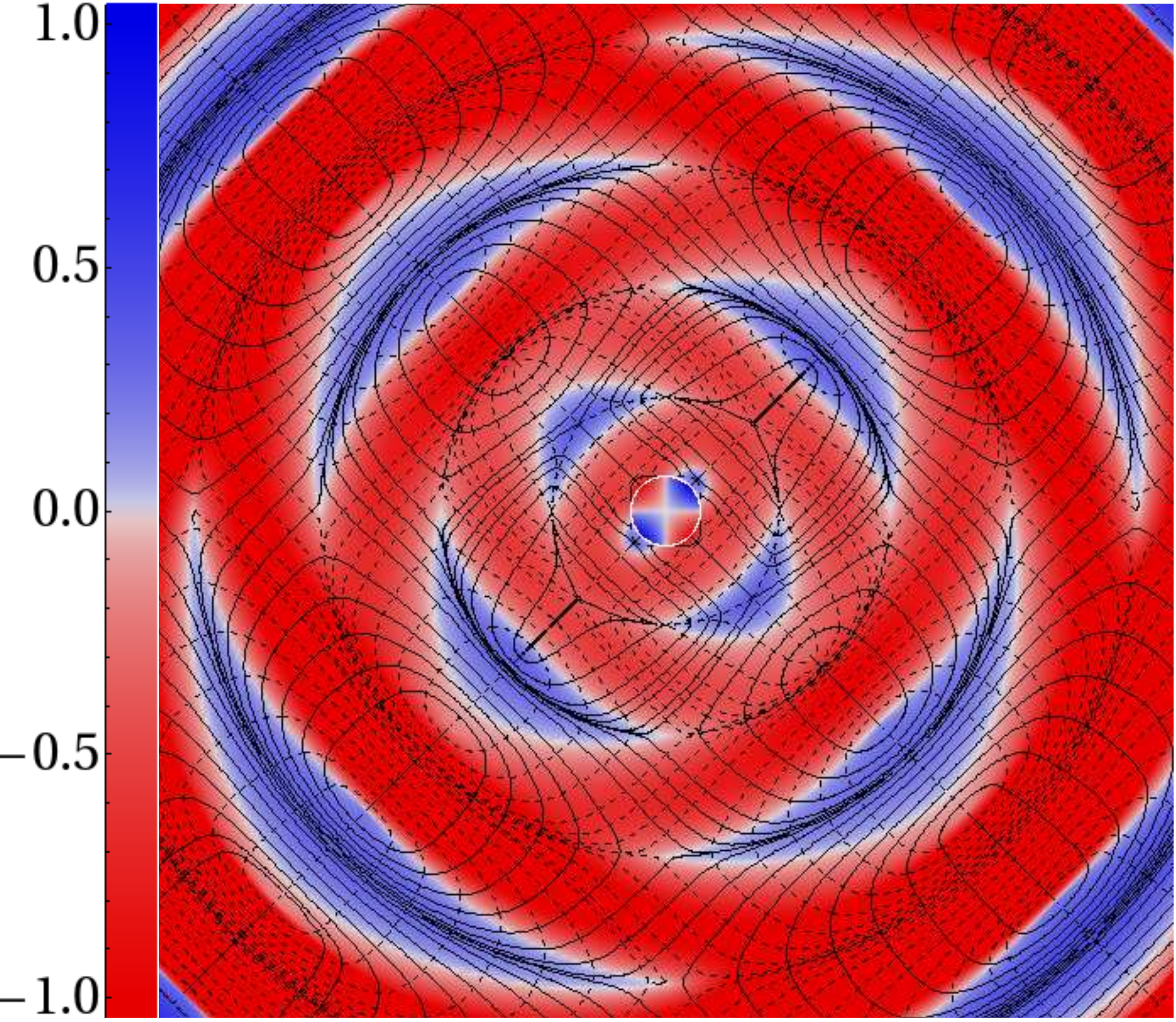}
\caption{(color online).
Equatorial vortex structure of the superposed $(2,2)$ and $(2,-2)$, magnetic-parity, fundamental modes of a Schwarzschild black hole.
The colors encode the vorticity of the dashed vortex lines.  
The vorticity of the solid lines is not shown, 
but can be inferred from the fact that under a
$90^{\circ}$ rotation, the dashed lines map into solid and the solid into
dashed.
}
\label{fig:22-2Smag}
\end{figure}

Figure \ref{fig:22-2Smag} is a snapshot of the two families of vortex
lines that lie in this 
mode's equatorial plane.  The plane is colored by the vorticity of the
dashed vortex lines; they are predominantly counterclockwise
(red), though in some regions they are clockwise (blue). 

The red (light gray) regions form \emph{interleaved rings} around the 
black hole, that expand outward at the speed of light, along with their dashed 
vortex lines.
These rings are not tori in three dimensions because [by contrast with the
$(2,0)$ mode] the frame drag field grows stronger as one moves up to the
polar regions, rather than weakening. As the mode oscillates, the longitudinal
near-zone frame-drag field $\boldsymbol{\mathcal B}^{\rm L}$, which drives
the mode, generates new interleaved rings, 
one after another and sends them outward.

During the oscillations, there are phases at which the longitudinal field
$\boldsymbol{\mathcal B}^{\rm L}$ 
threading the hole goes to zero, and so the hole has vanishing horizon vorticity.
The near-zone oscillation energy, at these phases, is locked up in the 
near-zone, longitudinal-transverse fields
$\boldsymbol{\mathcal B}^{\rm LT}$ and
$\delta \boldsymbol{\mathcal E}^{\rm LT}$, which, via the 
Maxwell-like Bianchi identities (and the propagation equation that they
imply), then feed energy into
the longitudinal near-zone frame-drag field $\boldsymbol{\mathcal B}^{\rm L}$, 
thereby generating new horizon-threading
vortex lines, which will give rise to the next ejected interleaved ring.

We explore these dynamics in greater detail in Sec.~\ref{sec:MagSuperposedDynamics}.

\subsection{This paper's organization}

The remainder of this paper is organized as follows:
In Sec.\ \ref{sec:SlicingDetails}, we introduce the time slicing and coordinates
used throughout this paper for the background Schwarzschild and Kerr spacetimes,
we introduce the two gauges that we use for Schwarzschild perturbations
(Regge-Wheeler-Zerilli and ingoing radiation gauges) and the one gauge (ingoing radiation)
we use for Kerr, we discuss how our various results are affected by 
changes of gauge, and we discuss how we perform our computations. 
In Secs.\ \ref{sec:22Perts}, 
\ref{sec:SuperposedPerts}, and \ref{sec:OtherPerts}, we present full
details of our results for the fundamental (most slowly damped)
quadrupolar modes of Schwarzschild and Kerr: (2,2) modes in 
Sec.\ \ref{sec:22Perts}; superposed (2,2) and (2,-2) modes in Sec.\ 
\ref{sec:SuperposedPerts}, and both (2,1) and  (2,0) modes in Sec.\ \ref{sec:OtherPerts}.
In Sec.\ \ref{sec:NumericalComparison}, we compare vortex lines computed
in a numerical-relativity simulation of a binary black hole at a late time,
when the merged hole is ringing down, with the vortex lines from this paper
for the relevant quasinormal mode; we obtain good agreement.
In Sec.\ \ref{sec:Conclusions}, we make a few concluding remarks.  And
in six appendices, we present mathematical details that underlie a number
of this paper's computations and results.

\section{Slicings, Gauges and Computational Methods}
\label{sec:SlicingDetails}

When calculating the tidal and frame-drag fields of perturbed black-hole spacetimes, 
we must choose a slicing and also spatial coordinates on each slice, for both the background
spacetime and at first order in the perturbations (``perturbative order'').  
The perturbative-order choices of slicing and spatial coordinates
are together called the chosen \emph{gauge}. We will always use the same choice of background slicing and coordinates in this study, but we will use different choices for our gauge.

This section describes the choices we make, how they influence the vortex and tendex
lines and their vorticities and tendicities (which together we call the 
``vortex and tendex structures''), and a few details of how, having made 
our choices, we compute the perturbative frame-drag and tidal fields and the vortex and
tendex structures.  
Most of the mathematical details are left to later sections and especially appendices. 

In Sec.~\ref{sec:BHPandGauge}, we describe our choices of slicing and spatial coordinates.
In Sec.\ \ref{sec:Methods}, we sketch how we calculate the perturbative frame-drag and 
tidal fields and visualize their vortex and tendex structures.
In Sec.\ \ref{sec:GaugePerts}, we explore how those structures
change under changes of gauge, i.e., changes of the perturbative slicing and 
perturbative spatial coordinates.

\subsection{Slicing, spatial coordinates, and gauge}
\label{sec:BHPandGauge}

Throughout this paper, for the background (unperturbed) Kerr spacetime, we use
slices of constant Kerr-Schild (KS) time $\tilde t$, which is related to the more
familiar Boyer-Lindquist time by 
\begin{equation}
\tilde t = t+ r_* - r\;, \quad {\rm where} \quad {dr_*\over dr} = {r^2+a^2 \over \Delta}\;
\label{eq:KStime}
\end{equation}
(Eq.\ (6.2) of Paper II \cite{ZhangPRD2}).  Here $t$ and $r$ are the Boyer-Lindquist 
time and radial coordinates, $a$ is the black hole's spin parameter (angular momentum per
unit mass), and $\Delta \equiv r^2 - 2Mr + a^2$, with $M$ the black-hole mass.
Our slices of constant $\tilde t$ penetrate the horizon smoothly, by contrast 
with slices of constant $t$, which are singular at the horizon.  
In the Schwarzschild limit $a\rightarrow 0$,
$t$ and $r$ become Schwarzschild's time and radial coordinates, and $\tilde t$ becomes
ingoing Eddington-Finkelstein time, $\tilde t = t + 2M \ln|r/2M-1|$.  

On a constant-$\tilde t$ slice in the background Kerr spacetime, we use Cartesian-like
KS (Kerr-Schild) spatial coordinates, when visualizing vortex and tendex structures; but in
many of our intermediary computations, we use Boyer-Lindquist spatial 
coordinates $\{r,\theta,\phi\}$
(which become Schwarzschild as $a \rightarrow 0$). The two sets of coordinates are related by
\begin{equation}
x+iy = (r+ia) e^{i \tilde\phi} \sin\theta\;, \quad z = r \cos\theta\;,
\label{eq:KSspace}  
\end{equation}
[Eq.\ (6.7) of Paper II]. Here 
\begin{equation}
\tilde\phi = \phi + \int_r^\infty {a\over \Delta} dr
\label{eq:tildephi}
\end{equation}
[Eq.\ (6.5) of Paper II]
is an angular coordinate that, unlike $\phi$, is well behaved at the horizon.  In the 
Schwarzschild limit, the KS $\{x, y,z \}$ coordinates become the quasi-Cartesian $\{x,y,z \}$ associated with Eddington-Finkelstein (EF) spherical coordinates $\{r,\theta, \phi\}$.

Our figures (e.g., \ref{fig:22MagVortex}--\ref{fig:22-2Smag} above) 
are drawn as though the KS $\{x,y,z\}$ were Cartesian coordinates in
flat spacetime---i.e., in the Schwarzschild limit, as though the EF 
$\{r,\theta,\phi \}$ were spherical polar coordinates in flat spacetime.

We denote by $g^{(0)}_{\mu\nu}$ the background metric in KS spacetime coordinates
[Eq.\ (6.8) of Paper II] (or EF spacetime coordinates in the Schwarzschild limit).  
When the black hole is perturbed, the metric acquires a perturbation 
$h_{\mu\nu}$ whose actual form depends on one's choice of gauge---i.e., one's 
choice of slicing and spatial coordinates at perturbative order.

For Schwarzschild black holes, we use two different gauges, as a way to assess
the gauge dependence of our results:
(i) \emph{Regge-Wheeler-Zerilli (RWZ) gauge}, in which 
$h_{\mu\nu}$ is a function of two scalars ($Q$ for magnetic parity and $Z$ for
electric parity) that obey separable wave 
equations in the Schwarzschild spacetime and that have spin-weight zero 
(see App.~\ref{sec:RWApp} for a review of this formalism), and 
(ii) \emph{ingoing radiation (IR) gauge}, in which $h_{\mu\nu}$ is computed
from the Weyl scalar $\Psi_0$ (or
$\Psi_4$) that obeys the separable Bardeen-Press equation.  The method used to
compute the metric perturbation from $\Psi_0$ is often called  
\emph{the Chrzanowski-Cohen-Kegeles (CCK) procedure of metric reconstruction} (see 
App. \ref{sec:CCKProc}).

In App.~\ref{sec:GaugeCompare}, we exhibit explicitly the relationship between 
the RWZ and IR gauges, for electric- and magnetic-parity perturbations.  
The magnetic-parity perturbations have different perturbative spatial 
coordinates, but the same slicing. (In fact,
\emph{all} gauges related by a magnetic-parity gauge transform have identically
the same slicing for magnetic-parity perturbations of Schwarzschild [although
the same is not true for Kerr]; see Sec.\ \ref{sec:GaugePerts}).
For electric-parity perturbations, the two gauges have different slicings and 
spatial coordinates.  

For all the perturbations that we visualize in this paper, the 
tendexes and vortexes show quite weak gauge dependence.
See, e.g., Sec.~\ref{sec:22Perts}, where we present 
results from both gauges.
The results in Secs.\ \ref{sec:SuperposedPerts} and \ref{sec:OtherPerts} are 
all computed in RWZ gauge.

For Kerr black holes, there is no gauge analogous to RWZ; but the IR gauge and
the CCK procedure that underlies it are readily extended from Schwarzschild to Kerr.  
In this extension, one constructs the metric perturbation from solutions to the Teukolsky equation
(see App. \ref{sec:Teukolsky}) for the perturbations to the Weyl scalars
$\Psi_0$ and $\Psi_4$, in an identical way to that for a Schwarzschild black 
hole described above.
Our results in this paper for Kerr black holes, therefore, come solely from the IR gauge.

\subsection{Sketch of computational methods}
\label{sec:Methods}

This section describes a few important aspects of how we calculate the tidal
and frame-drag fields, and their vortex and tendex structures
which are visualized and discussed in Secs.\ \ref{sec:22Perts}, 
\ref{sec:SuperposedPerts}, and \ref{sec:OtherPerts}.

We find it convenient to solve the eigenvalue problem in an orthonormal
basis (orthonormal tetrad) 
given by the four-velocities of the Kerr-Schild (KS) or 
Eddington-Finkelstein (EF) observers, and a spatial
triad, $\vec{e}_{\hat a}$, carried by these observers. 

The background EF
tetrad for the Schwarzschild spacetime, expressed in terms of 
Schwarzschild coordinates, is
\begin{eqnarray}
\vec u^{\,(0)} &=& \frac{1}{\sqrt{1+2M/r}}\left(\frac{1}{\alpha^2}
\frac{\partial}{\partial t} - \frac{2M}{r}\frac{\partial}{\partial r}\right)\;,
\nonumber \\
\vec e^{\,(0)}_{\hat r} &=& \frac{1}{\sqrt{1+2M/r}} \left( 
\frac{\partial}{\partial r} -\frac{2M}{\alpha^2 r} \frac{\partial}{\partial t}
\right) \;, \nonumber \\
\vec e^{\,(0)}_{\hat \theta} &=& \frac 1r \frac{\partial}{\partial\theta} \; ,
\qquad 
\vec e^{\,(0)}_{\hat \phi} = \frac{1}{r \sin\theta} 
\frac{\partial}{\partial\phi}\;
\label{EFbasis}
\end{eqnarray}
[cf.\ Eqs.\ (4.4) of Paper II, which, however, are written in terms of 
the EF coordinate basis rather than Schwarzschild].
The background orthonormal tetrad for KS observers (in ingoing Kerr 
coordinates $\{\tilde t, r ,\theta, \tilde \phi\}$; see Paper II, Sec.~VI C)
is
\begin{subequations}
\begin{align}
\vec{u}^{\,(0)} & = H \partial_{\tilde t} - \frac{2 M r}{H \Sigma } \partial_r 
\,, & \vec{e}^{\,(0)}_{\hat r} &=  \frac{\sqrt{A}}{H \Sigma} \partial_r + 
\frac{a H}{ \sqrt{A}} \partial_{\tilde \phi} \,, \notag \\
\vec{e}^{\,(0)}_{\hat \theta} & =  \frac{1}{\sqrt{\Sigma}} \partial_\theta 
\,, & \vec{e}^{\,(0)}_{\hat {\tilde\phi}}  &=  \sqrt{\frac{\Sigma}{A}}
\frac{1}{ \sin \theta} \partial_{\tilde \phi} \,,
\label{eq:KSbasis}
\end{align}
where we have defined
\begin{align}
\label{eq:SigmaDef}
\Sigma &= r^2 + a^2 \cos^2 \theta \,,\\
H &= 1 + \frac{2Mr}{\Sigma} \,, \\
A & =  (r^2 + a^2)^2 - a^2 (r^2 - 2Mr +a^2) \sin^2 \theta \,
\end{align}
\end{subequations}
[see Eq.\ (B2) of Paper II].

When the black hole is perturbed, the tetrad 
$\{\vec u^{(0)}, \vec e_{\hat r}^{\,(0)}, \vec e_{\hat \theta}^{\,(0)}, \vec e_{\hat{\tilde \phi}}^{\,(0)}\}$
acquires perturbative corrections that keep it orthonormal
with respect to the metric $g_{\mu \nu} = g^{(0)}_{\mu \nu} + h_{\mu \nu}$.
We choose the perturbative corrections to the observers' 4-velocity so as to 
keep it orthogonal to the space slices, i.e., so as to keep 
$\vec u = - \alpha \vec\nabla \tilde t$. (Here $\alpha = d\tau/d\tilde t$,
differentiating along the observer's world line, is the observer's lapse function.)
A straightforward calculation using the perturbed metric gives the following
contravariant components of this $\vec u$:
\begin{align}
u^\mu &= u^{\mu}_{(0)} + u^\mu_{(1)} \,, \notag \\
\label{u1BH}
u^\mu_{(1)} & =  -\frac12 h_{\hat 0 \hat 0} u^\mu_{(0)} - h^{\mu \nu} u^{(0)}_\nu \,.
\end{align}
where $h_{\hat 0\hat 0} = h_{\mu\nu} u^\mu_{(0)} u^\nu_{(0)}$, and 
$u^\mu_{(0)}$ is the four-velocity of the background observers. 

We choose the perturbative corrections to the spatial triad $\{\vec e_{\hat j}
\}$ so the
radial vector stays orthogonal to surfaces of constant $r$ in slices of constant
$\tilde t$, the $\hat \theta$ direction continues to run orthogonal 
to curves of 
constant $\theta$ in surfaces of constant $r$ and $\tilde t$, and the 
$\hat \phi$ vector changes only in its normalization.

When written in terms of the unperturbed tetrad and projections of the metric
perturbation into the unperturbed tetrad, the perturbation to the tetrad 
then takes 
the form
\begin{subequations}
\begin{eqnarray}
\label{eq:u1}
\vec u_{(1)} & = & \frac 12 h_{\hat 0\hat 0}  \vec u_{(0)}
- h_{\hat 0 \hat i} \vec e_{(0)}^{\, \hat i} \, ,\\
\vec e_{\hat r}^{\, (1)} & = & -\frac 12 h_{\hat r\hat r} 
\vec e_{\hat r}^{\, (0)} - h_{\hat r \hat A} \vec e_{(0)}^{\, \hat A} \, ,\\
\vec e_{\hat \theta}^{\, (1)} & = & -\frac 12 h_{\hat \theta\hat \theta} 
\vec e_{\hat \theta}^{\, (0)} - h_{\hat \theta \hat \phi} 
\vec e_{(0)}^{\, \hat \phi} \, ,\\
\vec e_{\hat \phi}^{\, (1)} & = & -\frac 12 h_{\hat \phi\hat \phi} 
\vec e_{\hat \phi}^{\, (0)} \, ,
\label{eq:e1}
\end{eqnarray}
\label{eq:PerturbedEFTetrad}
\end{subequations}
where $\hat i$ is summed over $\hat r$, $\hat \theta$, and $\hat \phi$, and 
$\hat A$ is summed over only $\hat \theta$ and $\hat \phi$.

In Appendices \ref{sec:RWApp} (RWZ gauge) and \ref{sec:CCKProc} (IR gauge), 
we give the details of how we compute the components 
\begin{equation}
\mathcal E_{\hat i\hat j} = \mathcal E_{\hat i\hat j}^{(0)} + 
\mathcal E_{\hat i\hat j}^{(1)} \, , \quad
\mathcal B_{\hat i\hat j} = \mathcal B_{\hat i\hat j}^{(0)} + 
\mathcal B_{\hat i\hat j}^{(1)} \, .
\end{equation}
of the tidal and frame-drag field in this perturbed tetrad.  
The background portions $\mathcal E_{\hat i\hat j}^{(0)}$ and
$\mathcal B_{\hat i\hat j}^{(0)}$ are the stationary fields of the unperturbed
black hole, which were computed and visualized in Paper II.
The perturbative pieces, $\mathcal E_{\hat i\hat j}^{(1)}$ and
$\mathcal B_{\hat i\hat j}^{(1)}$ are the time-dependent, perturbative parts, 
which carry the information about the quasinormal modes, their 
geometrodynamics, and their gravitational radiation.

As part of computing the perturbative $\mathcal E_{\hat i\hat j}^{(1)} \equiv
\delta \mathcal E_{\hat i\hat j}$ and
$\mathcal B_{\hat i\hat j}^{(1)} \equiv \delta \mathcal B_{\hat i\hat j}$ for 
a chosen quasinormal mode of a Kerr black hole, we have to solve for the mode's
Weyl-scalar eigenfunctions $\Psi_0^{(1)}$ and $\Psi_4^{(1)}$ and eigenfrequency $\omega$.
To compute the frequencies, we have used, throughout this paper, Emanuele 
Berti's elegant computer code \cite{BertiWebsite}, which is discussed in 
\cite{Berti2009} and is an implementation of Leaver's method \cite{Leaver1985}.
To compute the eigenfunctions, we our own independent code (which also uses the
same procedure as that of Berti).
In App. \ref{sec:CCKProc}, we describe how we extract the definite-parity
(electric or magnetic) eigenfunctions from the non-definite-parity functions.

To best visualize each mode's geometrodynamics and generation of gravitational 
waves in Secs.\ \ref{sec:22Perts}, \ref{sec:SuperposedPerts}, and 
\ref{sec:OtherPerts}, we usually plot the tendex and vortex structures of the 
perturbative fields $\mathcal E_{\hat i\hat j}^{(1)}$
and $\mathcal B_{\hat i\hat j}^{(1)}$. 
However, when we compare our results with numerical-relativity simulations, it
is necessary to compute the tendex and vortex structures of the full tidal and
frame-drag fields (background plus perturbation), because of the difficulty of
unambiguously removing a stationary background field from the
numerical simulations.
As one can see in Figs.~\ref{fig:NRComparison} and~\ref{fig:KerrTendexVortexFull}, in this case much of
the detail of the geometrodynamics and wave generation is hidden behind the 
large background field.

In either case, the tendex and vortex structure of the perturbative fields or the full fields,
we compute the field lines and their eigenvalues in the obvious way:  At selected points on
a slice, we numerically solve the eigenvalue problem
\begin{equation}
\mathcal E_{\hat i \hat j} V_{\hat j} = \lambda V_{\hat i}
\end{equation}
for the three eigenvalues $\lambda$ and unit-normed eigenvectors $V_{\hat i}$ 
of $\mathcal E_{\hat i \hat j}$,
and similarly for $\mathcal B_{\hat i \hat j}$; and we then compute the integral curve 
(tendex or vortex line) of
each eigenvector field by evaluating its coordinate components $V^j$ in the desired
coordinate system (KS or EF) and then 
numerically integrating the equation 
\begin{equation}
{dx^j \over ds} = V^j\;,
\label{eq:LineEq}
\end{equation}
where $s$ is the proper distance along the integral curve.

\subsection{Gauge changes: Their influence on tidal and frame-drag fields
and field lines}
\label{sec:GaugePerts}

For perturbations of black holes, a perturbative gauge change is a change
of the spacetime coordinates, $x^{\alpha '} = x^\alpha + \xi^\alpha$,
that induces changes of the metric that are of order the metric perturbation; when dealing with definite parity perturbation, we split the generator of the transform $\xi^\alpha$ into definite electric- and magnetic-parity components.
The gauge change has two parts: A change of slicing generated by $\xi^0$,
and a change of spatial coordinates
\begin{equation}
{\tilde t}'= \tilde t + \xi^0 \quad x^{j'}= x^{j}+ \xi^j\;.
\label{eq:gaugechange}
\end{equation}
Here all quantities are to be evaluated at the same event, $\mathcal P$,
in spacetime.

Because $\xi^0$ is a scalar under rotations in the Schwarzschild 
spacetime---and all scalar fields in Schwarzschild have electric parity---for 
a magnetic-parity $\xi^\alpha$, $\xi^0$ vanishes, and
\emph{the slicings for magnetic-parity quasinormal modes of Schwarzschild 
are unique.  For these modes, all gauges share the same slicing} 
(see Appendix \ref{sec:GaugeCompare}).\footnote{In the Kerr spacetime, however,
there are magnetic-parity changes of slicing, because $\xi^0$ no longer behaves
as a scalar under rotations.
To understand this more clearly, consider, as a concrete example, a vector in 
Boyer-Lindquist coordinates with covariant components 
$\xi_\mu = (0,0,X_\theta^{lm}, X_\phi^{lm})f(r)e^{-i\omega t}$, 
where $X_A^{lm}$ are the components of a magnetic-parity vector spherical 
harmonic [see Eq.\ (\ref{eq:BVectorHarmonic})].
This vector's contravariant components are 
$\xi^\mu = (g^{t\phi} X_\phi^{lm},0,g^{\theta\theta} X_\theta^{lm}, 
g^{\phi\phi} X_\phi^{lm})f(r)e^{-i\omega t}$, where $g^{t\phi}$, 
$g^{\theta\theta}$, and $g^{\phi\phi}$ are the contravariant components of
the Kerr metric (which have positive parity).
The vector $\xi^\mu$, has magnetic parity and a nonvanishing component $\xi^0$;
therefore, it is an example of a magnetic-parity gauge-change generator in the
Kerr spacetime that changes the slicing.}

\subsubsection{Influence of a perturbative slicing change}
\label{sec:PertSliceChange}

For (electric-parity) changes of slicing, the new observers, whose world lines 
are orthogonal to the new slices, $\tilde t' =$const, move at velocity 
\begin{equation}
\Delta \bm v = - \alpha \boldsymbol{\nabla} \xi^0
\label{eq:velchange}
\end{equation}
with respect to the old observers, whose world lines are orthogonal to the old 
slices $\tilde t =$const). Here $\boldsymbol{\nabla}$ is the gradient in
the slice of constant $\tilde t$, and $\alpha = (d\tau/d\tilde t) $ is the lapse function, evaluated along the observer's worldline.
In other words, $\Delta \bm v$ is the velocity of the boost that leads from an
old observer's local reference frame to a new observer's 
local reference frame.  Just as in electromagnetic theory, 
this boost produces a change in the observed
electric and magnetic fields for small $\Delta \bm v$ given by 
$\Delta \bm B = \Delta \bm v \times 
\bm E$ and $\Delta \bm E = - \Delta \bm v \times \bm B$, so also it produces a
change in the observed tidal and frame-drag fields given by
\begin{equation}
\Delta \boldsymbol{\mathcal B} = 
(\Delta \bm v \times \boldsymbol{\mathcal E})^{\rm S}\;, \quad
\Delta \boldsymbol{\mathcal E} = 
-(\Delta \bm v \times \boldsymbol{\mathcal B})^{\rm S}\; 
\label{eq:BoostedFields}
\end{equation}
(e.g., Eqs.\ (A12) and (A13) of \cite{maartens1999}, expanded to linear order in the boost velocity).
Here the superscript S means symmetrize.  

\subsubsection{Example: Perturbative slicing change for Schwarzschild black
hole}
\label{sec:ExSchwSlicingChange}

For a Schwarzschild black hole, because the unperturbed frame-drag field
vanishes, $\Delta \boldsymbol{\mathcal E}$ is second order in the perturbation
and thus negligible, so
\emph{the tidal field is invariant under a slicing change.}  By contrast,
the (fully perturbative) frame-drag field \emph{can} be altered by a slicing change;
$\Delta \boldsymbol{\mathcal B} =
(\Delta \bm v \times \boldsymbol{\mathcal E})^{\rm S} $ is nonzero at first order.  

Since the unperturbed tidal field is isotropic in the transverse ($\theta,
\phi$) plane, the radial part of $\Delta \bm v$ produces a vanishing
$\Delta \boldsymbol{\mathcal B}$.  The transverse part of $\Delta \bm v$,
by contrast, produces a radial-transverse $\Delta \boldsymbol{\mathcal B}$
(at first-order in the perturbation).
In other words, \emph{a perturbative slicing change in Schwarzschild gives rise to a 
vanishing $\Delta \boldsymbol{\mathcal E}$ and an electric-parity 
$\Delta \boldsymbol{\mathcal B}$ whose only nonzero components are}
\begin{equation}
\Delta \mathcal B_{\hat r \hat \theta} 
= \Delta \mathcal B_{\hat\theta \hat r} \quad \textrm{and}\quad
\Delta \mathcal B_{\hat r \hat \phi} 
= \Delta \mathcal B_{\hat\phi \hat r}\;.
\label{eq:DeltaB}
\end{equation} 

For a Schwarzschild black hole that is physically unperturbed, the first-order
frame-drag field is just this radial-transverse 
$\Delta \boldsymbol{\mathcal B}$, and its gauge-generated vortex lines make 45 degree angles
to the radial direction.

\subsubsection{Influence of perturbative change of spatial coordinates}

Because $\boldsymbol{\mathcal E}$ and $\boldsymbol{\mathcal B}$ are
tensors that live in a slice of constant $\tilde t$, the perturbative
change of spatial coordinates, which is confined to that slice, produces
changes in components that are given by the standard
tensorial transformation law, $\mathcal E_{i'j'}(x^{k'}[\mathcal P])
= \mathcal E_{pq}(x^{k}[\mathcal P])(\partial x^q/\partial x^{i'})
(\partial x^p/\partial x^{j'})$.  To first order in the gauge-change
generators $x^k$, this gives rise to the following perturbative change
in the tidal field
\begin{eqnarray}
\Delta \mathcal E_{ij} &=& - \mathcal E_{ij,k} \xi^k - \mathcal E_{ik} {\xi^k}_{,j}
- \mathcal E_{jk} {\xi^k}_{,i} \nonumber \\
&=& - \mathcal E_{ij|k} \xi^k - \mathcal E_{ik} {\xi^k}_{|j}
- \mathcal E_{jk} {\xi^k}_{|i}\;,
\label{eq:CoordChange}
\end{eqnarray}
and similarly for the frame-drag field $\mathcal B$.
Here the subscript ``$|$'' denotes covariant derivative with respect to
the background metric, in the slice of constant $\tilde t$.  The two
expressions in Eq.\ (\ref{eq:CoordChange}) are equal because 
the connection coefficients all cancel.

The brute-force way to compute the influence of a spatial coordinate change 
$x^{j'} = x^{j} + \xi^{j}$
on the
coordinate shape $x^j(s)$ of a tendex line (or vortex line) 
is to (i) solve the eigenequation to
compute the influence of $\Delta \mathcal E_{ij}$ [Eq.\ (\ref{eq:CoordChange})] on
the line's eigenvector, and then (ii) compute the integral curve of the altered
eigenvector field.  

Far simpler than this brute-force approach is to note that the tendex line, written
as location $\mathcal P(s)$ in the slice of constant $\tilde t$ as a function of spatial
distance $s$ along the curve, is unaffected by the coordinate change.  Therefore, if the
old coordinate description of the tendex line is $x^j(s) = x^j[\mathcal P(s)]$, then the
new coordinate description is $x^{j'}(s) = x^j[\mathcal P(s)] + \xi^j[\mathcal P(s)]$; i.e., 
$x^{j'}(s) = x^j(s) + \xi^j[x^j(s)]$.  In other words, \emph{as seen in the new (primed) 
coordinate system,
the tendex line appears to have been moved from its old coordinate location, along the
vector field $\xi^j$, from its tail to its tip; and similarly for any vortex line.}

\subsubsection{Example: Perturbative spatial coordinate change for a Schwarzschild black hole}
\label{sec:CoordChangeEx}

\emph{Because the frame-drag field of a perturbed Schwarzschild black hole 
is entirely perturbative, it is unaffected by a spatial coordinate change.
This, together with $\Delta \boldsymbol{\mathcal B} = 0$ for magnetic-parity
modes implies that the frame-drag field of any magnetic-parity mode of Schwarzschild
is fully gauge invariant!}

By contrast, a spatial coordinate change (of any parity) mixes some of the background tidal field
into the perturbation, altering the coordinate locations of the tendex lines.  

As an example, consider an electric-parity (2,2) mode of a Schwarzschild black hole.  In
RWZ gauge and in the wave zone, the tidal field is given by
\begin{eqnarray}
{\cal E}_{\hat\phi\hat\phi} &=& {M\over r^3} + {A\over r} \cos[2(\phi-\phi_o) -\omega (t - 
r_*)]\;, \nonumber \\
{\cal E}_{\hat\theta\hat\theta} &=& {M\over r^3} - {A\over r} \cos[2(\phi-\phi_o) -\omega( t - 
r_*)] \nonumber \\
{\cal E}_{\hat r\hat\phi} &=& {2A\over \omega r^2} \cos[2(\phi-\phi_o) -\omega (t - 
r_*)]\;,  \nonumber \\
{\cal E}_{\hat r \hat r} &=& -{2M\over r^3} + {\rm O}\left({A\over\omega^2 r^3}\right)\;,
\end{eqnarray}
where $A$ is the wave amplitude

Focus on radii large enough to be in the wave zone, but small enough that the wave's tidal
field is a small perturbation of the Schwarzschild tidal field.  Then 
the equation for the shape of the nearly circular tendex lines that lie in the equatorial plane,
at first order in the wave's amplitude, is
\begin{equation}
{1\over r}{dr\over d\phi} = { {\cal E}_{\hat r \hat \phi} \over ({\cal E}_{\hat \phi \hat \phi} - {\cal E}_{\hat r \hat r})}
= {2 rA\over 3M\omega} \cos[2(\phi-\phi_o) - \omega( t - r_*)] \;
\end{equation}
(an equation that can be derived using the standard perturbation theory of
eigenvector equations).
Solving for $r(\phi)$ using perturbation theory, we obtain for the tendex 
line's coordinate location
\begin{eqnarray}
r(\phi, t) &=& r_o + \rho(\phi, r_o,t)\;, 
\label{eq:E22tendexline}\\
\rho(\phi,r_o,t) &\equiv&  r_o {A r_o\over 3 M \omega } \sin[2(\phi-\phi_o) - \omega( t - 
r_{o *})]\;. \nonumber
\end{eqnarray}
Here $r_o$ is the radius that the chosen field line has when $\phi=\phi_o$.  Notice that
the field line undergoes a quadrupolar oscillation, in and out, as it circles around the
black hole, and it is closed---i.e., it is an ellipse centered on the hole.  The ellipticity
is caused by the gravitational wave.  As time passes,
the ellipse rotates with angular velocity $d\phi/dt=\omega/2$, and the phasing of successive
ellipses at larger and larger radii $r_o$ is delayed by an amount corresponding
to speed-of-light radial propagation.

Now, consider an unperturbed Schwarzschild black hole.  We can produce this
same pattern of elliptical oscillations of the equatorial-plane tendex lines, in the
absence of any gravitational waves, by simply changing 
our radial coordinate:  Introduce the new coordinate
\begin{equation}
r' = r + \xi^r, \quad \textrm{where} \quad \xi^r=\rho(\phi,r,t)\;,
\end{equation}
with $\rho$ the function defined in Eq.\ (\ref{eq:E22tendexline}).  In Schwarzschild coordinates,
the equatorial tendex lines are the circles $r=r_o = $constant.  In the new
coordinate system, those tendex lines will have precisely the same shape
as that induced by our gravitational wave [Eq.\ (\ref{eq:E22tendexline})]:
$r'=r_o + \rho(\phi, r_o,t)$.  Of course, a careful measurement of the radius 
of curvature of one of these tendex lines will show it to be constant
as one follows it around the black hole (rather than oscillating),
whereas the radius of curvature of the wave-influenced 
tendex line will oscillate. In fact, if we follow along with the tendex line and measure the tendicity \emph{along} the line, we find that the tendicity of the line is unchanged by the change in coordinates. To be explicit, consider the tendicity, which we denote $\lambda_\phi$, along one of the lines $r= r_o$. Enacting the coordinate transform on the tendicity but continuing to evaluate it along the perturbed line, we have the identity
\begin{align}
 \lambda_\phi (r)|_{r = r_o} & = \lambda_\phi (r' - \xi^r)|_{r' = r_o + \rho}  = \lambda_\phi(r_o  + \rho - \rho)  \notag \\
 &= \lambda_\phi (r_o) \,.
\end{align}
Nevertheless, if one just casually looks at the Schwarzschild tendex lines in the new, primed,
coordinate system, one will see a gravitational-wave pattern.

The situation is a bit more subtle for the perturbed black hole.  In this case,
the tendex lines are given by Eq.\ (\ref{eq:E22tendexline}), and we can change their 
ellipticity by again changing radial coordinates, say to
\begin{equation}
r'= r+\alpha \rho(\phi,r,t)\;.
\end{equation}
The radial oscillations $\Delta r'$ of the elliptical tendex lines in the new $(r',\phi,t)$ 
coordinate system will have amplitudes $1+\alpha$ times larger than in 
the original $(r,\phi,t)$ coordinates, and in the presence of the gravitational waves
it may not be easy to figure out how much of this amplitude is due to the physical
gravitational waves and how much due to rippling of the coordinates.

On the other hand, the tendicities of these tendex lines are unaffected by 
rippling of the coordinates.  They remain equal to 
$\lambda_\phi =$
${\cal E}_{\hat\phi\hat\phi}
= {M/ r^3} +(A/r) \cos[2(\phi-\phi_o) -\omega( t - r_*)]$ 
$={M/(r')^3} +(A/r') \cos[2(\phi-\phi_o) -\omega( t - r'_*)]$ at
leading order, which oscillates along each closed line by the amount 
$\Delta {\cal E}_{\hat\phi\hat\phi}=(A/r)\cos[2(\phi-\phi_o)-\omega(t - r_*)]$ 
that is precisely equal to the gravitational-wave contribution to the tendicity. Note that in this example, even without evaluating the tendicity along the perturbed lines to cancel the coordinate change, the change in the tendicity due to the coordinate change enters at a higher order than the contribution from the gravitational wave.

\emph{ Therefore, in this example, the tendicity and correspondingly the structures of tendexes
capture the gravitational waves cleanly, whereas the tendex-line shapes do not do so;
the lines get modified by spatial coordinate changes.}  This is why we pay significant
attention to tendexes and also vortexes in this paper, rather than focusing solely or
primarily on tendex and vortex lines.

\section{$(2,2)$ Quasinormal Modes of Schwarzschild and Kerr Black Holes}
\label{sec:22Perts}

In Sec.\ \ref{sec:Intro22}, we described the most important features
of the fundamental, (2,2) quasinormal modes of Schwarzschild black holes.
In this section, we shall explore these modes in much greater detail and
shall extend our results to the (2,2) modes of rapidly spinning 
Kerr black holes.
For binary-black-hole mergers, these are the dominant modes 
in the late stages of the merged hole's final ringdown (see, e.g., 
\cite{Schnittman2007}).  

\subsection{Horizon vorticity and tendicity}
\label{sec:HorizonQuant}

We can compute the horizon tendicity $\mathcal E_{NN}$ and vorticity 
$\mathcal B_{NN}$ [or equivalently $\Psi_2 = \frac12(\mathcal E_{NN}+i \mathcal
B_{NN})$] using two methods: first, we can directly evaluate them from 
the metric perturbations, and second, we can calculate them, via Eq.\ 
(\ref{eq:Psi0DrivePsi2}) in the form (\ref{eq:Psi2Mode}), from the 
ingoing-wave curvature perturbation $\Psi_0$, which obeys the Teukolsky
equation (App.\ \ref{sec:Teukolsky}).
For perturbations of Schwarzschild black holes, both methods produce simple 
analytical expressions for the horizon quantities; they both show that the 
quantities are proportional to a time-dependent phase times a scalar spherical
harmonic, $e^{-i\omega t} Y_{lm}$ [see, e.g., Eq.\ (\ref{eq:Psi2SchwPsi0})]. 
For Kerr holes, the simplest formal expression for the horizon quantities is
Eq.\ (\ref{eq:Psi2Mode}), and there is no very simple analytical formula.  
Nevertheless, from these calculations one can show that there is an exact 
duality between $\mathcal E_{NN}$ and $\mathcal B_{NN}$ in ingoing
radiation gauge for quasinormal
modes with the same order parameters $(n,l,m)$ but opposite parity, 
for both Schwarzschild and Kerr black holes; 
see App. \ref{App:HorizonFromPsi0}. 
For Schwarzschild black holes in RWZ gauge, there is also a duality for the 
horizon quantities, although it is complicated by a perturbation to the 
position of the horizon in this gauge; see Appendices.~\ref{sec:RWZTidalField} 
and~\ref{sec:RWZHorizon} for further discussion.

\begin{figure}
\includegraphics[width=0.95\columnwidth]{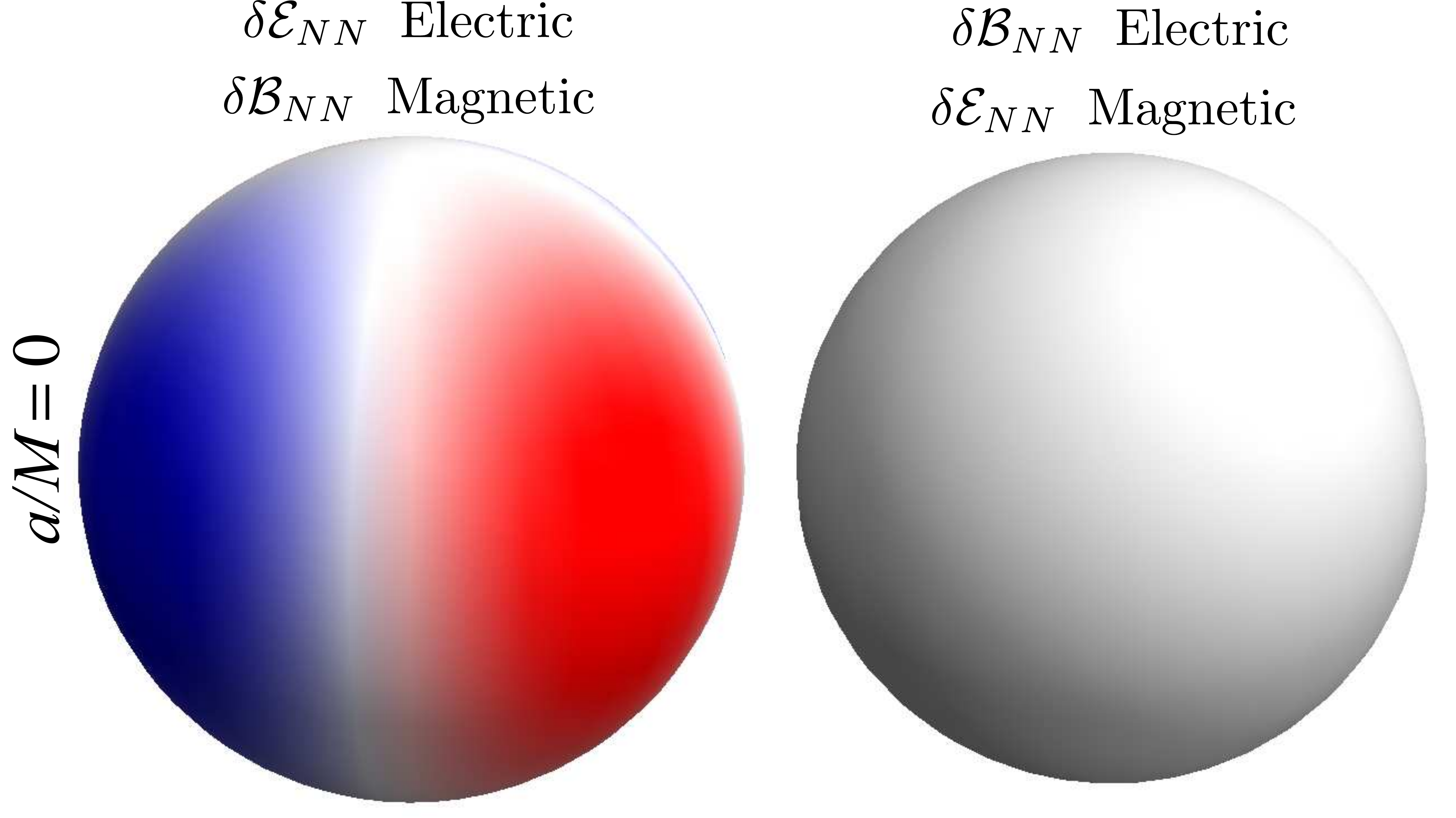}
\includegraphics[width=0.95\columnwidth]{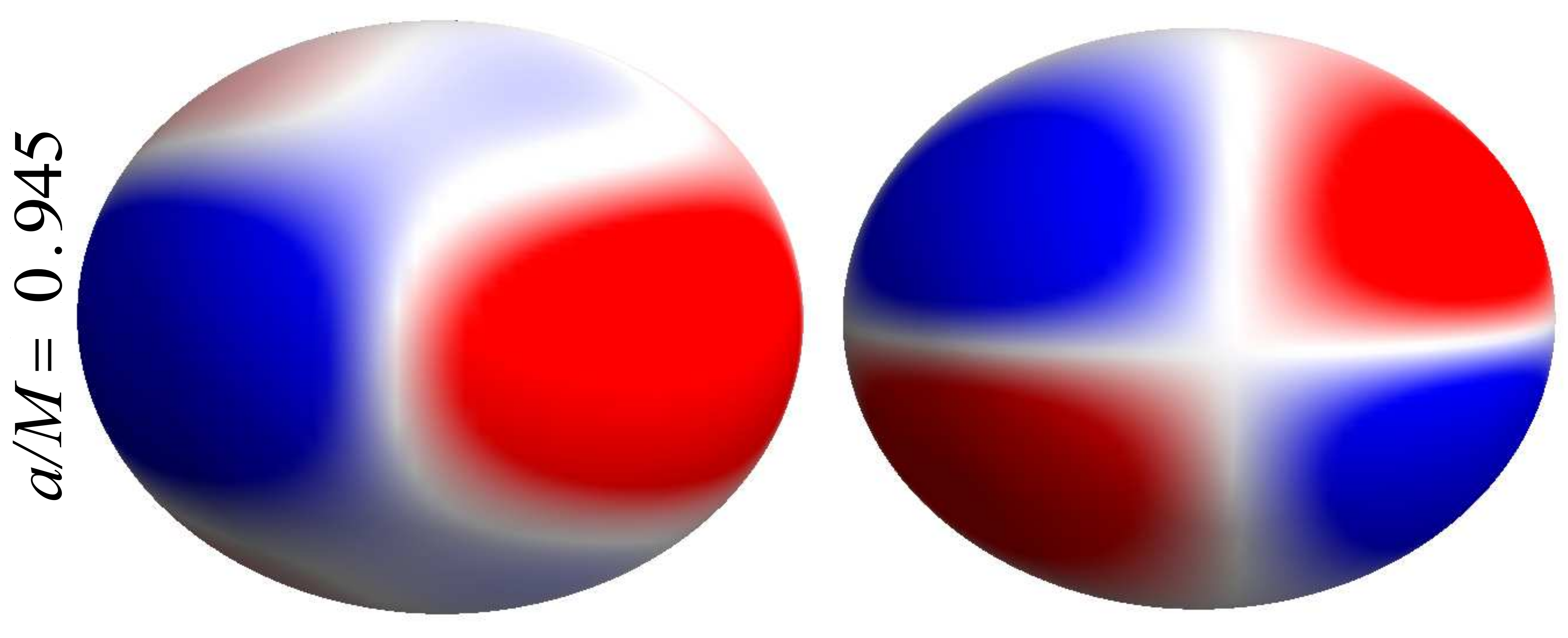}
\caption{(color online).
Perturbative horizon tendicities $\delta \mathcal E_{NN}$ and vorticities
$\delta \mathcal B_{NN}$ for the $(2,2)$ quasinormal modes with electric and magnetic
parities (see column labels at the top).  The top row is for a 
Schwarzschild black hole, $a=0$;
the bottom for a rapidly spinning Kerr black hole, $a/M=0.945$. 
The color intensity is proportional to the magnitude of the tendicity or
vorticity, with blue (dark gray) for positive and red (light gray) for 
negative.  For discussion, see Sec.\ \ref{sec:HorizonQuant} of the text.}
\label{fig:EvenPertHorizons}
\end{figure}

In Fig.~\ref{fig:EvenPertHorizons}, we show 
$\delta \mathcal E_{NN}$ and $\delta \mathcal B_{NN}$ for the $(2,2)$ modes with both parities,
of a Schwarzschild black hole (upper row) and a rapidly rotating
Kerr black hole (bottom row).  

The duality is explicit in the labels at the top: the patterns are identically
the same for $\delta \mathcal E_{NN}$ (tendexes) of electric-parity modes and 
$\delta \mathcal B_{NN}$ (vortexes) of magnetic-parity modes [left column]; and 
also identically the same when the parities are switched [right column]  
The color coding is similar to 
Fig.\ \ref{fig:22SmagEquator} above (left-hand scale).  The red (light gray) 
regions are stretching tendexes or counterclockwise vortexes (negative 
eigenvalues); the blue (dark gray), squeezing tendexes or clockwise vortexes 
(positive eigenvalues). 

For the Schwarzschild hole, the electric-parity tendex pattern and 
magnetic-parity vortex pattern (upper left) is that of the spherical harmonic 
$Y^{22}(\theta,\phi)$,
and the perturbative electric-parity vorticity and magnetic-parity tendicity
vanish (upper right).

For the rapidly spinning Kerr hole, the electric-parity tendexes and 
magnetic-parity vortexes (lower left) are concentrated  more tightly around 
the plane of reflection symmetry than they are for the Schwarzschild hole, and 
are twisted; but their patterns are still predominantly $Y^{22}$.  
And also for Kerr, the (perturbative)
electric-parity vorticity and magnetic-parity tendicity have become nonzero
(lower right),
they appear to be predominantly $Y^{32}(\theta,\phi)$ in shape, they are much
less concentrated near the equator and somewhat weaker than the 
electric-parity tendicity and
magnetic-parity vorticity (lower left).

\subsection{Equatorial-plane vortex and tendex lines, and vortexes and 
tendexes}
\label{sec:22EquatorialPlane}

As for the weak-field, radiative sources of Paper I, so also here,
the equatorial plane is an informative and simple region in which to 
study the generation of gravitational waves. 

For the (2,2) modes that we are studying, 
the $\delta \mathcal E_{jk}$ of an electric-parity perturbation and
the $\delta \mathcal B_{jk}$ for magnetic parity are
symmetric about the equatorial plane. This restricts two sets of field lines 
(tendex lines for electric-parity $\delta \mathcal E_{jk}$; vortex lines
for magnetic-parity $\delta \mathcal B_{jk}$) 
to lie in the plane and forces the third to be normal to the plane. 
By contrast, the electric-parity $\delta \mathcal B_{jk}$ and magnetic-parity
$\delta \mathcal E_{jk}$ are reflection antisymmetric.  This 
requires that two sets of field lines cross the equatorial
plane at $45^\circ$ angles, with equal and opposite eigenvalues (tendicities
or vorticities), and forces
the third set to lie in the plane and have zero eigenvalue; this third set of zero-vorticity vortex lines have less physical interest and so we will not illustrate them.

In this section, we shall focus on the in-plane field lines and their 
vorticities and tendicities.

\subsubsection{Magnetic-parity perturbations of Schwarzschild black holes}
\label{sec:MagPar22Sch}

In Sec.\ \ref{sec:Intro22} and Figs.\ \ref{fig:22MagVortex} 
and \ref{fig:22SmagEquator}, we discussed some equatorial-plane properties of the 
magnetic-parity $(2,2)$ mode.  Here we shall explore these and other properties
more deeply.  Recall that for the magnetic-parity mode,
the frame-drag field, and hence also the vortex lines
and their vorticities, 
are fully gauge invariant. 

\begin{figure*}
\includegraphics[width=0.575\columnwidth]{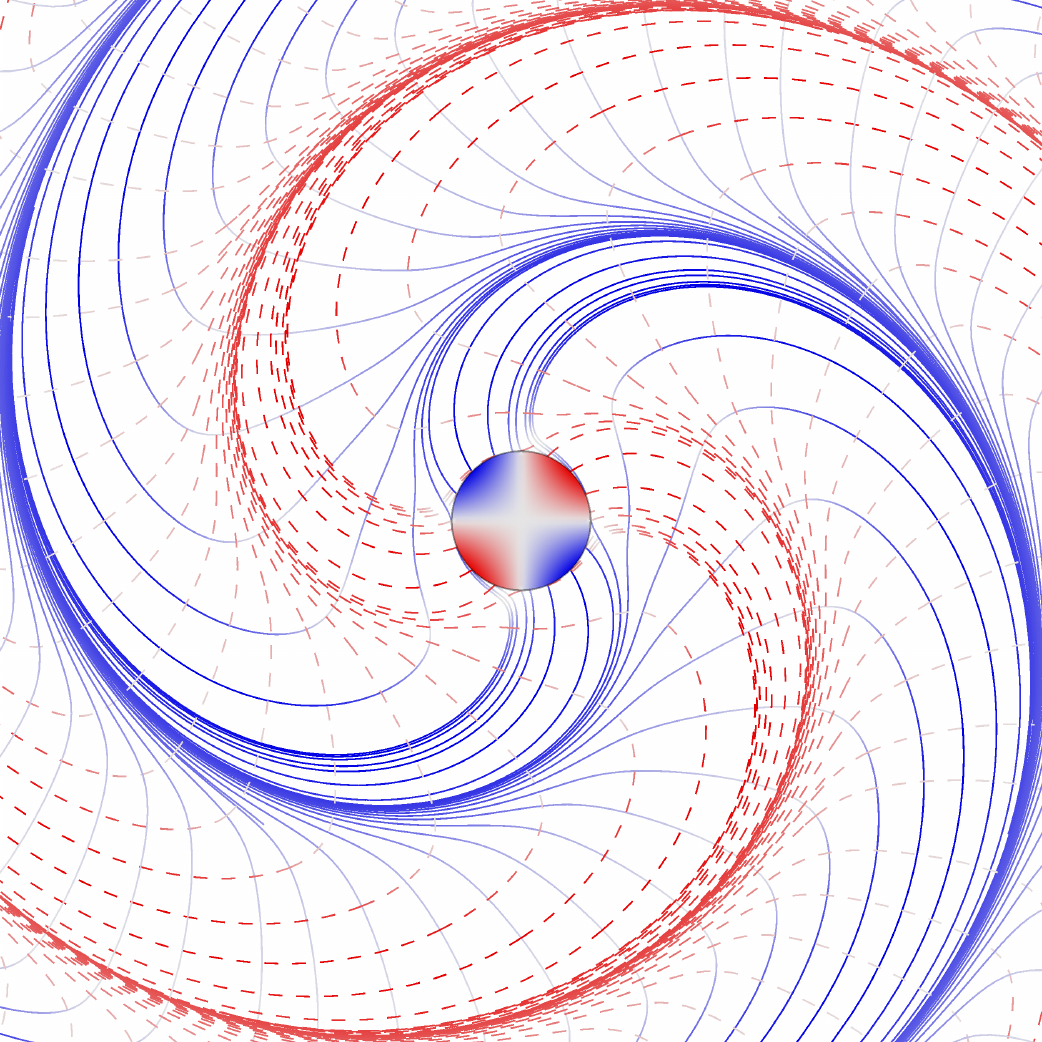}
\includegraphics[width=0.625\columnwidth]{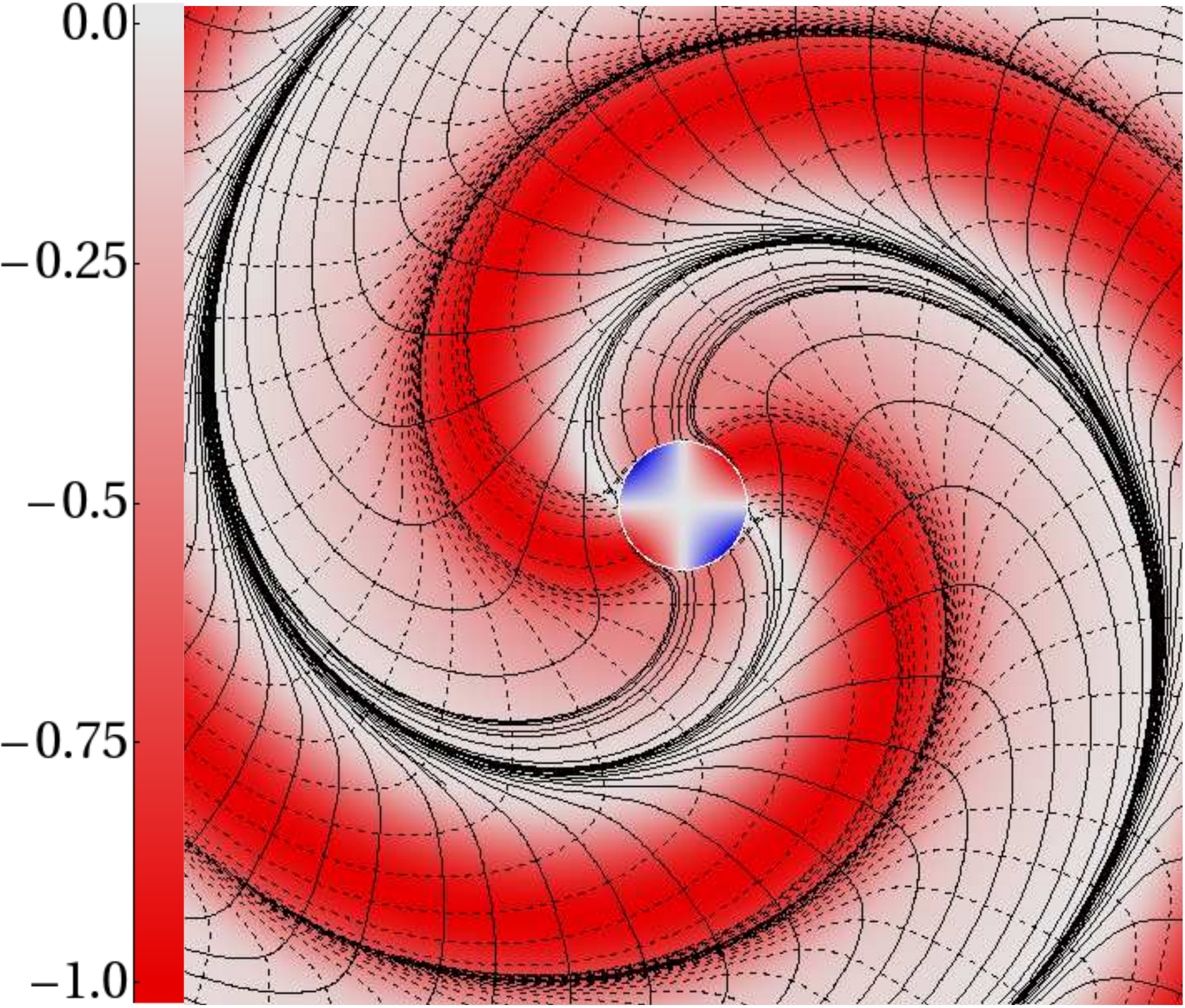}
\includegraphics[width=0.625\columnwidth]{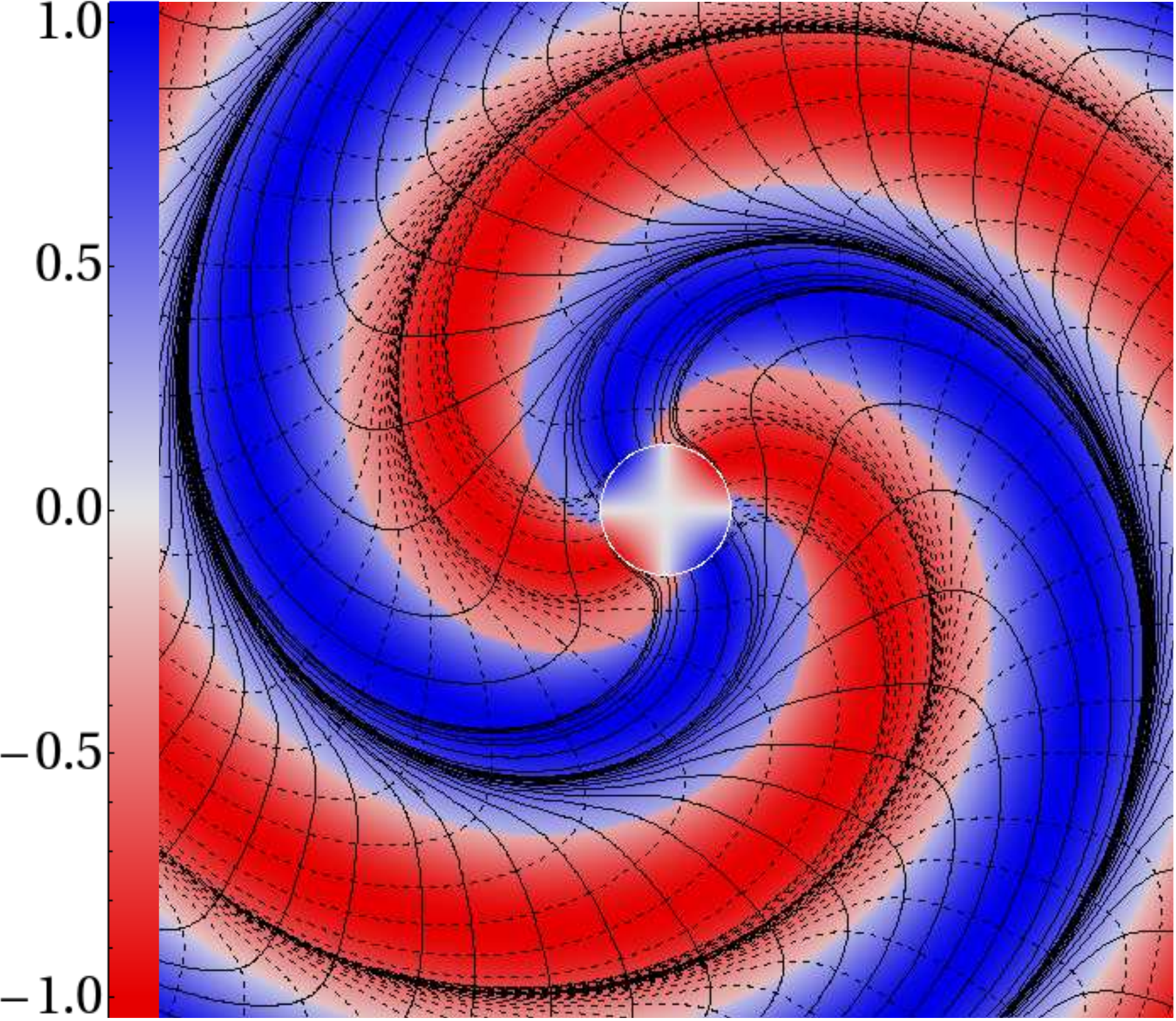}
\includegraphics[width=0.575\columnwidth]{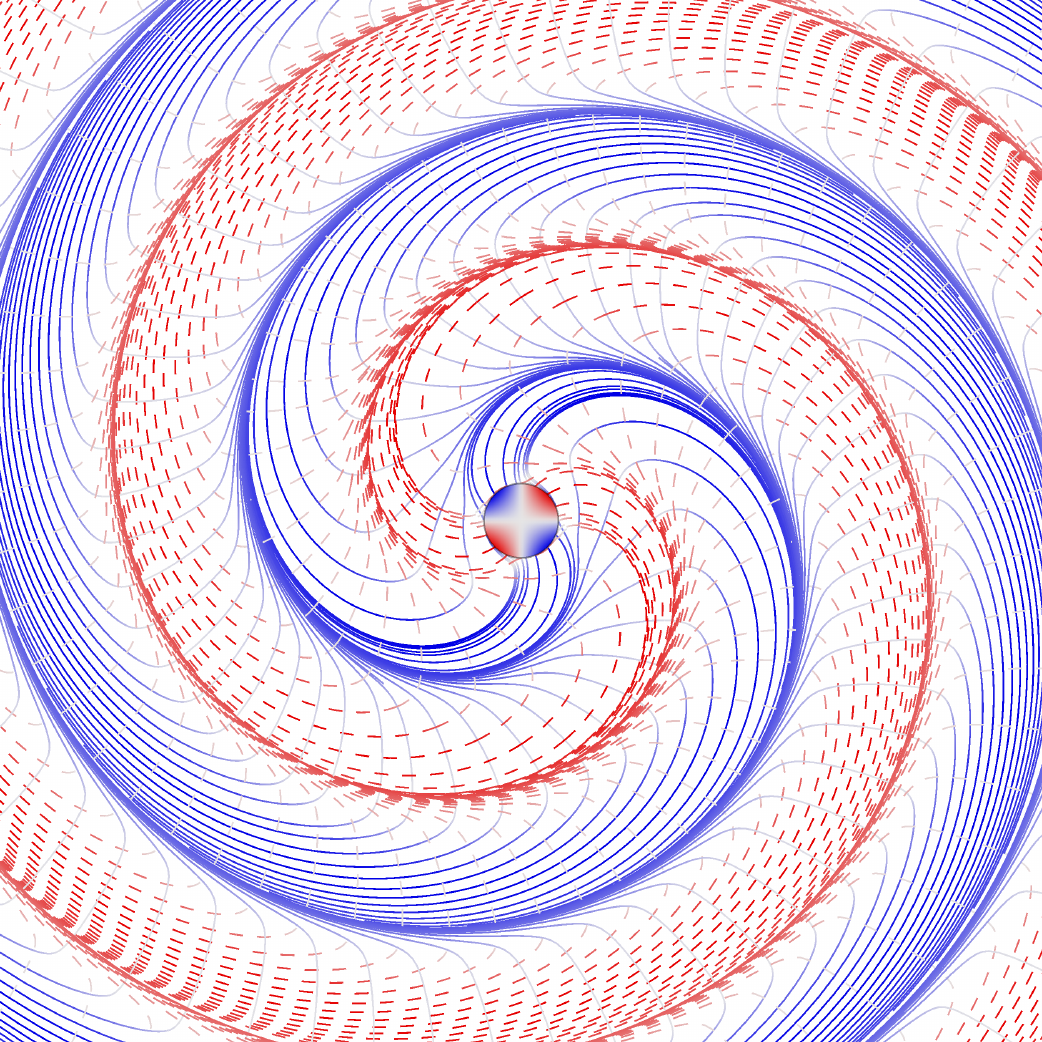}
\includegraphics[width=0.625\columnwidth]{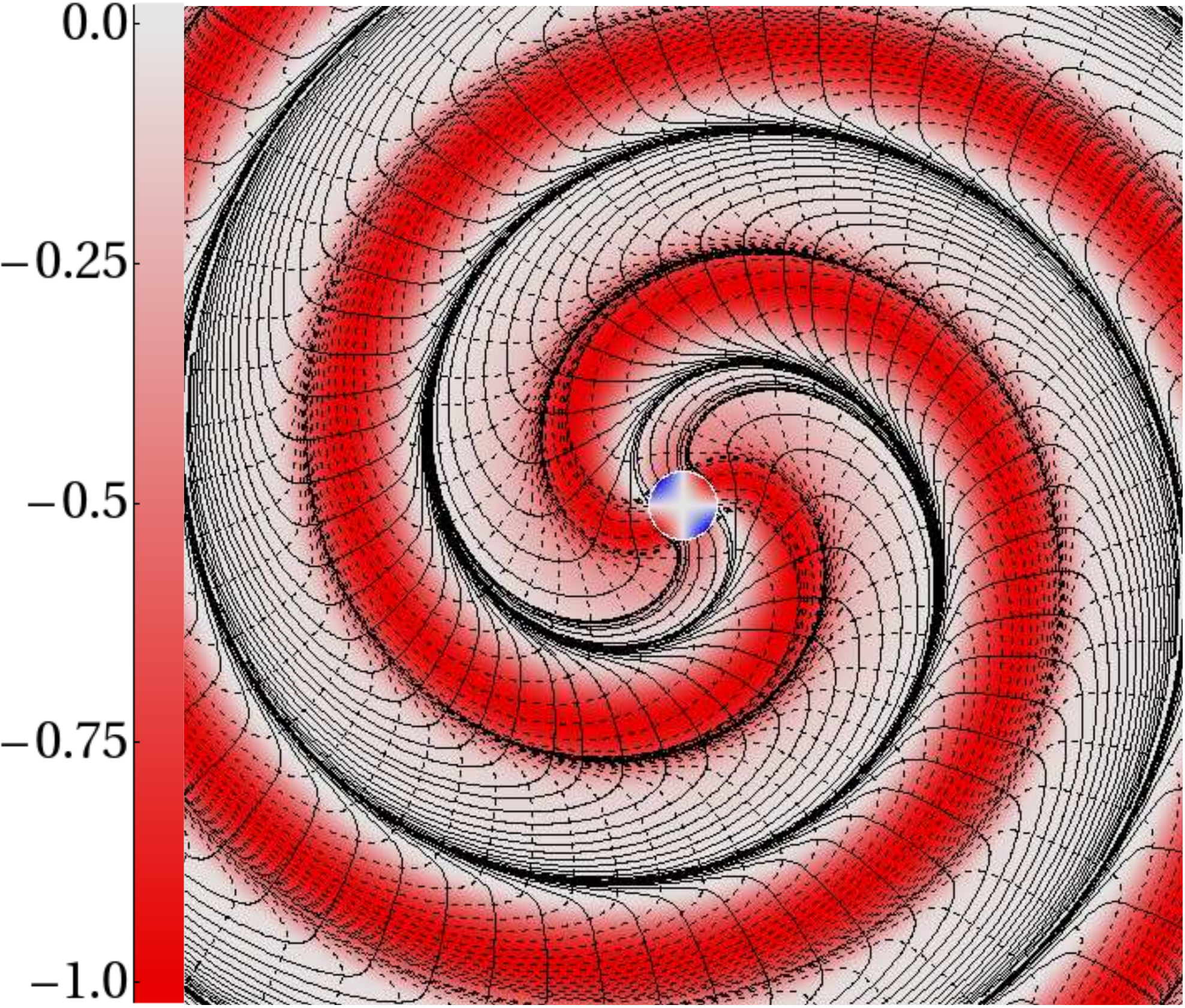}
\includegraphics[width=0.625\columnwidth]{RWBothVortexl2m2Far_2}
\caption{(color online).
Three representations of the vortex lines and vortexes in the equatorial
plane of a Schwarzschild black hole perturbed by a magnetic-parity $(2,2)$
quasinormal mode.
The bottom panels span a region $56M$ on each side, and the top panels are
a zoom-in of the lower panels, $30M$ on each side.
All panels show positive-vorticity lines as solid and negative-vorticity lines
as dashed.
In all panels, blue (dark gray) corresponds to positive and red (light gray) 
to negative; the intensity of the color indicates the strength of the vorticity
at that point normalized by the maximum of the vorticity at that radius (darker
shading indicates a larger strength and lighter, weaker). 
Similarly, in all panels, the central circle surrounded by a narrow white line
is the horizon colored by its vorticity as described above.
{\it Left column}: Vortex lines colored and shaded by their scaled vorticity.
{\it Middle column}: Negative vorticity coloring the plane with black vortex
lines.
{\it Right column}: Vorticity with the larger absolute value coloring the
plane and black vortex lines. For discussion of this figure, see Sec.\ 
\ref{sec:MagPar22Sch}.}
\label{fig:Sch22VortexAll}
\end{figure*}

In Fig.~\ref{fig:Sch22VortexAll}, we show six different depictions of the vortex
lines and their vorticities in the equatorial plane, each designed to highlight
particular issues.  See the caption for details of what is depicted. 

The radial variation of vorticity is not shown in this figure, only the angular
variation.
The vorticity actually passes through a large range of values as a 
function of radius: from the horizon to roughly $r=4M \simeq 1.5\lambdabar$
(roughly the outer edge of the near zone), the vorticity rapidly
decreases; between $r\simeq 4M$ and $12M$ 
(roughly the extent of the transition zone), 
it falls off as $1/r$; and at $r\gtrsim 12M$ (the wave zone),
it grows exponentially due to the damping of the quasinormal mode as time passes.
(The wave field at larger radii was emitted earlier when the mode was stronger.)
In the figure, we have removed these radial variations in order to highlight
the angular variations.

By comparing 
the left panels of Fig.~\ref{fig:Sch22VortexAll} with Fig~9 of Sec.~VI D
of Paper I, we see a strong  
resemblance between the vortex lines of our $(2,2)$, magnetic-parity 
perturbation of a Schwarzschild black hole, and those of a rotating current
quadrupole in linearized theory. 
As in linearized theory, when the radial (or, synonymously, longitudinal)
vortex lines in the near zone rotate, the effects of time retardation cause
the lines, in the transition and wave zones, to collect around four 
backspiraling regions of 
strong vorticity (the vortexes) and to acquire perturbative tendex lines as
they become transverse-traceless gravitational waves.
The most important difference is that, for the black-hole 
perturbations, the positive vortex lines emerge from the blue, clockwise 
horizon vortexes and spiral outward (and the negative vortex lines emerge from 
the counterclockwise horizon vortexes) rather than emerging from a 
near-zone current quadrupole. 

Although the left panels of Fig.~\ref{fig:Sch22VortexAll}
highlight most clearly the comparison with
figures in Paper I, the middle and right panels 
more clearly show the relationship between the vortex 
lines (in black) and the vorticities, throughout the equatorial plane.
In the middle panels (which show only the negative vorticity), the negative
vortex lines that emerge longitudinally from the horizon stay in the center of
their vortex in the near zone, and then collect onto the outer edge of the vortex in 
the transition and wave zones.
Interestingly, near the horizon, there are also two weaker regions of 
negative vorticity
between the two counterclockwise vortexes, regions associated with the 
tangential negative vortex lines that pass through this region without 
attaching to the horizon (and that presumably represent radiation traveling 
into the horizon).

In the right panels of Fig.\ \ref{fig:Sch22VortexAll} (which show the 
in-plane vorticity with the larger absolute value), a clockwise vortex
that extends radially from the horizon takes the place of the weaker region of
counterclockwise vorticity.
From these panels, it is most evident that the vortexes and vortex lines of 
opposite signs are identical, though rotated by $90^\circ$.
These panels also highlight that there are four spirals of nearly zero 
vorticity that separate the vortexes in the wave zone, which the spiraling 
vortex lines approach.
All three vorticities nearly vanish at these spirals; in the limit of infinite
radius, they become vanishing points for the radiation, which must exist 
for topological reasons \cite{Zimmerman2011}.

\subsubsection{Gauge dependence of electric-parity tendexes for a Schwarzschild black hole}
\label{sec:GaugeDep22}

\begin{figure*}[t]
\includegraphics[width=0.9\columnwidth]{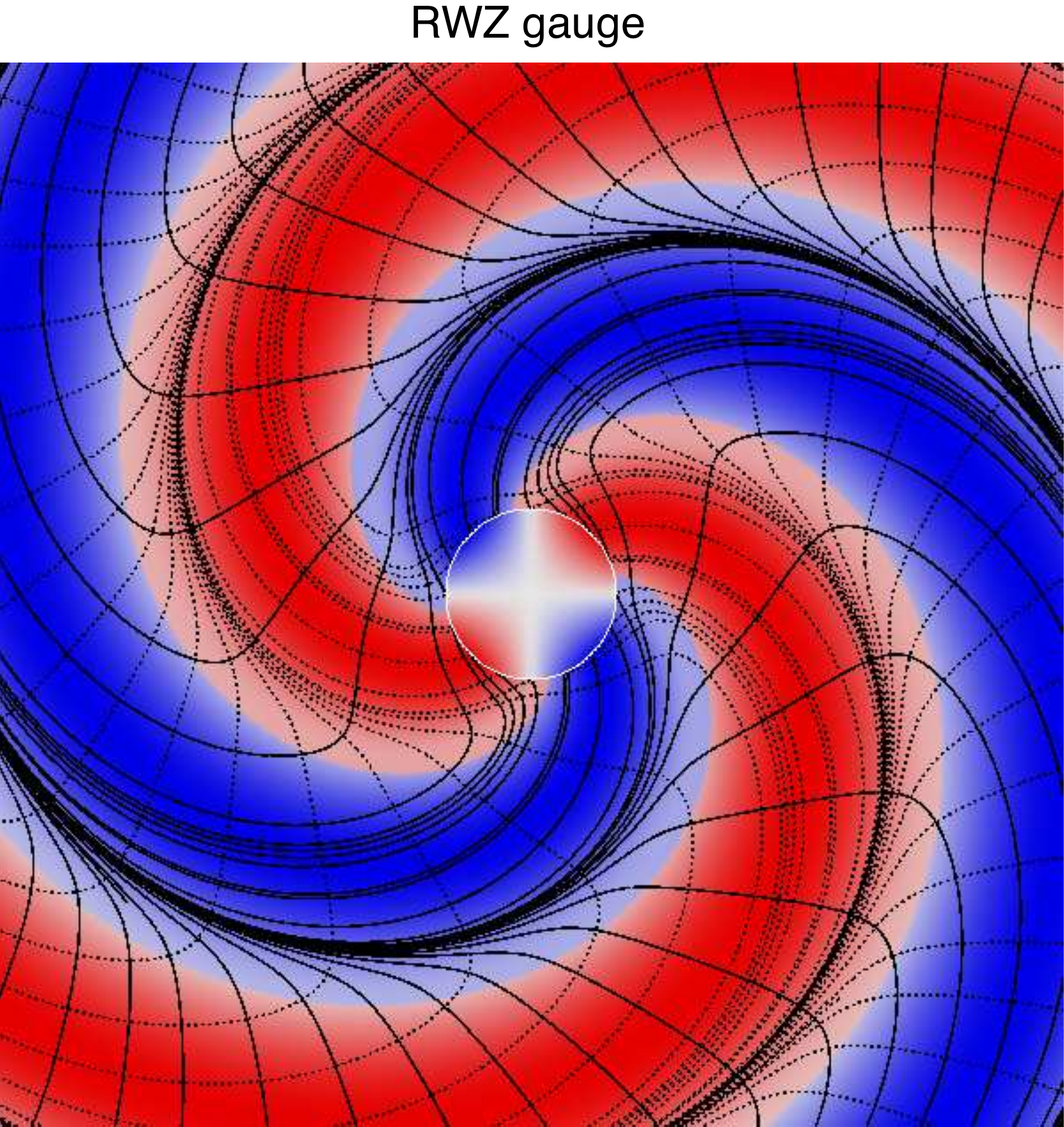}
\includegraphics[width=0.9\columnwidth]{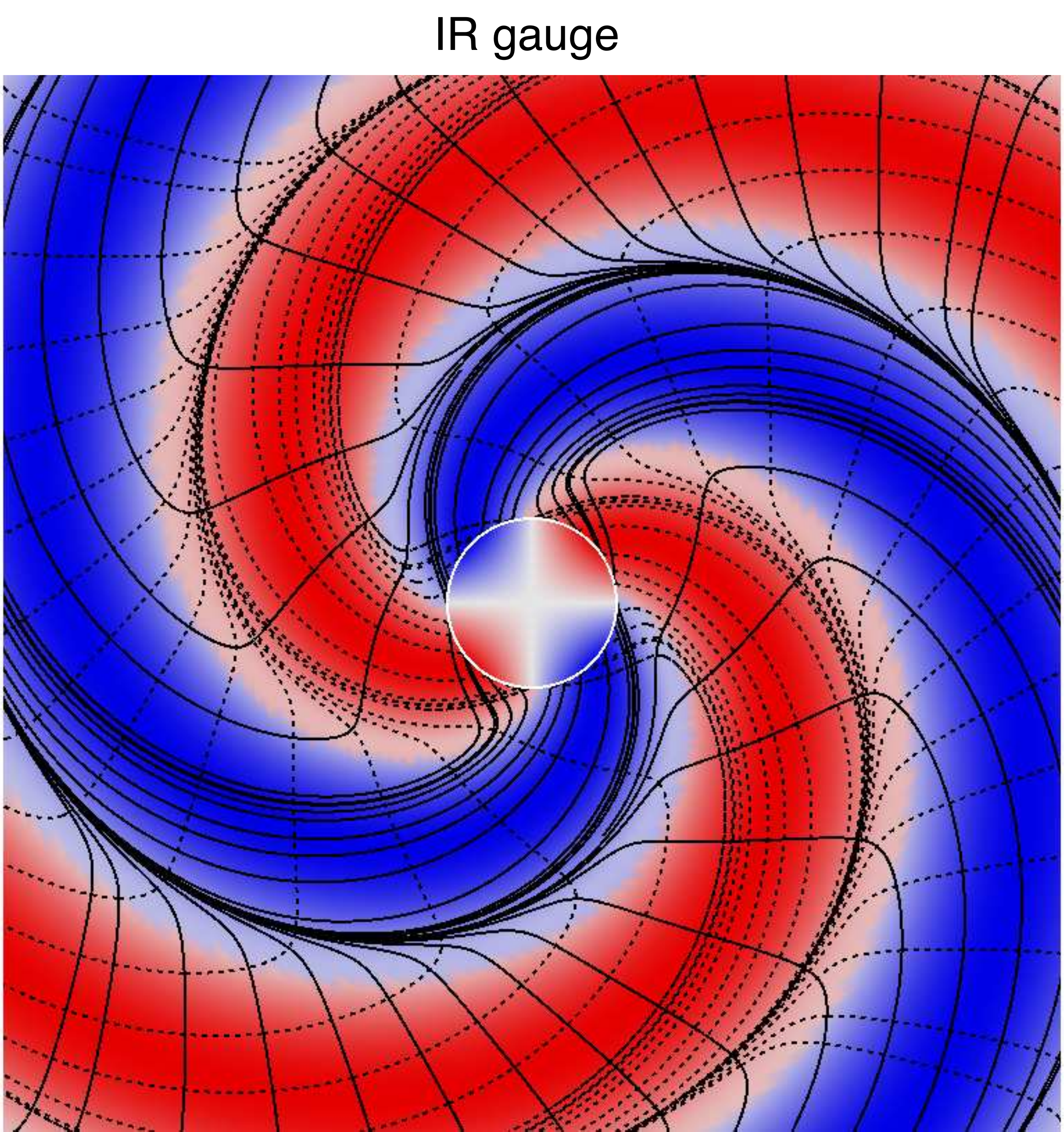}
\caption{(color online).
The equatorial-plane, electric-parity tendexes and tendex lines of a 
$(2,2)$ perturbation of a Schwarzschild black hole in RWZ gauge (left panel) 
and IR gauge (right panel). 
The conventions for the lines, the coloring and the shading are identical to 
those in the right panels of Fig.\ \ref{fig:Sch22VortexAll}.}
\label{fig:ElectricGaugeCompare}
\end{figure*}

In this subsection, we explore the gauge dependence of the (2,2) modes for
a Schwarzschild black hole. In Sec.\ \ref{sec:GaugePerts}, we showed that for magnetic-parity modes, all 
gauges share the same slicing.  Therefore, to maximize any gauge dependence 
that there might be, we focus on the electric-parity (2,2) mode. 

Because the frame-drag field of the unperturbed Schwarzschild black hole vanishes,
this mode's perturbative frame-drag field will be unaffected by 
perturbative changes of the spatial coordinates.  Therefore, we focus
on the perturbative tidal field $\delta \boldsymbol{\mathcal E}$
of the electric-parity mode, which is sensitive 
to \emph{both} perturbative slicing changes and perturbative spatial
coordinate changes.

In Fig.~\ref{fig:ElectricGaugeCompare}, we plot this field's 
perturbative equatorial 
tendexes and tendex lines for the electric-parity (2,2) mode 
in RWZ gauge (left panel) and IR gauge (right panel)---which differ,
for this mode, in both slicing and spatial coordinates. 
The tendex lines for the two gauges were seeded at the same coordinate points, 
so all the differences between the panels can be attributed to the 
gauge differences. 

The two panels are almost identical.  Therefore, these maximally sensitive
tendexes and tendex lines are remarkably unaffected by switching from
one gauge to the other.
The primary differences are that 
(i) the tendex lines of IR gauge which are near the black hole tend to be pulled closer to the horizon as compared to RWZ gauge (ii) the lines falling onto the 
attracting spiral are bunched even more tightly in IR gauge than in RWZ gauge; 
however, more lines reach the spiral in RWZ gauge in this figure, and 
(iii) the four tendex spirals wind more tightly in IR gauge, which is most 
easily seen by comparing the lower right and upper left corners of the two
panels.

One subtlety that must be remarked upon is that the central circle 
colored by the normal-normal component of the tidal field (surface
tendicity) in the RWZ gauge (left panel of
Fig.~\ref{fig:ElectricGaugeCompare}) is simply the surface $r=2M$, and
not the true event horizon. The location of the event horizon 
is affected by the
perturbations in a gauge-dependent manner, as discussed
by Vega, Poisson, and Massey \cite{Vega:2011}. We rely on the
results of this article in the brief discussion that follows. In RWZ
gauge, the horizon is at $r_H = 2M + \delta r(\tilde t, \theta, \phi)$,
where the function $\delta r$ can be solved for by ensuring that the
vector tangent to the perturbed generators (in our case, $l^\mu =
\partial x^\mu/\partial \tilde t$) remains null \cite{Vega:2011}. 
We give an expression for $\delta r$ in
Appendix~\ref{sec:RWZHorizon}. There we also discuss the correction to the
horizon tendicity in RWZ gauge. One key result is that the horizon
tendicity has the same angular distribution in RWZ gauge as in IR gauge
(given by the $Y^{22}$ spherical harmonic), so that the normal-normal
tendicity on the horizon and on the surface at $r=2M$ differ only by an
amplitude and phase in RWZ gauge. Meanwhile, in IR gauge the horizon remains
at $r_H= 2M$ and so the colored central circle is in fact the horizon, colored
by its horizon tendicity.

However, the bulk tendexes and tendex lines are determined completely 
independently of these horizon considerations, and so 
Fig.~\ref{fig:ElectricGaugeCompare} provides an accurate 
comparison of them in the two gauges.

\subsubsection{Duality and influence of spin in the equatorial plane}
\label{sec:TendVortSchwKerr}

\begin{figure*}
\includegraphics[width=0.65\columnwidth]{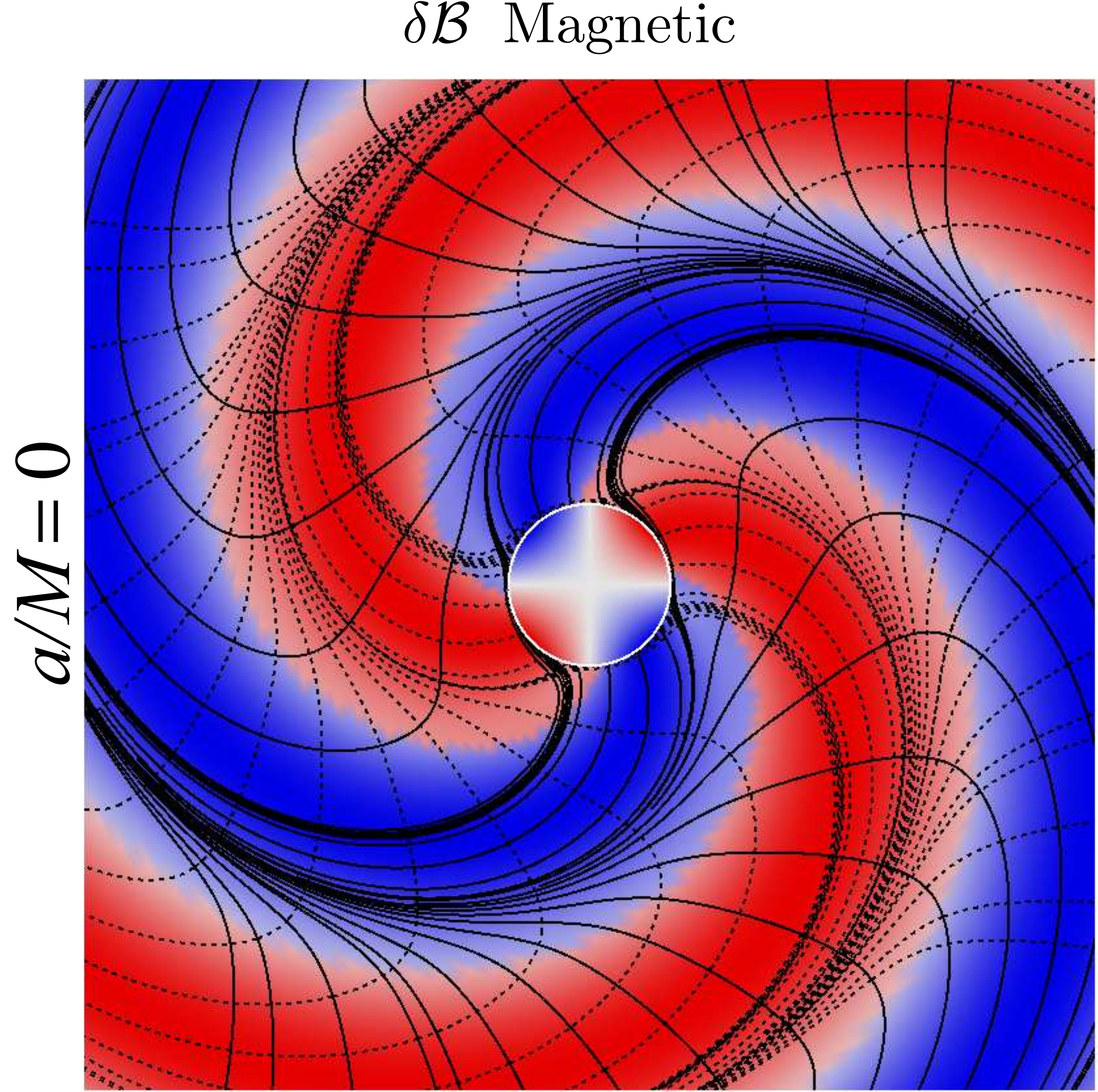}
\includegraphics[width=0.6\columnwidth]{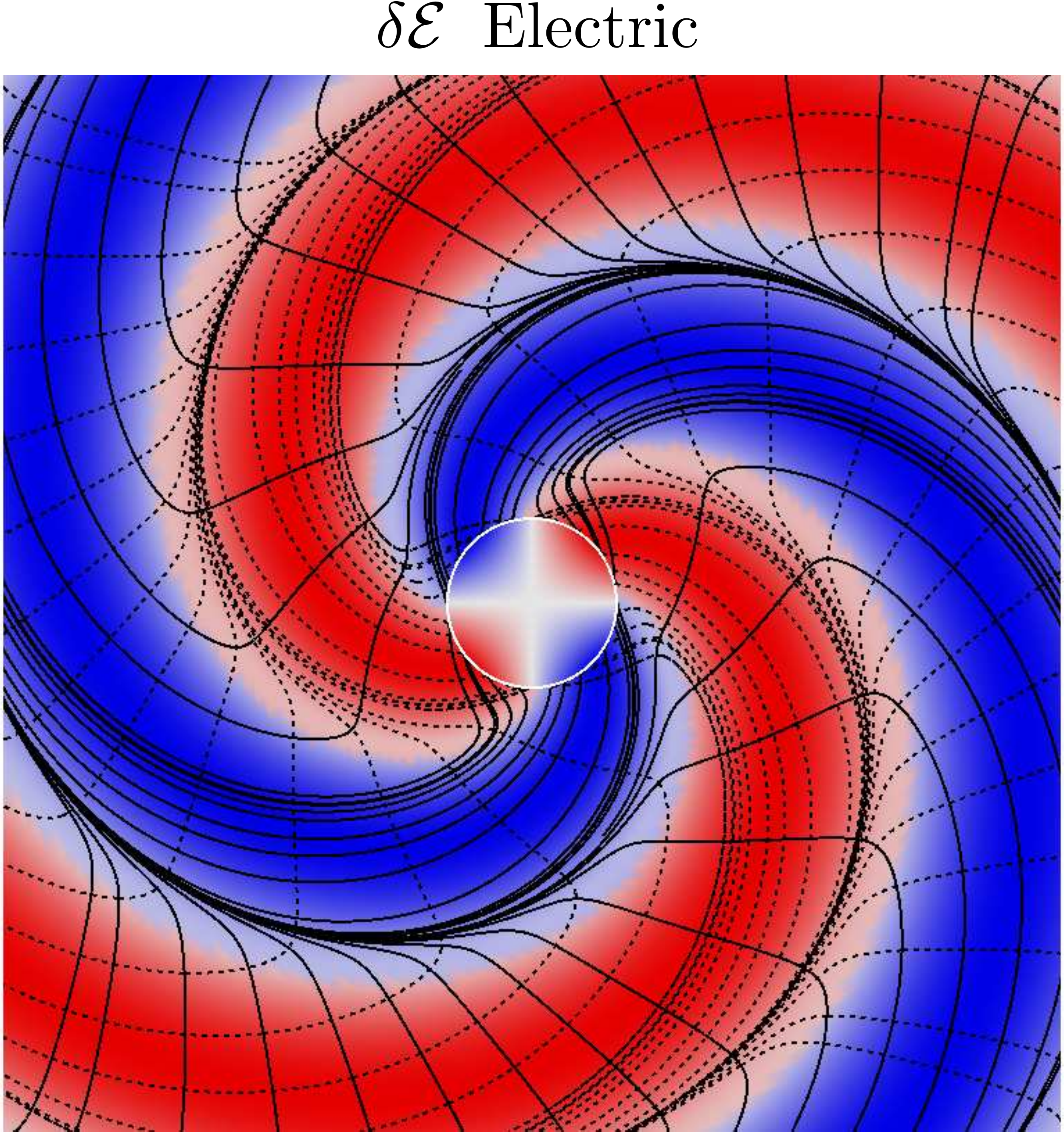}
\includegraphics[width=0.7\columnwidth]{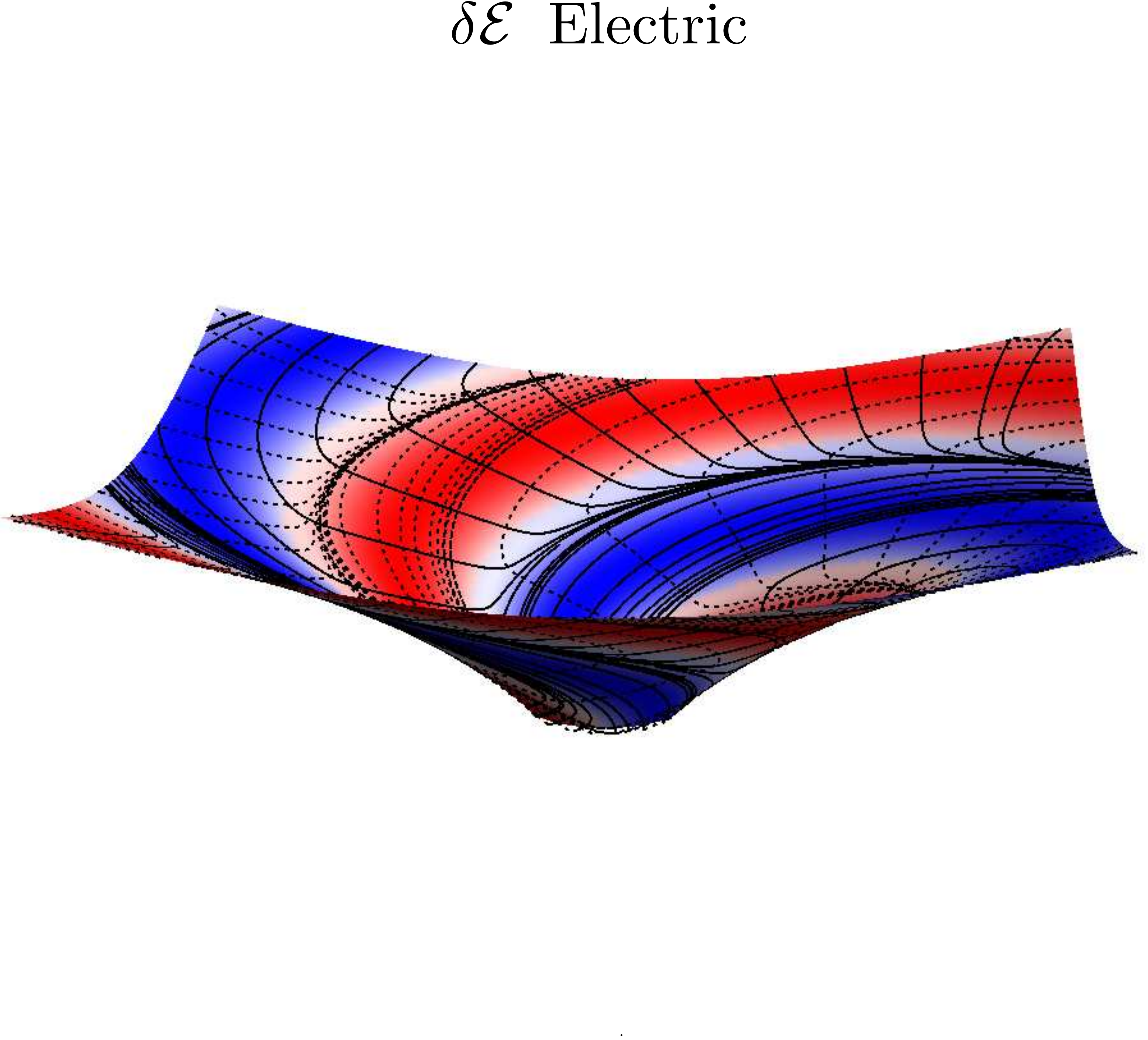}
\includegraphics[width=0.65\columnwidth]{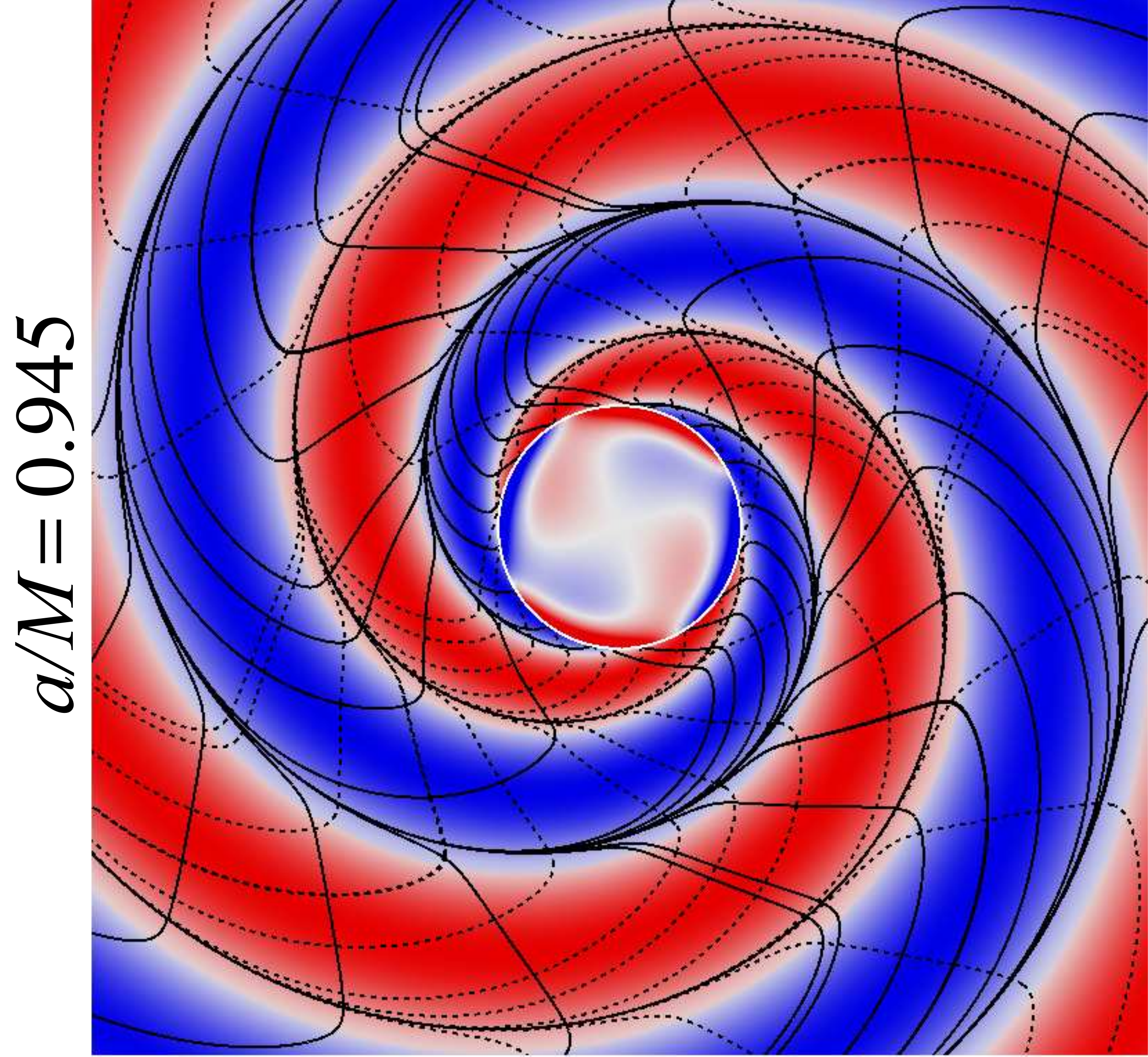}
\includegraphics[width=0.6\columnwidth]{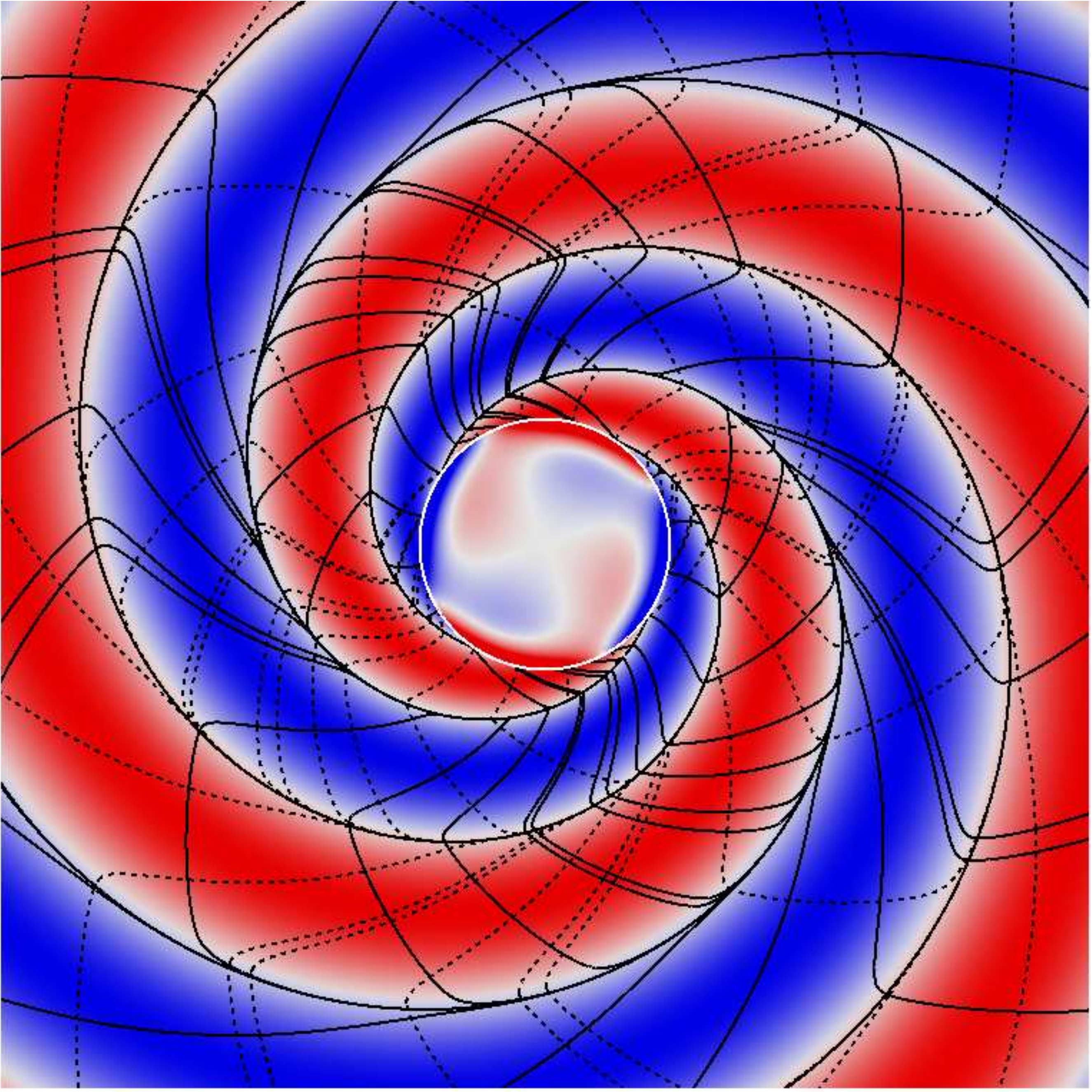}
\includegraphics[width=0.7\columnwidth]{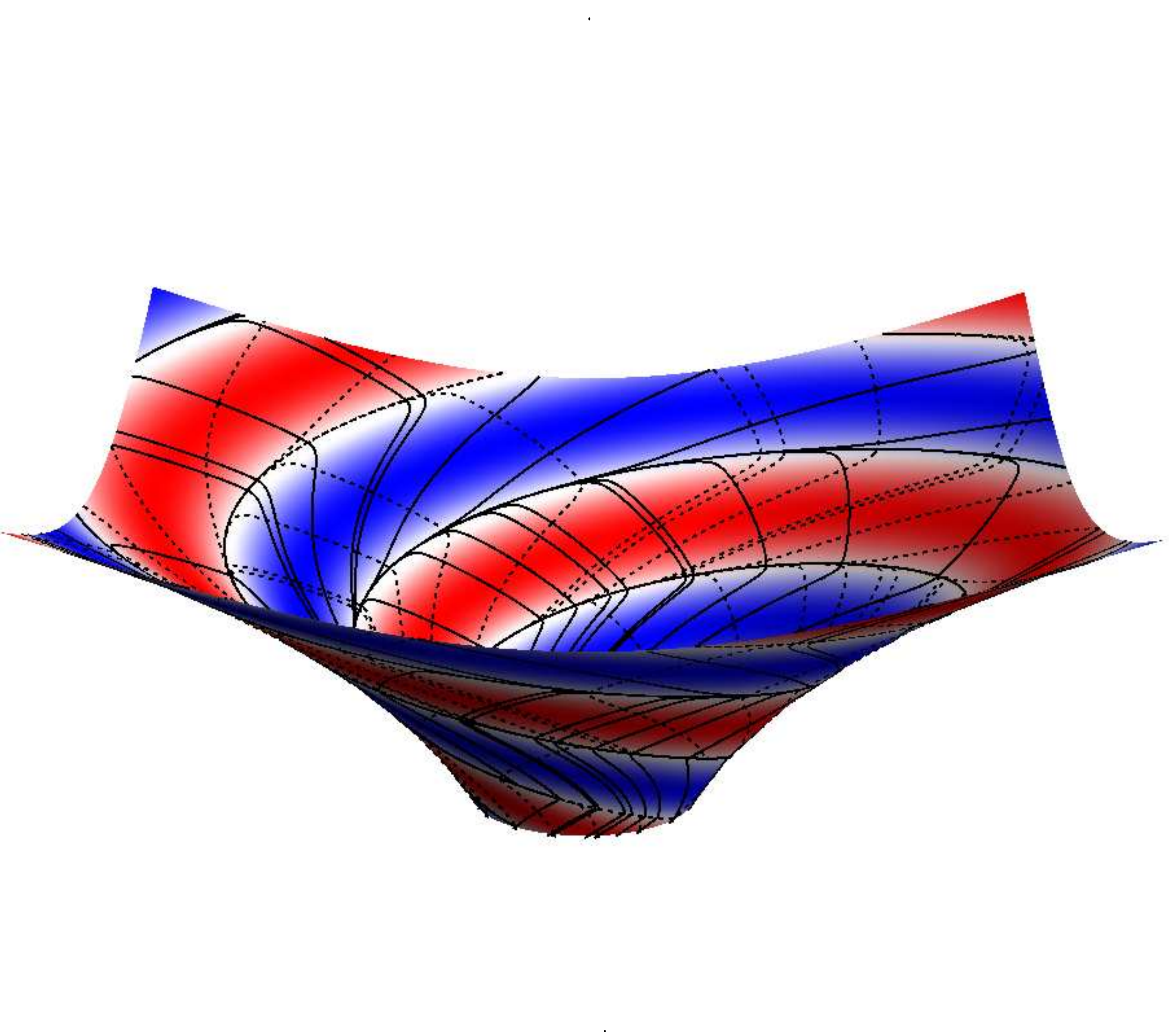}
\caption{(color online).
Vortexes and tendexes and their field lines in the equatorial plane 
for $(2,2)$ modes of Schwarzschild and Kerr black holes.
The lines, the coloring and the shading are identical to those in the right
panels of Fig.\ \ref{fig:Sch22VortexAll}.  The upper row is for a Schwarzschild
black hole $(a=0)$; the lower row, for a rapidly spinning Kerr black hole
$(a=0.945)$; see labels on the left. The left column shows the 
vortex lines and vorticities for magnetic parity (which are gauge
invariant for the perturbations of a Schwarzschild hole); the middle and right
columns show the tendex lines and tendicities for the electric-parity mode
in IR gauge; see labels at the top.  In the right column, the equatorial plane
is isometrically embedded in three-dimensional Euclidean space.  The top
panels are $24M$ across; the bottom, $14M$.  This figure elucidates 
duality and the influence of black-hole spin; see the discussion in Sec.\ 
\ref{sec:TendVortSchwKerr}.}
\label{fig:TendexVortexEq}
\end{figure*}

In this subsection, we use Fig.\ \ref{fig:TendexVortexEq} to explore 
duality and the influence of spin, for the fundamental $(2,2)$ mode.

By comparing the left and center panels in the top row of Fig.\
\ref{fig:TendexVortexEq}, we see \emph{visually} the near duality between the
electric- and magnetic-parity modes for a Schwarzschild hole.  This
near duality is explored \emph{mathematically} in App. \ref{sec:CCKProc}.
Specifically, the vortexes and their lines for the magnetic-parity mode
(left) are nearly identical to the tendexes and their lines for the
electric-parity mode.  The only
small differences appear in the size of the nearly zero-vorticity (or 
tendicity) regions, and the curvatures of the lines.

For the fast-spinning Kerr black hole (the bottom left and center panels of 
Fig.\ \ref{fig:TendexVortexEq}), the near duality is still obvious, especially 
in the colored vortexes and tendexes; but it is 
less strong than for Schwarzschild, especially in the field lines. 
The vortex lines (on the left) continue to look like those of a Schwarzschild
black hole, but the tendex lines (in the middle) curve in the opposite direction, 
which makes some lines reach out from the horizon and connect back to it 
instead of 
spiraling away from the horizon.

By comparing the top and bottom panels in the left and center columns of Fig.\ 
\ref{fig:TendexVortexEq}, we see the influence of the background  
black hole's spin on the dynamics of the perturbative vortexes and tendexes.
For fast Kerr (bottom), the vortexes and tendexes near the horizon look
more transverse (less radial) than for Schwarzschild, because the size of the 
near zone is 
much smaller. (The frequency
of the waves is nearly twice that for a perturbed Schwarzschild hole.) The
higher frequency also
explains why the spirals of the vortexes and tendexes are tighter.

In the isometric embedding diagrams in the right column of Fig.\ 
\ref{fig:TendexVortexEq}, we see that proper radial distance in the near 
zone is somewhat larger than it appears in the flat, planar drawing.
Taking this into account, we conclude that, aside from a few small 
differences, the qualitative ways in which waves are generated for fast Kerr
and for Schwarzschild are the same: two 
pairs of vortexes or tendexes emerge longitudinally from horizon
vortexes,  
and twist into backward spirals that eventually form the
transverse-traceless gravitational waves.

\subsubsection{Vortexes of electric-parity mode, and perturbative tendexes of magnetic-parity mode for a Schwarzschild black hole}
\label{sec:EparityVortexes22}

\begin{figure}
\includegraphics[width=0.95\columnwidth]{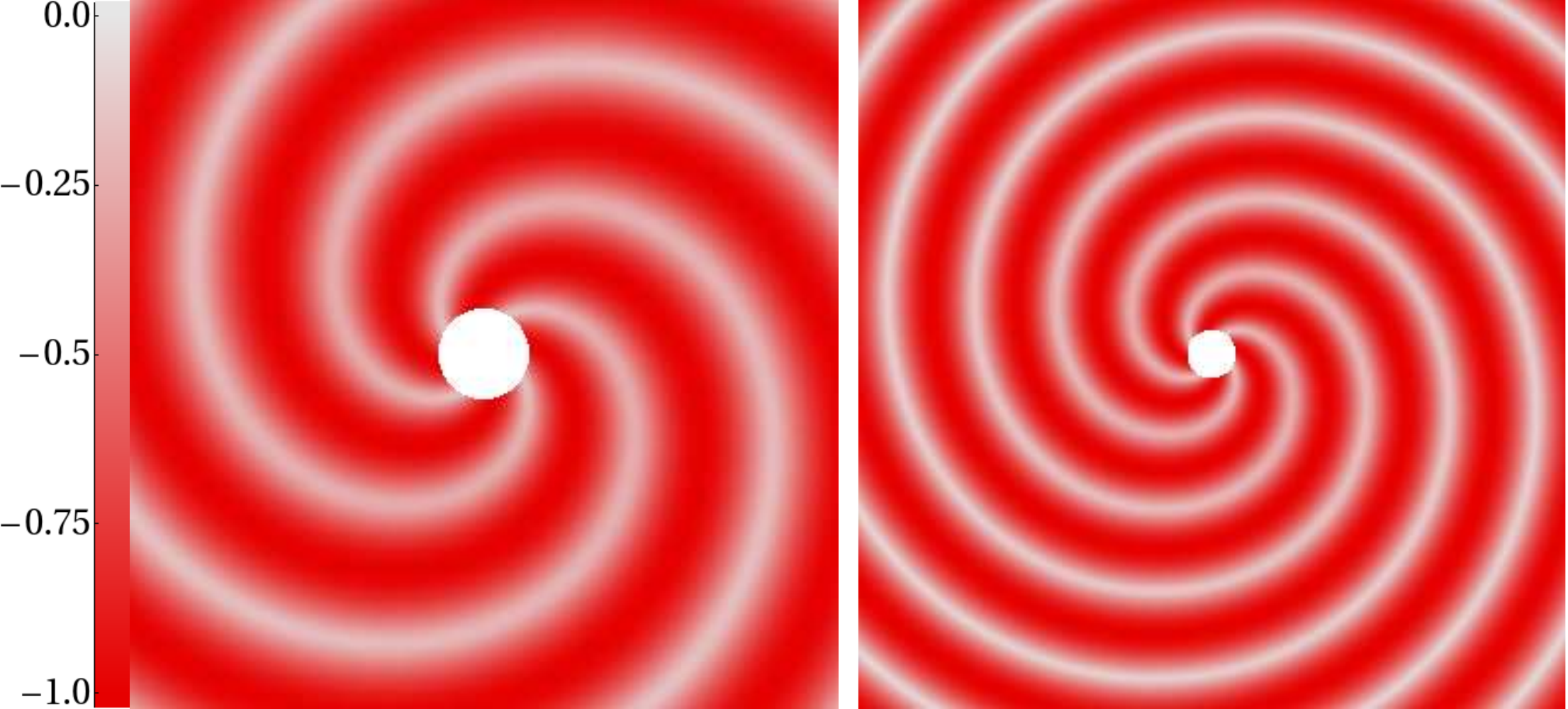}
\caption{(color online).
For an electric-parity (2,2) mode of a Schwarzschild black hole:
the vorticity of the counterclockwise vortex lines that pass through
the equatorial plane at a 45 degree angle.  The clockwise vortex lines 
that pass through the plane have equal and opposite vorticity.  
By near duality,
this figure also depicts the perturbative tendex structure for the
magnetic-parity (2,2) mode.
The conventions for coloring and shading are the same as in 
Fig.~\ref{fig:Sch22VortexAll}. 
Because the horizon vorticity is exactly 
zero for this mode, the horizon is shown as a white disk. 
The left panel, a region $30M$ across, is a zoom in of the right panel, which 
is $56M$ across.}
\label{fig:Schw22EvenVortexDensity}
\end{figure}

In Fig.\ \ref{fig:Schw22EvenVortexDensity}, we visualize
the vortexes of the electric-parity $(2,2)$ mode of a Schwarzschild hole. 
(By near duality, the perturbative tendexes of the magnetic-parity mode must 
look nearly the same.)

As noted at the beginning of Sec.\ \ref{sec:22EquatorialPlane}, 
reflection antisymmetry of the frame-drag field for this electric-parity
mode dictates
that through each point in the plane there will pass one zero-vorticity vortex 
line lying in the plane, and two vortex lines
with equal and opposite vorticities that pass through the plane at 45 degree angles
and are orthogonal to each other and to the zero-vorticity line.  
The vorticity plotted in Fig.\ 
\ref{fig:Schw22EvenVortexDensity} is that of the counterclockwise,
45 degree line.  For the clockwise line, the vorticity pattern is identically
the same, but blue instead of red.

There are again four regions of strong vorticity (four vortexes), which spiral 
outward from the horizon, becoming gravitational waves.  
In this case, the four regions look identical,
whereas for the tendexes of this same electric-parity mode (middle column of 
Fig.\ \ref{fig:TendexVortexEq}) there is an alternation between blue and red. 
There is actually an alternation here, too, though it does not show in the figure:
The relative tilt of the lines (in the sense of the $\phi$ direction) rotates, such that in one tendex, the red tendex lines pass through the plane with a forward
45 degree tilt on average, and in the next tendex, with a backward 45 degree tilt; and
conversely for the blue tendex lines.

\subsection{Three-Dimensional vortexes and tendexes }
\label{sec:3Dvortextendex}

\begin{figure*}
\includegraphics[width=0.55\columnwidth]{SchwarzschildOddVortex}
\includegraphics[width=0.47\columnwidth]{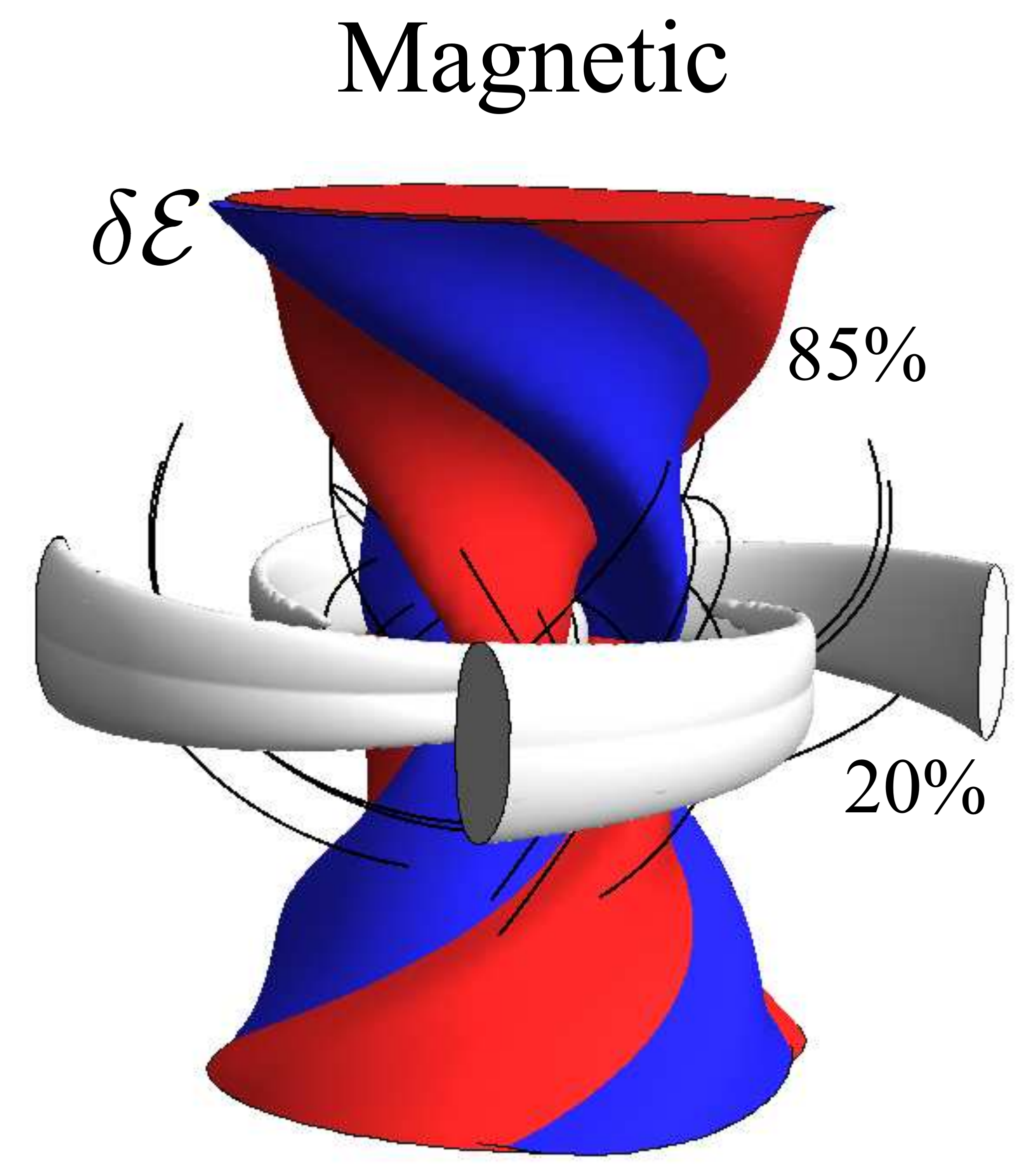}
\includegraphics[width=0.48\columnwidth]{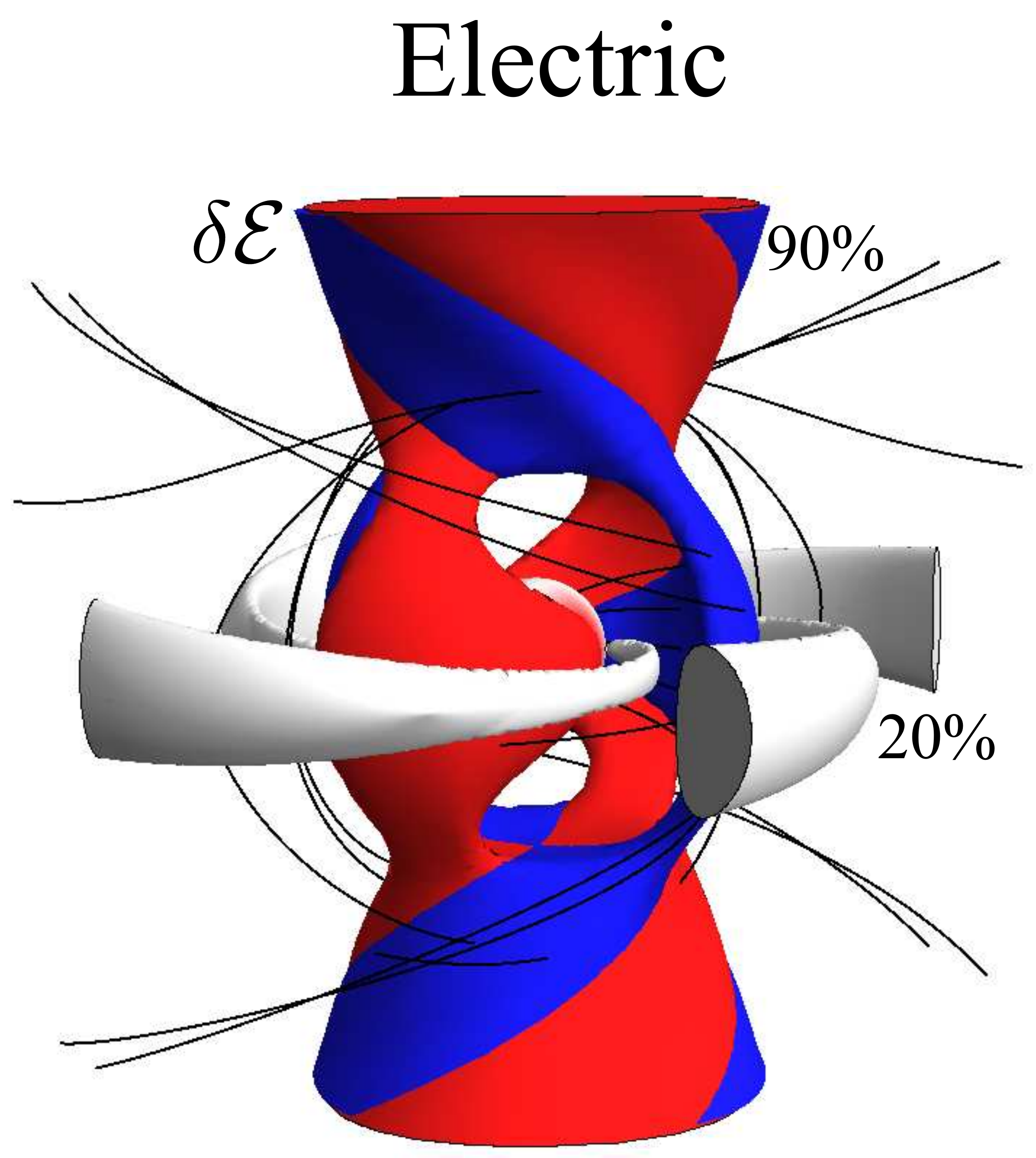}
\includegraphics[width=0.45\columnwidth]{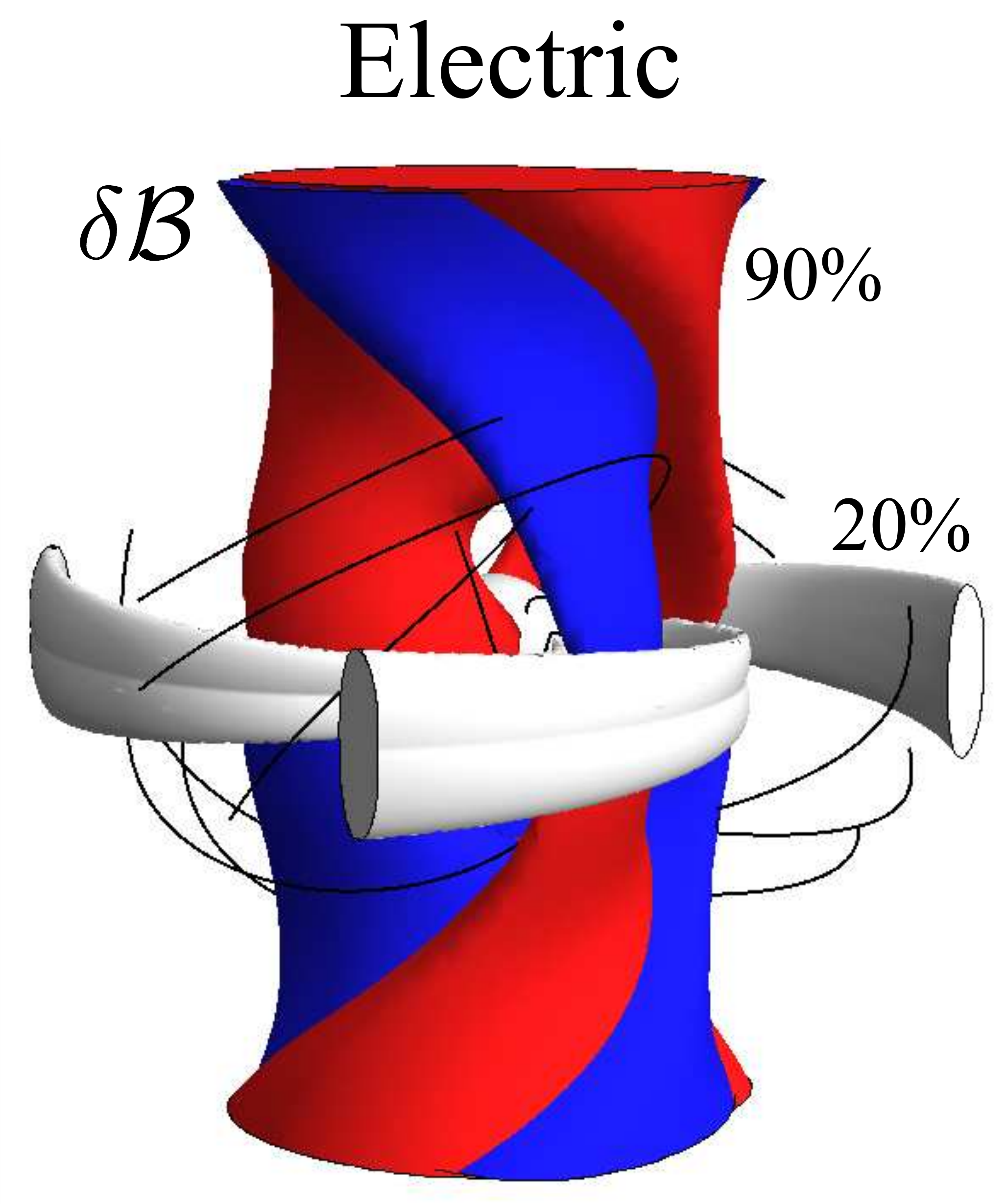}
\includegraphics[width=0.58\columnwidth]{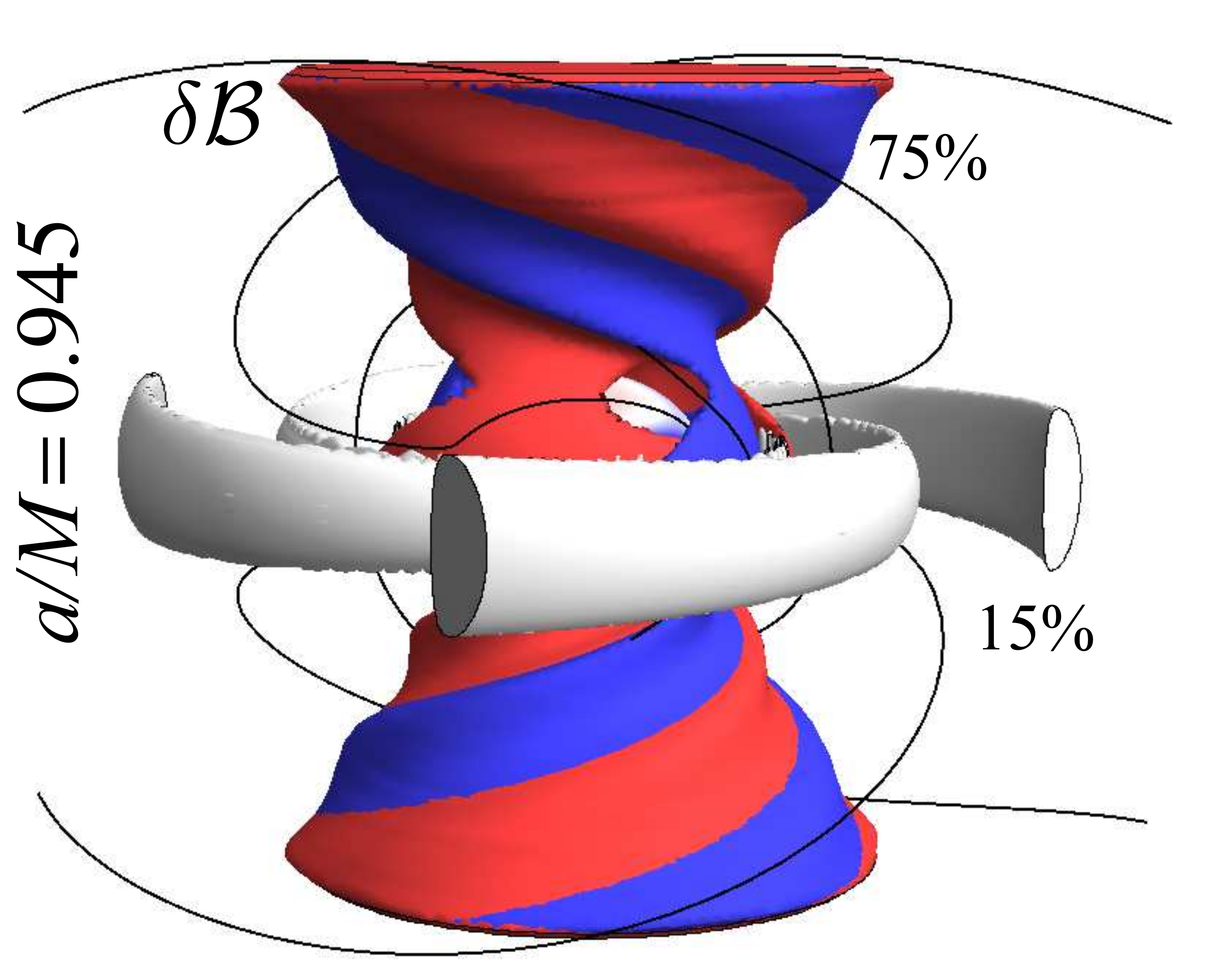}
\hskip-1pc \includegraphics[width=0.5\columnwidth]{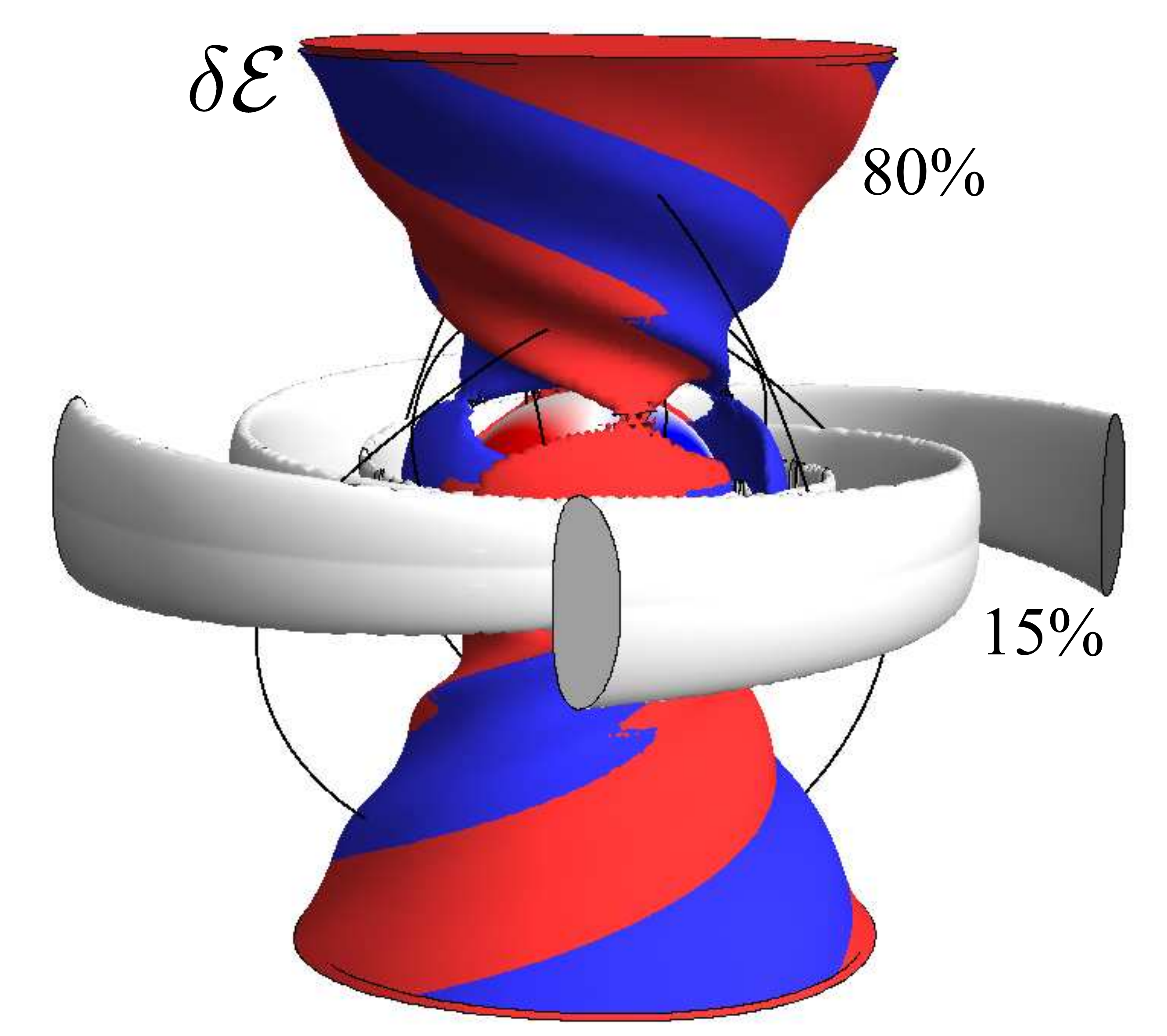}
\includegraphics[width=0.42\columnwidth]{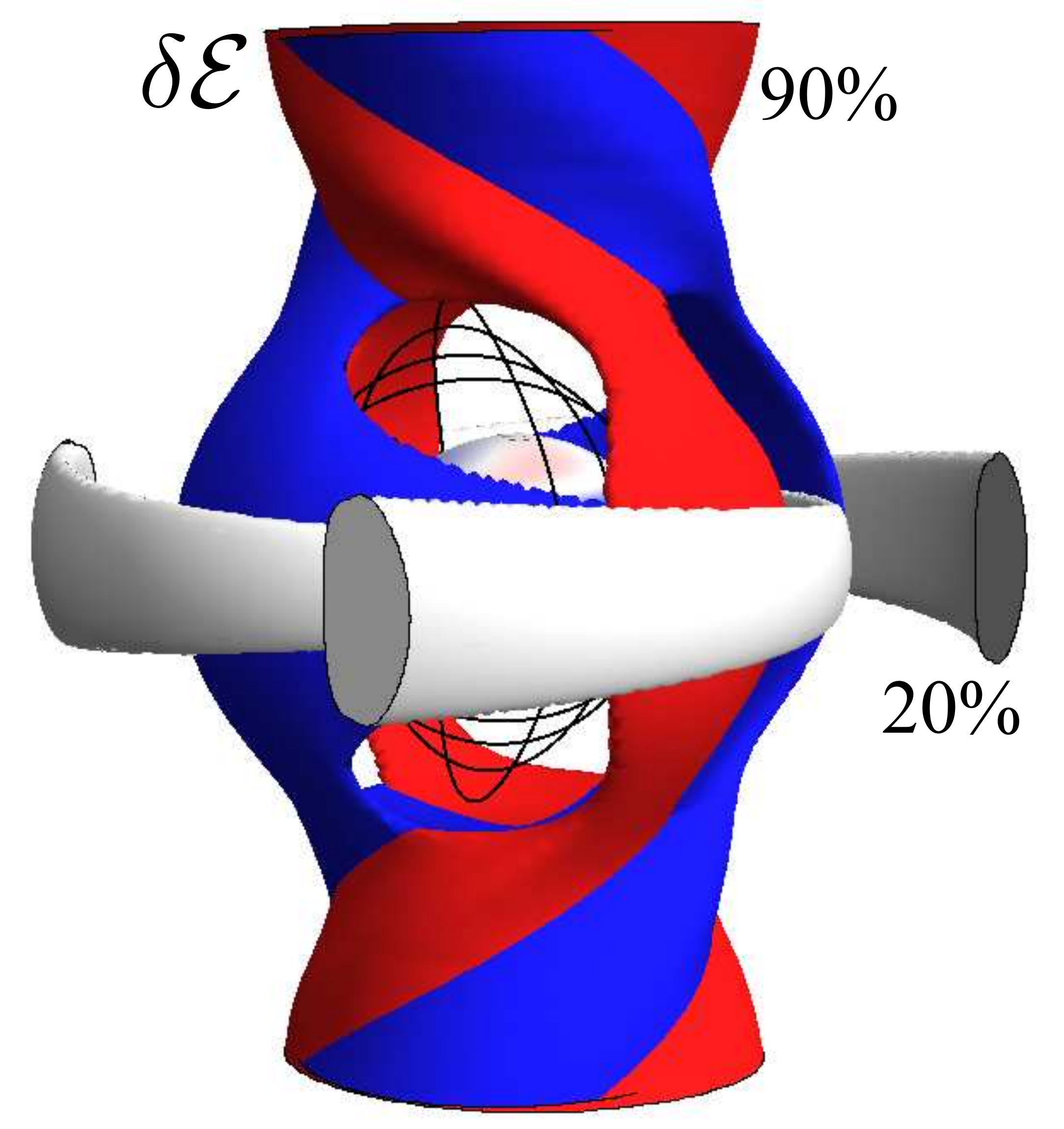}
\hskip1pc\includegraphics[width=0.42\columnwidth]{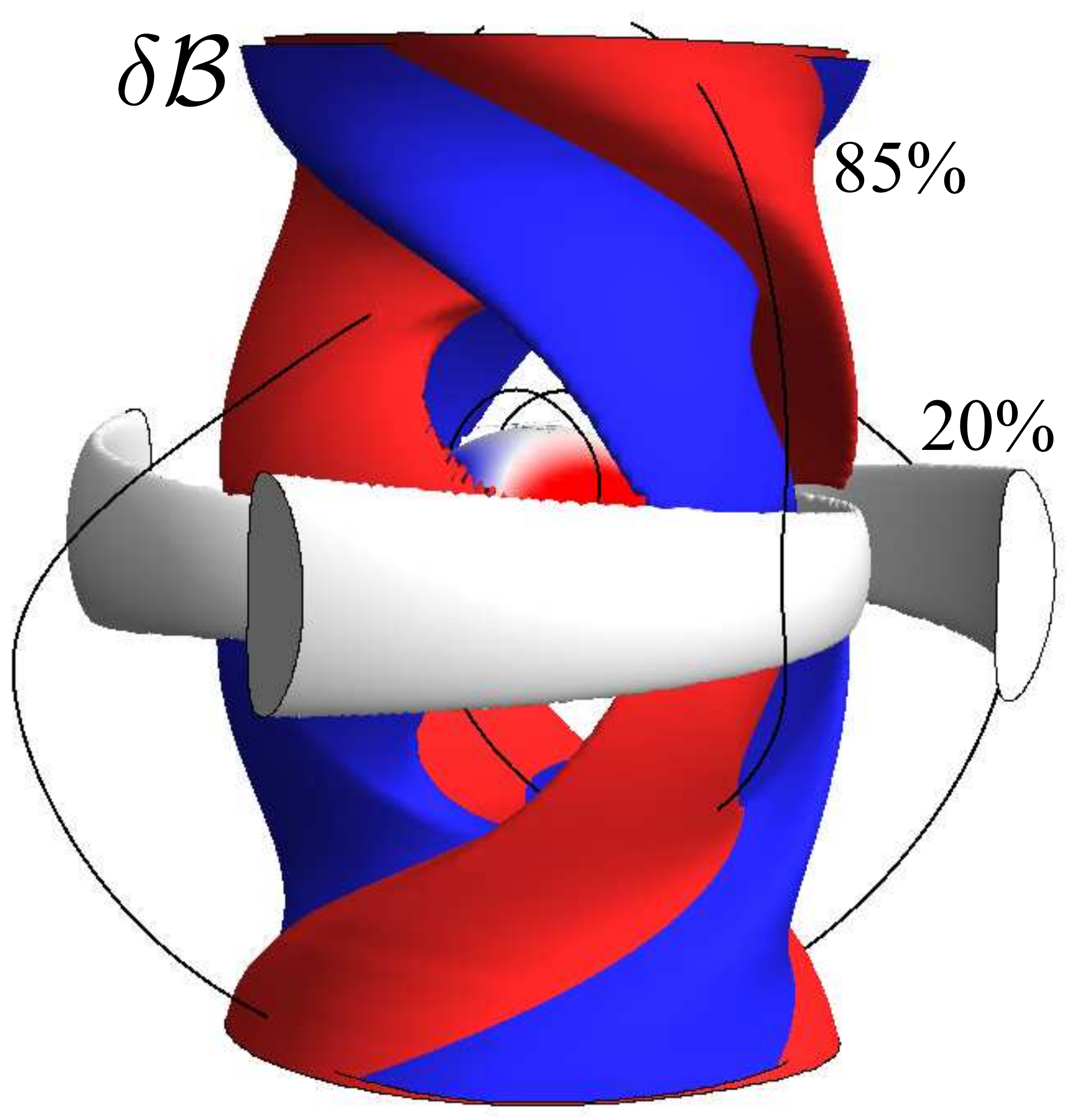}
\caption{(color online).
Three-dimensional vortexes and tendexes of the same four modes as
are shown in Fig.\ \ref{fig:TendexVortexEq}. 
As there, so here, the top
row is for a Schwarzschild black hole, and the bottom for fast-spin Kerr,
$a=0.945$; the left two columns (one in Fig.\ \ref{fig:TendexVortexEq})
are for a magnetic-parity (2,2) mode (with the vortexes in IR coordinates), 
and the right two columns (one in
Fig.\ \ref{fig:TendexVortexEq}) are for electric parity in IR gauge.
For each parity, the first column shows structures of the field that
generates the waves
($\delta \mathcal B$ for magnetic parity; $\delta\mathcal E$ for electric
parity) and the second column shows structures of the other field
(not included in the equatorial-plane drawings of
Fig.\ \ref{fig:TendexVortexEq}). 
In each panel, the colored surfaces
show the outer faces of vortexes (for $\delta \mathcal B$) or tendexes
(for $\delta \mathcal E$), defined as the locations,
for a given radius, where the largest-in-magnitude eigenvalue of the field
being plotted ($\delta \mathcal B$ or $\delta \mathcal E$) has dropped to a
certain percentage (90, 85, 80 or 75) of its maximum for that radius;
that percentage is shown alongside the colored surfaces.  
As in previous figures, the surface is red (light gray) if that 
largest-in-magnitude eigenvalue is negative and blue (dark gray) if positive. 
The off-white regions are surfaces where the largest-in-magnitude
eigenvalue has dropped to 15\%, 20\% or 25\% of the maximum at that radius.
In each panel the black lines are a few of the vortex lines (for $\delta \mathcal B$
panels) or tendex lines (for $\delta \mathcal E$ panels) that
become transverse when they reach large radii, and thereby produce the
tidal or frame-drag force of an emitted gravitational wave.  
For discussion of this figure, see Sec.\ \ref{sec:3Dvortextendex}.}
\label{fig:TendexVortex3D}
\end{figure*}

In this section, we shall explore the 3-dimensional vortexes and tendexes
of the $(2,2)$ modes of Schwarzschild and Kerr black holes, 
which are depicted in Fig.\ \ref{fig:TendexVortex3D}.
In the first subsection, we shall focus on what this figure tells us about
the generation of gravitational waves, and in the second, what
it tells us about duality.

\subsubsection{Physical description of gravitational-wave generation}
\label{sec:PhysDescGWgen}

In Sec.\ \ref{sec:Intro22} of the Introduction, we summarized in great detail
what we have learned about gravitational-wave generation from our vortex
and tendex studies.  There we focused on the $(2,2)$ magnetic-parity mode,
and among other things we scrutinized the upper left panel of 
Fig.\ \ref{fig:TendexVortex3D} (which we reproduced as Fig.\ \ref{fig:22Smag3D}).
Here, instead, we shall focus on the $(2,2)$ electric-parity mode as 
depicted in the right half of Fig.\ \ref{fig:TendexVortex3D}. 

We begin with the perturbative tendexes of the electric-parity, (2,2) mode
of a Schwarzschild black hole 
(third panel on top row of Fig.\ \ref{fig:TendexVortex3D}).
The 3-D tendexes emerge from the horizon as four deformed-cylinder structures,
two red (light gray) and two blue (dark gray).  
These are the extensions into the third dimension of the
four near-zone, equatorial-plane tendexes of the center panel in Fig.\ 
\ref{fig:TendexVortexEq} above.  As we enter the transition zone, the four 
3D tendexes lengthen vertically (parallel to the poles), and then as we 
enter the wave zone, they spiral upward and downward around the poles; they
have become gravitational waves.
They are concentrated near the poles because the (2,2)--mode 
gravitational waves are significantly stronger in 
polar directions than in the equator.

In this panel, we also see black tendex lines
that emerge from the horizon and spiral upward and downward alongside the
polar-spiraling tendexes, becoming nearly transverse at large radii---part
of the outgoing gravitational waves.  Of course, there are similar tendex 
lines, not shown, inside the spiraling tendexes.  
In addition, we also see tendex lines 
in the inner part of the wave zone that are approximately polar circles;
these are also part of the outgoing waves.

The top rightmost panel depicts the vortexes associated with this 
electric-parity mode. 
The horizon vorticity vanishes, so the horizon is white. The vortexes near the 
horizon are dominated by the longitudinal-transverse part of the 
frame-drag field $\boldsymbol{\mathcal B}^{\rm LT}$, which interacts with
$\delta \boldsymbol{\mathcal E}^{\rm L}$ 
and $\delta \boldsymbol{\mathcal E}^{\rm LT}$ 
to maintain their joint near-zone structure as they
rotate (cf.\ the description of the dual magnetic-parity mode in
Sec.\ \ref{sec:Intro22}). However, of course, there is also a 
$\boldsymbol{\mathcal B}^{\rm TT}$ associated with the ingoing gravitational
waves.
At large radii, in the outgoing-wave zone, the vortexes, like the tendexes
of the third panel top row, spiral upward and downward around the polar axis;
they have joined with the tendexes to form the full gravitational-wave
structure. 

For insight into how (we think) the near-zone tendexes of this electric-parity
mode,
extending radially out of the horizon, generate these outgoing gravitational waves,
and how the ingoing waves, that they also generate, act back on them and drive
their gradual decay, see the description of this mode's dual in
Sec.\ \ref{sec:Intro22}.

For the rapid-spin Kerr black hole, the tendex and vortex structures 
(last two panels of second row of Fig.\ \ref{fig:TendexVortex3D}) are
quite similar to those for the Schwarzschild black hole. 
The detailed differences are similar to those in the equatorial plane
(see discussion in Sec.\ \ref{sec:TendVortSchwKerr} above): 
smaller near zone and tighter
spiraling for the tendexes because of the higher eigenfrequency; 
nonvanishing horizon vorticity with a predominantly $Y^{32}(\theta,\phi)$
angular structure. 
In the near zone, the 3D vortexes seem to have acquired 
a longitudinal (radial) part, emerging from the $Y^{32}$ horizon vortexes
(though this is largely hidden behind the off-white structures).
Thus, for a Kerr black hole, one might intuitively describe the generation of
gravitational waves as being produced by a superposition of near-zone tendexes 
that induce vortexes by their motions, and near-zone vortexes that induce 
tendexes by their motions.  However, because the near-zone vortexes are weaker 
than the tendexes, the tendexes still play the dominant role for 
gravitational-wave generation in this electric-parity mode.

\subsubsection{Approximate duality}
\label{sec:ApproximateDual}

By comparing the magnetic-parity left half of Fig.\ \ref{fig:TendexVortex3D}
with the electric-parity right half, we can visually assess the degree to 
which there is a duality between the modes in three dimensions.
For the perturbations of Schwarzschild black holes (top row), the most 
notable difference
between the magnetic- and electric-parity perturbations is that the transition
between the longitudinal near-zone and spiraling wave-zone vortexes of the 
magnetic-parity perturbation is more abrupt, and happens closer to the horizon
than it does in the electric-parity perturbations.  
The reason for this is encoded in Eqs.\ (\ref{eq:IRG22FieldsE}), 
(\ref{eq:IRG22Fields}), and (\ref{eq:IRG22Duality}), but we do not have a 
simple physical explanation for why this occurs. 
This difference is magnified for perturbations of the rapidly rotating
Kerr black hole (bottom row). 
Thus, the small breaking of duality quantified in App. \ref{sec:CCKProc}
for Schwarzschild black holes seems to be more pronounced in three-dimensions
than in two, and stronger for rapidly rotating black holes than for 
non-rotating ones.

Nevertheless, the qualitative picture of wave generation by longitudinal
near-zone tendexes and vortexes is essentially dual for perturbations of
the two parities.

\subsection{Comparing vortex lines of a perturbed Kerr black hole and a
binary-black-hole-merger remnant}
\label{sec:NumericalComparison}

\begin{figure*}[t!]
\includegraphics[width=0.9\columnwidth]{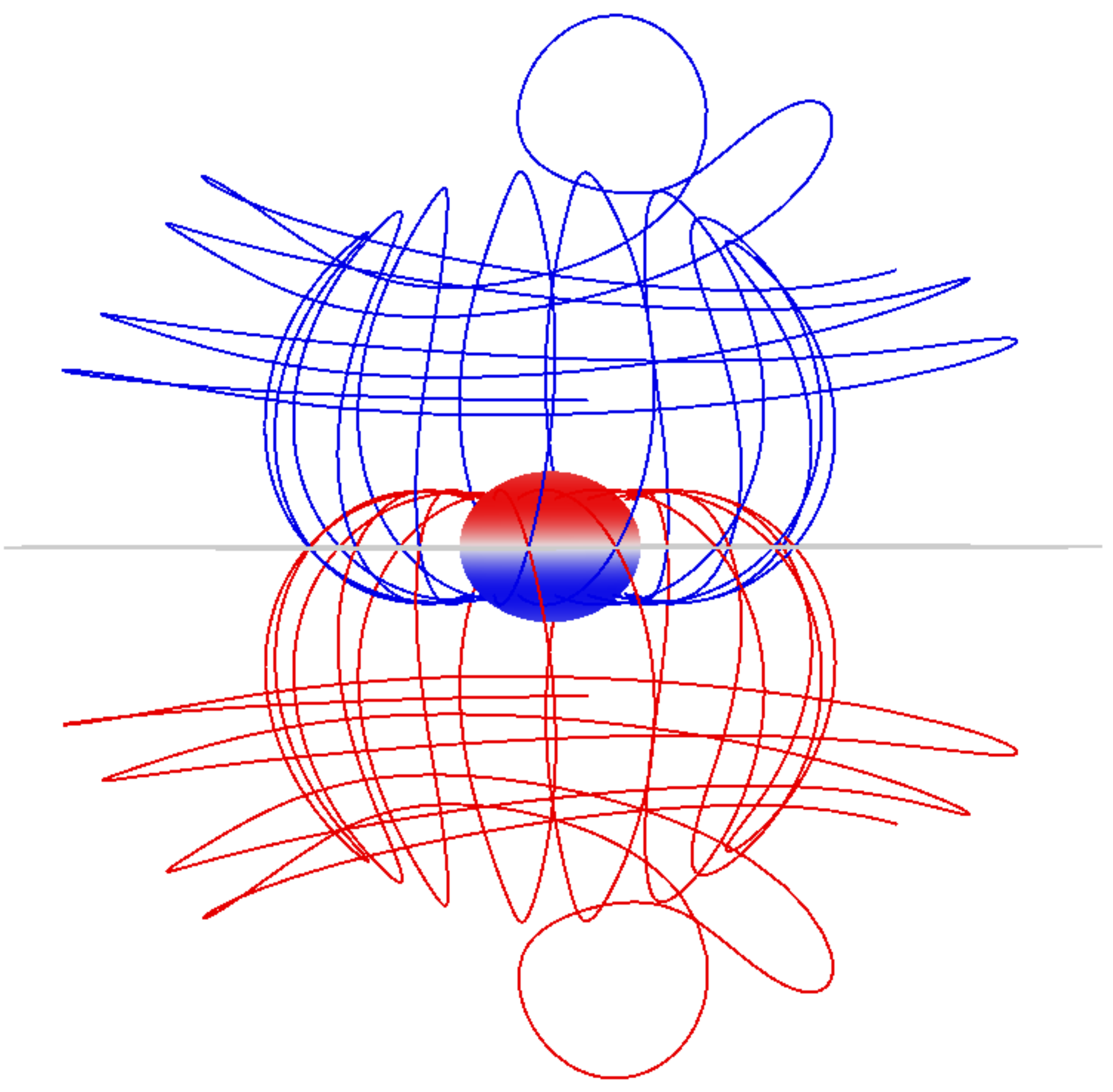}
$\quad$
\includegraphics[width=1.1\columnwidth]{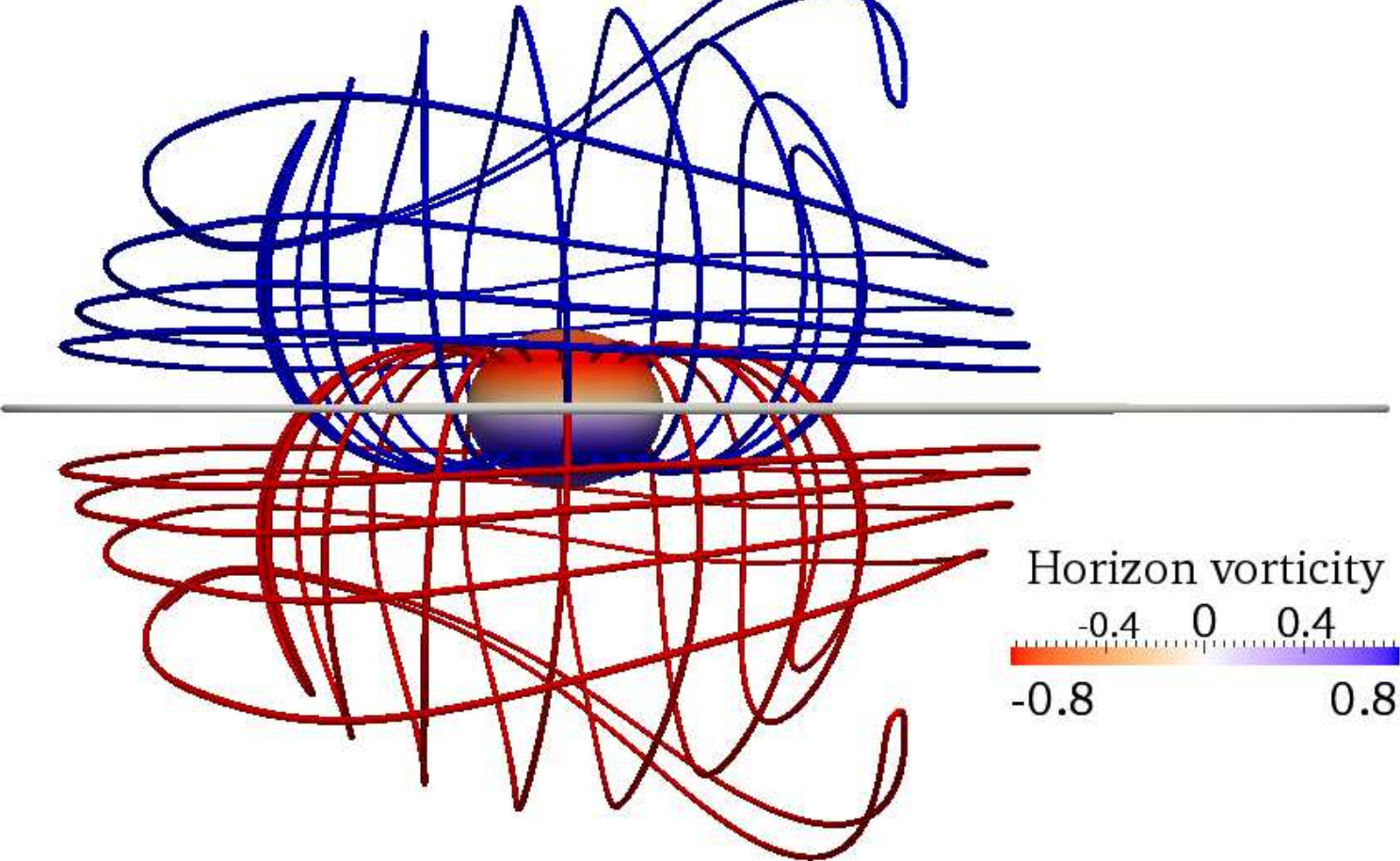}
\caption{(color online).
\emph{Left:} Vortex lines of a $a/M = 0.945$ Kerr black hole perturbed
by an electric-parity $(2,2)$ quasinormal-mode in IR gauge. 
\emph{Right:} Vortex lines from a $a/M=0.945$ ringing down Kerr black hole 
obtained from a numerical 
simulation~\cite{Lovelace:2011nu} of two identical merging black holes with 
spins of magnitude $0.97$ aligned parallel to the orbital angular momentum.
In the simulation, we chose a late enough time that the common
apparent horizon is essentially that of a single, perturbed black hole,
and we computed the vortex lines using methods summarized in Ref.\ 
\cite{OwenEtAl:2011}.}
\label{fig:NRComparison}
\end{figure*}

As a conclusion to this section and a prelude to future work, in Fig.\ 
\ref{fig:NRComparison} we compare the vortex lines found using our analytic 
methods to those found in a numerical ringdown of a fast-spinning Kerr black 
hole. 

More specifically,
we compare an electric-parity, $(2,2)$ quasinormal-mode perturbation 
of a Kerr black hole with dimensionless spin $a/M=0.945$, 
to a ringing-down Kerr black hole of the same spin 
formed in a numerical simulation~\cite{Lovelace:2011nu} 
of the merger of two equal-mass black holes with equal spins 
of magnitude 0.97 aligned with the orbital angular momentum. (Note that 
because of the symmetry of this configuration, during the ringdown there 
is no magnetic-parity $(2,2)$ mode excited.)

For both the analytical and numerical calculations, the vortex lines
are those of the full frame-drag tensor.
To describe the magnitude of the perturbation in the analytical calculation,
we write the frame-drag field as $\mathcal B_{\hat i \hat j}
= \mathcal B_{\hat i \hat j}^{(0)} + \mathcal B_{\hat i \hat j}^{(1)}$, 
including the background part $\mathcal B_{\hat i \hat j}^{(0)}$ and the 
perturbation
$\mathcal B_{\hat i \hat j}^{(1)} \equiv \delta \mathcal B_{\hat i \hat j}$
(as in App. \ref{sec:ExtraFigures}).
We choose the ratio of the maximum of the perturbation of
the horizon 
vorticity, $\mathcal B_{NN}^{(1)}$ to the background horizon vorticity 
$\mathcal B_{NN}^{(0)}$ to be of order $10^{-3}$.
This amplitude of the perturbation produces lines that agree 
qualitatively with those from
the numerical simulation.

The lines of the full frame-drag field look quite different from those for 
just $\delta \mathcal B_{\hat i \hat j}$ depicted in Figs. \ref{fig:TendexVortexEq} and 
\ref{fig:TendexVortex3D}. 
Near the equator, the vortex lines in both panels look like those of an 
unperturbed Kerr black hole (see Paper II). 
Closer to the axis of rotation, the background vortex lines become degenerate, 
and the perturbations break the degeneracy by picking the principal axes
of the perturbative field.  
Correspondingly, near the rotation axis and at large enough radii to be
in or near the wave zone, the vortex lines resemble those of 
transverse-traceless gravitational waves, which are emitted symmetrically 
above and below the hole.

Although the vortex lines from these similar physical situations were 
computed using very different methods and gauge conditions, the results are
qualitatively similar (see Fig.\ \ref{fig:NRComparison}). 
The lines in the two figures are not identical, but they were selected to
intersect the horizon in approximately the same places;
a careful inspection shows there are small differences, for example very
near the horizon.

This comparison ultimately gives us confidence that
our analytical methods can guide our understanding of the vortexes and
tendexes in the late stages of numerical simulations.

\section{Superposed $(2,2)$ and $(2,-2)$ Quasinormal Modes of Schwarzschild}
\label{sec:SuperposedPerts}

\subsection{Magnetic-parity superposed modes}
\label{sec:MP22sup}

\begin{figure*}[tb]
\includegraphics[width=0.275\textwidth]{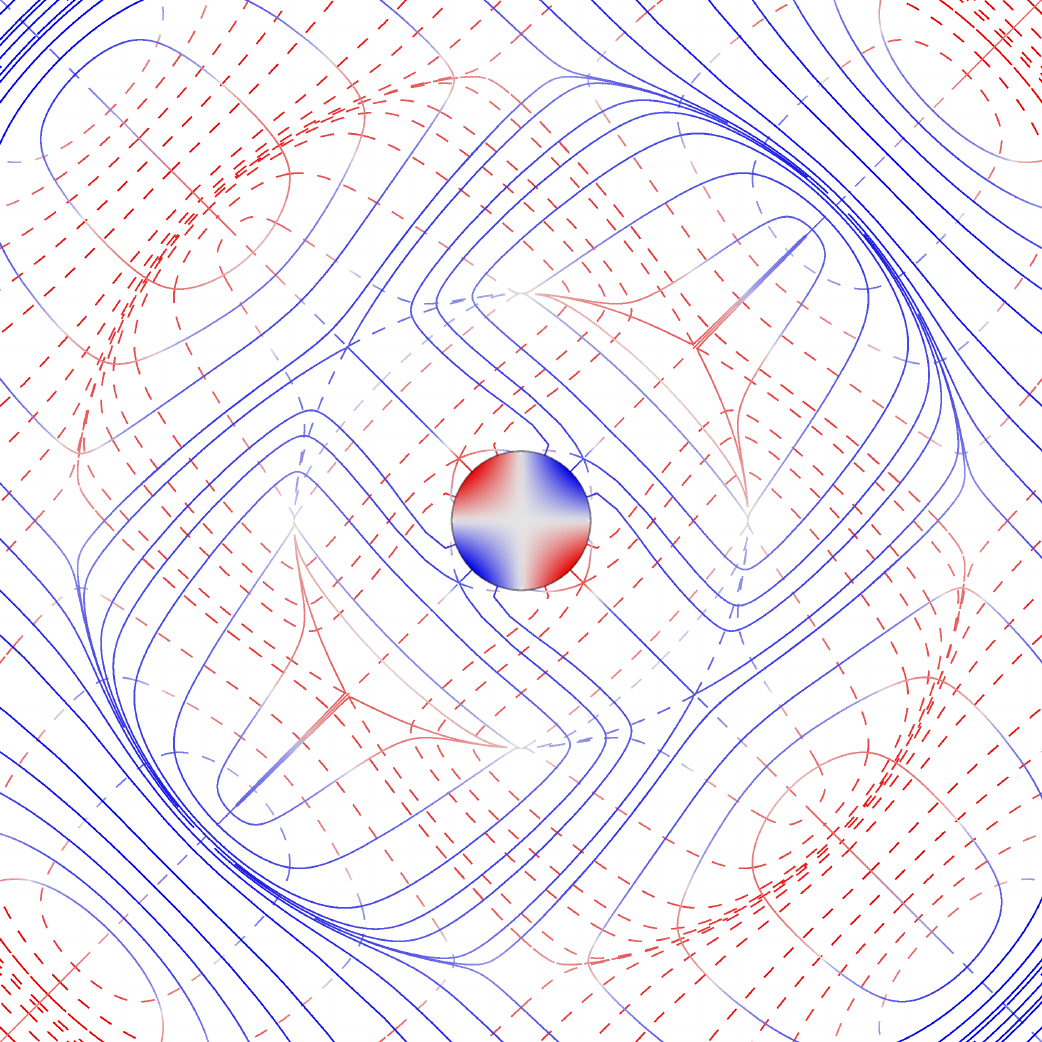}
\includegraphics[width=0.325\textwidth]{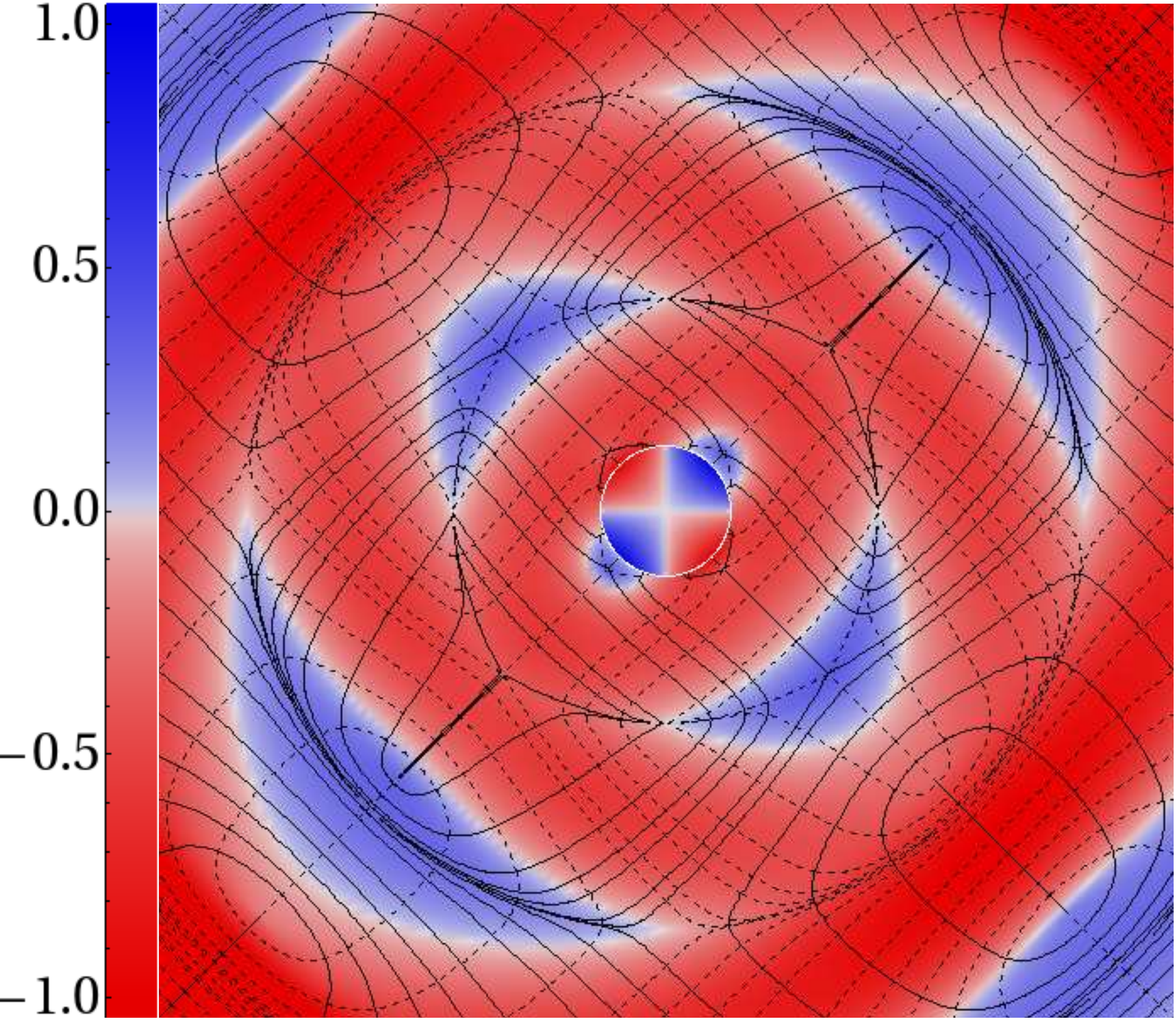}
\includegraphics[width=0.325\textwidth]{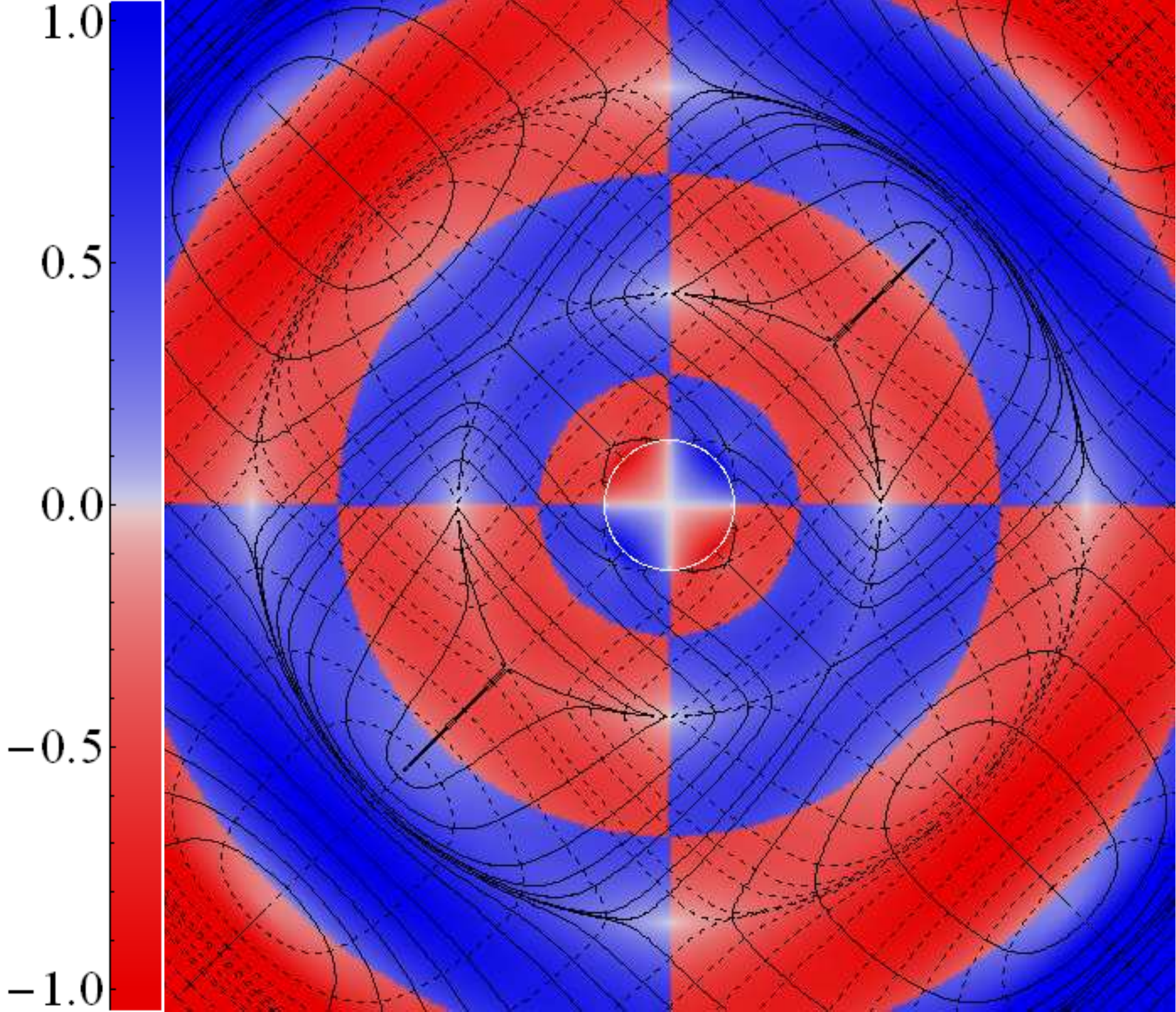}
\includegraphics[width=0.275\textwidth]{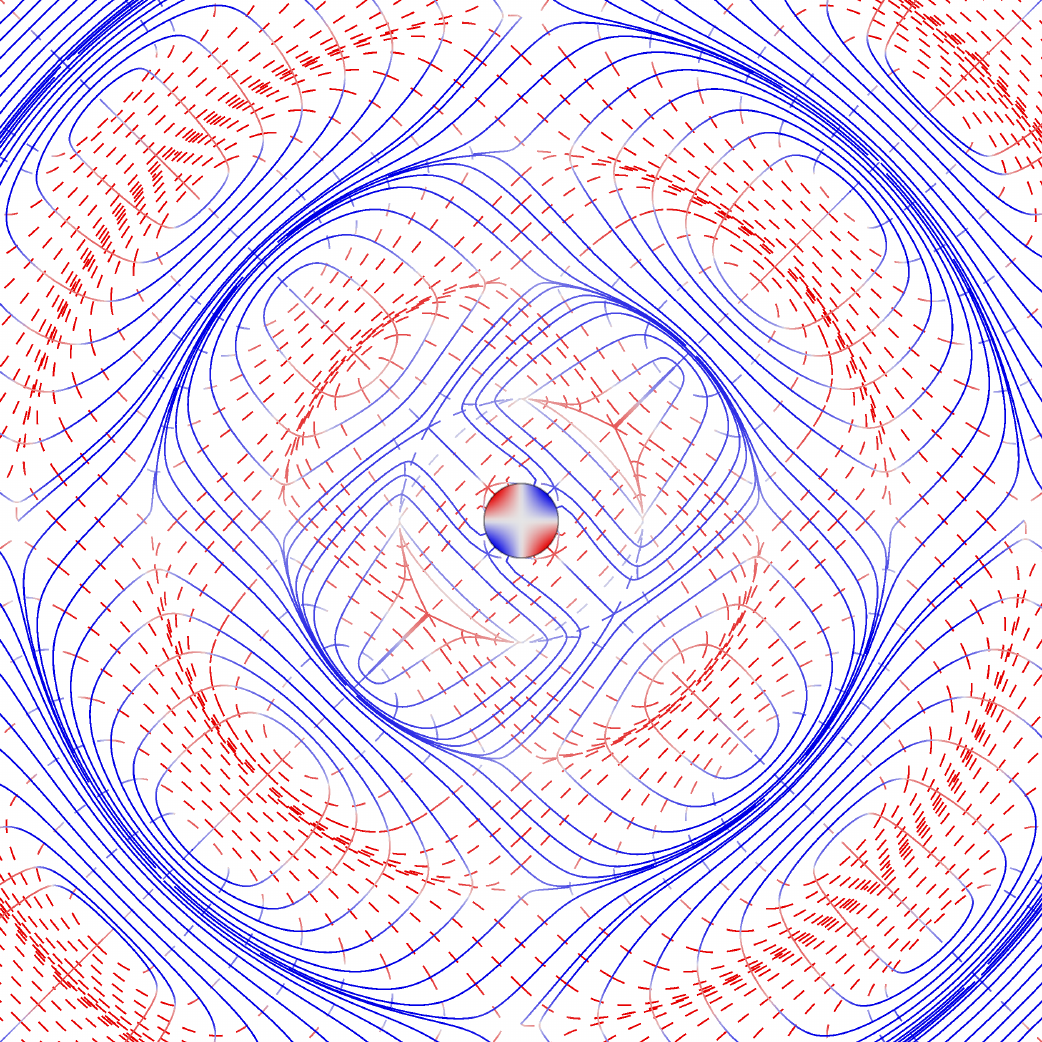}
\includegraphics[width=0.325\textwidth]{RWNegVortexDensityl2oscFarBoth}
\includegraphics[width=0.325\textwidth]{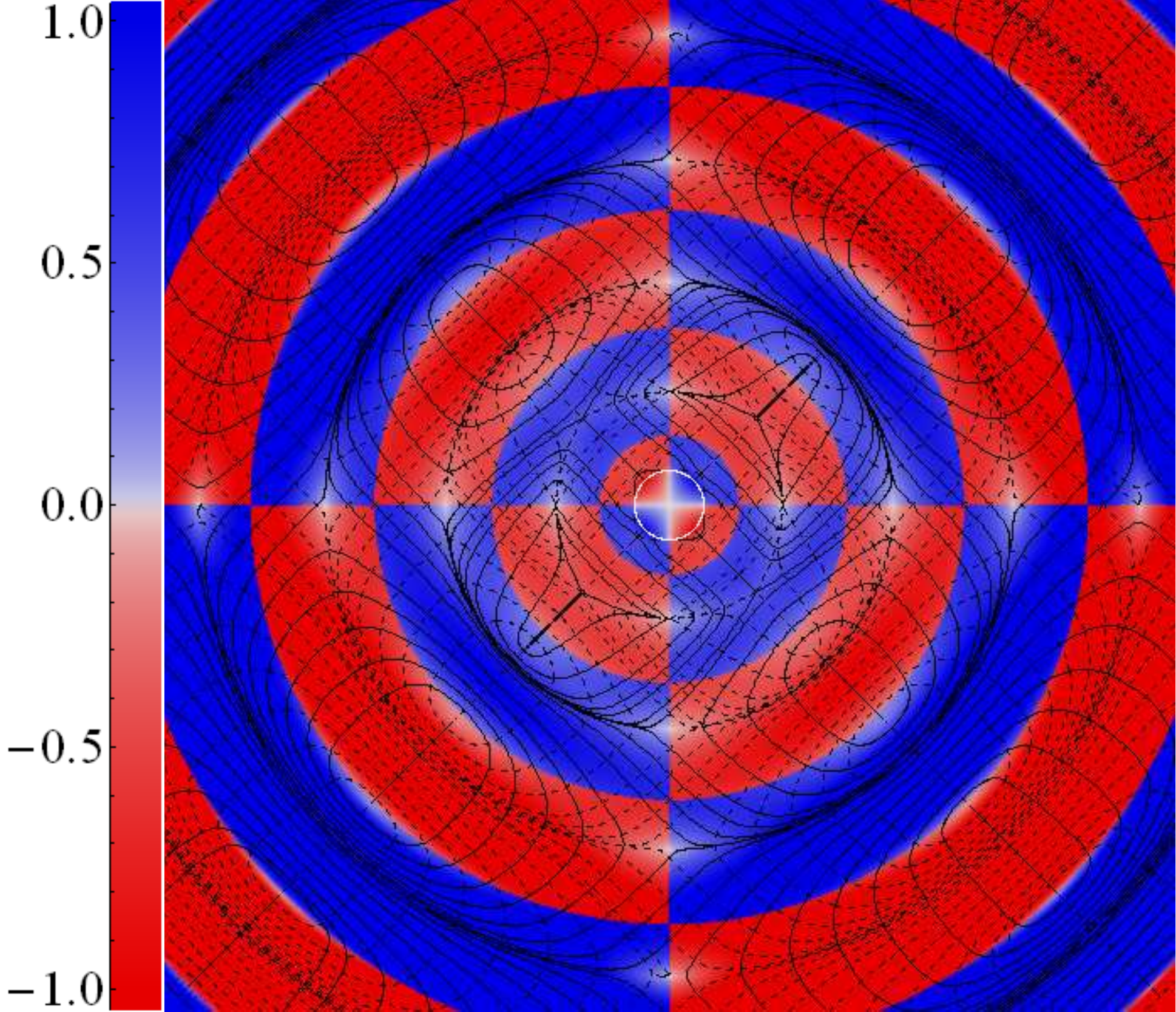}
\caption{(color online).
The vorticities and vortex lines in the equatorial plane of a 
Schwarzschild black hole, for the fundamental magnetic-parity $(2,2)$ mode 
superposed on the fundamental magnetic-parity $(2,-2)$ mode, depicted using 
the same three visualization techniques as in Fig.\ \ref{fig:Sch22VortexAll}.
Here, however, we do not scale the vorticity by any function, but the numbers 
on the vorticity scale on the left of the panels are equal to 
$\sqrt{{\rm vorticity}\times r}$ (where $r$ is radius), in units of the 
maximum value of this quantity, which occurs on the horizon at $\phi=3\pi/4$ 
and $\phi=7\pi/4$.
The top panels cover a region $30M$ across, and the bottom panels are a 
zoom-out of the upper panels, $56M$ across.
The central circle in all panels is 
the horizon as viewed from the polar axis, colored by its vorticity.
{\it Left column}: The two families of vortex lines (one shown dashed, the 
other solid) with each line colored, at each point, by the sign of its 
vorticity (blue [dark gray] for positive, i.e., clockwise; red (light gray) 
for negative, i.e., counterclockwise), and each line has an intensity 
proportional to the magnitude of its vorticity. 
{\it Center column}:  The same vortex lines are colored black,
and the equatorial plane is colored by the vorticity of the dashed family
of lines. 
{\it Right column}: The same as the center column, but the equatorial plane is
colored by the vorticity with the larger magnitude.}
\label{fig:22sup2-2VortexDensity}
\end{figure*}

In Sec.\ \ref{sec:magneticSuperposed}, we summarized the properties of
the quasinormal mode of Schwarzschild that is obtained by superposing
the magnetic-parity $(2,2)$ and $(2,-2)$ modes.  Here we give details.
The vortex lines and vorticities for this superposed mode are depicted in
Fig.\ \ref{fig:22sup2-2VortexDensity} using the three types of visualizations
in Fig.\ \ref{fig:Sch22VortexAll}.

The left column of Fig.\ \ref{fig:22sup2-2VortexDensity} shows the two families
of vortex lines that lie in the equatorial plane, color coded by their 
vorticities.
The solid-line family has predominantly positive (clockwise) vorticity, 
but in some
regions its vorticity becomes weakly negative (counterclockwise).  
The dashed-line family has predominantly negative vorticity, but in some
regions it is weakly positive.  A rotation around the hole by angle
$\pi/2$ maps each family into the other.   

In the center column of Fig.\ \ref{fig:22sup2-2VortexDensity}, the vortex 
lines are drawn black and the equatorial plane is colored by the vorticity of 
the dashed lines.
To deduce the coloring for the solid lines, just rotate the colored plane 
(but not the lines) by $\pi/2$ and interchange red (light gray) and blue (dark
gray). 
By contrast with most previous figures, the radial variation of the vorticity 
is not scaled out of this figure; so in the wave zone (roughly, the outer half of 
right panel) the coloring oscillates radially, in color and intensity, in the
manner of a gravitational wave.  At large radii, there is also a growth 
of intensity (and saturation of the color scale) due to the waves emitted
earlier having larger amplitude.  

In the right column of Fig.\ \ref{fig:22sup2-2VortexDensity}, the vortex lines
are again drawn black, and the equatorial plane is now colored by the larger 
of the two vorticities in amplitude.

Together, the columns of Fig.\ \ref{fig:22sup2-2VortexDensity} provide the 
following picture: 
For each family of lines, the 
equatorial-plane vortexes form interleaved
rings (dashed lines and red [light gray] vortexes for center column).
Most of the family's vortex lines form closed, distorted ellipses that,
when tangential, lie in a single vortex (red for dashed lines), and when 
more nearly radial, travel from one vortex to another.   
In the wave zone, these line and vortex structures grow longer tangentially 
as they
propagate outward, and they maintain fixed radial thickness.  When one
looks at both families simultaneously, focusing on the strongest at each
point (right column), one sees vortexes of alternating red and blue vorticity
(light and dark gray).
The angular oscillations are those of a quadrupolar structure; the radial
oscillations are those of a propagating wave.

In the near zone, the vortex lines have rather sharp,  
right angled features associated 
with the quadrupolar nature of the near-zone perturbation; this is to 
be compared to the oscillating current quadrupole in linearized gravity 
(Paper I, Sec.~VI, Fig.~15). There are multiple singular points in 
the lines, degenerate points where three lines cross with sharp bends, 
and where both families of vortex lines take on the same eigenvalue. 
(The third eigenvalue, that of the lines perpendicular to the plane, 
must then be minus twice the vorticity of these lines, in order for 
the sum of the eigenvalues to vanish). 

We shall discuss the dynamics of these vortex lines and vortexes in
Sec.\ \ref{sec:MagSuperposedDynamics} below, 
after first gaining insight into the 
electric-parity superposed mode (whose vortexes will teach us
about this magnetic-parity mode's tendexes through the near duality).

\subsection{Electric-parity superposed mode}
\label{sec:EP22sup}

For the mode constructed by superposing electric-parity (2,2) and (2,-2)
modes of Schwarzschild, as for the electric-parity (2,2) mode itself
(Sec.\ \ref{sec:EparityVortexes22}), 
symmetry considerations dictate that: (i) one family
of vortex lines lies in the equatorial plane and has vanishing vorticity,
(ii) two families pass through the equatorial plane at $45^\circ$
angles, with equal and opposite vorticities, and (iii) the horizon
vorticity vanishes.

\begin{figure}[tb]
\includegraphics[width=0.95\columnwidth]{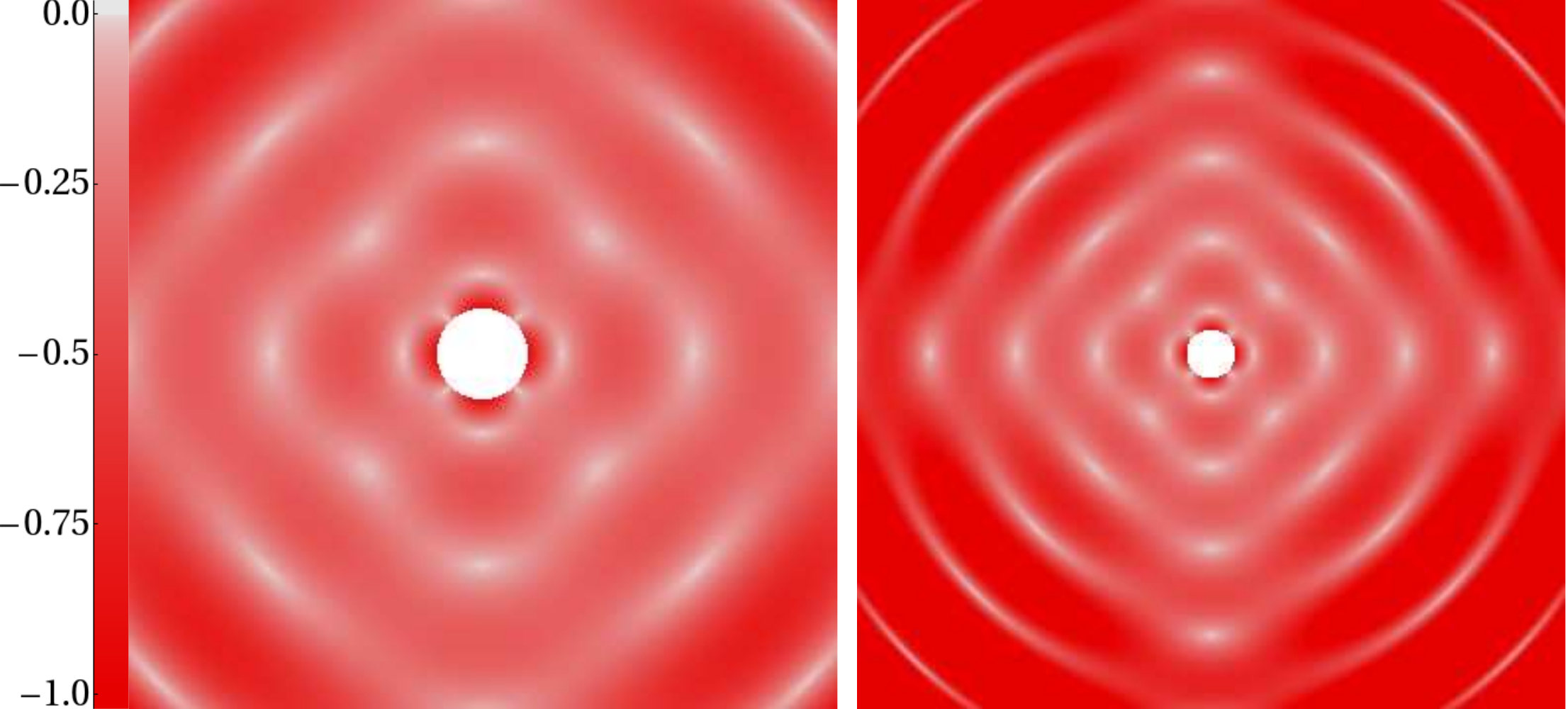}
\caption{(color online).
For the electric-parity, superposed (2,2) and (2,-2) fundamental modes
of Schwarzschild: the vorticity of the counterclockwise vortex lines
that pass through the equatorial plane at 
$45^\circ$ angles. 
By near duality,
this figure also depicts the perturbative tendex structure for the
magnetic-parity superposed mode.
The intensity scale of the red color (left edge of
figure) is the same as that in the center column of 
Fig.\ \ref{fig:22sup2-2VortexDensity}.  
The left panel, a region $30M$ across, is a zoom-in of the 
right panel, which is $56M$ across. 
This figure is the superposed-mode
analog of Fig.\ \ref{fig:Schw22EvenVortexDensity}.
}
\label{fig:22sup2-2VortexDensityEven}
\end{figure}

In Fig.\ \ref{fig:22sup2-2VortexDensityEven} [analog of Fig.\ 
\ref{fig:Schw22EvenVortexDensity} for the (2,2) electric-parity mode],
we show the vorticity of the family of counterclockwise vortex lines, 
as they pass through the equatorial plane.  

In the near zone of this figure,
we see again a distinct quadrupolar structure, with four lobes of strong vorticity present near the horizon (four near-zone vortexes). Beyond these near-zone lobes, there is a ring of vanishing vorticity, followed by an annulus where the cast-off vortexes of a previous cycle have begun to deform into an annulus of stronger vorticity. In the wave zone, the vortexes have transitioned into outward traveling transverse waves, with regions of vanishing vorticity between the crests and troughs of each wave. The waves are strongest along the diagonals, though in the near zone the (LT) frame-drag field is strongest in the up, down, left and right directions. 

By (near) duality, the tendexes of the magnetic-parity superposed mode will have the same form as these
electric-parity-mode vortexes.  Accordingly, in the next section, we 
will use this figure to elucidate the magnetic-parity mode's dynamics --- 
and by duality, also the dynamics of this electric-parity mode.

\subsection{Dynamics of the magnetic-parity superposed mode}
\label{sec:MagSuperposedDynamics}

We now turn to the dynamics of the magnetic-parity superposed mode,
which we studied in Sec.\ \ref{sec:MP22sup}

\begin{figure*}[tb]
\includegraphics[width = 0.95 \textwidth]{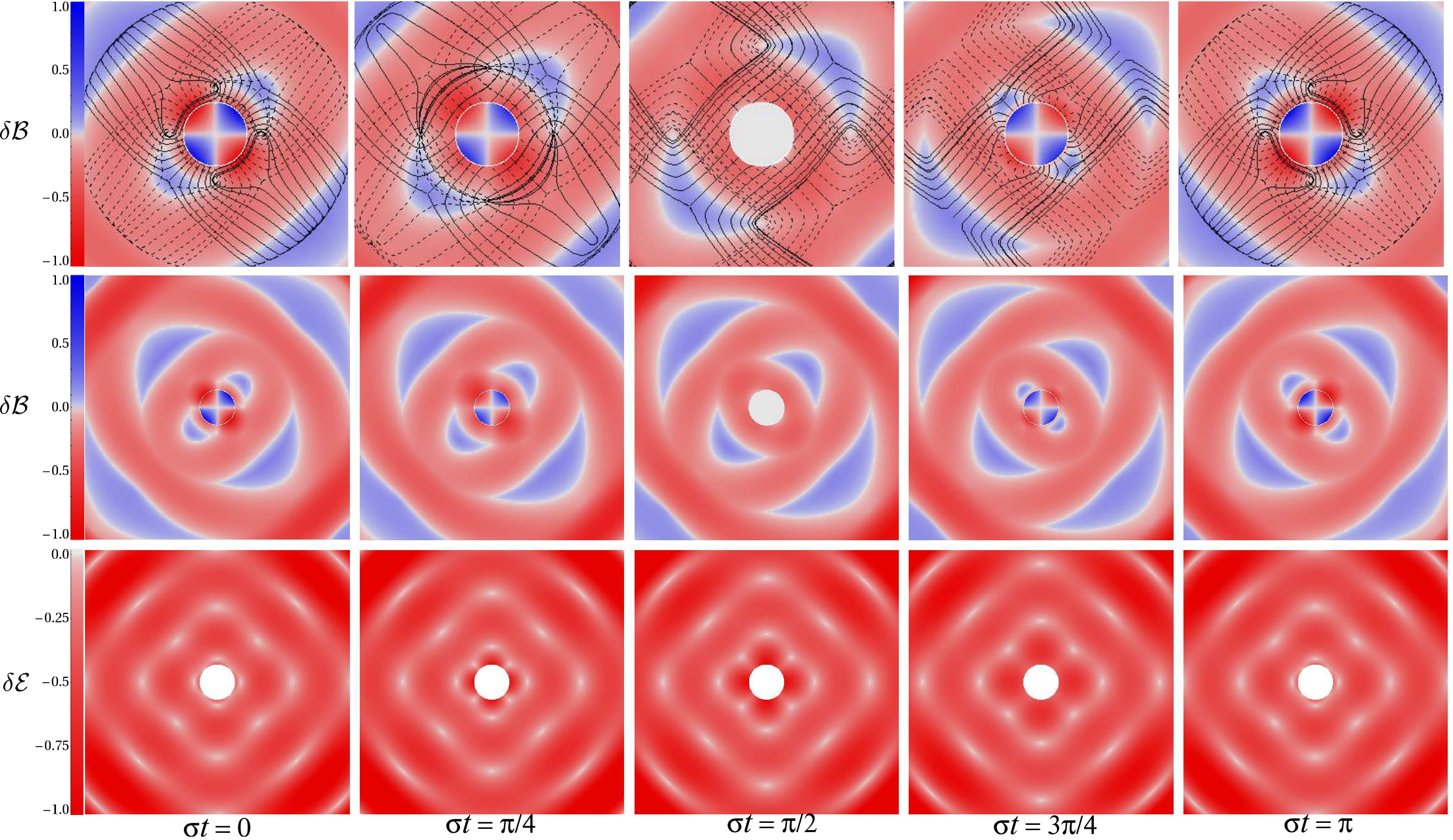}
\caption{(color online).
Time evolution of the equatorial vortexes (top and middle rows) and
equatorial perturbative tendexes (bottom row) for the superposed $(2,2)$ and
$(2,-2)$
magnetic-parity mode of Schwarzschild in RWZ gauge. The color scale is the
same as the center column of Fig.\ \ref{fig:22sup2-2VortexDensity}, and the
gravitational-wave-induced exponential decay of the vorticity and
tendicity has been removed.  {\it Top row}: Equatorial vortex lines and their
vorticity plotted in a region near the horizon ($16M$ across) followed over
time $t$.  The real part of the
eigenfrequency is denoted $\sigma$, so the successive panels, left
to right cover half a cycle of the mode's oscillation. {\it Middle row}: The
vorticity of the equatorial vortex lines in the near, intermediate and 
beginning of wave
zone ($30M$ across) at the same time steps as the top row. {\it Bottom row}:
Tendicity of the counterclockwise tendex lines passing through the equatorial plane
(which is dual to the left panel of Fig.\ \ref{fig:22sup2-2VortexDensityEven}),
plotted at the same time steps as the top row.
}
\label{fig:TimeSeries}
\end{figure*}  

In Fig.~\ref{fig:TimeSeries} for this mode 
we show, in the equatorial plane, the time evolution of 
(i) the vortex lines and their vorticities near the black hole 
(top row), and (ii) on a larger scale
that extends into the wave zone, the mode's 
vortexes (middle row) and perturbative tendexes (bottom row).
The five panels in the top row are stills from
a movie at \cite{OscillatingVortexMovie}. 
To be absolutely clear, the vortexes and tendexes 
are those of the same 
magnetic-parity mode.  As in the center column of Fig.\ 
\ref{fig:22sup2-2VortexDensity}, the top and middle rows show
only one family
of vortexes, that for the dashed vortex lines which have
predominantly negative vorticity; 
and as in Fig.\ \ref{fig:22sup2-2VortexDensityEven},
the bottom row shows only the tendicity of the negative-tendicity perturbative
tendexes that pass
through the equatorial plane at $45^\circ$.  Time $t=0$ (left
panels of Fig.\ \ref{fig:TimeSeries}) is chosen at a moment when the 
horizon vorticity is
maximum, whereas Figs.\ \ref{fig:22sup2-2VortexDensity} and
\ref{fig:22sup2-2VortexDensityEven} are snapshots at the
slightly earlier time $\sigma t \approx -\pi /3$ (which gives nearly the
same vortex structures as the fourth column, after a rotation by $\pi/2$).  

In interpreting Fig.~\ref{fig:TimeSeries}, especially the top row, we emphasize that there is no unique way of following a single vortex or tendex line in time. The same is true of electric and magnetic field lines in Maxwell's theory (cf.\ \cite{Belcher2003}). While we hope to elucidate this issue in future work, here, in constructing the panels in the top row of Fig.~\ref{fig:TimeSeries}, we have simply started the integration of the vortex lines from the same points at each time step, \emph{making no attempt to identify and follow individual lines from moment to moment}. Correspondingly, 
in order to interpret Fig.~\ref{fig:TimeSeries} and gain insight into the dynamics of the superposed mode, instead of trying to follow individual lines, we will 
focus on the lines' evolving shapes, and the structures of the 
vortexes and tendexes and the equations governing their evolution on the horizon.

As a foundation for understanding the near-zone dynamics 
depicted in this figure, 
we write down explicit expressions for the longitudinal
and longitudinal-transverse parts of the frame-drag and tidal fields
\emph{on the horizon}:
\begin{subequations}
\begin{eqnarray}
\mathcal B_{\hat r\hat r} &=& \Re\left[{3\over 2i\omega M^3}
e^{-i\omega (\tilde t + 2M)}\right] \mathcal Y(\theta,\phi)\;,
\label{eq:BrrHsuperposed} \\
\mathcal B_{\hat r \hat A} &=& \Re\left[{1\over 2\sqrt2 M^2}
(1-\beta)
e^{-i\omega (\tilde t + 2M)}\right] D_{\hat A} \mathcal Y\;,
\label{eq:BrAHsuperposed} \\
\mathcal \delta E_{\hat r \hat A} &=& \Re\left[{1\over 2\sqrt2 M^2}
(1+\beta)
e^{-i\omega (\tilde t + 2M)}\right] \left( - {\epsilon_{\hat A}}^{\hat B}
D_{\hat A}\right) \mathcal Y\;, 
\nonumber \\
\label{eq:ErAHsuperposed}  
\end{eqnarray}
where
\begin{equation}
\beta \equiv {3\over 2i\omega M(1-4i\omega M)}\;.
\end{equation}

Here the normalization is that of App.\ \ref{sec:RWApp}, 
$\omega = (0.37367-0.08896i)/M$
is the mode's eigenfrequency, ${\epsilon_{\hat A}}^{\hat B}$ is the 
Levi-Civita tensor on the horizon,
\begin{equation}
\mathcal Y \equiv Y^{22} + Y^{2\, -2} = \sqrt{15/8\pi}  \sin^2\theta \cos2\phi
\label{eq:SHsuperposed}
\end{equation}
\label{eq:SuperposedHorizon}
\end{subequations}
is this mode's scalar spherical harmonic, and $D_{\hat A}$ is the 
covariant derivative on the unit 2-sphere 
(related to the covariant derivative on the horizon by
$ D_{\hat A} = 2M \boldsymbol{\nabla}_{\hat A}$).
[Equations (\ref{eq:SuperposedHorizon}) follow from Eqs.\ (\ref{eq:FD22mag}),
(\ref{Qanalytic}), (\ref{eq:magneticSuperposedTidal}) of App.\ 
\ref{sec:RWApp}, the vector-spherical-harmonic definitions
(\ref{eq:YlmAdef}) and (\ref{eq:BVectorHarmonic}), and definition
(\ref{eq:KStime}) of the EF time coordinate.]

Equations (\ref{eq:SuperposedHorizon}) are the fields measured by 
Eddington-Finkelstein observers.  The
conservation law (\ref{eq:DPsiCons}) for longitudinal field lines 
threading the 
horizon (which
we shall need below) involves, by contrast, the LT frame-drag field measured
by Schwarzschild observers on the ``stretched horizon'' (very close to the
event horizon). Since the Schwarzschild observers
move outward with velocity $\bm v = (2M/r) \bm N \simeq \bm N$ with respect to
the EF observers, and with $\gamma \equiv 1/\sqrt{1-v^2} \simeq 1/(\sqrt2 \alpha)$,
the field they measure is
$\boldsymbol{\mathcal B}^{\rm LT}_{\rm Sch} = \gamma
(\boldsymbol{\mathcal B}^{\rm LT} - \bm N \times
\delta \boldsymbol{\mathcal E}^{\rm LT})$.  This field diverges as $1/\alpha$
as the stretched horizon is pushed toward the event horizon; to remove that 
divergence, in the Membrane Paradigm \cite{Thorne-Price-MacDonald:Kipversion})
we renormalize by multiplying with $\alpha$:
\begin{eqnarray}
{\mathcal B}^{\rm H}_{\hat r \hat A} 
&\equiv& 
\alpha \boldsymbol{\mathcal B}^{\rm LT}_{\rm Sch} = \alpha \gamma
(\boldsymbol{\mathcal B}^{\rm LT} - \bm N \times
\delta \boldsymbol{\mathcal E}^{\rm LT}) \nonumber \\
&=& \Re\left[{1\over 4M^2}
e^{-i\omega (\tilde t + 2M)}\right] D_{\hat A} \mathcal Y\;.
\label{eq:BrAHHsuperposed}
\end{eqnarray}
The second line is obtained by inserting the EF fields 
(\ref{eq:BrrHsuperposed}) and (\ref{eq:BrAHsuperposed}),
and $\gamma = 1/\sqrt2 \alpha$, into the first line. 
The conservation law for longitudinal vortex lines threading 
the horizon (actually,
one of the Maxwell-like Bianchi identities in disguise) says that 
\begin{equation}
\partial \mathcal B_{NN} / \partial \tilde t + \nabla_{\hat A}
(-\mathcal B^{\rm H}_{\hat r \hat A}) = 0\;;
\label{eq:LineConservationSuperposed}
\end{equation}
cf.\ Eq.\ (\ref{eq:DPsiCons}) and subsequent discussion.
(In this Schwarzschild-perturbation-theory case, there are no small 
spin-coefficient terms to spoil the perfection of the conservation law.)
The vortex-line density and flux expressions
(\ref{eq:BrrHsuperposed}) and (\ref{eq:BrAHHsuperposed}) do, indeed, satisfy 
this conservation law, by virtue of the 
fact that the 2-dimensional Laplacian acting on the quadrupolar spherical
harmonic $\mathcal Y$ gives 
$D_{\hat A} D^{\hat A} \mathcal Y = - 6 \mathcal Y$.

Equations (\ref{eq:SuperposedHorizon}) 
and (\ref{eq:LineConservationSuperposed})
tell us the following:
(i) On and near the horizon, the LT fields $\boldsymbol{\mathcal B}^{\rm LT}$ [Eq.\ 
(\ref{eq:BrAHsuperposed})] and 
$\delta \boldsymbol{\mathcal E}^{\rm LT}$ [Eq.\ 
(\ref{eq:ErAHsuperposed})], and also $\mathcal B^{\rm H}_{\hat r \hat A}$
[Eq.\ (\ref{eq:BrAHHsuperposed})], 
all oscillate approximately out of phase with the longitudinal field
$\boldsymbol{\mathcal B}^{\rm L}$ [Eq.\ (\ref{eq:BrrHsuperposed})].\footnote{
The longitudinal-transverse frame-drag field $\boldsymbol{\mathcal B}^{\rm LT}$ lags approximately $1.04 \approx \pi/3$ radians behind $\mathcal B_{NN}$ on the horizon, while the LT component of the tidal field, $\boldsymbol{\mathcal E}^{\rm LT}$, lags approximately $1.21\approx 2 \pi/5$ radians behind $\mathcal B_{NN}$. Most importantly for interpreting Fig.~\ref{fig:TimeSeries}, the two nonzero tendicities of the tidal field are (in the equatorial plane)  $\pm \sqrt{\mathcal E_{\hat r \hat \theta}^2 + \mathcal E_{\hat \theta \hat \phi}^2 }$, and on the horizon they lag nearly $\pi/2$ radians behind $\mathcal B_{NN}$; the damping of the perturbation adds slightly to the phase lag, so that it is actually $\pi/2 +  \arctan[\Im(\omega)/\Re(\omega)] \simeq \pi/2 + 0.234$ out of phase with $\mathcal B_{NN}$. }
Therefore, near-zone energy is fed back and forth between the L and LT fields as
the black hole pulsates.

(ii) The conservation law (\ref{eq:LineConservationSuperposed}) says that,
if we regard $\mathcal B_{NN}$ as the density of vortex lines of
$\boldsymbol{\mathcal B}^{\rm L}$ threading the horizon, 
and $-\mathcal B^{\rm H}_{\hat r \hat A} $ as the
flux of vortex lines (number crossing a unit length in the horizon per
unit time), then these horizon-threading vortex lines of 
$\boldsymbol{\mathcal B}^{\rm L}$ are conserved during the pulsation. 
More specifically:

(iii) As the mode evolves
in Fig.\ \ref{fig:TimeSeries} from $\sigma t=0$ to $\sigma t=\pi$, the conserved vortex lines
are pushed away from the center of each horizon vortex toward its white
edges, and there the conserved lines from the red region (counterclockwise)
annihilate with the conserved lines from the blue region (clockwise).
The pushing is embodied in the vortex-line flux 
$-\mathcal B^{\rm H}_{\hat r \hat A} $, which grows stronger during this
evolution. 

(iv) As the mode evolves further from $\sigma t=\pi/2$ to $\sigma t=\pi$,
conserved vortex lines of $\boldsymbol{\mathcal B}^{\rm L}$ are created in
pairs (one clockwise, the other counterclockwise) at the white edges of the 
horizon vortexes, and move inward toward the center of each vortex.

Turn, now, from the conserved vortex lines of 
$\boldsymbol{\mathcal B}^{\rm L}$ piercing the horizon to the 
3D vortex lines outside the horizon, depicted
in the top row of Fig.\ \ref{fig:TimeSeries}.  Because these 
are lines of the full
3D frame-drag field $\boldsymbol{\mathcal B}$ and not its longitudinal part
$\boldsymbol{\mathcal B}^{\rm L}$,
they do not obey a conservation law and there is no unique way of following 
individual lines from one panel to the next. However, their evolving shapes
teach us much about the geometrodynamics of this superposed mode:

At time $\sigma t=0$ (upper left panel), 
the horizon-piercing vortex lines of the full frame-drag
field $\boldsymbol{\mathcal B}$  
are almost perfectly radial, with clockwise (solid) tendex lines emerging from
the two blue horizon tendexes, and counterclockwise (dashed) tendex lines
emerging from the two red horizon tendexes.  As time passes, the 
horizon piercing lines become less radial and the horizon vorticity decreases
($\sigma t=\pi/4$) until the lines' angles to the horizon are almost all near
$45^\circ$ and the horizon vorticity vanishes ($\sigma t=\pi/2$).  
Note that the lines that lie precisely on the diagonals, and which contact the 
horizon radially in the middle panel ($\sigma t = \pi/2$), have zero vorticity 
where they strike the horizon.
This latter fact allows them to have a more radial angle of intersection than 
almost all other lines.
The near-horizon frame-drag field has evolved at this time from being 
predominantly longitudinal, 
$\boldsymbol{\mathcal B}^{\rm L}$, to being predominantly
longitudinal-transverse, $\boldsymbol{\mathcal B}^{\rm LT}$, but with some
small admixture of transverse-traceless ingoing waves, $\boldsymbol{\mathcal B}^{\rm
TT}$. As time moves onward from $\sigma t=\pi/2$ to $\sigma t = \pi$, the
vortex lines in the outer part of each panel reach around on the horizon
and attach to a quadrant on the side rather than directly below 
themselves---a quadrant that has newly acquired the color corresponding
to the lines' own vorticity (blue for solid lines, red for dashed lines).

At $\sigma t=0$ (upper left panel), the near-horizon, 
nearly circular vortex lines in each quadrant
represent, predominantly, the transverse-isotropic part
of $\boldsymbol{\mathcal B}^{\rm L}$ and keep it trace free. As time passes
and $\boldsymbol{\mathcal B}^{\rm L}$ decreases,we 
can regard these lines as traveling
outward, forming the distorted ellipses which become gravitational waves in 
the far zone. 
The manner in which these circular vortex lines are restored each cycle appears
to be as follows: 
As discussed above, as the horizon vorticity oscillates through zero, formerly longitudinal lines are pushed away from their respective vortexes and become first more longitudinal-transverse, and then attach to a different quadrant of the horizon; meanwhile the $\boldsymbol{ \mathcal B}^{\rm L} $ is being regenerated with opposite sign in each quadrant, which requires new transverse-isotropic lines of opposite vorticity. These lines run tangent to the horizon, hugging it while they cross through a vortex of strong opposite vorticity. At each edge of the vortex these isotropic-transverse lines link up with predominantly radial lines which have the same sign of vorticity (it appears that it is the degenerate points at these edges that allow for such a reconnection). 
This deforms the highly distorted, nearly circular arcs, which then lift off the horizon and propagate away as the cycle progresses. 

This entire evolution is being driven by the
oscillatory turn-off and turn-on of the longitudinal part of the frame-drag
field $\boldsymbol{\mathcal B}^{\rm L}$.  

Turn attention to the evolution of the mode's equatorial vortexes
(middle row) and tendexes (bottom row).  In accord with our discussion
above of the evolution of the horizon fields, 
Eqs.\ (\ref{eq:SuperposedHorizon}), these panels
reveal (see below) that the vortexes and tendexes oscillate out of phase with each other.  Near-zone
energy (see footnote 1 in Sec.~\ref{sec:20Mag}) gets fed back and forth between vortexes and 
tendexes in an oscillatory manner (though during this feeding, some
of it leaks out into the transition zone and thence into 
gravitational waves).  This oscillatory feeding enables the near-zone
perturbative tendexes to store half of the oscillation energy 
(while the LT vortexes store the other half) 
when the longitudinal frame-drag field $\boldsymbol{\mathcal B}^{\rm L}$
is temporarily zero; and then use that energy to regenerate $\boldsymbol{\mathcal B}^{\rm L}$.

The evidence for this near-zone feeding, in Fig.\
\ref{fig:TimeSeries}, is the
following: (i) The near-horizon vortexes
are strongest along the diagonals, while the regions of strong near-zone 
tendicity always occur along the vertical and horizontal
directions. Thus, the vortexes and tendexes tend to occupy
different regions, with a $\pi/4$ rotation between the patterns [as
one should expect from the angular dependences in Eqs.\ 
(\ref{eq:BrrHsuperposed}) and (\ref{eq:ErAHsuperposed})].
(ii) There is a $\simeq \pi/2$ phase difference in the time evolution
of the vortexes and tendexes. 
At those times when the horizon vorticity and near-horizon 
vortexes are strongest, the near-horizon tendexes are weak.
As the horizon oscillates through zero vorticity, 
the tendexes are reaching their maximum strength.

A careful study of the phases of these time behaviors
reveals that the dynamics are not precisely $\pi/2$ out of phase,
as can be seen clearly in the first panel in the bottom row of
Fig.~\ref{fig:TimeSeries}: though the horizon vorticity is at it's
maximum, the tendicities have just oscillated through zero in this
region and are beginning to regenerate. 
As mentioned in footnote 3 above, this additional phase lag is due to the mode's damping, and in radians its
magnitude is $\arctan[\Im(\omega)/ \Re(\omega)] \simeq  0.234$. 

The $\pi/4$ differences in spatial phase and 
$\simeq\pi/2$ differences in temporal phase are lost as the 
frame-drag and tidal fields travel outward through 
the intermediate zone and into the wave zone; an inspection of the 
outer edges of the
time series plots shows bands of strong tendicity and vorticity
\emph{in phase} in time and space, propagating outward in synch. This must be
the case, since for plane waves in linearized gravity, the vortex
and tendex lines are in phase temporally and spatially (though the
lines are rotated by $\pi/4$ with respect to each other at each
event; see Paper I, Sec VI A).

By scrutinizing the middle and bottom rows of Fig.\ \ref{fig:TimeSeries},
(which extend from the near zone through the intermediate zone and into the
inner parts of the wave zone), one can see visually how the oscillatory feeding of
energy between near-zone vortexes and tendexes gives rise to outgoing
vortexes and tendexes that represent gravitational waves.

\section{$(2,1)$ and $(2,0)$ Quasinormal Modes of Schwarzschild}
\label{sec:OtherPerts}

In this section we will complete our study of the quadrupolar perturbations of Schwarzschild black holes. Specifically, we will explore the vortex 
and tendex
structures and the dynamics of the $(2,1)$ and $(2,0)$ magnetic- and 
electric-parity perturbations of a Schwarzschild hole, in RWZ gauge.

\subsection{Vortexes of $(2,1)$ magnetic-parity mode and perturbative tendexes of $(2,1)$ electric-parity mode}
\label{sec:21Mag}

In Sec.\ \ref{sec:21MagSch}, we summarized the most important
properties of the
(2,1) magnetic-parity mode of a Schwarzschild black hole.  In this
subsection and the next, we shall give additional details about this mode and its
electric-parity dual.  We begin with the vortex structure for magnetic
parity.

\begin{figure*}
\includegraphics[width=0.6 \columnwidth]{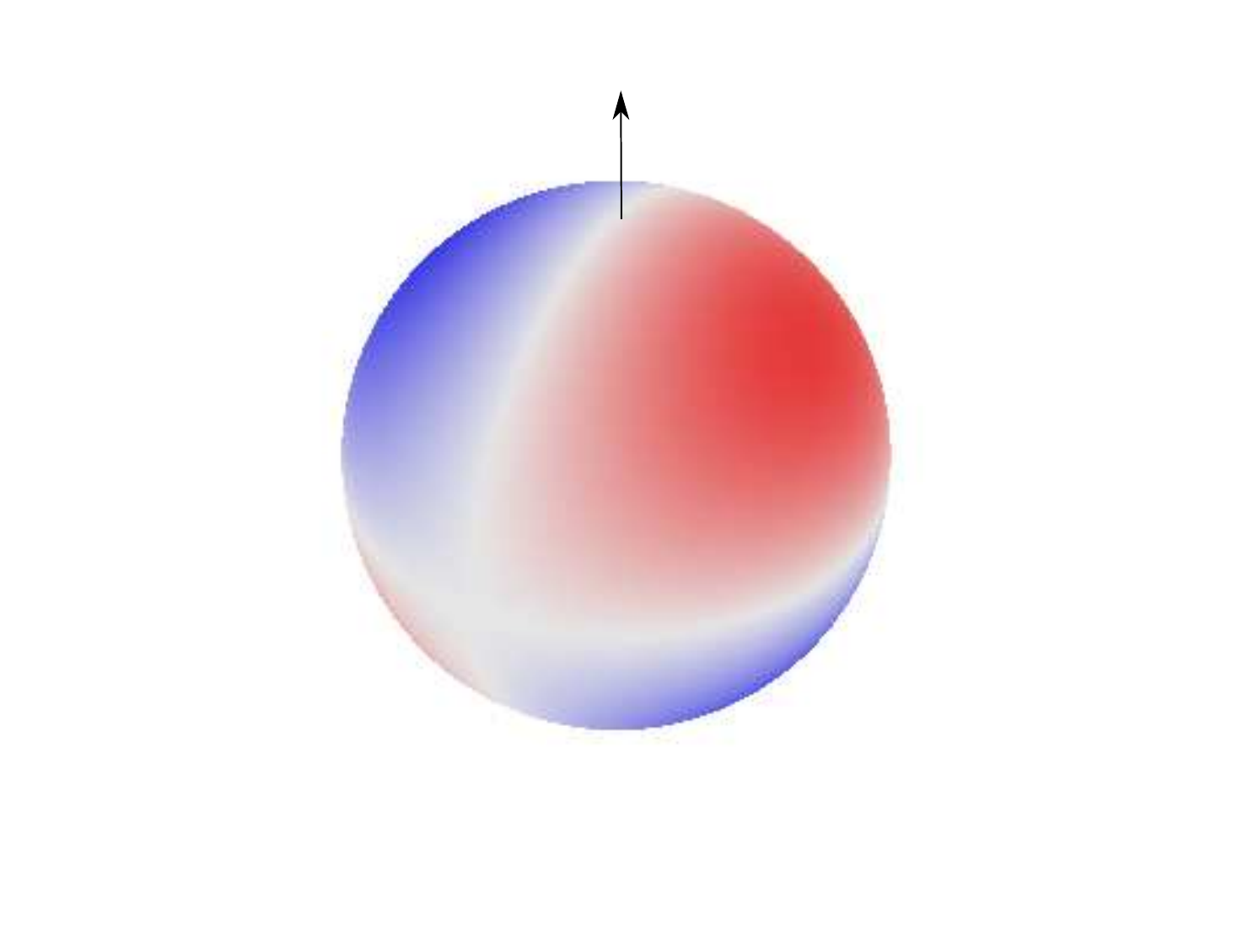}
\includegraphics[width=0.6\columnwidth]{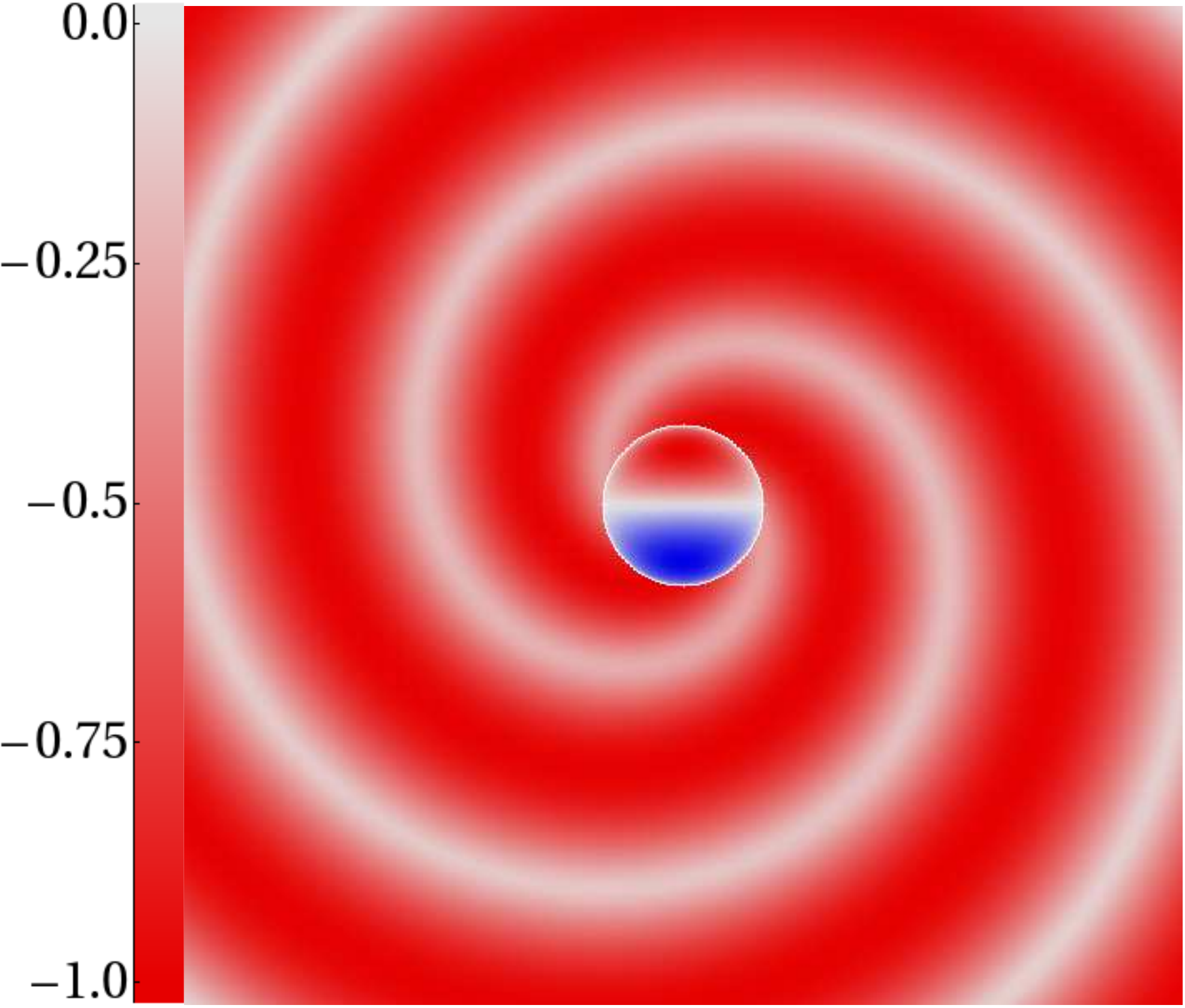}
\includegraphics[width=0.6\columnwidth]{SchwOddRW21BothVort}
\caption{(color online).
The $(2,1)$ magnetic-parity horizon vorticity and vortexes. 
{\it Left panel}: The horizon vorticity for the $(2,1)$ magnetic-parity
perturbation of Schwarzschild, colored as in Fig.\
\ref{fig:EvenPertHorizons}. There are four 
horizon vortexes, two clockwise (blue [dark gray]) and two counterclockwise 
(red [light gray]). 
The horizon vorticity vanishes at the equator and the poles. 
{\it Middle panel}: Vorticity of the counterclockwise 
vortex lines passing through the equatorial plane, colored and normalized as in 
the middle and right hand columns of Fig.~\ref{fig:Sch22VortexAll} 
and plotted in a region $24M$ across. The (blue [dark gray]) vorticity of the 
clockwise vortex lines has precisely this same pattern, because the two 
families of lines pass through the equatorial plane with the same magnitude of
vorticity at each point. {\it Right panel}: Three dimensional vortexes
colored and labeled as in Fig.\ \ref{fig:TendexVortex3D}.  By near duality,
this figure also represents (to good accuracy) the tendicity and tendex
structure of the (2,1) electric-parity mode.}
\label{fig:21MagneticVorts}
\end{figure*}

The horizon vorticity of the magnetic-parity (2,1) mode has an angular
dependence given by the spherical harmonic 
$Y^{21}(\theta,\phi)$ (of course). We display
this horizon vorticity in the left panel of
Fig.~\ref{fig:21MagneticVorts}. There are four horizon vortexes, two of
each sign, and vanishing horizon vorticity all along the equator. 

As we noted in Sec.\ \ref{sec:21MagSch}, this mode's symmetry dictates that 
the frame-drag field be reflection antisymmetric through the equatorial
plane.  As for the electric-parity (2,2) frame drag field, which also has
this property (second paragraph of Sec.\ \ref{sec:22EquatorialPlane};
also Sec.\  \ref{sec:EparityVortexes22} and Fig.\
\ref{fig:Schw22EvenVortexDensity}), this implies that one family of vortex 
lines lies in the equatorial plane with vanishing vorticity,
and two cross through that plane at $45^\circ$ with equal and opposite 
vorticities.  The negative
vorticities of the crossing lines are plotted in the middle panel of
Fig.~\ref{fig:21MagneticVorts}, along with the projected horizon
vorticity, as if the horizon were viewed from above. 
The positive-vorticity pattern of the other family of crossing lines is
identical to this negative-vorticity pattern, since at each point the two
lines have the same vorticity magnitude.

The fact that there are just two spiraling vortexes in this figure, by
contrast with four for the (2,2) modes, is
guaranteed by the modes' azimuthal orders, $m=1$ here and $m=2$ for 
(2,2).

The vortex structure outside the equatorial plane, depicted in the right
panel of Fig.\ \ref{fig:21MagneticVorts}, was discussed in Sec.\ 
\ref{sec:21MagSch}.  The two red (light gray), 3D vortexes are the same ones 
depicted in the middle panel.  They actually extend across the equatorial plane
(via the $45^\circ$ vortex lines) into the region occupied by the blue (dark 
gray) vortexes; but we do not see them there in the 3D drawing because the blue
vortexes have larger vorticity and we have chosen to show at each point
only the largest-vorticity vortex.

\subsection{Vortexes of $(2,1)$ electric-parity mode and perturbative tendexes of $(2,1)$ magnetic-parity mode}
\label{sec:21Elec}

\begin{figure*}
\includegraphics[width=0.65\columnwidth]{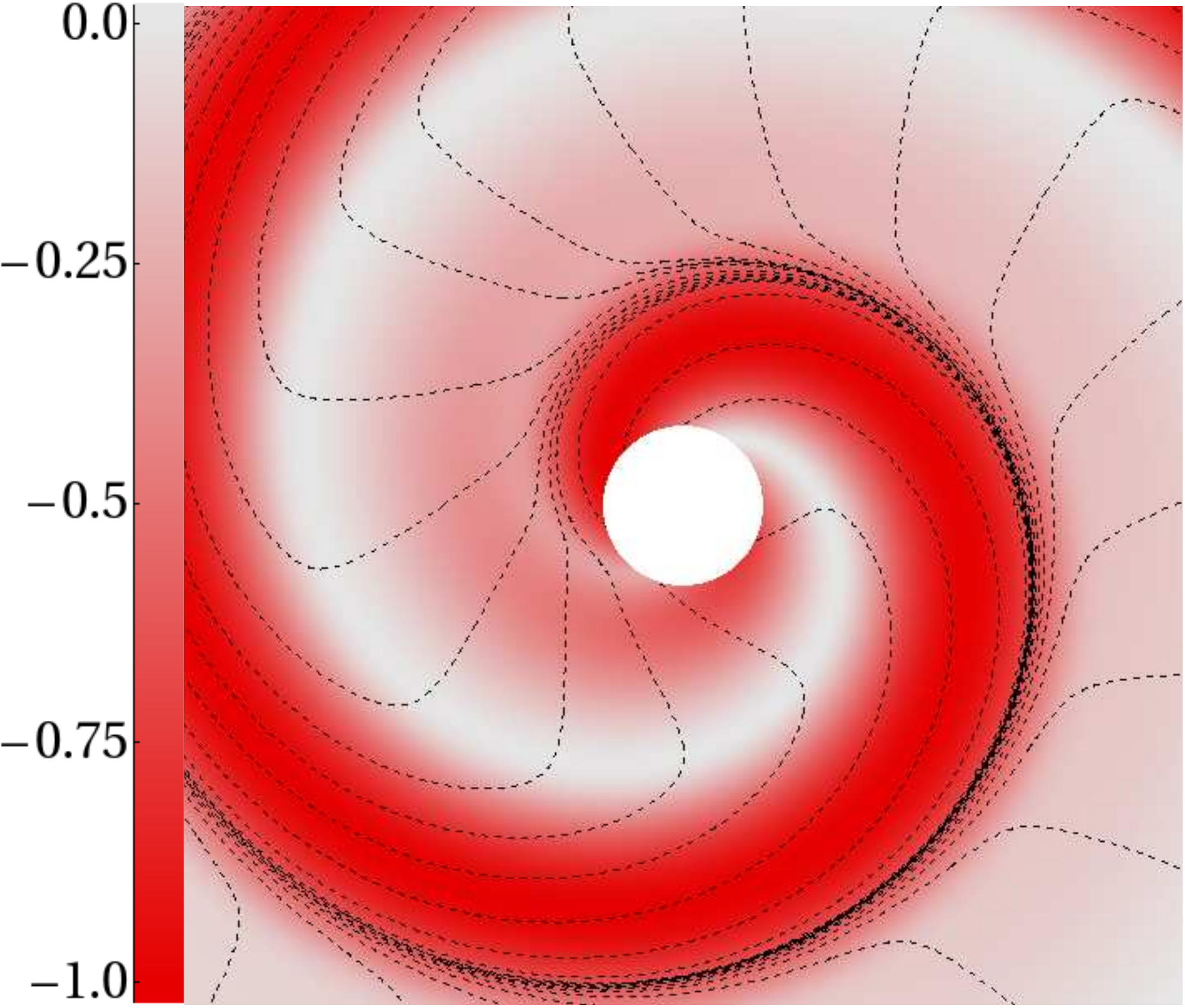}
\includegraphics[width=0.65\columnwidth]{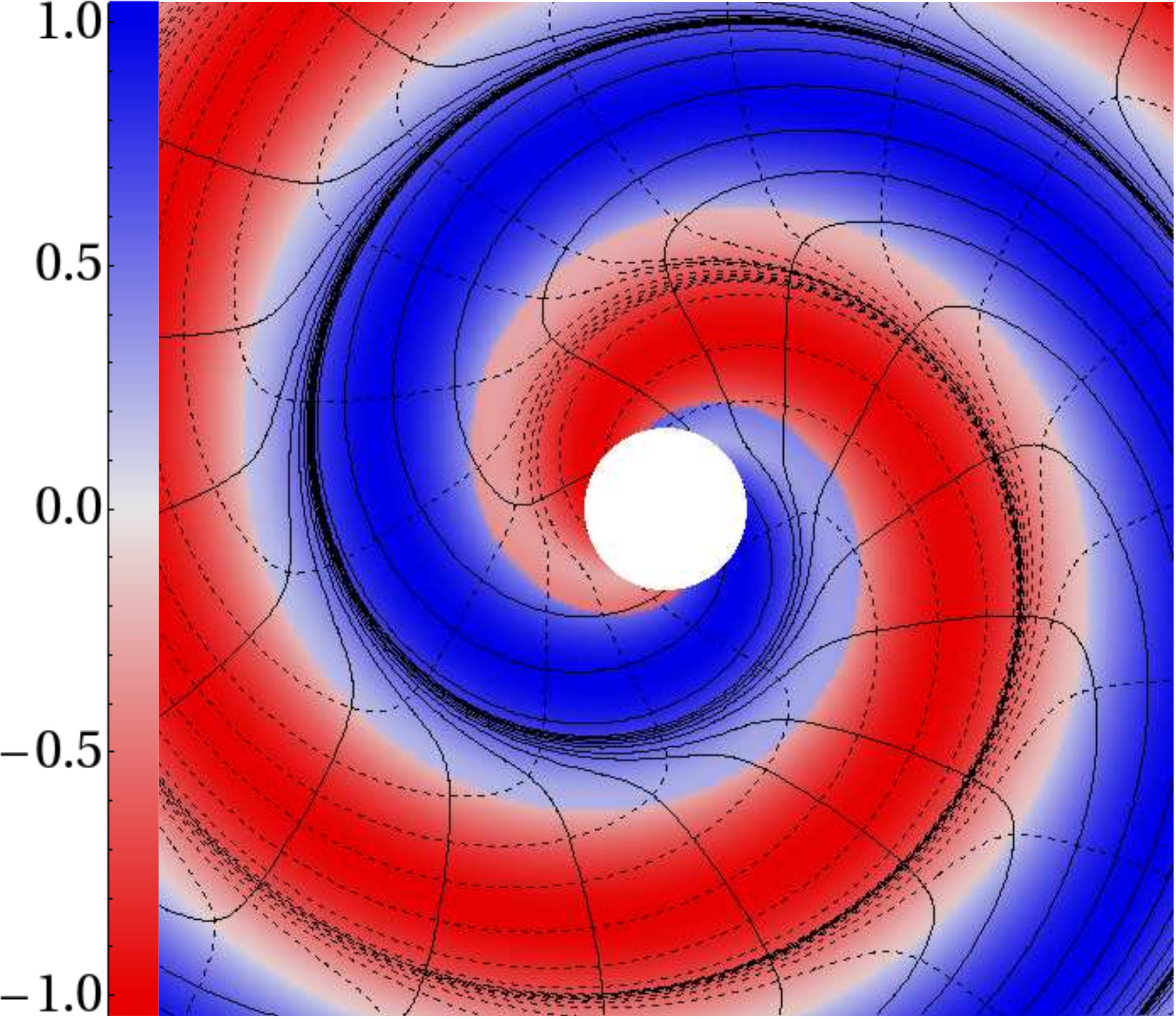}
\includegraphics[width=0.65\columnwidth]{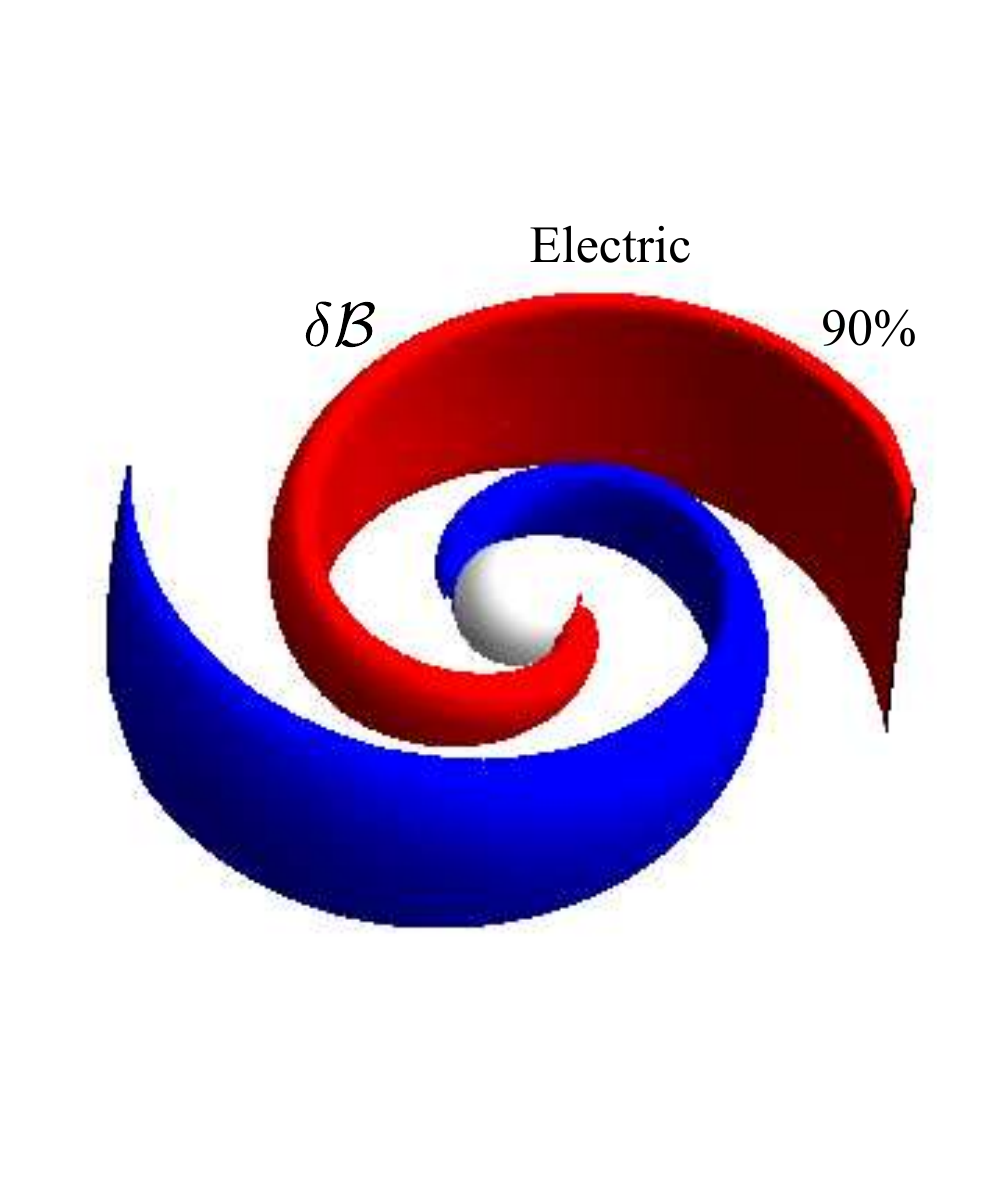}
\caption{(color online).
The $(2,1)$ electric-parity vortex lines, vorticities and vortexes in the 
equatorial plane. For this mode the horizon's vorticity vanishes, 
so the horizon is plotted as a white disk or sphere. 
{\it Left panel}: Counterclockwise vortex lines (dashed) and their vorticity 
(red [light gray] color) normalized as in
Fig.~\ref{fig:Sch22VortexAll}, and plotted in a region $24M$ across. {\it
Middle panel}: Both clockwise (solid) and counterclockwise (dashed)
vortex lines, and the vorticity (color) of the line with the larger
magnitude of vorticity, 
in a region $24M$ across.  {\it Right Panel}: Three-dimensional vortexes
colored and labeled as in Fig.\ \ref{fig:TendexVortex3D}.
By near duality, this figure also represents to good accuracy
the perturbative tendex lines, tendicities and 
tendexes for the magnetic-parity (2,1) mode.
}
\label{fig:21ElectricVorts}
\end{figure*}

Turn, next, to the vortex lines and vortex structure of the (2,1) 
electric-parity mode.  
[By near duality, the perturbative tendex lines and tendex
structure of the (2,1) magnetic-parity mode will be the same.] 

For this mode, with the parity reversed from
the previous section, the frame-drag field is symmetric under reflection 
through the equatorial plane rather than antisymmetric.  Therefore, 
there are two sets of vortex lines that remain
in the equatorial plane, with the third set normal to it.  In this sense
the vortexes structures are analogous to those of the magnetic-parity, $(2,2)$
mode; and in fact they are strikingly similar,
aside from having two arms rather than four.  

We show the
vortexes and vortex lines in Fig.~\ref{fig:21ElectricVorts}.
The left and middle panels of Fig.~\ref{fig:21ElectricVorts} show the
lines that remain in the equatorial plane, along with color-intensity
plots depicting the lines' vorticities. 

The left panel of
Fig.~\ref{fig:21ElectricVorts} shows the counterclockwise lines and 
their vorticities. As
in the case of the $(2,2)$ mode, we see a spiraling region of strong
vorticity which contacts the horizon, and an accompanying spiral of low
vorticity. At the horizon, the frame-drag field is primarily 
longitudinal-transverse, and correspondingly its vortex lines enter
the horizon at a (nearly) $45^\circ$ angle. 
As for the (2,2) mode, there is a limiting spiral that all the outspiraling 
vortex lines approach, near the edge of the vortex. 

There is also a small
region of strong vorticity near the horizon which forms a second spiral,
opposite the primary spiral, although it quickly becomes weak; 
this second vortex coincides with the region of strong positive 
vorticity, as we see in the
middle panel of Fig.~\ref{fig:21ElectricVorts}, and we think its
existence is due to the frame-drag field at the horizon being primarily
longitudinal-transverse.
It also should be compared to the similar regions of strong negative vorticity near the positive horizon vortexes of the magnetic-parity $(2,2)$ mode in Fig.~\ref{fig:Sch22VortexAll}.

In the middle panel, we plot both the
counterclockwise (dashed) and clockwise (solid) vortex lines, and we
color each point by the vorticity that is strongest.  We see two strong
vortexes spiraling out to form gravitational waves, and we see that 
under a rotation through $180^\circ$ the clockwise and counterclockwise
vortex lines map into each other. 

Finally, in the right panel of Fig.~\ref{fig:21ElectricVorts}, we
show the vortexes in three dimensions using the same conventions as in
Fig.\ \ref{fig:TendexVortex3D}:  
the red (light gray) and blue (dark gray) surfaces are the locations where
the vorticity of largest magnitude has fallen to $90\%$ of its maximum at each
radius. By contrast with the (2,1) magnetic-parity mode, where the 
3D vortexes are antisymmetric through the equator (and so they flip colors; 
see Fig.\ \ref{fig:21MagneticVorts}),
here they are symmetric and so have the same color above and below the
equatorial plane.

By duality, for the (2,1) magnetic-parity mode, with its antisymmetric
3D vortexes (Fig.\ \ref{fig:21MagneticVorts}), the 3D perturbative 
tendexes are symmetric through the
equatorial plane and have the form shown in this right panel of Fig.\ 
~\ref{fig:21ElectricVorts}.

\subsection{Vortexes of $(2,0)$ magnetic-parity mode and perturbative
tendexes of (2,0) electric-parity mode}
\label{sec:20MagVortexes}

In Sec.\ \ref{sec:20Mag}, we described in detail the dynamics of the axisymmetric
(2,0) magnetic-parity mode of Schwarzschild and the gravitational
waves it emits---waves in which the vortex and tendex lines wrap around
deformed tori.  In this section and the next, we shall discuss some
other details of this mode and its dual, the (2,0) electric-parity mode.

\begin{figure}
\includegraphics[width=0.95\columnwidth]{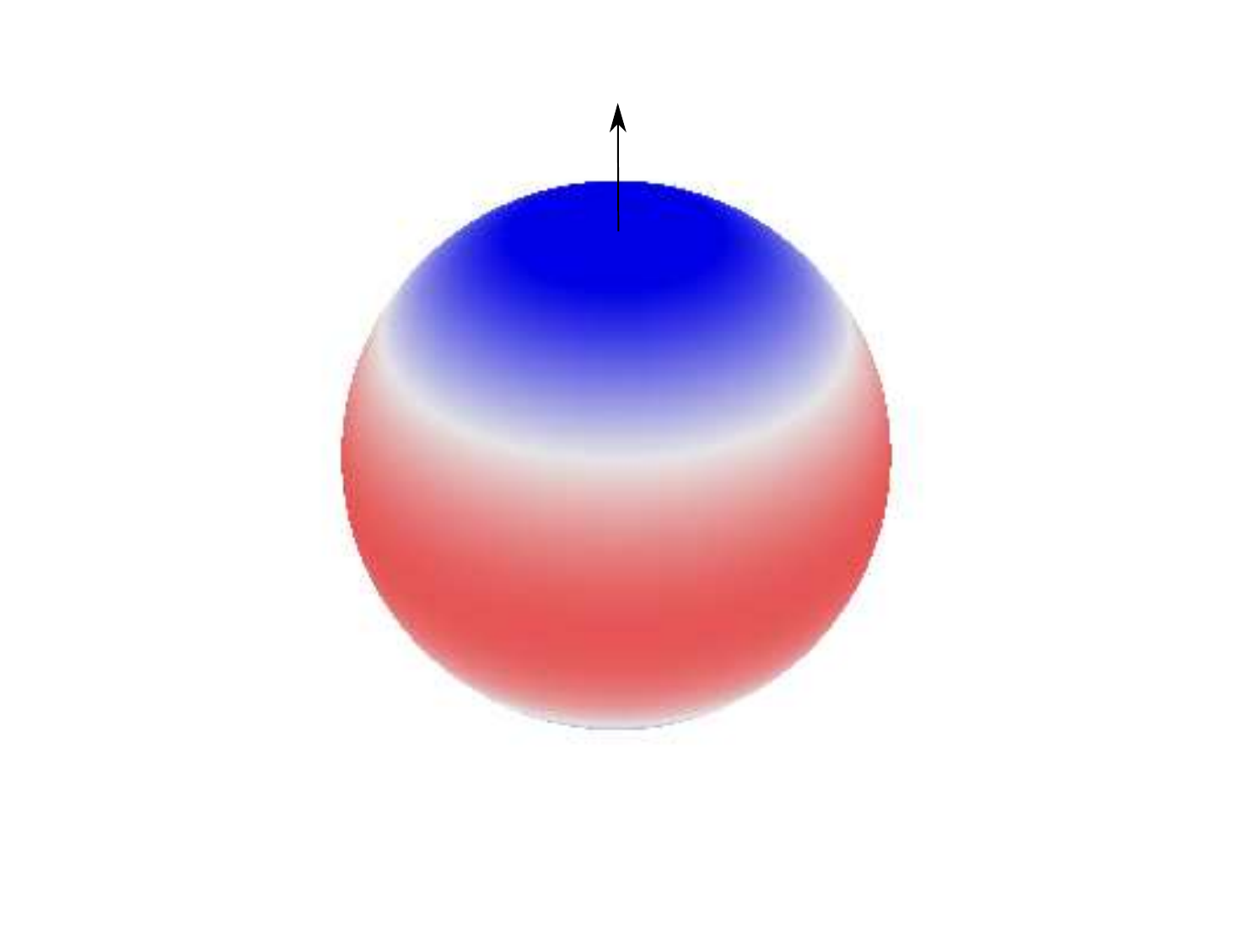}
\caption{(color online).
The horizon vorticities  $(\delta \mathcal B_{NN})$ of the
quadrupolar, $(2,0)$, magnetic-parity mode. 
As in Fig.\ \ref{fig:EvenPertHorizons}, the color intensity is proportional 
to the magnitude of the vorticity with blue (dark gray) for positive and red 
(light gray) for negative.
The arrow points along the polar axis. 
The vorticity oscillates sinusoidally in time, causing $\delta \mathcal
B_{NN}$ first to vanish and then to change sign.}
\label{fig:SchwHorVort20}
\end{figure}

In Fig.~\ref{fig:SchwHorVort20}, we show the horizon vorticity for
this magnetic-parity
mode.  Of course, it is proportional to the scalar spherical harmonic
$Y^{20}(\theta,\phi)$.
At this moment of time, there are clockwise  
vortexes (blue [dark gray]) in the northern and southern hemispheres, 
and a band-shaped counterclockwise vortex (red [light gray]) in the equatorial 
region.  As time passes, the 
horizon vorticity oscillates, with red vortexes becoming blue
and blue becoming red in each half cycle, while also decaying
exponentially.  The cause of these oscillations, as we discussed in
Sec.\ \ref{sec:20Mag}, is exchange of energy between 
$\delta {\boldsymbol{\mathcal B}}^{\rm L}$ (whose normal-normal component
is the horizon vorticity) and $\delta {\boldsymbol{\mathcal E}}^{\rm LT}$ 
(which we will visualize in the next section).

\begin{figure*}[tb]
\includegraphics[width=0.65\columnwidth]{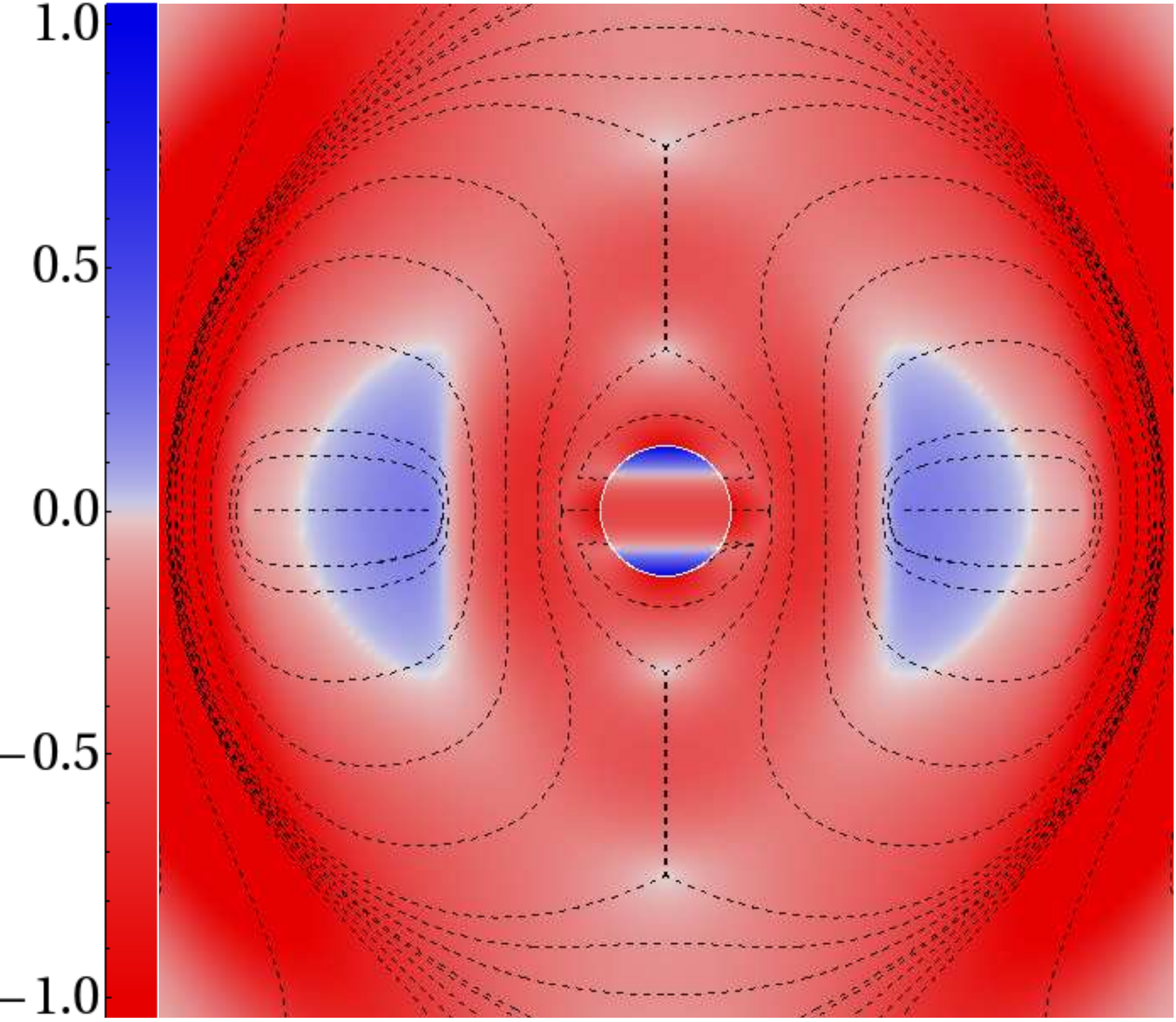}
\includegraphics[width=0.65\columnwidth]{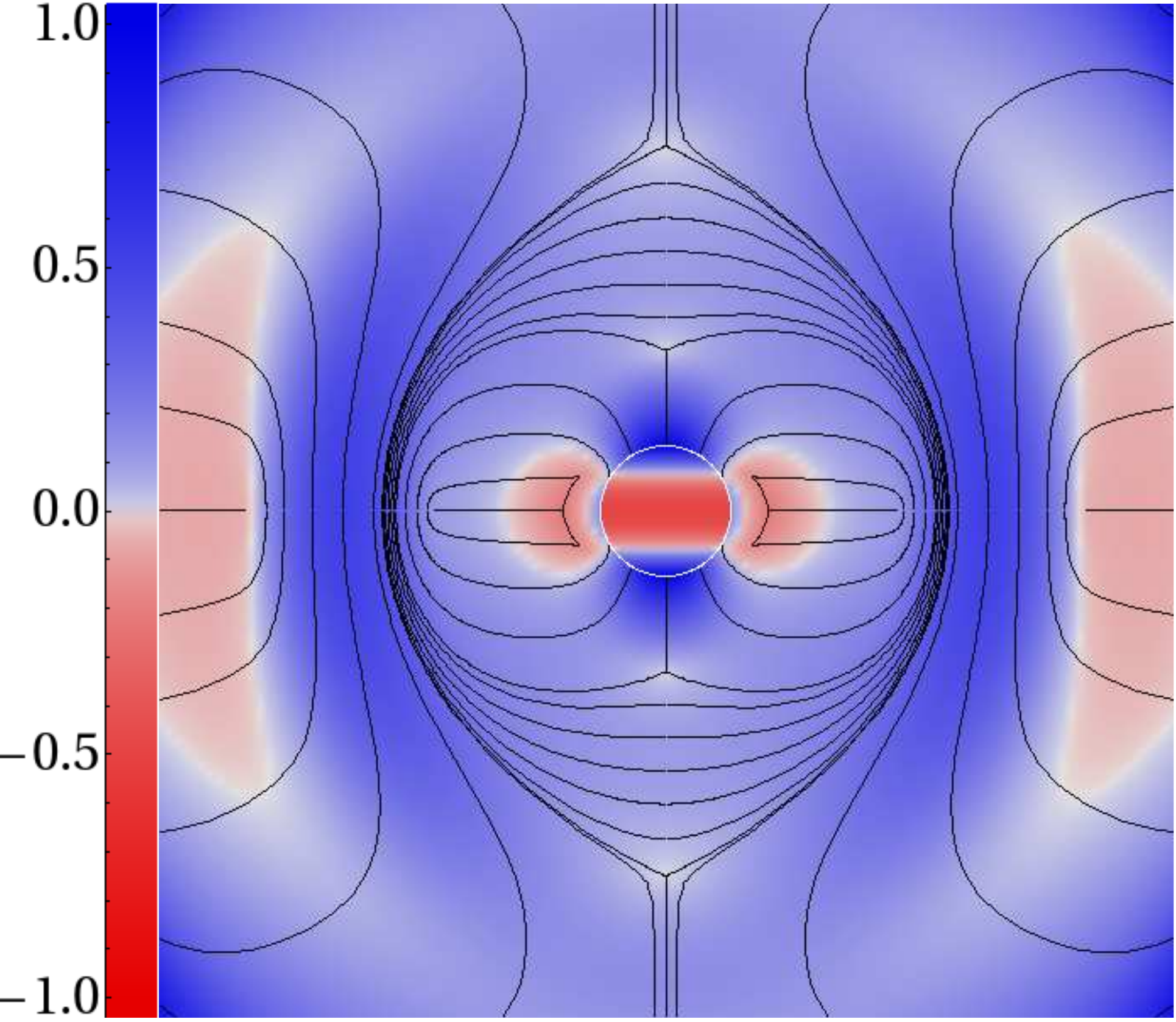}
\includegraphics[width=0.65\columnwidth]{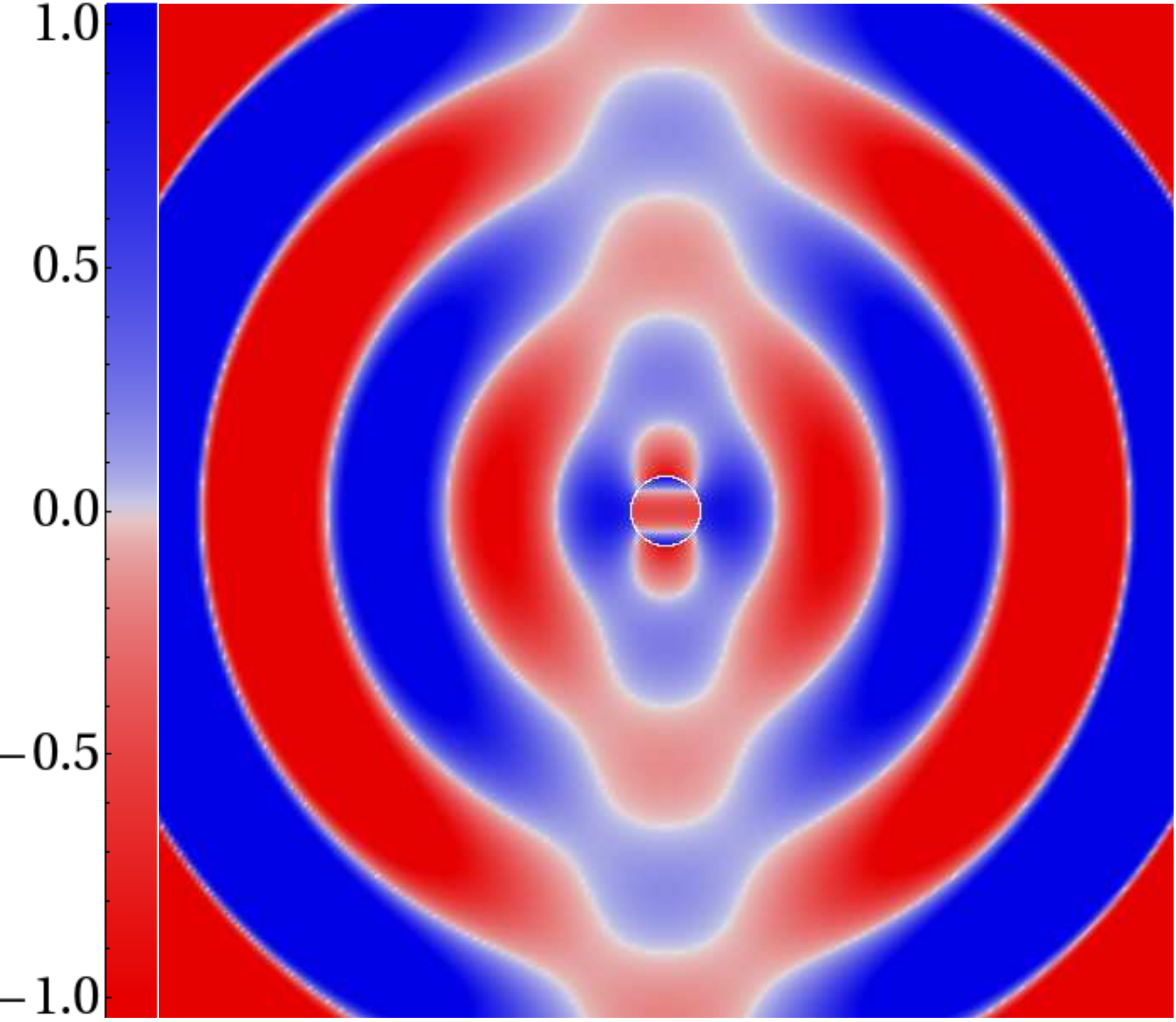}
\caption{(color online).
{\it Left and middle panels:} Near- and transition-zone vortex lines and their vorticities in an
$\mathcal S_\phi$ plane of constant $\phi$, for the
axisymmetric (2,0) magnetic-parity mode of Schwarzschild in the near and
transition zones ($30M$ across).
{\it Left panel}: The predominantly counterclockwise family of vortex lines.
{\it Middle panel}: The predominantly clockwise family of vortex lines.
{\it Right panel}: Vorticity of the axial lines normal to an $\mathcal S_\phi$ plane, in near and wave zones ($56 M$ across).
The color intensity in each panel gives the vorticity of the lines, scaled as 
in Fig.~\ref{fig:22sup2-2VortexDensity}.}
\label{fig:Schw20NearZone}
\end{figure*}

As we discussed in Sec.\ \ref{sec:20Mag}, symmetries dictate that this 
mode have two families of vortex lines lying in planes $\mathcal S_\phi$
of constant $\phi$ and a third family consisting of azimuthal circles of constant
$r$ and $\theta$.  In Fig.\ \ref{fig:20Sphi}, we explored in detail the
wave-zone wrap-around-torus shapes of the $\mathcal S_\phi$ vortex lines,
and their vorticity patterns.  In the near zone, the line shapes and 
vorticities are
somewhat more complex.  We elucidate them in Fig.\
\ref{fig:Schw20NearZone}, where, to make the figure more understandable
and preserve some features lost in Fig.\ \ref{fig:20Sphi},
we show the two families of vortex lines in separate panels, left and center.  

As for the superposed $(2,2)$ and $(2,-2)$ perturbations of 
Sec.~\ref{sec:SuperposedPerts}, each family takes on both positive
and negative vorticities, but is predominantly one or the other. 
And unlike the (2,2) mode and the superposed mode, the (2,0) line families
do not map into each other after 
a $90^\circ$ rotation; rather, they have
distinct patterns (as one might expect, since their plane is $\mathcal S_\phi$
rather than the equatorial plane). On the other hand, because of this 
mode's oscillating nature, 
the predominantly positive lines are the same as the predominantly negative
lines a half-cycle previous (with signs reversed). For this reason, we
illustrate the two families at the same
moment in time, the moment when the horizon vorticity reaches a maximum with
blue (dark gray) near the poles and red (light gray) near the equator. 

One striking feature of Fig.\ \ref{fig:Schw20NearZone} 
is a set of isolated points where
six lines meet, three from each family (three in each panel).  
These are nodes (zeros) of the frame-drag field, as one
can see from the fact that the coloring there is white.  These are also points
where, dynamically, the field lines can reconnect, changing their
topologies.

Let us focus on the near-horizon, predominantly negative vortex lines 
(dashed lines) 
in the left panel of Fig.\ \ref{fig:Schw20NearZone}. 
The lines that emerge from the counterclockwise horizon vortex in the
equatorial region loop over the north or
south pole of the black hole, and reconnect to the opposite side of that
horizon vortex. We think that, as the
mode oscillates, these lines will merge at the equator then slide off the 
horizon and form
closed loops surrounding the hole, of the sort that we see in the 
outer parts of the lenticular blue (dark gray) region of the center panel, and 
these will then expand and deform and reconnect to form
the set of wrap-around-deformed-torus lines of the left panel, which lie
in the outer part of the transition zone and are becoming outgoing
gravitational waves.  

Next focus on the near-horizon, predominantly positive vortex lines (solid
lines) in the middle panel of Fig.\ \ref{fig:Schw20NearZone}. 
The lines, that emerge from the clockwise horizon vortex in the north polar
region, swing around the equator and descend into the south polar horizon vortex.
We think that, as the oscillation proceeds, these lines will slide
off the horizon and immediately form closed loops that wrap around deformed 
tori, which expand to become like those near the left and right edges of the 
left panel (outer part of transition zone), and then continue their expansion,
becoming the gravitational-wave wrap-around-torus lines whose inner parts
are at the left and right edges of the right panel.

Notice, in the middle panel near the equator, two 
regions of weakly negative (pink [light gray]) vorticity, and their
near-zone lines that appear to have just disconnected from the horizon but
are mostly radially directed.  And notice similarly the pink regions near
the left and right edges of this panel, again with vortex lines that are
traveling roughly radially.  These pink regions are actually toroidal, because
of the rotation symmetry around the vertical axis. In the outer transition
zone and the wave zone, they are the regions
in which this family's wrap-around-torus, gravitational-wave vortex lines 
are crossing over from one clockwise vortex (wave crest) to another.  
This feature of crossover lines with weakly reversed vorticity 
appears to be a robust feature of oscillatory
modes. For other examples, see the weakly blue regions in the left panel, and 
see the superposed (2,2) and (2,-2) mode in 
Fig.\ \ref{fig:22sup2-2VortexDensity},  where the dashed vortex lines, with 
predominantly counterclockwise (red [light gray]) vorticity, become weakly 
blue (darker gray) in the crossover regions.  

The right panel of Fig~\ref{fig:Schw20NearZone} shows the vorticity of the axial lines
(constant $r,\theta$ circles) in both near zone and wave zone.  
Near the horizon, these lines are largely part of the transverse,
isotropic piece of the
longitudinal field $\delta {\boldsymbol{\mathcal B}}^{\rm L}$; they have
opposite color to the horizon vortexes at the horizon, as they must, 
in order to keep
$\delta {\boldsymbol{\mathcal B}}^{\rm L}$ trace-free. Near the horizon,
these lines also contain a smaller component 
of the ingoing-wave transverse-traceless field 
$\delta {\boldsymbol{\mathcal B}}^{\rm TT}$. 
In the wave zone, they
are fully outgoing-wave $\delta {\boldsymbol{\mathcal B}}^{\rm TT}$.

\subsection{Vortex lines of $(2,0)$ electric-parity mode and perturbative
tendex lines of (2,0) magnetic-parity mode}
\label{sec:20ElectricVortexes}

For the (2,0) electric-parity mode of Schwarzschild in RWZ gauge,
the only nonzero components of the frame-drag field are
$\delta \mathcal B_{\hat r \hat \phi}$ and $\delta \mathcal
B_{\hat \theta \hat \phi}$. Near the horizon, where decomposition
into longitudinal, longitudinal-transverse, and transverse-traceless parts
is meaningful, 
$\delta \boldsymbol{\mathcal B}^{\rm L}$ vanishes (and hence the horizon vorticity
vanishes), 
$\delta \mathcal B_{\hat r \hat \phi}$ is the sole component
of $\delta \boldsymbol{\mathcal B}^{\rm LT}$, and $\delta \mathcal
B_{\hat \theta \hat \phi}$ is the sole component of $\delta \boldsymbol{\mathcal
B}^{\rm TT}$. 

By (near) duality, the same is true for the (2,0) magnetic-parity mode
of the last subsection, with $\delta \boldsymbol{\mathcal B}$ replaced by
$\delta \boldsymbol{\mathcal E}$.

\begin{figure}[tb]
\includegraphics[width=0.95\columnwidth]{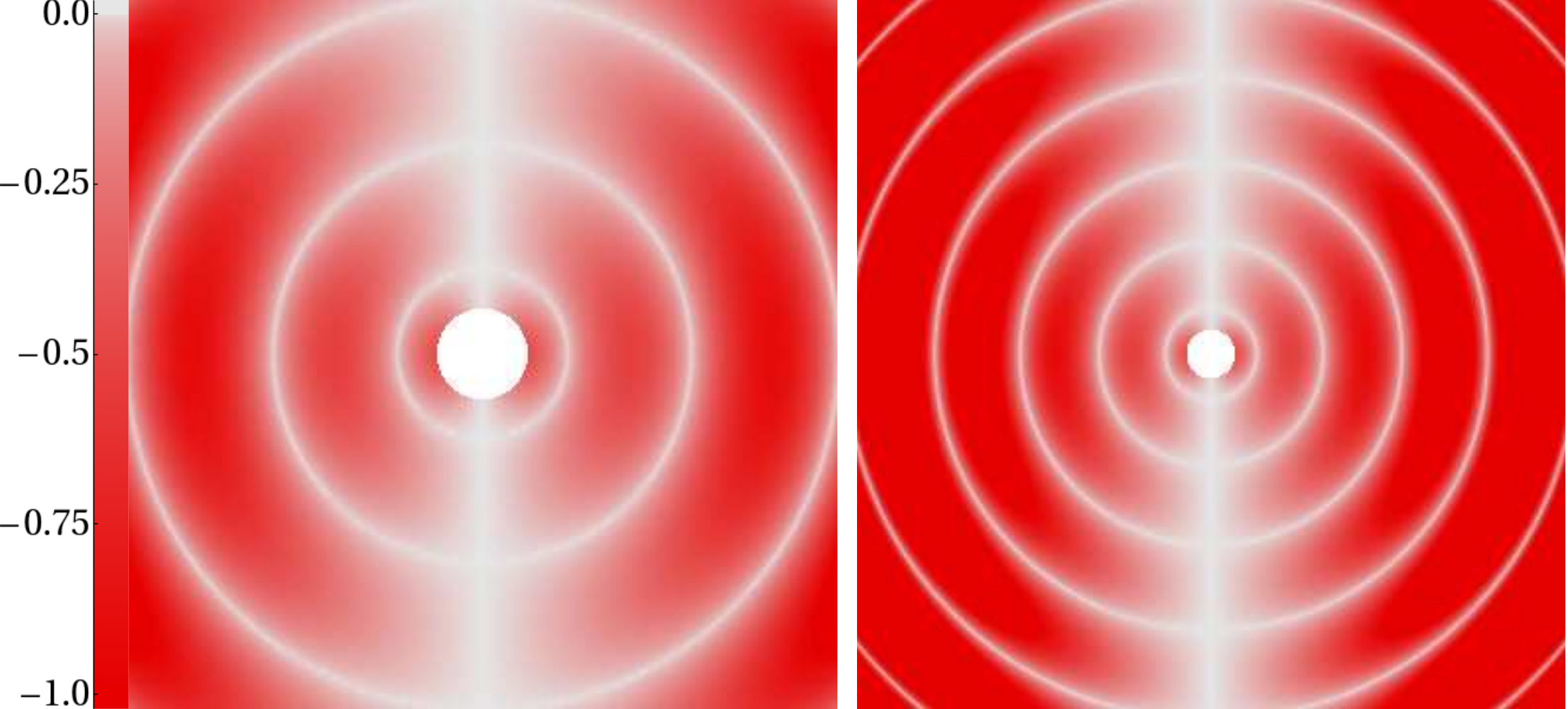}
\caption{(color online).
For the (2,0) electric-parity mode of Schwarzschild in RWZ gauge: 
the vorticity of the 
counterclockwise vortex lines that pass through the plane $\mathcal S_\phi$
of constant $\phi$.  The color intensity (scale on left) is scaled as in Fig.~\ref{fig:22sup2-2VortexDensity}.
{\it Left panel}: $30M$ across, showing the near and intermediate zones and beginning
of the wave zone. {\it Right panel}: $56M$ across.}
\label{fig:Schw20EvenNegativeVort}
\end{figure}

Because the only nonzero components are 
$\delta \mathcal B_{\hat r \hat \phi}$ and $\delta \mathcal
B_{\hat \theta \hat \phi}$ and because of the axisymmetry, 
there is a family of zero-vorticity vortex lines which lie in a plane 
$\mathcal S_\phi$ 
of constant $\phi$, and the other two sets of vortex lines have equal and
opposite vorticity and pass through $\mathcal S_\phi$ at 45 degree angles. 
In Fig.~\ref{fig:Schw20EvenNegativeVort}, we show in $\mathcal S_\phi$
the vorticity of the counterclockwise lines that pass through it.  A plot
for the clockwise lines would be identical, but with blue changed into red.

Notice the remarkable absence of structure in the near zone.  All we see
is toroidal vortexes separated by circular null surfaces and a polar
null line. (Recall the 
axisymmetry around the vertical polar axis).  The absence of structure
is presumably due to the fact that this mode is sourced by the longitudinal
perturbative tendex field, and not by this frame-drag field (though its
longitudinal-transverse part plays a key role of periodically storing
near-zone energy during the oscillations; cf.\ the discussion of the dual 
mode below).  The vorticity
vanishes along the polar axis because of axisymmetry and the fact that
the radial-radial part of the frame-drag field vanishes. 

\begin{figure}[tb]
\includegraphics[width=0.45\columnwidth]{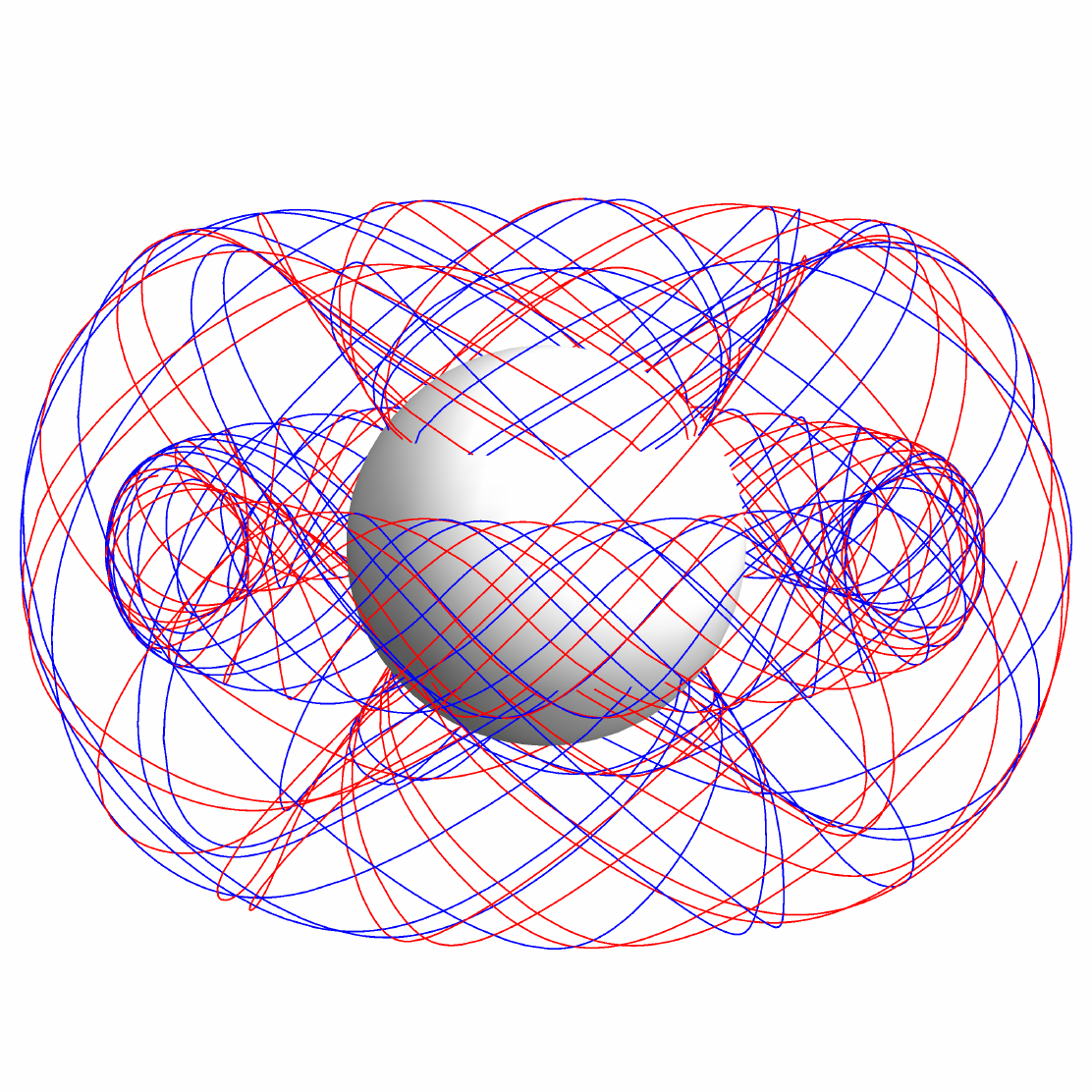}
\includegraphics[width=0.45\columnwidth]{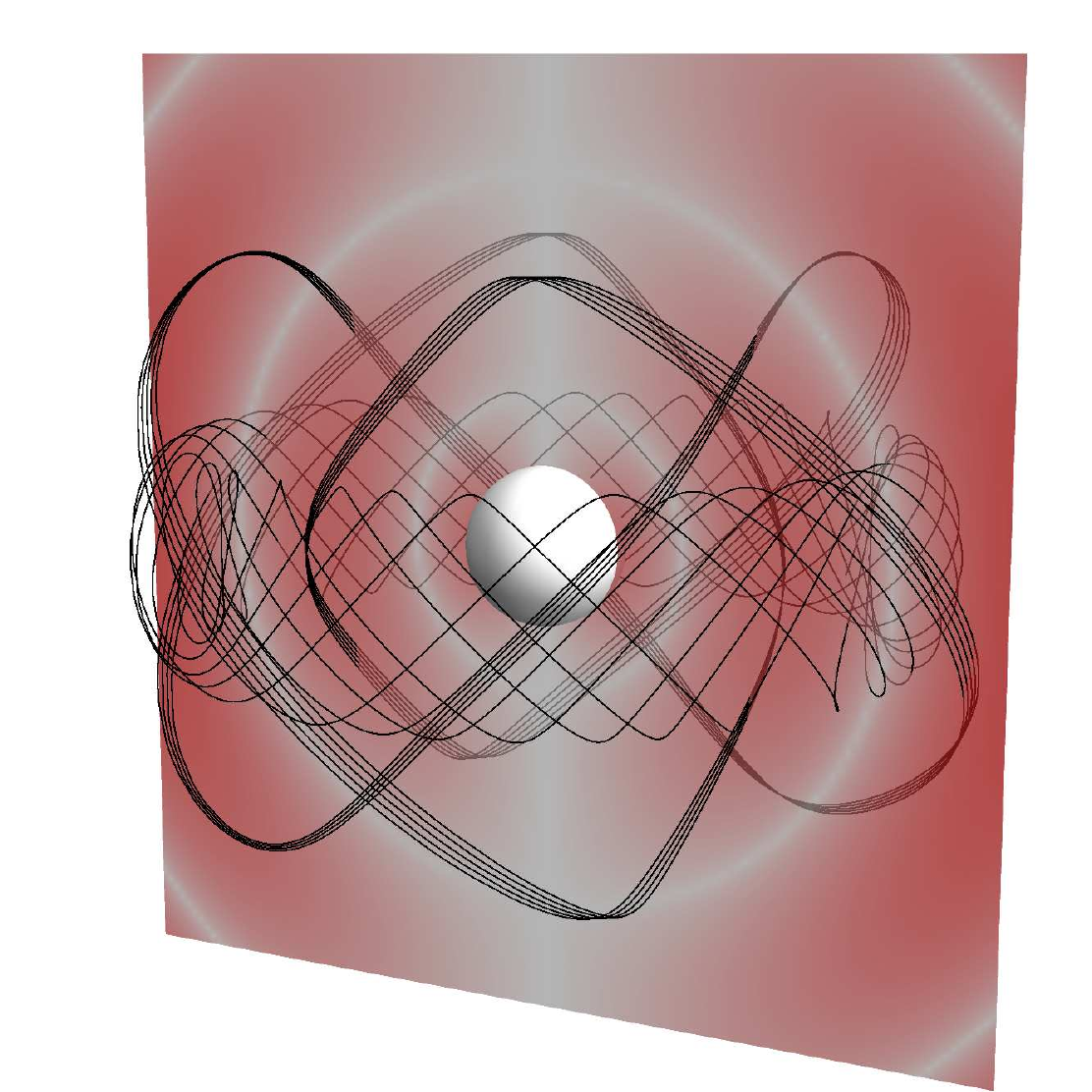}
\caption{(color online).
Vortex lines for the $(2,0)$ electric-parity mode of Schwarzschild in RWZ
gauge. 
{\it Left panel}: Two vortex lines of each sign 
in the near zone and innermost part of transition zone.
The positive (blue lines) and
negative (red) lines are identical but wind their tori in opposite directions.
{\it Right panel}: Two counterclockwise vortex lines in the transition zone,
with the vorticity shown in a semitransparent slice $\mathcal S_\phi$
of constant $\phi$ as a density plot. The vortex lines are plotted in black rather than red in this panel to aid the eye.}
\label{fig:Schw20EvenVortLines}
\end{figure}

For greater insight into this frame-drag field, we show in the left panel of  
Fig.~\ref{fig:Schw20EvenVortLines} several of its three-dimensional 
vortex lines in the near zone and innermost part of the transition zone.  
These vortex lines wind densely around axisymmetric deformed tori. 

Note that the large torus is attached to the horizon. Its vortex
lines intersect the horizon at the approximately $45^\circ$ angles 
characteristic of the longitudinal-transverse part of the field, which is
these lines' dominant component.  

For the dual, magnetic-parity (2,0) mode, 
this torus depicts the perturbative tendex lines of the near zone,
and those lines predominantly belong to the longitudinal-transverse
part of the tidal field, 
$\delta \boldsymbol{\mathcal E}^{\rm LT}$.  This is the part that
stores the mode's near-zone oscillation energy when 
$\delta \boldsymbol{\mathcal B}^{\rm L}$ is passing through zero and its
perturbative vortex lines are detached from the horizon
(see the discussion of this mode's dynamics in Sec.\ \ref{sec:20Mag}).  
Immediately after this snapshot, these tendex lines' $\delta
\boldsymbol{\mathcal E}^{\rm LT}$ begin to regenerate the near-horizon
longitudinal frame-drag field $\delta \boldsymbol{\mathcal B}^{\rm L}$
and its horizon vorticity.  As it does so, these tendex lines 
and their torus (presumably) detach from the horizon and expand outward 
into the transition then wave zone, becoming the tendex-line component
of a gravitational-wave torus like those displayed in Fig.\ 
\ref{fig:Schw20EvenWaveZone} above.  

The small torus in the left panel of Fig.~\ref{fig:Schw20EvenVortLines} 
encircles the equatorial point on the innermost node of the field
(innermost white circle in Fig.\ \ref{fig:Schw20EvenNegativeVort}).
It is also the innermost torus shown in Fig.\
\ref{fig:Schw20EvenWaveZone} above.

In the right-hand panel of
Fig.~\ref{fig:Schw20EvenVortLines}, for the (2,0) electric-parity mode
we show two counterclockwise vortex
lines in the transition zone. By chance, the larger of the selected lines nearly
forms closed orbits, and so even after wrapping its torus four times it
appears as a thin ribbon. While it is difficult to tell with this nearly
closed line, extended integration reveals that it does wrap a (deformed)
torus. 

Notice that both lines (both tori) in the right panel
straddle the second null of the frame-drag field (second-from-center
white circle in Fig.\ \ref{fig:Schw20EvenNegativeVort}).

\section{Conclusions}
\label{sec:Conclusions}

Although the theory of black-hole quasinormal modes is roughly 
half a century old, most past studies of them have focused on their
mathematical properties, their eigenfrequencies and emitted 
gravitational waves, and their excitation by various physical processes.
Aside from a geometric-optics interpretation of high-frequency modes 
(see, e.g., \cite{FerrariMashhoon1984,Dolan2010,Yang2012}),
little was known, before this paper, about their geometrodynamic
properties---e.g., the structure and dynamics of their spacetime
curvature in the near zone and transition zone, and how the
near-zone curvature generates gravitational waves.

In this paper we have used our new vortex and tendex tools to 
explore, in depth, the geometrodynamics of the quadrupolar modes
for Schwarzschild and Kerr black holes.  Most importantly, we have
discovered that: 

(i) There is a near duality between
electric-parity modes and magnetic-parity modes, in which tendexes
get mapped into vortexes and vortexes into tendexes.  

(ii) The electric-parity $(l,m) = (2,2)$ and (2,1) modes 
are generated by near-zone, longitudinal vortexes that extend out 
of the horizon and rotate (four tendexes for $m=2$; two for $m=1$).
The vortexes' rotation generates outgoing and ingoing gravitational
waves; the ingoing waves act back on the longitudinal vortexes, 
gradually pushing them off the horizon, which results in the mode's
exponential attenuation.  

(iv) By (near) duality, the electric-parity (2,2) and (2,1) modes
are generated and attenuated in the same way, but with near-zone,
longitudinal tendexes rather than vortexes playing the central role.

(v) The magnetic-parity (2,0) mode and superposed (2,2) \& (2,-2) mode 
are generated by near-zone, longitudinal vortexes that extend out
of the horizon and oscillate between clockwise and
counterclockwise vorticity. In these oscillations, energy is fed
back and forth between the longitudinal vortexes and 
longitudinal-transverse,
near-zone tendexes that do not penetrate the horizon.  
In each 
oscillation, as the horizon vorticity passes through zero, the
longitudinal vortex lines slide off the horizon and reconnect to form
toroidal vortexes that travel outward, becoming gravitational waves;
and the near-zone tendexes then regenerate the longitudinal vortexes
(with reversed vorticity), thereby triggering the next half cycle
of oscillation.

(vi) The electric-parity (2,0) mode and superposed (2,2) \& (2,-2) mode
exhibit these same geometrodynamics, but with the roles of the vortexes
and tendexes reversed.

In future papers, these quasinormal-mode 
insights will be a foundation as we explore the geometrodynamics of 
merging binary black holes using numerical simulations. 

While all analytic 
approximations fail near the time of merger, black-hole perturbation theory 
does approximate a binary-black-hole spacetime well in some epochs: 
the merged hole during its ringdown, 
each tidally deformed hole during inspiral, and each perturbed hole during 
the initial relaxation that causes spurious ``junk'' gravitational 
radiation. 
Before exploring the fully nonlinear vortex and tendex structures in 
simulations, 
we are likely first to
compare numerical vortex and tendex structures 
during these epochs
with the corresponding perturbative results 
(as in Fig.~\ref{fig:NRComparison}). 
Such comparisons will allow us to determine to what degree the insights 
we have gained from our perturbative studies can also be applied to 
numerical simulations---particularly the relative insensitivity of vortexes 
and tendexes to changes in gauge and slicing.

Building on these comparisons, our future work will then include 
initial explorations of the fully nonlinear 
geometrodynamics of the warped spacetimes present in binary-black-hole 
simulations. For example, Kamaretsos, Hannam, and 
Sathyaprakash~\cite{Kamaretsos:2012bs} 
have recently observed relationships 
between the properties (masses and spins) of the initial holes 
in a binary-black-hole merger and the particular quasinormal modes that 
are excited in the remnant (modes which generate the ringdown portion of the 
gravitational waves). By examining the vortex and tendex structures of a 
variety of binary-black-hole mergers, we hope to gain insight into the origin 
of such relationships. 
Also, following Dennison and Baumgarte's recent 
exploration~\cite{Dennison2012} of the vortex and tendex structures in 
approximate, perturbative initial data, we intend to explore the vortex and tendex 
structures of constraint-satisfying binary-black-hole initial data, which 
could give insight into the initial perturbations (and the corresponding 
spurious ``junk'' gravitational radiation) that 
appear in all currently used, binary-black-hole initial-data schemes.

Ultimately, we plan to use vortexes and tendexes to explore the 
geometrodynamics of binary-black-hole spacetimes throughout the entire 
simulated inspiral, merger, and ringdown. We expect that these tools will 
provide insights into the behavior of these spacetimes and perhaps 
also motivate new ways of constructing phenomenological waveform templates for 
use in gravitational-wave data analysis.

\acknowledgments
We thank John Belcher, Jeandrew Brink, and Richard Price for helpful discussions. We thank Jeff Kaplan for helpful discussions and web assistance. We thank Mark Scheel for helpful discussions and for version control assistance. A.Z. would also like to thank the National Institute for Theoretical Physics of South Africa for hosting him during a portion of this work. Some calculations in this paper were performed using the Spectral Einstein Code ({\tt SpEC})~\cite{SpECwebsite}.
This research was supported by NSF grants PHY-0960291, PHY-1068881 and CAREER grant PHY-0956189 at Caltech, by NSF grants PHY-0969111 and PHY-1005426 at Cornell, by NASA grant NNX09AF97G at Caltech, by NASA grant NNX09AF96G at Cornell, and by the Sherman Fairchild Foundation at Caltech and Cornell, the Brinson Foundation at Caltech, and the David and Barbara Groce fund at Caltech.

\appendix

\section{Quasinormal Modes of a Schwarzschild Black Hole in Regge-Wheeler Gauge}
\label{sec:RWApp}

In this appendix, we review the Regge-Wheeler-Zerilli (RWZ) formalism for black-hole perturbations, and we discuss the calculations that underlie the results reported in Secs.~\ref{sec:22Perts},~\ref{sec:SuperposedPerts}, and~\ref{sec:OtherPerts} for quadrupolar perturbations of non-spinning black holes in the RWZ
gauge.

\subsection{Regge-Wheeler-Zerilli formalism}
\label{sec:ReggeWheeler}

Here we review the equations governing quasinormal modes for a non-rotating black hole in the 
Regge-Wheeler-Zerilli gauge~\cite{ReggeWheeler1957,Zerilli1970c}.\footnote{
\label{fn1}There are many errors in the original paper of Regge and Wheeler 
\cite{ReggeWheeler1957}, most of which were corrected by Edelstein and Vishveshwara
\cite{Edelstein1970}. We use the corrected equations without further comment.}
We write the metric in Schwarzschild coordinates with a small perturbation $h_{\mu\nu}$,
\begin{eqnarray}
ds^2 &=& -\alpha^2 dt^2 + \frac{dr^2}{\alpha^2} + r^2(d\theta^2 + \sin^2\theta d\phi^2) + h_{\mu\nu} 
dx^\mu dx^\nu \:, \nonumber\\
 \alpha^2&=&1-2M/r \:.
\label{RWmetric}
\end{eqnarray}
The components of $h_{\mu \nu}$ obey separable differential equations, and importantly $h_{\mu \nu}$ can be split into definite-parity perturbations (electric and magnetic) which do not couple to each other.

For magnetic-parity perturbations, the only nonzero components of 
$h_{\mu \nu}$ in Regge-Wheeler gauge are
\begin{equation}
h_{tA} = h_0(r) e^{-i\omega t} X^{l m}_A(\theta,\phi)\;, \quad
h_{rA} = h_1(r) e^{-i\omega t} X^{l m}_A(\theta,\phi)\;.
\label{h0h1def}
\end{equation}
Here $\omega$ is the mode's complex QNM eigenfrequency, and $X^{l m}_A$ is the magnetic-parity vector spherical harmonic on the unit two-sphere,
\begin{equation}
X^{l m}_\theta  = - \csc \theta \, {Y^{l m}}_{,\phi}\;, 
\qquad 
X^{l m}_\phi  =\sin \theta \,  {Y^{l m}}_{,\theta}\;,
\label{Phiadef}
\end{equation}
with $Y^{l m}(\theta,\phi)$ denoting the scalar spherical harmonics. Regge and Wheeler \cite{ReggeWheeler1957,Edelstein1970} showed that the radial parts of the metric perturbation, $h_0(r)$ and $h_1(r)$, can be expressed in terms of a single scalar radial eigenfunction $Q(r)$ as
\begin{equation}
h_0 = - \frac{\alpha^2}{i\omega}(rQ)_{,r}\;, \quad
h_1 =  \frac{rQ}{\alpha^2}\;,
\label{h01Q}
\end{equation}
which 
satisfies the eigenequation
\begin{equation}
Q_{,r_*r_*} + \omega^2 Q = \mathcal V_Q(r) Q\;, \;\; \mathcal V_Q(r) = \alpha^2\left(\frac{l(l+1)}{r^2} - 
\frac{6M}{r^3}\right)\;.
\label{Qdeq}
\end{equation}
Here $r_*$ is the tortoise coordinate
\begin{equation}
dr_* = \frac{dr}{\alpha^2}\;, \quad r_* = r + 2M \ln(\alpha^2 r/2M)\;,
\label{rstardef}
\end{equation}
which goes to $+\infty$ far from the hole and $-\infty$ at the hole's horizon.  
This eigenequation must be solved subject to the boundary conditions of outgoing waves
at infinity, $Q \sim e^{i\omega r_*}$ as $r_* \rightarrow + \infty$, and ingoing waves at the horizon, $Q \sim e^{-i\omega r_*}$ as $r_* \rightarrow - \infty$.

For electric-parity modes,
the nonzero components of $h_{\mu\nu}$
in RWZ gauge are~\cite{Zerilli1970c}
\begin{eqnarray}
h_{tt} = \alpha^2 H_0(r)e^{-i\omega t} Y^{l m} \;,&& h_{rr} = \frac{H_0(r)}{\alpha^2} e^{-i\omega t} Y^{l m}\;, \nonumber \\
h_{tr} = H_1(r)e^{-i\omega t} Y^{l m}\;, && h_{AB} = r^2 \Omega_{AB} K(r)
e^{-i\omega t} Y^{l m}\;. \nonumber \\
\label{EPmetric}
\end{eqnarray}
Here $\Omega_{AB}$ denotes the metric on the unit 2-sphere. We can write the metric perturbation functions in terms of the Zerilli function $Z(r)$ 
as\footnote{In $H_1$ we have corrected a term in the numerator of the fraction: the last term, $-3M^2$, was
incorrectly written as $-3M$ by Zerilli, an error that should be obvious on dimensional grounds. }
\begin{eqnarray}
K&=&\left[\frac{\lambda(\lambda+1)r^2 + 3\lambda M r + 6M^2}{r^2(\lambda r + 3M)}\right]
 Z + \alpha Z_{,r}\;,
\nonumber \\
H_1&=&-i\omega \left[\frac{\lambda r^2 -3\lambda M r -3 M^2}{(r-2M)(\lambda r+3M)}\right]
Z - i\omega r Z_{,r}\;,
\nonumber \\
H_0 &=& \left[\frac{\lambda r(r-2M) - \omega^2 r^4 + M(r-3M)}{(r-2M)(\lambda r+3M)}\right]
 K \nonumber\\
&&+ \left[\frac{(\lambda +1)M - \omega^2 r^3}{i\omega r(\lambda r+ 3M)}\right] H_1\;.
\label{KH0H1inZ}
\end{eqnarray}
Here we have used Zerilli's notation
\begin{equation}
\lambda = \frac12 (l-1)(l+2)\;.
\label{lambdadef}
\end{equation}
The Zerilli function satisfies the eigenequation
\begin{equation}
Z_{,r_*r_*} + \omega^2 Z = \mathcal V_z(r) Z\;,
\label{ZerilliEqn}
\end{equation}
where
\begin{equation}
\mathcal V_z(r) = \alpha^2\left[\frac{ 2\lambda^2(\lambda+1)r^3 + 6\lambda^2 Mr^2 + 18\lambda M^2 r
+18M^3}{r^3(\lambda r+3M)^2}      
\right].
\label{Vzdef}
\end{equation}

The slices of constant Schwarzschild time $t$ do not intersect the black hole's horizon, so 
in performing our 3+1 split, we use slices of constant Eddington-Finklestein
time $\tilde t = t +2M \ln (r/2M-1)$. 
Written in Schwarzschild coordinates, the perturbed tetrad for the EF observers
is given by Eqs.~(\ref{EFbasis}) and~ (\ref{eq:PerturbedEFTetrad}).
For any chosen mode, we compute the frame-drag and tidal fields by 
(i) computing, from the metric-perturbation components $h_{\mu\nu}$, 
the perturbation $\delta R_{\alpha\beta\gamma\delta}$
to the Riemann tensor (same as Weyl) in Schwarzschild coordinates;
(ii) projecting the total Riemann tensor $R_{\alpha\beta\gamma\delta}
= R^{(0)}_{\alpha\beta\gamma\delta} + \delta R_{\alpha\beta\gamma\delta}$
(where $R^{(0)}_{\alpha\beta\gamma\delta}$ is the unperturbed Riemann
tensor)
onto the perturbed EF tetrad; (iii) reading off $\mathcal E_{\hat a \hat b}
= R_{\hat a \hat 0 \hat b \hat 0}$ and $\mathcal B_{\hat a \hat b}
= \tfrac 12\epsilon_{\hat a\hat p\hat q} R_{\hat p \hat q \hat 0 \hat b}$ and 
splitting them into
their unperturbed and perturbed parts.

\subsection{Magnetic-parity $(2,m)$ mode: Frame-drag field}
\label{sec:l2m2Vortexes}

We first focus on the $(2,m)$ quadrupolar modes for magnetic-parity 
perturbations. Carrying out the above computation, expressing the
answer for the frame-drag field in terms of the Regge-Wheeler
function $Q(r)$ and the electric-parity scalar, vector, and tensor
harmonics (see discussion in App.\ \ref{sec:IRGSchwarzschild}), 
and simplifying using Eq.\ (\ref{Qdeq}), we obtain:

\begin{subequations}
\label{eq:Bij22RWZ}
\begin{align}
\label{Bij22}
\mathcal B^{(1)}_{\hat r \hat r} &= \Re \left[B_{1(\rm{m})}   e^{-i\omega t} Y^{2m} \right]\;, 
\\
\mathcal B^{(1)}_{\hat r \hat A} &=  \Re \left[B_{2(\rm{m})} e^{ -i \omega t}  Y^{2m}_{\hat A} \right]\;,  \\
\mathcal B^{(1)}_{\hat A \hat B} & = \Re \left [\left(-\frac12 B_{1(\rm{m})} \delta_{\hat A \hat B} Y^{2m} +  B_{3(\rm{m})} Y^{2m}_{\hat A \hat B} \right)e^{ -i \omega t} \right], \\
\label{eq:RWB1m}
B_{1(\rm{m})} (r)& = -\frac{12 Q}{ i \omega r^3 } \:,\\
\label{eq:RWB2m}
 B_{2(\rm{m})} (r)& = -\frac{4i M \omega Q +2 \alpha^2 r Q' }{i \omega r^3 \alpha^2\sqrt{1+2M/r}} \,,
\end{align}
  
\begin{widetext}
\begin{flalign}
\label{eq:B3mRW}
B_{3(\rm{m})} (r)=  -\frac{1}{ i \omega r^5 \alpha^4 (r+2M)} \Big( 
&\left[3\alpha^2(r-M)(r^2+4 M^2) + 4 i M \omega r^2(r-3M) - r^3 \omega^2 (r^2 + 4 M^2) \right] Q 
\notag \\ &
+ r \alpha^2 \left[(r-3M)(r^2+4M^2)+4 i M \omega r^3\right]Q' \Big) \;,
\end{flalign}   
\end{widetext}
\label{eq:FD22mag}
\end{subequations}
where a prime denotes a derivative with respect to $r$, $Y^{lm}_{A}$ and $Y^{lm}_{AB}$ are given by Eqs.~\eqref{ElectricHarmonics}, and $\delta_{\hat A \hat B}$ is the Kronecker delta.

We have solved the Regge-Wheeler equation (\ref{Qdeq}) numerically for the most slowly
damped, quadrupolar normal mode.  When the numerical solution is inserted into the above expressions for $\mathcal B^{(1)}_{\hat a \hat b}$, numerical errors cause problems
with delicate cancellations in the
transverse-traceless and radial-transverse
components near the horizon.  To deal with this, we have derived the following asymptotic
formula for $Q(r)$ near the horizon, $ r_*/M \ll -1$:
\begin{eqnarray}
Q&=&Y^{-2 i M \omega } 
\left[
1
+\frac{3 Y}{ (1-4 iM \omega )e} \right. \nonumber \\
&&
+\frac{9 i M \omega Y^2}{(1- 4 i M \omega)(1 - 2 i M \omega)e^2}
\nonumber \\
&&
\left.
-\frac{3(1+12 i M \omega + 40 M^2 \omega^2)Y^3}{2(1-4 i M \omega)(1-2 M i\omega)(3-4i M \omega)e^3}
+ O(Y^4) \right]\; \nonumber\\
\label{Qanalytic}
\end{eqnarray}
where $Y=e^{r_*/2M}$. Inserting this into Eqs.\ (\ref{eq:Bij22RWZ}), we find,
of course, that all components of $\mathcal B_{\hat a \hat b}$ are finite 
at the horizon. 

Using Eqs.~(\ref{eq:Bij22RWZ}) for the frame-drag field, our analytic formula (\ref{Qanalytic}) for
$Q(r)$ near the horizon, and our numerical solution for $Q(r)$ at larger radii, and the $(2,2)$ harmonics, we compute the vortex lines and their vorticities for the fundamental $(2,2)$ quasinormal mode. We illustrate them in Figs.~\ref{fig:22MagVortex},
\ref{fig:22SmagEquator}, \ref{fig:22Smag3D} and
\ref{fig:Sch22VortexAll}. 

For our superposition of the $(2,2)$ and $(2,-2)$ modes we can simply sum the 
$(2,2)$ and $(2,-2)$ harmonics in the above expressions. 
We plot the vortex lines for the resulting frame-drag field 
in Figs.~\ref{fig:22-2Smag},
\ref{fig:22sup2-2VortexDensity},
and the top row of Fig.\ \ref{fig:TimeSeries}. 
We use the $(2,1)$ harmonics for generating the vortexes of the 
magnetic-parity, $(2,1)$ perturbations that are illustrated in 
Fig.~\ref{fig:21MagneticVorts}. 
Finally, we use the $(2,0)$ harmonics to produce the vortexes and vortex lines 
of the $(2,0)$ magnetic-parity perturbation. 
We note that $Y^{20}_{\hat \phi} = Y^{20}_{\hat \theta \hat \phi} = 0$ for 
this mode. 
This means that $\mathcal B^{(1)}_{\hat a \hat b}$ is block-diagonal, and the 
vortex lines split into a pair of lines which remains in a slice of constant 
$\phi$ and a single, axial line that runs in circles of constant $(r,\theta)$. 
In a slice of constant $\phi$, we illustrate the two sets of vortex lines in 
the slice and their voticity together in Fig.~\ref{fig:20Sphi} and separately 
in the left and middle panels of Fig.~\ref{fig:Schw20NearZone}. 
We also plot the vorticity of the axial lines in a slice in the right panel of Fig.~\ref{fig:Schw20NearZone}.

\subsection{Electric-parity $(2,m)$ modes: Frame-drag field}

Carrying out the calculation described at the end of
Sec.\ \ref{sec:ReggeWheeler} using the electric-parity metric perturbation
(\ref{EPmetric}), expressing the result in terms of the Zerilli 
function $Z$ with
the aid of Eqs.~\eqref{KH0H1inZ}, and 
simplifying using Zerilli's differential equation 
\eqref{ZerilliEqn}, we obtain for the frame-drag field of a $(2,m)$ electric-parity perturbation
\begin{subequations}
\begin{align}
\mathcal B^{(1)}_{\hat r \hat r } & = 0 \,, \\
\mathcal B^{(1)}_{\hat r \hat A} & = \Re \left[ B_{1(\rm{e})} e^{- i\omega t} X^{2m}_{\hat A} \right]\,, \\
\mathcal B^{(1)}_{\hat A \hat B} & = \Re \left [  B_{2(\rm{e})}e^{- i\omega t} X^{2m}_{\hat A \hat B} \right] \,, 
\end{align}
where
\begin{widetext}
\begin{align}
\label{eq:RWB1e}
B_{1(\rm{e})} =& \frac{ \left[ 6 M^2 \alpha^2 - i \omega r^2 (2r + 3M) \right] Z - 2 M r \alpha^2 (2r+ 3 M) Z' }{ 2 r^5 \alpha^2 \sqrt{1+2M/r}} \,, \\
\label{eq:RWB2e}
B_{2(\rm{e})}  =& \frac{1}{2r^4 \alpha^4(2r+3M) (r+2M)}
\Big( \left[ -12 M \alpha^2(M^2 + 4 r \beta_1) - i \omega (r^2 + 4M^2)\beta_2 + 4 M \omega^2 r^3 (2r+ 3M) \right] Z  \notag \\ 
& \qquad \qquad \qquad  \qquad \qquad \qquad - r \alpha^2 \left[ 4 M \beta_2 +i \omega r (2r+3M) (r^2 + 4 M^2) \right] Z' \Big) \,.
\end{align}
\end{widetext}
Here a prime denotes a derivative with respect to $r$, and $X^{lm}_{A}$ and $X^{lm}_{AB}$ are the magnetic-parity vector and tensor spherical harmonics
given by Eqs.~\eqref{MagneticHarmonics}
(see discussion in App.\ \ref{sec:IRGSchwarzschild}). 
We have defined here for convenience the functions
\begin{align}
\beta_1 &= \frac{r^2 + M r + M^2}{2r + 3M} \,, & \beta_2 & = (2r^2 - 6Mr - 3M^2) \,.
\end{align}
\label{eq:FD22electric}
\end{subequations}
We note again that the horizon vorticity, $\mathcal B^{(1)}_{NN}$, vanishes.
With this $\mathcal B^{(1)}_{\hat a \hat b}$ we can again compute the 
eigenvector fields and eigenvalues for the perturbed spacetime, and from them 
compute the vortex lines. 
We use these expressions to calculate the vortex lines and their vorticities 
generated by electric-parity perturbations. 
In order to compute the vortex lines for these modes, once again we expand $Z$ 
around the horizon in terms of $Y = e^{r_*/2M}$ up to $O(Y^3)$.
We use this series to match to a numerical solution of the Zerilli equation 
subject to ingoing-wave boundary conditions. 
Because the Zerilli potential $\mathcal V_z$ is more complicated than the 
Regge-Wheeler potential $\mathcal V_Q$, the coefficients of the expansion of 
$Z$ in powers of $Y$ are lengthy, but easily computed using algebraic computing
software such as Mathematica. 
For this reason, we do not give the coefficients here.

For an electric-parity $(2,2)$ perturbation, the only set of vortex lines that 
are confined to the equatorial plane have vanishing vorticity (and are of less
physical interest).
Instead, we used the above frame-drag field to compute, and then plot in 
Fig.~\ref{fig:Schw22EvenVortexDensity}, the vorticity of one of the
sets of vortex lines that pass through the equatorial plane at a 45 degree
angle: the set with negative vorticity. 
Just as with the magnetic-parity modes, we superpose a $(2,2)$ perturbation 
with a $(2,-2)$ perturbation by a simple sum of the harmonics.
The vorticity of these lines passing through the equatorial plane (the analog 
of Fig.~\ref{fig:Schw22EvenVortexDensity}) 
is plotted in Fig.~\ref{fig:22sup2-2VortexDensityEven}.

For the vortex lines of the $(2,1)$ mode, there is a reflection symmetry about 
the equatorial plane, which implies that there are two sets of vortex lines 
confined to the plane, with a third normal to it. 
We illustrate the vortex lines and 3D vortexes of this mode in 
Fig.~\ref{fig:21ElectricVorts}. 
Finally, when we use the $(2,0)$ harmonics, we note that 
$X^{20}_{\hat \theta} = X^{20}_{\hat \theta \hat \theta} = X^{20}_{\hat \phi \hat \phi} = 0$. 
While this means that the frame drag field is simple, it is not block-diagonal
and its nonzero vortex lines pass through all three dimensions. 
There is a single set of axial vortex lines with zero vorticity, and two sets 
with equal and opposite vorticity that wind around deformed tori. 
We illustrate the vorticity in a slice of constant $\phi$ for the negative 
lines in Fig.~\ref{fig:Schw20EvenNegativeVort}. 
In addition, we illustrate some 3D vortex lines in 
Figs.~\ref{fig:Schw20EvenWaveZone} and~\ref{fig:Schw20EvenVortLines}.

\subsection{Electric-parity $(2,2)$ mode: Tidal field}
\label{sec:RWZTidalField}

To help understand the slicing dependence of our results, we compare fields 
generated by electric-parity perturbations, because the slicings are identical 
for all magnetic-parity perturbations.
In particular, we focused on the perturbed tidal field for the electric-parity,
$(2,2)$ mode. 
Carrying out the calculation of this mode as above when using the 
electric-parity metric perturbation (\ref{EPmetric}), we obtain 
\begin{subequations} 
\begin{align}
\label{eq:RWE1e22}
\mathcal E^{(1)}_{\hat r \hat r} & = \Re\left[ E_{1(\rm{e})} e^{- i \omega t} Y^{22} \right] \, \\
\label{eq:RWE2e22}
\mathcal E^{(1)}_{\hat r \hat A} & =\Re \left[  E_{2(\rm{e})} e^{-i \omega t} Y_{\hat A}^{22} \right] \,, \\
\label{eq:RWE3e22}
\mathcal E^{(1)}_{\hat A \hat B} & =\Re \left[ \left(-\frac12 E_{1(\rm{e})} \delta_{\hat A \hat B} Y^{22} + E_{3(\rm{e})} Y_{\hat A \hat B}^{22}\right ) e^{- i \omega t}\right],
\end{align}
\begin{widetext}
\begin{flalign}
E_{1(\rm{e})} (r) =& -\frac{3 \mathcal Z}{2 r^3} \,, \\
E_{2(\rm{e})} (r)=&  \frac{\left[3 M \alpha^2 - 2 i M \omega (2r+3M) \right] Z -  \alpha^2 r (2r+ 3M) Z'}{2 r^4 \alpha^2 \sqrt{1 + 2M /r}} \,,\\
E_{3(\rm{e})} (r) = &\frac{1}{2 r^4 \alpha^4(r + 2M)}\biggl( \biggl[ -\frac{3\alpha^2(3M^3 + 6 M^2r + 4 M r^2 + 4r^3)(4M +r \alpha^2)}{ (2r + 3M)^2} + \frac{4 i M \omega r (3M^2 + 6 M r - 2r^2)}{ (2r + 3M)} \notag \\ & \qquad
 +\omega^2 r^3(4M + r \alpha^4)  \biggr] Z  + \alpha^2 r \biggl[  \frac{ (3M^2 + 6 M r - 2r^2)(4M +r \alpha^2)}{2r+3M} - 4 i M \omega r^2 \biggr] Z'
\biggr)\,,
\end{flalign}
\end{widetext}
\end{subequations}
where $\mathcal Z$ is a function which obeys the same Regge-Wheeler equation
\eqref{Qdeq} as $Q$ and can be built from the Zerilli function $Z$ as, (see, 
e.g., Ch.~4, Eq.~(156) of \cite{ChandrasekharBook})
\begin{align}
\mathcal Z & = \left[ \frac{\lambda^2(\lambda+1)}{3M} + \frac{3M\alpha^2}{r(\lambda r + 3M)} \right] Z - \frac{Z_{,r_*}}{\lambda} \,,
\end{align}
for integers $l\geq 2$. 
This implies there is an exact duality between $\mathcal E_{\hat r \hat r} $ 
for electric-parity perturbations and $\mathcal B_{\hat r \hat r}$ 
magnetic-parity perturbations [in fact for any $(l,m)$ mode] in RWZ gauge. 
This follows from the facts that these radial-radial components have the same 
time, radial, and angular dependence (but not necessarily the same amplitude 
and phase). 
However, we can fix the relative normalization of the Regge-Wheeler function 
$Q$ and Zerilli function $Z$ such that $ Q = - \omega \mathcal Z /8$, in which 
case we have for Eqs.~\eqref{eq:RWB1m} and~\eqref{eq:RWE1e22}
\begin{align}
B_{1(\rm{m})} (r)  = i  E_{1(\rm{e})}(r) \,.
\label{eq:RWZradialDuality}
\end{align}
Substituting $\mathcal Z$ into Eqs.~\eqref{eq:RWE2e22}---\eqref{eq:RWE3e22} does 
not illustrate the near-duality of the other components of 
$\boldsymbol{\mathcal E}^{(1)}$ and $\boldsymbol{\mathcal B}^{(1)}$ in an 
obvious manner, so we leave these equations in terms of $Z$.

As we discuss in the next section, however, the exact duality of 
$\mathcal E^{(1)}_{\hat r \hat r} $ and  $\mathcal B^{(1)}_{\hat r \hat r} $ 
does not immediately correspond to an exact duality of the horizon tendicity 
and vorticity.
This happens because in RWZ gauge, the electric-parity perturbations deform the
horizon, which changes the horizon tendicity.

For the electric-parity, $(2,2)$ perturbation, the tidal field is symmetric 
about the equatorial plane, and there are two sets of tendex lines that remain 
in the equatorial plane (just as the vortex lines of the $(2,2)$ 
magnetic-parity mode did).  
The tendex lines are illustrated in the left-hand panel of 
Fig.~\ref{fig:ElectricGaugeCompare}

\subsection{Perturbed horizon and horizon tendicity for electric-parity modes}
\label{sec:RWZHorizon}

We discuss here the correction to the position of the horizon and its influence
on the perturbed horizon tendicity for the electric-parity $(2,2)$ modes. 
First, we calculate the correction to the horizon position $\delta r$ using
the same procedure as that of Vega, Poisson, and Massey \cite{Vega:2011}. 
The horizon generators, $\vec l$, for the perturbed spacetime are given by
\begin{align}
l^\mu = \frac{\partial x^\mu}{\partial \tilde t} = (1 + \delta \dot t, \delta \dot r, \delta \dot \theta, \delta \dot \phi) \,,
\end{align}
with an overdot represents a derivative with respect to $\tilde t$. 
The functions $\delta \theta$ and $\delta \phi$ change the location of 
individual generators, but do not alter the shape of the surface defined by the 
instantaneous horizon.
We will not treat them here, but they are described in \cite{Vega:2011}. 
By requiring that the generators remain null to first order in the 
perturbation, we find
\begin{align}
\label{eq:equationdeltar}
\delta r - 4 M \delta \dot r = 2 M h_{ll} \,.
\end{align}
For IR gauge, $h_{ll} =0$ and the only physical solution of Eq.\ 
\eqref{eq:equationdeltar} is $\delta r  = 0$. 
Magnetic-parity RWZ perturbations also have $h_{ll}=0$, and, therefore, the 
coordinate location of the horizon does not change in this gauge either. 
For electric-parity perturbations in RWZ gauge, we use the fact that 
$h_{ll} = h_{\tilde t \tilde t}$ on the horizon to solve for the perturbation
to the horizon's shape.
For a general electric-parity perturbation of indices $(l,m)$, Eqs.\ 
\eqref{EPmetric} allow us to write
\begin{align}
\label{eq:deltarsolve}
\delta r = \Re\left[\frac{e^{-2 i M \omega}}{\kappa + i \omega}e^{-i \omega \tilde t} Y_{lm} \lim_{r \to 2M} \left( \frac{H_1 - H_0}{r - 2 M} e^{i \omega r^*} \right) \right] \,,
\end{align}
where $\kappa = (4 M)^{-1}$ is the horizon's unperturbed surface gravity. 
We evaluate these quantities on the horizon using the near-horizon expansion of
the Zerilli function $Z$, and they are finite.

The perturbation to the position of the horizon corrects the perturbative 
horizon tendicity in two ways: 
First, the background horizon tendicity $\mathcal E^{(0)}_{\hat r \hat r}$,
when evaluated at $r = 2M + \delta r$, becomes, through first order in 
$\delta r$,
\begin{align}
\label{eq:ENNposition}
\mathcal E^{(0)}_{\hat r \hat r} (r = r_H) &  = - \frac{1}{4 M^2} + \frac{3}{8M^3} \delta r \,.
\end{align}
Next, we recall that 
$\vec{e}_{\hat r}=\vec{e}^{\,(0)}_{\hat r}+\vec{e}^{\,(1)}_{\hat r}$ is normal 
to surfaces of constant $r$ through perturbative order.
Now that the horizon's surface is deformed, however, the normal to the horizon 
$\vec N$ is no longer precisely the same as $\vec{e}_{\hat r}$. 
It receives a correction such that
\begin{align}
N^\mu =&\frac{1}{N} \gamma^{\mu \nu} \nabla_\nu ( r + \delta r) \notag \\
=& e_{\hat r}^{(0)\mu} + e_{\hat r}^{(1)\mu} + \delta N^\mu  - \left(\delta N_{\nu} e^{(0)\nu}_{\hat r} \right) e^{(0)^\mu}_{\hat r} \,,  
\end{align}
where $N = N^{(0)} + N^{(1)}$ is a normalization factor and 
$\delta N^\mu = (\gamma_{(0)}^{\mu \nu} \nabla_{\nu} \delta r)/N^{(0)}$ 
deforms $\vec N$ away from $\vec{e}_{\hat r}$. 
Note that the leading-order normal remains 
$\vec{N}^{(0)} = \vec{e}^{\,(0)}_{\hat r}$. 
The deformation of the horizon normal produces additional modifications to the 
horizon tendicity,
\begin{align}
\label{eq:RWZENNfull}
\mathcal E_{N N}  =& \mathcal E_{\mu \nu} (2M+\delta r) N^\mu N^\mu \notag \\
 =& \mathcal E^{(0)}_{\hat r \hat r}  + \mathcal E^{(1)}_{\hat r \hat r} +  \frac{3}{8M^3} \delta r + 2 \mathcal E^{(0)}_{\hat r \mu} \delta N^{\mu} \notag \\
& - 2 \mathcal E^{(0)}_{\hat r \hat r}  \delta N_{\nu} e^{(0)\nu}_{\hat r} \,,
\end{align}
where, as usual, $\mathcal E^{(1)}_{\hat r \hat r}$ includes the effects of 
both the perturbation to the tidal field and to $\vec{e}^{\,(1)}_{\hat r}$ (and
where all quantities are evaluated at the unperturbed 
horizon position $r= 2M$). 
The new contributions [the last three terms on the right-hand side of 
Eq.~\eqref{eq:RWZENNfull}] come from the displacement of the position of the 
horizon $\delta r$ and the deformation to the normal $\delta \vec{N}$. 

In RWZ gauge, the $(\tilde t, r)$ components of $\delta \vec{N}$ vanish, 
although $\delta \vec{N}$ does have angular components; this means that the 
deformation to the normal to the horizon $\delta \vec N$ does not affect the 
horizon tendicity in RWZ gauge. 
[To show this, note first that  when the deformation to the normal has no 
$(\tilde t,r)$ components $\delta N_{\nu} e^{(0)\nu}_{\hat r}=0$.
Then observe that the (projected) spatial tidal field 
$\mathcal E^{(0)}_{\alpha \beta}$ is diagonal and that $\delta \vec{N}$ has 
only angular components; therefore, the term 
$\mathcal E^{(0)}_{\hat r \mu} \delta N^{\mu} =0$ and all terms involving 
$\delta \vec{N}$ in Eq.~\eqref{eq:RWZENNfull} vanish as well.]
Only the shifted coordinate location of the horizon, changes the horizon 
tendicity, and we find
\begin{align}
\label{eq:RWZENNpert}
\mathcal E^{(1)}_{NN} & = \mathcal E^{(1)}_{\hat r \hat r} +  \frac{3}{8M^3} \delta r \,.
\end{align}
From Eq.~\eqref{eq:deltarsolve}, we see that the angular distribution of 
$\mathcal E^{(1)}_{NN}$ in RWZ gauge is the same as in IR gauge 
[it is $Y^{22}(\theta, \phi)$].

\begin{figure}
\includegraphics[width=0.95\columnwidth]{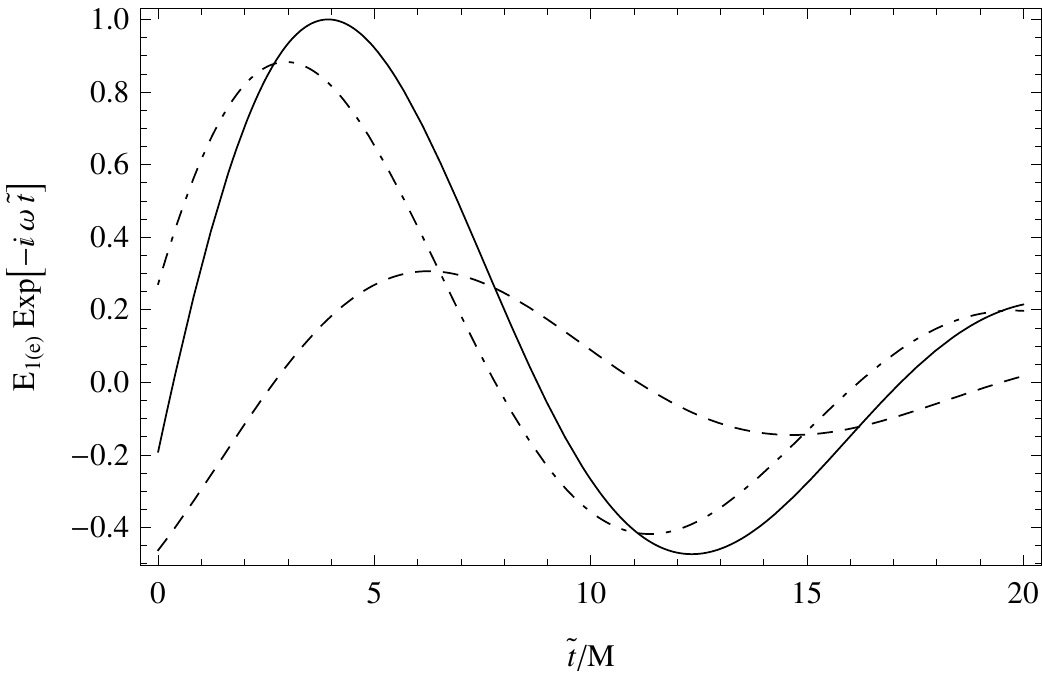}
\caption{Plot illustrating the contributions to the amplitude 
$E_{1(e)}e^{-i \omega \tilde t}$ [Eq.~\eqref{eq:ENNAmpSplit}] of the perturbed 
horizon tendicity $\mathcal E^{(1)}_{NN}$, in RWZ gauge for the 
electric-parity, (2,2) perturbation. 
Plotted agianst $\tilde t$ are the amplitude contributions from 
$\mathcal E^{(1)}_{\hat r \hat r}$ (dashed line), and from the perturbative 
shift of the horizon generators [dot-dashed line; see 
Eqs.~\eqref{eq:ENNposition} and~\eqref{eq:RWZENNpert}]. 
The time-dependent amplitude of the total perturbed horizon tendicity is the 
solid line.}
\label{fig:ENNElectricPert}
\end{figure}

With the angular dependence of the horizon tendicity well understood, let us 
focus on the amplitude and time dependence of the horizon tendicity. 
Using a notation analogous to that in Eq.~\eqref{eq:RWE1e22}, we write the 
horizon tendicity in the form
\begin{align}
\label{eq:ENNAmpSplit}
\mathcal E^{(1)}_{NN} & = E_{1(\rm{e})} (r=2M) e^{-i \omega \tilde t} Y^{22} \,,
\end{align}
for some amplitude $E_{1(\rm{e})}$ [we can do this because both terms in 
Eq.~\eqref{eq:RWZENNpert} have the same time dependence]. 
This amplitude has two contributions: one from the amplitude (and phase) of 
$\mathcal E^{(1)}_{\hat r \hat r} $, and the other from the correction to the
radial perturbation of the generators [second term on the right hand side of 
Eq.~\eqref{eq:RWZENNpert}]. 
We plot these contributions to $E_{1(\rm{e})}e^{-i \omega \tilde t}$ of 
Eq.~\eqref{eq:ENNAmpSplit} in Fig.~\ref{fig:ENNElectricPert}, as a function of 
$\tilde t$ and normalized by the maximum of the (perturbed) horizon tendicity. 
We also plot the amplitude of total perturbation to the horizon tendicity, 
$E_{1(\rm{e})}e^{-i \omega \tilde t}$ (the sum of the two contributions). 
The two contributions are of roughly the same magnitude, but are out of phase. 
The influence of the change in horizon position (dot-dashed line) is slightly 
larger than $\mathcal E^{(1)}_{\hat r\hat r}$ (dashed line).

That $\mathcal E^{(1)}_{NN}$ differs from $\mathcal E^{(1)}_{\hat r \hat r}$ 
only by an amplitude and phase means that, in some sense, the duality between 
the horizon tendicity and vorticity (which is exact in IR gauge) is still
intact; however, they are no longer related by the simple phase shift of $i$. 
In fact, we could choose a different normalization between the Regge-Wheeler 
function $Q$ and the Zerilli function $Z$ than we did in 
App.~\ref{sec:RWZTidalField} to restore this duality relation, but this would 
only hold for the horizon tendicity and vorticity [and the duality in Eq.\ 
(\ref{eq:RWZradialDuality}) would be more complicated, with a complex amplitude replacing the factor of $i$].

\subsection{Magnetic-parity, superposed $(2,2)$ and $(2,-2)$ modes:
Tidal field}

In order to understand better the dynamics and interaction of the tendexes and 
vortexes, we compute the perturbations to the tidal field that must accompany 
the frame-drag field of a magnetic-parity perturbation for the superposed 
$(2,2)$ and $(2,-2)$ modes. 
These tidal perturbations are much like the frame-drag perturbations of an electric-parity metric perturbation, as expected by the near-duality. 
Their odd parity ensures that they must be reflection antisymmetric about the 
equatorial plane.
If we consider the tendex lines of $\boldsymbol{\mathcal E}^{(1)}$ alone, there
must be a zero tendicity set of tendex lines in the plane, and two sets which 
pass through the plane at $45^\circ$ with equal and opposite tendicity. 
We are also assured that $\mathcal E^{(1)}_{NN} = 0$. 
From the calcultion described at the beginning of Sec.\ \ref{sec:ReggeWheeler} 
above, we obtain:
\begin{subequations}
\begin{align}
\mathcal E^{(1)}_{\hat r \hat r} & = 0 \,, \\
\mathcal E^{(1)}_{\hat r \hat A} & = \Re \left[ E_{1({\rm m})} e^{-i \omega t} \left( X^{22}_{\hat A} + X^{2-2}_{\hat A} \right) \right]\,, \\
\mathcal E^{(1)}_{\hat A \hat B} & = \Re\left[ E_{2({\rm m})} e^{-i \omega t} \left( X^{22}_{\hat A \hat B} + X^{2-2}_{\hat A \hat B}\right) \right], \\
E_{1({\rm m})} & = \frac{2i \omega r Q + 4 M \alpha^2 Q'}{i \omega  r^3 \alpha^2 \sqrt{1 + 2M/r}} \,,
\end{align}
\begin{widetext}
\begin{align}
E_{2({\rm m})} & = \frac{\left[6 M \alpha^2(\alpha^2+1) +i \omega (r^2 - 3 M r - 2M^2 + 6 M^2 \alpha^2) - 4 M \omega^2 r^2 \right] Q +\alpha^2 \left[ -12M^2 + 4r(M + i M^2 \omega ) + i \omega r^3 \right] Q'}{i\omega r^4 \alpha^4 ( 1 + 2M/r)}\,.
\end{align}
\end{widetext}
\label{eq:magneticSuperposedTidal}
\end{subequations}
We illustrate the tendicity of the predominantly negative tendex lines in the equatorial plane in the time series of Fig.~\ref{fig:TimeSeries} (bottom row), 
which shows the evolution over a half period of oscillation of the metric perturbation.

\section{Teukolsky's Equation and Black-Hole Perturbations in the 
Newman-Penrose Formalism}
\label{sec:Teukolsky}

The results in this appendix appear in many places in the literature (see, for
example, Teukolsky's paper \cite{Teukolsky}). We summarize them here 
because we will need them in Apps.~\ref{sec:CCKProc} and \ref{sec:HorizonDetails}.

Teukolsky's equation relies on the Newman-Penrose (NP) formalism using 
Kinnersley's tetrad, which is
the principal complex null tetrad in the Schwarzschild and Kerr spacetimes. 
For Kerr, in the Boyer-Lindquist coordinate basis 
$\{\partial_t,\partial_r,\partial_\theta,\partial_\phi\}$ [Eq. (6.1) of Paper II],
this tetrad's contravariant components are given by
\begin{eqnarray}
\label{eq:KinnM}
l^\mu &=& \frac 1\Delta (r^2+a^2,\Delta,0,a)\, ,\notag \\
n^\mu &=& \frac 1{2\Sigma}(r^2+a^2,-\Delta,0,a)\, , \notag\\
m^\mu &=& \frac 1{\sqrt 2(r+ia\cos\theta)} (i a\sin\theta, 0, 1, i\csc\theta)\, ,
\end{eqnarray}
with the final leg given by $\vec {m}^*$, the complex conjugate of $\vec m$.
Here
\begin{equation}
\Delta = r^2 + 2Mr + a^2\;, \quad
\Sigma = r^2 +a^2 \cos^2\theta\;.
\label{eq:DeltaSigmaDef}
\end{equation}
When $a$ is taken to zero, we recover the Kinnersley tetrad for 
Schwarzschild spacetime in the Schwarzschild coordinate basis [Eq.\ (4.1)
of Paper II]. 
The Teukolsky equation also requires the NP spin coefficients, 
certain contractions of covariant 
derivatives of the tetrad above given by Eq.~(4.1a) of \cite{Newman1962} 
(though with the opposite signs because of differing metric-signature 
conventions).
The nonzero spin coefficients in this tetrad are 
\begin{align}
\rho & = -\frac 1{r-ia\cos\theta}\, , &
\pi &= \frac{i a}{\sqrt{2}}\rho^2\sin\theta\, , \notag \\
\beta &= -\frac 1{2\sqrt 2} \rho^* \cot\theta\, , &
\alpha &= \pi - \beta^* \, ,\notag \\
\mu &= \frac{\rho^*\Delta}{2\Sigma}\, , &
\gamma &= \mu + \frac{r-M}{2\Sigma}\, , \notag \\
\tau &= -\frac{ia}{\sqrt 2 \Sigma}\sin\theta\, .
\end{align}

The Weyl scalars $\Psi_0$ and $\Psi_4$ are defined in terms of the Weyl tensor
by 
$\Psi_0 = C_{\mu\nu\rho\sigma}l^\mu m^\nu l^\rho m^\sigma$ and 
$\Psi_4 = C_{\mu\nu\rho\sigma}n^\mu m^{*\nu} n^\rho m^{*\sigma}$ . 
These both vanish in the background when using the Kinnersley tetrad, and 
are gauge invariant at first order in the perturbation theory 
\cite{Teukolsky}, consequently. 
At that perturbative order, they satisfy decoupled, linear, second-order 
partial-differential equations. 
Teukolsky's big breakthrough \cite{Teukolsky}
was to show that, when those equations are re-expressed in terms of
\begin{equation}
\psi_{2} \equiv \Psi_0 \quad {\rm and} \quad 
\psi_{-2}\equiv\rho^{-4} \Psi_4\;,
\label{eq:psi2def}
\end{equation} 
they take on a unified form (the \emph{Teukolsky equation})
that depends on the spin-weight $s=+2$ for $\psi_2$
and $s=-2$ for $\psi_{-2}$, and that is separable; i.e., it has a solution
of the form  
$\psi_{s}^{lm\omega} = {}_{s}R_{lm\omega}(r) 
{}_{s}S_{lm\omega}(\theta) e^{i(m\phi-\omega t)}$. The Teukolsky equation
implies for the radial function 
${}_{s}R_{lm\omega}$ the following ordinary differential equation 
(in vacuum)
\begin{eqnarray}
\label{eq:TeukolskyRadial}
0 &=& \Delta^{-s} \frac{d}{dr}\left(\Delta^{s+1}
\frac{d \, {}_{s}R_{lm\omega}}{dr} \right) + \bigg(\frac{K^2-2is(r-M)K}{\Delta}
\nonumber \\
&& + 4is\omega r +2 a m \omega -a^2 \omega^2 - {}_s A_{lm}\bigg) 
{}_{s}R_{lm\omega} \, ,
\label{eq:RlmwEqn}
\end{eqnarray}
where ${}_s A_{lm}$ is a separation constant 
that is a function of $a\omega$ [i.e., ${}_sA_{lm} = {}_s A_{lm}(a\omega)$],
and $K\equiv(r^2+a^2)\omega -am$. 
The radial function has the symmetries
${}_s R_{lm\omega} = (-1)^m {}_s R_{l-m-\omega^*}$.
The angular function, ${}_{-2}S_{lm\omega}(\theta)$, (called the
spin-weighted spheroidal harmonic) satisfies the ordinary differential
equation (in vacuum)
\begin{eqnarray}
0 &=& \csc\theta \frac{d}{d\theta}\left(\sin\theta\frac{d \, {}_sS_{lm\omega}}
{d\theta}\right) + (a^2\omega^2 \cos^2\theta - m^2\csc^2\theta \nonumber \\
&-& 2a\omega s\cos\theta -2ms\cot\theta\csc\theta - s^2\cot^2\theta + s
+ {}_sA_{lm}) \nonumber \\
&\times& {}_sS_{lm\omega} \, .
\label{eq:SlmwEqn}
\end{eqnarray}
This angular function has the symmetries 
${}_sS_{lm\omega}(\pi-\theta) = (-1)^{(m+l)}{}_{-s}S_{lm\omega}(\theta)$ and
${}_sS_{lm\omega}^*(\theta) = (-1)^{m+s}{}_{-s}S_{l-m-\omega^*}(\theta)$, 
where we are using a phase convention such that the angular functions agree
with the usual convention for spin-weighted spherical harmonics in the limit
that spin parameter, $a$, goes to zero.

It is often useful, in working with the perturbation equations, to
change variables from the separation constants ${}_s A_{lm}$
to Chandrasekhar's \cite{ChandrasekharBook}
\begin{equation}
{}_s\lambdabar_{lm} \equiv {}_s A_{lm} +s + |s| -2 a m \omega+ a^2 \omega^2\;,
\label{eq:lambdabar}
\end{equation} 
which are the same for positive and negative spin weights, $\pm s$.

\section{The Chrzanowski-Cohen-Kegeles Procedure and the 
Ingoing-Radiation-Gauge Metric}
\label{sec:CCKProc}

In this appendix, we will review the formalism used for computing
the ingoing-radiation-gauge (IR gauge) metric, using what is known as the
Chrzanowski-Cohen-Kegeles (CCK) procedure.
We will also connect the CCK procedure to Chrzanowski's original calculation
of definite-parity harmonics, which we find useful for our calculations.

Although Chrzanowski conjectured that ``the conceptual benefits of having 
found the perturbed Kerr metric potentials surpass the usefulness of these
potentials for doing future computations'' \cite{Chrzanowski1975}, the
procedure he helped to formulate has found several applications in the
past few years.
Lousto and Whiting \cite{Lousto2002} revisited Chrzanowski's construction 
and found explicit expressions for computing the Hertz potential corresponding
to specific perturbations of the Weyl curvature scalars $\Psi_0$ and
$\Psi_4$ in the Schwarzschild spacetime.
Ori then derived a similar result for Kerr black holes, using a frequency-domain
calculation \cite{Ori2003}.
Yunes and Gonzalez  were the first to explicitly compute the metric of a 
perturbed Kerr black hole from the Hertz potential \cite{Yunes2006}, and Keidl,
Friedman, and Wiseman were the first to use the procedure to calculate the 
metric perturbation from a static point particle in the Schwarzschild 
spacetime \cite{Keidl2007}.
More recently, Keidl, Shah, and their collaborators 
articulated a formalism for computing the gravitational self-force of a point
particle in the Schwarzschild or Kerr spacetimes using the metric constructed
from a Hertz potential \cite{Keidl2010}.
They were then able to compute the conservative piece of the self-force from 
this metric perturbation in the Schwarzschild spacetime \cite{Shah2011}.
In the first article \cite{Keidl2010}, they gave a concise summary of 
constructing metric perturbations from a Hertz potential, and they called this 
process the Chrzanowski-Cohen-Kegeles (CCK) procedure or formalism (names we 
will also adopt).

In the first part of this appendix, we will review the CCK formalism in a similar manner to how Keidl summarized it in \cite{Keidl2010}.
While the metric we ultimately compute in this paper is nearly identical to 
that described by Chrzanowski \cite{Chrzanowski1975}, we find it helpful to 
put Chrzanowski's original calculation into the context of the more recent 
work on the CCK procedure.
Furthermore, we review the CCK procedure here, rather than simply referring
the interested reader to \cite{Keidl2010}, because there are several
differences between our calculation and that set forth in \cite{Keidl2010}:
we use a metric of the opposite signature, we calculate the metric 
corresponding to quasinormal modes with complex frequencies, we construct the
metric in a different radiation gauge, and (like Chrzanowski's original 
calculation) we are interested in metric perturbations of definite parities. 

Because the CCK formalism relies heavily on the Newman-Penrose formalism 
and Teukolsky's equation for perturbations of Weyl curvature scalars, we 
review these in Appendix \ref{sec:Teukolsky}.
In the second part of this appendix, we will describe how to use the CCK 
procedure to compute definte-parity metric perturbations corresponding to 
quasinormal modes.
In the third part of this appendix, we compute the metric perturbations in a
notation in which they can be compared more easily with those of the RWZ 
formalism (a calculation originally performed by \cite{Chrzanowski1975}), and
we also give explicit analytial expressions for the tidal and frame-drag fields
for $(2,2)$ perturbations, which highlight a near duality between the 
perturbative pieces of these fields for perturbations of opposite parities.
In the final part, we summarize how we numerically calculate the IR gauge metric
perturbations that we use in the visualizations in Figs.\ \ref{fig:22Smag3D},
\ref{fig:TendexVortexEq}, \ref{fig:TendexVortex3D}, \ref{fig:NRComparison},
and \ref{fig:KerrTendexVortexFull}.

\subsection{The CCK procedure}
\label{sec:Reconstruction}

The purpose of the CCK procedure is to construct a metric perturbation,
$h_{\mu\nu}$, from a given solution to Teukolsky's equation, either 
$\psi_2=\Psi_0$ or $\psi_{-2} = \rho^{-4} \Psi_4$ (see Appendix 
\ref{sec:Teukolsky} for a summary of the Teukolsky formalism).
As part of the calculation, it is necessary to relate the solutions of the 
Teukolsky equation to a Hertz potential from which the metric perturbation
is directly constructed [see Eq.\ (\ref{eq:Psi4Hertz}) for the general 
relationship, Eq.\ (\ref{eq:RadialHertz}) for the relationship for the radial 
functions for their harmonics, and Eq.\ (\ref{eq:RadialHertzParity}) for the 
relationship of the radial functions of definite-parity perturbations].

The CCK procedure can construct a metric in either ingoing-radiation (IR)
gauge, 
\begin{equation}
h_{\mu\nu}l^\nu=0 \, , \quad h_{\mu\nu}g_{(0)}^{\mu\nu} = 0 \, ,
\end{equation}
or outgoing-radiation gauge 
\begin{equation}
h_{\mu\nu}n^\nu=0 \, , \quad h_{\mu\nu}g_{(0)}^{\mu\nu} = 0 \, ,
\end{equation}
for Schwarzschild and Kerr black holes.
Here $l^\nu$ and $n^\nu$ are two vectors of a Newman-Penrose null tetrad [for 
our calculations, we will use the Kinnersley tetrad, Eq. (\ref{eq:KinnM})], 
and $g_{(0)}^{\mu\nu}$ is the background Schwarzschild or Kerr metric.
Because our goal is to compute vacuum perturbations of Kerr that are regular
on the future event horizon, we will construct the metric perturbation in IR gauge, 
and we will be able to compute it by algebraically inverting a differential
relationship between the harmonics of the Hertz potential and those of 
$\psi_{-2}$ [the result is in Eq.\ (\ref{eq:RadialHertz})].

The Hertz potential is tensor with the same symmetries as the Riemann tensor,
whose double coordinate divergence is a harmonic coordinate metric.
Stewart \cite{Stewart1979} showed that in Type D spacetimes, there is 
sufficient gauge freedom that one can represent the independent degrees of 
freedom of the perturbative part of the Hertz potential as a single complex
scalar; furthermore, if one applies a coordinate transformation from harmonic 
gauge into IR gauge, the Hertz potential, which we will denote by $\Psi_H$, is 
a solution of the vacuum Teukolsky equation for scalars of spin weight $s=-2$ 
(the same as $\psi_{-2}$).
One can then construct a metric perturbation from the Hertz potential 
by applying several differential operators to $\Psi_H$, 
\begin{eqnarray}
\label{eq:MetricPert}
h_{\mu\nu} &=& \left\{ -l_\mu l_\nu (\bm\delta + \alpha^* + 3\beta - \tau)
(\bm \delta +4\beta + 3\tau) \right.  \nonumber \\
&& -m_\mu m_\nu ({\bf D} -\rho + 3\epsilon -\epsilon^*)
({\bf D} + 3\rho + 4\epsilon) \nonumber \\
&& + l_{(\mu} m_{\nu)} \left[({\bf D} + \rho^* - \rho + \epsilon^* +3\epsilon)
(\bm \delta + 4\beta + 3\tau) \right. \nonumber \\
&& \left. \left . + (\bm \delta +3\beta -\alpha^* -\pi^* -\tau)
({\bf D} + 3\rho + 4\epsilon)\right]\right\}
\Psi_H \nonumber \\
&& + {\rm c.c.}
\end{eqnarray}
(see, e.g., Eqs.\ (93) and (94) of \cite{Keidl2007}).
The differential operators are defined by ${\bf D} = l^\mu \nabla_\mu$ and 
$\bm \delta = m^\mu\nabla_\mu$.
The last term in Eq.\ (\ref{eq:MetricPert}), denoted by ``c.c.,'' means
to take the complex conjugate of the entire expression, so that the metric
perturbation is real.

When computing perturbations of black holes, it is helpful to be able to 
relate a given Hertz potential $\Psi_H$ to a specific perturbation of the 
Weyl scalar $\Psi_4$.
It is possible to do this by computing the components of the perturbative 
Riemann tensor from the metric perturbations (\ref{eq:MetricPert} that 
correspond to the Weyl scalar
\begin{equation}
\Psi_4 = C_{\alpha\beta\mu\nu} n^\alpha m^{*\beta} n^\mu m^{*\nu} \, .
\end{equation}
The result can be expressed compactly as
\begin{equation}
\psi_{-2} = \frac 18 \left( \mathcal L^{\dag 4} \Psi_H^* - 12 M \partial_t 
\Psi_H \right) 
\label{eq:Psi4Hertz}
\end{equation}
[see, e.g., Table I of the paper by Keidl \cite{Keidl2010}, where our 
$\mathcal L^\dag$ is their $\tilde{\mathcal L}$].
We have used the shorthand that 
$\mathcal L^{\dag 4} = \mathcal L_{-1}^\dag \mathcal L_0^\dag 
\mathcal L_{1}^\dag \mathcal L_{2}^\dag$
where
\begin{equation}
\mathcal L_s^\dag = -\left(\partial_\theta + s\cot\theta - i\csc\theta 
\partial_\phi\right) + i a\sin\theta \partial_t \, . 
\end{equation}
In general, solving for the Hertz potential $\Psi_H$ that corresponds to a 
perturbed Weyl scalar $\Psi_4$ involves inverting the fourth-order partial 
differential equation (\ref{eq:Psi4Hertz}); however, when $\Psi_H$ and 
$\psi_{-2}$ are expanded in harmonics in the frequency domain, it is 
possible to perform the inversion algebraically.

The algebraic inversion can be completed by expanding $\psi_{-2}$ in harmonics,
\begin{equation}
\psi_{-2} = \sum_{lm\omega} \psi_{-2}^{(l m \omega)} = \sum_{lm\omega} \, 
{}_{-2} R_{l m \omega}(r) \, {}_{-2} S_{l m \omega}(\theta)
e^{i(m\phi-\omega t)} \, ,
\label{eq:Psi4Harmonics}
\end{equation}
where 
${}_{-2} R_{l m \omega}(r)$ and ${}_{-2} S_{l m \omega}(\theta)$ 
satisfy Eqs.\ (\ref{eq:RlmwEqn}) and (\ref{eq:SlmwEqn}).
In the IR gauge, the Hertz potential is a solution to the Teukolsky equation with 
spin $s = -2$; consequently, it can also be expanded in the same harmonics 
\begin{equation}
\Psi_H = \sum_{lm\omega} \Psi_H^{(l m \omega)} = \sum_{lm\omega} \, 
{}_{-2} X_{l m \omega}(r) \,{}_{-2} S_{l m \omega}(\theta)
e^{i(m \phi-\omega t)} \, .
\label{eq:HertzHarmonics}
\end{equation}
The radial function of the Hertz potential's harmonics 
${}_{-2} X_{l m \omega}(r)$ also satisfies the vacuum Teukolsky radial 
equation, but because it is not the same radial function as in the harmonics of 
$\psi_{-2}$, we denote it with a different function.
The radial functions of the harmonics of $\Psi_H$ and $\psi_{-2}$ can be 
related by substituting Eqs.\ (\ref{eq:Psi4Harmonics}) and 
(\ref{eq:HertzHarmonics}) into Eq.\ (\ref{eq:Psi4Hertz}) and using the 
Teukolsky-Starobinsky identity
\begin{equation}
\mathcal L^{\dag 4} {}_{2} S_{l m \omega} = D {}_{-2} S_{l m \omega} 
\end{equation}
(Eq.\ (59) of Ch.\ 9 of \cite{ChandrasekharBook}, after noting that our 
$\mathcal L^\dag_s$ is equivalent to $-\mathcal L_s$ there), and the identity
\begin{equation}
{}_{-2} S_{l m \omega}^* = (-1)^m {}_2 S_{l -m -\omega^*} \, .
\end{equation}
Then, it is necessary to equate the full radial function for a given angular 
and time harmonic of the Hertz potential to the radial functions 
${}_{-2} R_{l m \omega}$ of $\psi_{-2}$.
After this relationship is inverted, the individual radial harmonics of the 
Hertz potential can be written as
\begin{equation}
{}_{-2} X_{l m \omega} =8\frac{(-1)^m D^* \,{}_{-2} R^*_{l -m -\omega^*} 
- 12 i M \omega \,{}_{-2} R_{l m \omega}}{D^{*2} + 144 M^2 \omega^2} \, .
\label{eq:RadialHertz}
\end{equation}
The constant $D^* \equiv D_{lm\omega}^* = D_{l-m-\omega^*}$ is defined by 
\begin{eqnarray}
D^{2} & =& \lambdabar^2(\lambdabar+2)^2 - 8\lambdabar(5 \lambdabar+6)
(a^2 \omega^2 - a m \omega) +96\lambdabar a^2 \omega^2 \nonumber \\ 
&& + 144 (a^2 \omega^2 - a m \omega)^2 \, ,
\end{eqnarray}
where $\lambdabar$ is the separation constant (\ref{eq:lambdabar}) 
used by Chandrasekhar 
\cite{ChandrasekharBook} (a choice of the separation constant that is the same 
for both the angular and the radial equations).
Although the Teukolsky-Starobinsky identities are usually derived assuming
real frequencies, they have been shown to hold for complex frequencies as
well (for a recent derivation, see \cite{Fiziev2009}).

The general description of the CCK formalism is now complete: (i) For a vacuum 
perturbation of $\Psi_4$, we can find the Hertz potential $\Psi_H$ that 
corresponds to this perturbation by expanding $\Psi_H$ in harmonics 
[Eq.\ (\ref{eq:HertzHarmonics})], and expressing the radial functions of this 
expansion in terms of those of $\psi_{-2} = \rho^{-4}\Psi_4$ via Eq.\ 
(\ref{eq:RadialHertz}). 
(ii) From the resulting $\Psi_H$, we can then compute the metric perturbations 
via Eq.\ (\ref{eq:MetricPert}).

Next, we will show that if we choose the radial function of the Hertz potential
to correspond to a perturbation of $\psi_{-2}$ with definite parity, then the
result of this calculation is equivalent to Chrzanowski's original calculation 
of definite-parity metric perturbations \cite{Chrzanowski1975}.

\subsection{Definite-parity harmonics and Chrzanowski's calculation} 
\label{sec:DefiniteParity}

Let us now connect this CCK procedure with Chrzanowski's original calculation 
of metric perturbations of definite parities.
We shall begin by showing that, if the perturbations have definite parity 
(electric or magnetic), then the Hertz potential must itself transform as 
$\Psi_H \rightarrow \pm (-1)^l \Psi_H^*$ under parity (the plus and minus 
correspond to electric- and magnetic-parity perturbations, respectively), and 
its radial harmonics must satisfy 
${}_{-2} X_{l-m-\omega^*}^* = \pm (-1)^m {}_{-2} X_{l m \omega}$.
In turn, this implies that the radial harmonics of $\psi_{-2}$ must satisfy 
the same relationship 
${}_{-2} R_{l-m-\omega^*}^* = \pm (-1)^m {}_{-2} R_{l m \omega}$.

To deduce these relationships, we discuss the parity of the terms that appear 
in Eq.\ (\ref{eq:MetricPert}). 
The Newman-Penrose tetrad and spin coefficients of the background spacetime
transform in several different ways under parity:
$\vec l$ and $\vec n$ have positive parity, and $\vec m$ does not have a 
definite parity, $\vec m \rightarrow -\vec m^*$.
Similarly, the differential operator ${\bf D} = l^\mu \nabla_\mu$ has positive
parity, and $\bm \delta = m^\mu \nabla_\mu$ again does not have a definite 
parity, $\bm \delta \rightarrow - \bm \delta^*$.
Three of the nonzero spin coefficients map to their complex conjugates 
under parity ($\rho \rightarrow \rho^*$, $\mu \rightarrow \mu^*$, and
$\gamma \rightarrow \gamma^*$), and the remaining four spin coefficients
become minus their complex conjugates under parity 
($\alpha \rightarrow -\alpha^*$, $\beta \rightarrow -\beta^*$, 
$\pi \rightarrow -\pi^*$, and $\tau \rightarrow -\tau^*$).
These relationships hold true for both Schwarzschild and Kerr, although in 
the former case, the spin coefficients are real and, therefore, have definite
parity.

When applying a parity transformation to the perturbative metric tensor,
$h_{\mu\nu}dx^\mu dx^\nu$, where $h_{\mu\nu}$ is given by Eq.\ 
(\ref{eq:MetricPert}), we can show that the tensor differential operator 
in Eq.\ (\ref{eq:MetricPert}) becomes its complex conjugate by using the 
parity transformations for the spin coefficients, NP tetrad, and differential
operators above.
As a result, the metric perturbation will have either electric or magnetic
parity when the Hertz potential transforms as 
\begin{equation}
\Psi_H \rightarrow \pm (-1)^l \Psi_H^*
\end{equation}
under parity. 
The plus sign corresponds to an electric-parity perturbation, and the minus 
sign describes a magnetic-parity perturbation.
The condition this implies on the harmonics is also quite simple, which we
can determine by applying a parity transformation to the Hertz potential 
expanded in harmonics [Eq.\ (\ref{eq:HertzHarmonics})] and equating it to its 
complex conjugate.
Then using the properties of the Teukolsky angular functions 
\begin{eqnarray}
{}_sS_{lm\omega}(\pi-\theta) &=& (-1)^{m+l}{}_{-s}S_{lm\omega}(\theta)\\
{}_sS_{lm\omega}^*(\theta) &=& (-1)^{m+s}{}_{-s}S_{l-m-\omega^*}(\theta)
\end{eqnarray}
(see Appendix \ref{sec:Teukolsky}) and equating the radial function of each 
time and angular harmonic, we obtain the following condition on its radial 
functions,
\begin{equation}
{}_{-2} X_{l-m-\omega^*}^* = \pm (-1)^m {}_{-2} X_{l m \omega} \, .
\end{equation}
Similarly by substituting Eq.\ (\ref{eq:RadialHertz}) into the expression
above, we find an analogous relationship for the radial function of the Weyl 
scalar $\psi_{-2}$, 
\begin{equation}
{}_{-2} R_{l-m-\omega^*}^* = \pm (-1)^m {}_{-2} R_{l m \omega} \, .
\end{equation}
For these definite-parity perturbations, the relationship between the radial 
functions of the Hertz potential and $\psi_{-2}$, Eq.\ (\ref{eq:RadialHertz})
also simplifies,
\begin{equation}
{}_{-2} X_{l m \omega} = \pm 8 (D^*\pm 12 i M\omega)^{-1} 
{}_{-2} R_{l m \omega} \, ;
\label{eq:RadialHertzParity}
\end{equation}
namely, for definite-parity perturbations, the radial functions of 
$\psi_{-2}$ and $\Psi_H$ differ by only a complex constant.
Because Eq.\ (\ref{eq:RadialHertzParity}) shows that the two radial functions
${}_{-2} X_{l m \omega}$ and ${}_{-2} R_{l m \omega}$ differ only by a constant
multiple, we will express both $\Psi_H$ and $\psi_{-2}$ in terms of the radial
function of $\psi_{-2}$, ${}_{-2} R_{l m \omega}$, for simplicity.

In the next part (and also for all other IR gauge calculations in this paper), 
we will compute a metric perturbation that corresponds to a perturbation of 
$\psi_{-2}$ of the form
\begin{eqnarray}
\psi_{-2} &=& \pm \frac 18 (D^*\pm 12 i M\omega) {}_{-2} R_{lm\omega} 
e^{i(m\phi-\omega t)} {}_{-2} S_{lm\omega} \nonumber \\
&& + \frac 18 (-1)^m (D \mp 12 i M\omega^*) {}_{-2} R_{lm\omega}^*
e^{-i(m\phi-\omega^* t)} \nonumber \\
&& \times {}_{-2} S_{l-m-\omega^*} \, .
\end{eqnarray}
The corresponding Hertz potential is
\begin{eqnarray}
\label{eq:HertzParity}
\Psi_H &=& {}_{-2} R_{lm\omega} e^{i(m\phi-\omega t)} {}_{-2} S_{lm\omega} \\
\nonumber 
&& \pm (-1)^m {}_{-2} R_{lm\omega}^* {}_2 e^{-i(m\phi-\omega^* t)} 
{}_{-2} S_{l-m-\omega^*}\, .
\end{eqnarray}
We choose the prefactors on the modes of $\psi_{-2}$ so as to make the Hertz
potential (and, therefore, the metric) as simple as possible.
Furthermore, this choice gives the same definite-parity metric as that of
Chrzanowski (when we take the real part of his expressions).

\subsection{Definite-parity CCK metric perturbations and tidal and frame-drag 
fields for Schwarzschild black holes}
\label{sec:IRGSchwarzschild}

In the first two parts of this section, we will calculate electric- and 
magnetic-parity perturbations of Schwarzschild black holes in IR gauge.
Because Chrzanowski performed this calculation in Table III of reference
\cite{Chrzanowski1975}, and our results agree with his, we do not go into 
great detail describing the calculations; instead, we aim show the results
here so as to be able to compare with the RWZ formalism in Appendix 
\ref{sec:RWApp}.
In the third part, we will compute the tidal and frame-drag fields 
corresponding to these metric perturbations and show a near duality of the 
tidal and frame-drag fields of opposite parity perturbations for the $(2,2)$
mode.

\subsubsection{Electric-parity metric perturbations}

We begin this part by comparing the metric produced by the CCK procedure
to that of the RWZ formalism.
We will write the RWZ metric using the covariant notation described by Martel 
and Poisson \cite{Martel2005}.
Martel and Poisson write the electric-parity perturbations as
\begin{subequations}
\begin{align}
h_{ab}^{\rm (e)} &= \sum_{lm} h_{ab}^{lm} Y^{lm} \, , \\
h_{aB}^{\rm (e)} &= \sum_{lm} j_{a}^{lm} Y^{lm}_B \, , \\
h_{AB}^{\rm (e)} &= r^2 \sum_{lm} \left(K^{lm} \Omega_{AB} Y^{lm}
+ G^{lm} Y^{lm}_{AB} \right) \, ,
\end{align}
\end{subequations}
where the lowercase indices run over the radial and time coordinates (e.g., 
$a,b=t,r$), and uppercase indices run over the angular coordinates as before,
$A,B=\theta,\phi$.
The angular functions $Y^{lm}$ are scalar spherical harmonics, $Y^{lm}_B$
are the electric-parity Regge-Wheeler harmonics, and $Y^{lm}_{AB}$ are 
transverse-traceless, electric-parity tensor harmonics; the term $\Omega_{AB}$ 
is the metric on a 2-sphere.
The vector and tensor harmonics are defined by
\begin{subequations}
\begin{align}
Y^{lm}_A &= D_A Y^{lm} \, , 
\label{eq:YlmAdef}\\
Y^{lm}_{AB} &= \left[D_A D_B + \frac 12 l(l+1) \Omega_{AB}\right] Y^{lm} \, ,
\end{align}
\end{subequations}
where $D_A$ is the covariant derivative on a 2-sphere.

Because the Schwarzschild spacetime is spherically symmetric, we can see, 
intuitively, that the CCK metric, Eq.\ (\ref{eq:MetricPert}), corresponding to 
an electric-parity quasinormal-mode perturbation [the plus sign in Eq.\ 
(\ref{eq:HertzParity})] will have a relatively simple form.
The angular operators acting on the Hertz potential in Eq.\
(\ref{eq:MetricPert}) become the spin-weight raising and lowering operators,
and the angular functions become the spin-weighted spherical harmonics;
furthermore, and when the spin-weighted harmonics are combined with
the appropriate factors of $\vec m$ and $\vec m^*$ the angular functions
become proportional to the scalar, vector, and tensor harmonics described
above.
When performing the calculation, we will need to use the following 
identities, which can be found, for example, by adapting Eqs.\ (2.22a) and \
(2.38e) in the review by Thorne \cite{thorne80} to the notation used here,
\begin{subequations}
\label{ElectricHarmonics}
\begin{align}
Y^{lm}_A &= \sqrt{\frac{l(l+1)}2} ({}_{-1}Y_{lm} m_A - {}_1Y_{lm} m^*_A) \, ,\\
Y^{lm}_{AB} &= \frac{\sqrt D}2 ({}_{-2}Y_{lm} m_A m_B + {}_2Y_{lm} m^*_A m^*_B)
\, .
\end{align}
\end{subequations}
The Teukolsky-Starobinsky constant for spin-weighted spherical harmonics
is $D=(l+2)!/(l-2)!$.
We can then find that the metric coefficients are given by 
\begin{subequations}
\begin{align}
h_{tt}^{\rm (e)} & = -\alpha^2 h_{tr}^{\rm (e)} = \alpha^4 h_{rr}^{\rm (e)} =
-\frac{2\sqrt D}{r^2} \Re[{}_{-2}R_{lm} e^{-i\omega t} Y^{lm}] \, , \\
\label{eq:EvenVectorPert}
h_{tA}^{\rm (e)} & = -\alpha^2 h_{rA}^{\rm (e)} = \frac{\sqrt D }
{2 l(l+1)\alpha^2}\Re\left\{\left[\frac{d}{dr_*} {}_{-2}R_{lm} \right.
\right. \nonumber \\
& \qquad \qquad \left.\left. - \left(i \omega + \frac{2\alpha^2}r \right)
{}_{-2}R_{lm} \right] Y_A^{lm}e^{-i\omega t} \right\} \, , \\
h_{AB}^{\rm (e)} &= \frac{2}{\sqrt D\alpha^4} \Re\left\{\left[ 
(i\omega r^2 -M) \frac{d}{dr_*} {}_{-2}R_{lm} - [\tfrac 12 \mu^2\alpha^2 
\right.\right. \nonumber \\
& - i\omega (-3r+7M) - r^2\omega^2]{}_{-2}R_{lm}\bigg]
Y_{AB}^{lm}e^{-i\omega t} \bigg\} \, . 
\label{eq:EvenTensorPert}
\end{align}
\end{subequations}
In the last equation we have used the radial Teukolsky equation to eliminate
the second-derivative term, and we have defined $\mu^2=(l-1)(l+2)$
(which is also equal to $l(l+1)-s(s+1)$ for $s=-2$).

There are a few noteworthy differences between the IR gauge electric-parity 
perturbations, and the electric-parity RWZ-gauge metric.
The CCK metric has a strictly angular part of the perturbation which is
proportional to the transverse-traceless harmonics, and the trace portion of 
the angular block vanishes; conversely, the angular block of the RWZ metric 
perturbation has a trace part, but no transverse-traceless perturbation.
The $h_{tr}^{\rm (e)}$ part of the metric perturbation also has a simpler 
relationship with the $h_{tt}^{\rm (e)}$ and $h_{rr}^{\rm (e)}$ components in
IR gauge than in RWZ gauge; one reason for this is that the IR gauge metric has 
electric-parity vector perturbations, whereas the RWZ metric sets these to 
zero.
Finally, the IR gauge metric is finite on the future event horizon for 
ingoing radiation. 
One can see this by noting that both ${}_{-2}R_{lm}$ and 
$d{}_{-2}R_{lm}/dr_*$ scale as $\alpha^4 e^{-i\omega r_*}$ near the
horizon, which will cancel any negative powers of $\alpha^2$ in the 
expressions for the metric coefficients.
The same is not as manifest for the RWZ perturbations (see Appendix 
\ref{sec:RWApp} for more details on the RWZ formalism).

\subsubsection{Magnetic-parity metric perturbations}

The magnetic-parity perturbations are given by
\begin{subequations}
\begin{align}
h_{ab}^{\rm (m)} &= 0 \, , \\
h_{aB}^{\rm (m)} &= \sum_{lm} h_{a}^{lm} X^{lm}_B \, , \\
h_{AB}^{\rm (m)} &= \sum_{lm} h_2^{lm} X^{lm}_{AB} \, ,
\label{eq:MagneticTT}
\end{align}
\end{subequations}
where the magnetic-parity harmonics are defined by 
\begin{subequations}
\label{MagneticHarmonics}
\begin{align}
\label{eq:BVectorHarmonic}
X^{lm}_A &= -\epsilon_A{}^B D_B Y^{lm} \, ,\\
\label{eq:BTensorHarmonic}
X^{lm}_{AB} &= -\frac 12 (\epsilon_A{}^C D_B + \epsilon_B{}^C D_A )D_C 
Y^{lm} \, ,
\end{align}
\end{subequations}
and $\epsilon_{AB}$ is the Levi-Civita tensor on a unit 2-sphere.
As in the previous part, we can compute the CCK metric (\ref{eq:MetricPert}), 
which is relativitely simple for a Schwarzschild black hole.
The reason for the simplification is the same, but we will need the following
two identities that relate the spin-weighted spherical harmonics to
magnetic-parity vector and tensor harmonics
\begin{subequations}
\begin{align}
X^{lm}_A &= -i\sqrt{\frac{l(l+1)}2} ({}_{-1}Y_{lm} m_A
+ {}_1Y_{lm} m^*_A) \, , \\
X^{lm}_{AB} &= -i\frac{\sqrt D}2 ({}_{-2}Y_{lm} m_A m_B
- {}_2Y_{lm} m^*_A m^*_B) \, .
\label{eq:TTMagneticHarmonic}
\end{align}
\end{subequations}
These relationships can found in Eqs.\ (2.22b) and (2.38f) of \cite{thorne80}.
The magnetic-parity metric perturbations have the same radial and time 
dependence as the electric-parity perturbations for the vector and tensor 
parts, 
\begin{subequations}
\begin{align}
-h_{tA}^{\rm (m)} & = \alpha^2 h_{rA}^{\rm (m)} = \frac{\sqrt D }{2 l(l+1)
\alpha^2} \Im\left\{\left[\frac{d}{dr_*} {}_{-2}R_{lm} \right.\right. 
\nonumber \\
& \qquad \qquad \left.\left. - \left(i \omega + \frac{2\alpha^2}r \right)
{}_{-2}R_{lm} \right] X_A^{lm}e^{-i\omega t} \right\} \, . \\
h_{AB}^{\rm (m)} &= -\frac{2}{\sqrt D\alpha^4} \Im\left\{\left[ 
(i\omega r^2 -M) \frac{d}{dr_*} {}_{-2}R_{lm} - [\tfrac 12\mu^2\alpha^2 
\right.\right.\nonumber \\
& - i\omega (-3r+7M) - r^2\omega^2]{}_{-2}R_{lm}\bigg]
X_{AB}^{lm}e^{-i\omega t} \bigg\} \, .
\label{eq:IRGpAB}
\end{align}
\end{subequations}
Because they have the same radial dependence as the electric-parity metric, the
magnetic-parity peturbations will also be well-behaved on the future event 
horizon.

The major difference between the RWZ formalism's magnetic-parity metric and 
the IR gauge metric is that in IR gauge, the transverse-traceless metric 
perturbation is no longer required to be zero.

\subsubsection{Tidal and frame-drag fields of the (2,2) mode}

In this part, we calculate the tidal and frame-drag fields for a $(2,2)$ mode
in IR gauge of both electric and magnetic parities.
We find an interesting near duality between the tidal and frame-drag fields of
opposite-parity peturbations that we noted in Secs.\ \ref{sec:modeDuality},
\ref{sec:TendVortSchwKerr} and \ref{sec:ApproximateDual}.

We compute the tidal and frame-drag fields from the metric by evaluating the 
components of the Weyl tensor and its dual in the tetrad (\ref{EFbasis}) 
including the perturbative corrections to the tetrad 
(\ref{eq:u1})--(\ref{eq:e1}).
We find that for an electric-parity mode, the tidal and frame-drag fields 
can be written as 
\begin{subequations}
\label{eq:IRG22FieldsE}
\begin{align}
\mathcal E_{\hat r\hat r}^{\rm (1,e)} &= 2\Re[E_{{\rm I(e)}}(r) Y^{22}
e^{-i\omega t}] \, ,\\
\mathcal E_{\hat r\hat A}^{\rm (1,e)} &= 2\Re[E_{{\rm II(e)}}(r) 
Y^{22}_{\hat A} e^{-i\omega t}] \, , \\
\mathcal E_{\hat A\hat B}^{\rm (1,e)} &= 2\Re\left[\left(-\frac 12 
E_{{\rm I(e)}}(r) \delta_{\hat A\hat B} Y^{22} \right.\right. \nonumber \\
& + E_{{\rm III(e)}}(r) Y^{22}_{\hat A\hat B}\bigg) e^{-i\omega t}\bigg] \, , 
\label{eq:AngularTidalFieldEven} \\
\mathcal B_{\hat r\hat r}^{\rm (1,e)} &= 0 \, ,\\
\mathcal B_{\hat r\hat A}^{\rm (1,e)} &= 2\Re[B_{{\rm I(e)}}(r) X^{22}_{\hat A} 
e^{-i\omega t}] \, , \\
\mathcal B_{\hat A\hat B}^{\rm (1,e)} &= 2\Re[B_{{\rm II(e)}}(r)
X^{22}_{\hat A\hat B}
e^{-i\omega t}] \, .
\end{align}
\end{subequations}
The symbol $\delta_{\hat A\hat B}$ is the Kronecker delta function, and the 
traceless property of $\mathcal E$ requires that the radial function in front 
of the Kronecker delta must be minus one-half that of
$\mathcal E_{\hat r\hat r}^{(1)}$ [i.e., $-(1/2) E_{I{\rm (e)}}(r)$].

For the magnetic-parity perturbation, the frame-drag and tidal fields are
\begin{subequations}
\label{eq:IRG22Fields}
\begin{align}
\label{eq:Bij22IRG}
\mathcal B_{\hat r\hat r}^{\rm (1,m)} &= 2\Re[B_{{\rm I(m)}}(r) Y^{22}
e^{-i\omega t}] \, ,\\
\mathcal B_{\hat r\hat A}^{\rm (1,m)} &= 2\Re[B_{{\rm II(m)}}(r) 
Y^{22}_{\hat A} e^{-i\omega t}] \, , \\
\mathcal B_{\hat A\hat B}^{\rm (1,m)} &= 2\Re\bigl[\bigl(-\frac 12 
B_{{\rm I(m)}}(r) \delta_{\hat A\hat B} Y^{22} \nonumber \\
& + B_{{\rm III(m)}}(r) Y^{22}_{\hat A\hat B}\bigr) e^{-i\omega t}\bigr] \, ,\\
\mathcal E_{\hat r\hat r}^{\rm (1,m)} &= 0 \, ,\\
\mathcal E_{\hat r\hat A}^{\rm (1,m)} &= 2\Re[E_{{\rm I(m)}}(r) X^{22}_{\hat A} 
e^{-i\omega t}] \, , \\
\mathcal E_{\hat A\hat B}^{\rm (1,m)} &= 2\Re[E_{{\rm II(m)}}(r)
X^{22}_{\hat A\hat B} e^{-i\omega t}] \, .
\label{eq:AngularTidalFieldOdd}
\end{align}
\end{subequations}

Interestingly, the radial functions of the tidal and frame-drag fields of the
opposite-parity perturbations are nearly identical
\begin{subequations}
\label{eq:IRG22Duality}
\begin{align}
B_{{\rm I(m)}}(r) &= i E_{{\rm I(e)}}(r) \\
B_{{\rm II(m)}}(r) &= i E_{{\rm II(e)}}(r) - i\frac{M\sqrt{3(r+2M)}}
{r^5\alpha^4\sqrt{2r}} \nonumber \\
&\times \left[-(2\alpha^2+i\omega r){}_{-2}R_{22}
+r\frac{d}{dr_*}{}_{-2}R_{22}\right] \, , \\
B_{{\rm III(m)}}(r) &= i E_{{\rm III(e)}}(r) + \frac{\sqrt 3 (r+2M)}
{r^5\alpha^4\sqrt 2} M \omega {}_{-2}R_{22}\, , \nonumber \\
\label{eq:B3mIRG} \\
E_{{\rm I(m)}}(r) &= -i B_{{\rm I(e)}}(r) +i \frac{M\sqrt{3(r+2M)}}
{r^5\alpha^4\sqrt{2r}} \nonumber \\
&\times \left[-(2\alpha^2+i\omega r){}_{-2}R_{22}
+r\frac{d}{dr_*}{}_{-2}R_{22}\right] \, , \\
E_{{\rm II(m)}}(r) &= -i B_{{\rm II(e)}}(r) - \frac{\sqrt 3 (r+2M)}
{r^5\alpha^4\sqrt 2} M\omega {}_{-2}R_{22}\, . 
\end{align}
\end{subequations}
In fact, there is an exact duality of the radial-radial components, which 
implies that the horizon vorticity of a magnetic-parity perturbation is the
same as the horizon tendicity of an electric-parity perturbation.
For completeness, we list the expressions for the radial functions for the
electric-parity perturbations, which are lengthy, but will be needed in the
next appendix.
\begin{widetext}
\begin{subequations}
\label{eq:IRG22Radial}
\begin{align}
E_{{\rm I(e)}}(r) &=-\frac{2\sqrt 6}{r^6 \alpha^4}\left\{r^2(r-3M+i\omega r^2)
\frac{d}{dr_*}{}_{-2}R_{22} + [-5r^2 + 16M r -12M^2 - i\omega r^2(4r-9M) + 
r^4\omega^2]{}_{-2}R_{22}\right\} \, , \\
E_{{\rm II(e)}}(r) &= \frac 1{r^6\sqrt{6r(r+2M)}\alpha^4}\bigg\{
r^2[3r^2+6M^2 + i\omega r^2(r-3M) + r^4\omega^2 ]
\frac{d}{dr_*}{}_{-2}R_{22} \nonumber \\
& + [(-9r^3+18Mr^2-12M^2r+24M^3) - i\omega r^2(8r^2-16Mr+18M^2) + 2 \omega^2
r^4 (4r-9M) + ir^6\omega^3]{}_{-2}R_{22}\bigg\} \, , \\
E_{{\rm III(e)}}(r) &= \frac{1}{r^5(r+2M)\alpha^4\sqrt 6}\bigg\{
ir^2\omega(-2r^2 + 3Mr +3M^2 + r^4\omega^2)\frac{d}{dr_*}{}_{-2}R_{22} 
+[6(r^2 + 4M^2) \nonumber \\
& +i\omega(4r^3-11Mr^2+12M^2r+12M^3)-\omega^2 r^2(4r^2-4Mr-9M^2)
-3i\omega^3 r^5\alpha^2+r^6\omega^4]{}_{-2}R_{22}\bigg\} \, , \\
B_{{\rm I(e)}}(r) &=  \frac{\sqrt 2}{r^5\sqrt{3r(r+2M)}\alpha^4}\bigg\{
r^2[9M -i\omega r(r-3M)+ r^3 \omega^2]\frac{d}{dr_*}{}_{-2}R_{22} \nonumber \\
& + [-24Mr\alpha^2 +i\omega r(12M^2-25Mr + 5r^2) - \omega^2 r^3(4r-9M)
- i r^5\omega^3]{}_{-2}R_{22} \bigg\} \, , \\
B_{{\rm II(e)}}(r) &= \frac{1}{r^4(r+2M)\alpha^4\sqrt 6}\bigg\{ i\omega r
(-2r^2 + 3Mr + 3M^2 + r^4\omega^2) \frac{d}{dr_*}{}_{-2}R_{22} \nonumber \\
& + [-24M + 2i\omega r(2r - 7M) +\omega^2 r (-4r^2 + 4Mr + 9M^2)
-3i\omega^3 r^4 \alpha^2 + r^5\omega^4]{}_{-2}R_{22} \bigg\} \, . 
\end{align}
\end{subequations}
\end{widetext}
From these expressions, it is clear that the tidal and frame drag-fields are
regular on the horizon, because, as noted above ${}_{-2}R_{lm}$ and 
$d{}_{-2}R_{lm}/dr_*$ scale as $\alpha^4 e^{-i\omega r_*}$ near the horizon; 
consequently, they will cancel the corresponding powers of $\alpha$ in the
denominators of these functions.

\subsection{Analytical and numerical methods for computing metric perturbations
and tidal and frame-drag fields in IR gauge}
\label{sec:ComputationMethods}

The procedures for calculating the metric perturbations and their tidal and 
frame-drag fields are identical for Schwazschild and Kerr black holes; however,
because the analytical expressions for the Newman-Penrose quantities, the 
angular Teukolsky function, and the metric derived from these mathematical 
objects are significantly simpler for Schwarzschild black holes, the amount of 
work we can perform analytically differs for rotating and non-rotating black 
holes.
Even for Schwarzschild black holes, however, we will not be able to compute
all aspects of the metric perturbation analytically.
We calculate the the least-damped $l=2$, $m=2$ quasinormal-mode frequencies 
for both Schwarzschild and Kerr black holes using the Mathematica notebook 
associated with \cite{Berti2009}, an implementation of Leaver's method 
\cite{Leaver1985}.
Similarly, we compute the radial Teukolsky functions ${}_{-2} R_{lm\omega}$
corresponding to a quasinormal-mode solution for both Schwarzschild and
Kerr black holes numerically.
We compute it in two ways, which give comparable results: we solve the 
boundary-value problem for a quasinormal mode solution to the radial Teukolsky 
equation, Eq.\ (\ref{eq:TeukolskyRadial}), using a shooting method, and we 
compare the result with a series solution given by Leaver \cite{Leaver1985} 
(as is also done in the notebook of \cite{Berti2009}).
For Kerr black holes, the numerical solution requires the angular eigenvalue,
${}_s A_{lm}$ associated with the quasinormal mode frequency, which we again 
compute from the implementation of Leaver's method in \cite{Berti2009}.

The most significant difference between the calculations of quasinormal modes
of Schwarzschild and Kerr black holes arises from differences in the 
Teukolsky angular function, and the angular operators used in computing the
metric (\ref{eq:MetricPert}).
First, the spin-weighted spheroidal harmonics in the expression for 
the Hertz potential, Eq.\ (\ref{eq:HertzParity}), reduce to spin-weighted
spherical harmonics for Schwarzschild black holes. 
Second, the angular operators in Eq.\ (\ref{eq:MetricPert}) reduce to 
spin-weight lowering operators, in the non-spinning limit.
As a result, the metric perturbation can be expressed, analytically, in terms 
of electric- or magnetic-parity scalar, vector, and tensor spherical harmonics 
of a single $l$, for Schwarzschild black holes.
For perturbations of Kerr black holes, there are not these additional 
simplifications.
First, we must calculate the spin-weighted spheroidal harmonics numerically,
which we do using a series solution put forward by Leaver \cite{Leaver1985}
(the same method as that implemented in \cite{Berti2009}).
Second, the angular operators are no longer the spin-weight lowering operators.
The metric perturbation computed from these functions, therefore, is not
nearly as simple as that of the Schwarzschild limit.  
In fact, for our calculations with spinning black holes, we find it easier to
work with a numerical fit to the analytical expression for the metric.

Once we calculate the metric perturbation, we construct the perturbation to
the Weyl tensor in the same way for both rotating and non-rotating black 
holes.
We can then calculate the tetrad components of the tidal field, 
$\mathcal E_{\hat a \hat b}$, and frame-drag field, 
$\mathcal B_{\hat a \hat b}$, using the background tetrad in Eq.\ 
(\ref{EFbasis} for Schwarzschild holes or Eq.\ (\ref{eq:KSbasis}) for Kerr 
holes and its perturbative corrections in Eqs.\ (\ref{eq:u1})--(\ref{eq:e1}).
From these fields, we can solve the eigenvalue problem and compute tendex and
vortex lines, and their corresponding tendicities and vorticities.

\section{Relationship Between Regge-Wheeler-Zerilli and Ingoing-Radiation 
Gauges}
\label{sec:GaugeCompare}

In this appendix, we construct generators of infinitesimal coordinate 
transformations between RWZ and IR gauges, for both magnetic- and 
electric-parity perturbations of Schwarzschild black holes.

\subsection{Magnetic-parity gauge transformation}

In this part, we compute the gauge-change generator that transforms the 
magnetic-parity metric in IR gauge to the same metric in Regge-Wheeler gauge.
We show, as noted in Sec.\ \ref{sec:GaugePerts}, that this infinitesimal 
magnetic-parity coordinate transformations does not change the time function 
that specifies the slicing (into surfaces of constant $\tilde t$).
In addition, perturbative changes of the spatial coordinates will not alter 
the coordinate (or tetrad) components of the frame-drag field; therefore, the
fields in both gauges will be equal.

The calculation that shows these facts is relatively straightforward.
Regge and Wheeler showed in Eq.\ (17) of \cite{ReggeWheeler1957} that, 
beginning in any gauge, it is 
possible to remove the transverse-traceless part of the magnetic-parity metric 
perturbation
[Eq.\ (\ref{eq:MagneticTT}) in the notation used in Appendix 
\ref{sec:IRGSchwarzschild}] by an infinitesimal coordinate transformation of the form
\begin{equation}
\vec \xi^{\, \rm (m)} = -\frac 12 \sum_{lm} h_2^{lm} (0,0,{\bf X}^{lm}) \, ,
\end{equation}
where ${\bf X}^{lm}$ is a magnetic-parity, vector spherical harmonic.
This follows from the fact that the perturbation to the metric transforms under
this change of coordinates by 
\begin{equation}
h_{\mu\nu} \rightarrow h_{\mu\nu} + 2\xi_{(\mu|\nu)} \, ,
\end{equation}
(where $|$ denotes a covariant derivative with respect to the background 
metric, and parenthesis around the indices means the expression is symmetrized)
and from the definition of the magnetic-parity, transverse-traceless tensor 
harmonics (\ref{eq:BTensorHarmonic}).
The result can also be found from Eqs.\ (5.5) and (5.6) of \cite{Martel2005}.

For a multipolar perturbation with indices $(l,m)$ in IR gauge, the function
$-\tfrac12 \sum_{lm} h_2^{lm}/2$ is given by the radial function in Eq.\ 
(\ref{eq:IRGpAB}) multiplied by $e^{-i\omega t}$, and the full coordinate 
transformation vector is therefore
\begin{subequations}
\begin{align}
\xi_t^{\, \rm (m)} & = \xi_r^{\, \rm (m)} = 0 \, ,\\
\xi_A^{\, \rm (m)} &= \frac{1}{\sqrt D\alpha^4} \Im\left\{\left[
(i\omega r^2 -M) \frac{d}{dr_*} {}_{-2}R_{lm} - [\tfrac 12\mu^2\alpha^2 
\right.\right. \nonumber \\
& - i\omega (-3r+7M) - r^2\omega^2]{}_{-2}R_{lm}\bigg]
X_A^{lm}e^{-i\omega t} \bigg\} \, .
\end{align}
\end{subequations}
A short calculation can verify that $h_{tA}$ and $h_{rA}$ are the
only nonzero components of the metric after this transformation (the same as
in RWZ gauge) and they are given by
\begin{subequations}
\begin{align}
h_{tA} & = \Im\bigg\{\frac 1{\sqrt{l(l+1)D} r\alpha^4}\bigg[[-\tfrac 12 \mu^2 
r\alpha^2 + (iM\omega + r^2\omega^2)] \nonumber \\
& \times \frac{d}{dr_*} {}_{-2}R_{lm} + [\alpha^2(\alpha^2+i\omega r)
- \omega^2 r (3r-7M) \nonumber \\
& - i\omega^3 r^3] {}_{-2}R_{lm} \bigg] X^{lm}_A e^{-i\omega t}\bigg\} 
\, ,\\
h_{rA} & = \Im\bigg\{\frac{-i\omega}{\sqrt{l(l+1)D}\alpha^6} \bigg[
(r-3M+i\omega r^2)\frac{d}{dr_*} {}_{-2}R_{lm} \nonumber \\
& + [-\alpha^2(\tfrac 12 \mu^2 + 3\alpha^2) -i\omega (4r-9M) + \omega^2 r^2]
{}_{-2}R_{lm} \bigg] \nonumber \\
& \times X^{lm}_A e^{-i\omega t}\bigg\} \, .
\end{align}
\end{subequations}
It is not immediately apparent, however, that this gauge is RWZ
gauge, because it is expressed in terms of the radial function of $\psi_{-2}$,
(${}_{-2}R_{lm}$), rather than the Regge-Wheeler function $Q$.

To show that this transformation did bring the metric into Regge-Wheeler gauge,
it is necessary to use the relationship between $Q$ and ${}_{-2}R_{lm}$ given 
in, e.g., Eq.\ (319) of Ch.\ 4 of \cite{ChandrasekharBook}\footnote{Aside from 
several differences in notation (the radial function used by Chandrasekhar, 
$Z^{(-)}$, is related to the Regge-Wheeler function by $Q = i\omega Z^{(-)}$, 
and his radial function for the Teukolsky equation, $Y_{-2}$ is related to that
of this paper by ${}_{-2}R_{lm}=r^3 Y_{-2}$), there is one additional subtle
point about using this equation.
This equation is expressed as a relationship between $Y_{+2}$ (proportional to 
the radial function of $\Psi_0$) and $Z^{(-)}$.
Because the time dependence of $\Psi_0$ is given by $e^{+i\sigma t}$ in 
\cite{ChandrasekharBook}, then the $Y_{+2}$ there is equivalent to $Y_{+2}^*$
of $\Psi_0$ with a time dependence given by $e^{-i\omega t}$.
In addition, because $Y_{-2}$ satisfies the same equation as $Y_{+2}^*$, then 
this equation is valid for $Y_{-2}$ when $\sigma$ is replaced by $\omega$.} 
\begin{align}
Q =& \frac{-2i\omega}{r \alpha^4 (D-12iM\omega)}\bigg\{(r-3M+i\omega r^2)
\frac{d}{dr_*}{}_{-2}R_{lm} \nonumber \\
& + [-\alpha^2 (\tfrac 12 \mu^2 +3\alpha^2) - i\omega (4r-9M) 
\nonumber \\
& + r^2\omega^2]{}_{-2}R_{lm}\bigg\} \, .
\label{eq:QandR}
\end{align}
After substituting this relationship into Eqs.\ (\ref{h01Q}) and 
(\ref{h0h1def}) and taking its imaginary part---so that the RWZ
metric is real and is expressed in terms of ${}_{-2}R_{lm}$---it becomes 
apparent that the transformation brings the IR gauge metric into RWZ
gauge.

Because the gauge change from IR to RWZ is generated by a strictly spatial
$\vec \xi^{\, \rm (m)}$ and because $\boldsymbol{\mathcal B}$ is a strictly 
first-order quantity for perturbations of Schwarzschild holes, the frame-drag 
analog of Eq.\ (\ref{eq:CoordChange}) guarantees that the frame-drag field 
must be identically the same in the two gauges:
\begin{equation}
\mathcal B^{\rm IRG}_{\hat i\hat j} = \mathcal B_{\hat i \hat j}^{\rm RW} \, .
\end{equation}
This can be confirmed explicitly for the $(2,2)$ mode by substituting Eq.\ 
(\ref{eq:QandR}) into the tetrad components of the RWZ frame-drag field in 
Eqs.\ (\ref{Bij22})--(\ref{eq:B3mRW}) and finding that they are identical to 
the IR gauge frame-drag field of Eqs.\ (\ref{eq:IRG22Fields}), 
(\ref{eq:IRG22Duality}), and (\ref{eq:IRG22Radial}).

When thought of as abstract tensors without reference to any coordinate system,
it is also the case that the tidal fields are equal, 
\begin{equation}
\boldsymbol{\mathcal E}^{\rm IRG} = \boldsymbol{\mathcal E}^{\rm RW} \, .
\end{equation}
Because there is a background tidal field, perturbative differences in
the coordinates enter into the components of the tidal field and the 
components are no longer equal; see Eq.\ (\ref{eq:CoordChange}).
Therefore, visualizations of the tendex lines or tendicity in the two 
coordinate systems (when the coordinates are drawn as though they were flat)
look different.

\subsection{Electric-parity gauge transformation}

In this part, we construct an infinitesimal generator of an electric-parity 
coordinate transformation that brings the electric-parity IR gauge to RWZ 
gauge.
The transformation changes the time function (and hence how we define the 
slicing) in addition to the spatial coordinates.
This implies that neither the frame-drag fields nor the coordinate components 
of the tidal field will equal in the two gauges (but the tidal field written
without coordinates will be); see Eqs.\ (\ref{eq:DeltaB}) and 
(\ref{eq:CoordChange}), respectively.

The gauge-change generator that connects the two gauges is somewhat more 
complex for the electric-parity perturbations than it was for the 
magnetic-parity ones.
The transformation can be found by using Eq.\ (19) of \cite{ReggeWheeler1957}
or Eqs.\ (4.6)--(4.9) of \cite{Martel2005}.
The general approach to find the transformation is to use $\xi_A^{\, \rm (e)}$ 
to remove the transverse-traceless part of the IRG metric, and then use 
$\xi_t^{\, \rm (e)}$ and $\xi_r^{\, \rm (e)}$ to annul the transverse metric 
coefficients.
After a short calculation, it is possible to express the generator as
\begin{subequations}
\begin{align}
\xi_t^{\, \rm (e)} &= \frac{1}{2\mu^2 r^2\alpha^4} \Re\bigg\{\bigg[
-r^2[\mu^3\alpha^2 -4i\omega(M-r^2\omega)] \nonumber \\
& \times \frac{d}{dr_*} {}_{-2}R_{lm} + [\mu^3 r\alpha^4
 +i\omega r^2 \mu^2\alpha^2(\mu-2) \nonumber \\
& -4rM\omega^2r^2(3r-7M) -4i\omega^3 r^4] {}_{-2}R_{lm} \bigg] Y^{lm} 
e^{-i\omega t}\bigg\} \, \\
\xi_r^{\, \rm (e)} &= \frac{1}{2\mu^2 r^3\alpha^6} 
\Re\bigg\{\bigg[r^3[\alpha^2 \mu^2(\mu+2) + 4i\omega (r-3M) \nonumber \\
& -4r^2\omega^2] \frac{d}{dr_*} {}_{-2}R_{lm} + 
\{-r\mu^2\alpha^2(\mu+2)(2r\alpha^2+ir^2\omega) \nonumber \\
& -2r^2(\mu+2) -2i\omega r[(\mu^2+\mu-2)r^2+2Mr(\mu-8) \nonumber \\
& +24M^2] +4\omega^2r^3(4r-9M) + 4i\omega^3 r^5\} {}_{-2}R_{lm}\bigg] 
\nonumber \\
& \times Y^{lm} e^{-i\omega t} \bigg\} \\
\xi_A^{\, \rm (e)} &= \frac{-1}{\sqrt D\alpha^4} \Re\bigg\{\bigg[
(i\omega r^2 -M) \frac{d}{dr_*} {}_{-2}R_{lm} - [\tfrac 12\mu^2\alpha^2 
\nonumber \\
& - i\omega (-3r+7M) - r^2\omega^2]{}_{-2}R_{lm}\bigg]
Y_A^{lm}e^{-i\omega t} \bigg\} \, ,
\end{align}
\end{subequations}
where we used Eq.\ (\ref{eq:TeukolskyRadial}) to reduce second-order radial 
derivatives to first-order ones.

To confirm that this gauge-change generator does bring the IR-gauge metric to
the RWZ metric, we again use the relation between the Zerilli function and 
the radial Teukolsky function encoded in Eq.\ (319) of Ch.\ 4 of 
\cite{ChandrasekharBook}:
\begin{align}
Z &= \frac 1{r^2\alpha^4(D+12iM\omega)(\mu^2r+6M)}\bigg\{2r[i\omega r^2
(\mu^2 r+6M) \nonumber \\
& + (\mu^2 r^2 -3\mu^2 M r -6M^2)] \frac{d}{dr_*} {}_{-2}R_{lm} + \{ \alpha^2
(\mu^2 r+ 6M)^2 \nonumber \\
& + 2(3\alpha^2+i\omega r)[\mu^2 r^2-3\mu^2 Mr -6M^2 \nonumber \\
& + i\omega r^2(\mu^2 r+ 6M)] \} {}_{-2}R_{lm} \bigg\} \, .
\end{align}
This allows us to confirm that the IR metric was brought to RWZ gauge through
the transformation vector $\vec \xi^{\, \rm (e)}$.

With this expression, we can also compare the frame-drag fields in the two 
gauges for the $(2,2)$ mode.
By expressing the radial functions for the frame-drag field of a $(2,2)$, 
electric-parity mode in RWZ gauge [$B_{1 {\rm (e)}}(r)$ and 
$B_{2 {\rm (e)}}(r)$ of Eqs.\ (\ref{eq:RWB1e}) and (\ref{eq:RWB2e})] in terms
of the radial Teukolsky function ${}_{-2}R_{22}$, we find
\begin{subequations}
\begin{align}
B_{1 {\rm (e)}}(r) & = B_{\rm I (e)}(r) + \frac{\sqrt 3}{2 r^5\alpha^4
\sqrt{2r(r+2M)}}\bigg\{[-4r\alpha^2 \nonumber \\
& -2i\omega(2r^2-5Mr+6M^2) + 3r^2\omega^2(r-3M) \nonumber \\
& + ir^4\omega^3]{}_{-2}R_{22} - r^2\alpha^2(r^2\omega^2 + 3iM\omega-2) 
\nonumber \\
& \times \frac{d}{dr_*} {}_{-2}R_{22}\bigg\} \, , \\
B_{2 {\rm (e)}}(r) & = B_{\rm II (e)}(r) \, ,
\end{align}
\end{subequations}
where $B_{\rm I (e)}(r)$ and $B_{\rm II (e)}(r)$ are the equivalent radial
functions of the IR-gauge frame-drag field in Eqs.\ (\ref{eq:IRG22Radial}).
Because the functions $B_{1 {\rm (e)}}(r)$ and $B_{\rm I (e)}(r)$ determine the
radial dependence of the transverse part of the frame-drag field (and 
$B_{2 {\rm (e)}}(r)$ and $B_{\rm II (e)}(r)$ do the same for the 
transverse-traceless part), we see a particular illustration of the result of
Eq.\ (\ref{eq:DeltaB}) of Sec.\ \ref{sec:ExSchwSlicingChange}: namely, a change
in slicing from an electric-parity gauge change will induce a change in the 
longitudinal-transverse components of the frame-drag field (but not the 
longitudinal or transverse-traceless parts).

\section{Horizon Tendicity and Vorticity Calculated from the Weyl Scalar
$\Psi_0$}
\label{sec:HorizonDetails}

In this appendix, we modify a calculation by Hartle \cite{Hartle:1974} to 
compute the horizon tendicity and vorticity by applying differential operators
of the background spacetime to a perturbation of $\Psi_0 \equiv \psi_{2}$ 
(where $\psi_{2}$ satisfies Teukolsky's equation; see Appendix
\ref{sec:Teukolsky}).
Using this result, we derive the duality between the horizon vorticity and
tendicity of opposite-parity perturbation mentioned in Sec.\ 
\ref{sec:HorizonQuant}, for both Schwarzschild and Kerr black holes.
We also relate the horizon quantities to the complex curvature and show that 
they are proportional for Schwarzschild holes and differ only by the product 
of spin coefficients $\lambda^{(0)} \sigma^{(1)}$ for Kerr holes.
This proves these claims made in Sec.\ \ref{sec:NewTools}.

\subsection{Constructing a hypersurface-orthogonal tetrad on the horizon}

As in Hartle's calculation, we must work in a NP tetrad in which the null 
vector $\vec l$ is tangent to the horizon, $\vec n$ is normal to the horizon,
and $\vec m$ and its complex conjugate lie in the instantaneous horizon 
(constant $v=\tilde t + r$ for Hartle, though we will use constant $\tilde t$).
This NP tetrad must also satisfy additional constraints
\begin{subequations}
\label{eq:NullAndOrtho}
\begin{align}
\vec u & = \frac 1{\sqrt 2} (\vec l + \vec n) \, \\
\vec N & = \frac 1{\sqrt 2} (\vec l - \vec n) \, \\
\vec m & = \frac 1{\sqrt 2} (\vec e_2 + i \vec e_3) \, 
\end{align}
\end{subequations}
[with the associated non-null tetrad given by Eqs.\ (\ref{eq:KSbasis}) and 
(\ref{eq:PerturbedEFTetrad})], which ensure that the slicing vector $\vec u$ 
associated with this NP tetrad is hypersurface-orthogonal on the horizon and 
the spatial basis vectors are tied to our coordinate system in the desired way.

To describe the unperturbed NP tetrad, it is useful to first construct
Hartle's tetrad, which can be obtained from Kinnersley's tetrad 
(\ref{eq:KinnM}), by a boost followed by a null rotation about $\vec l$ (also 
called class III and I transformations, respectively):
\begin{subequations}
\begin{align}
\vec l_{\rm H} & = \frac{\Delta}{2(r^2+a^2)} \vec l_{\rm K} \, , \\ 
\vec m_{\rm H} & = \vec m_{\rm K} - 
\frac{ia\sin\theta}{\sqrt 2(r+ia\cos\theta)} \vec l_{\rm H} \, , \\
\vec n_{\rm H} & = \frac{2(r^2+a^2)}\Delta \vec n_{\rm K} 
+\frac{ia\sin\theta}{\sqrt 2(r+ia\cos\theta)} \vec m_{\rm K} \nonumber \\
& -\frac{ia\sin\theta}{\sqrt 2(r+ia\cos\theta)} \vec m^*_{\rm K}
+ \frac{a^2\sin^2\theta}{2 \Sigma} \vec l_{\rm H} \, .
\end{align}
\end{subequations}
The quantites $\Delta$ and $\Sigma$ are defined in Eq.\ 
(\ref{eq:DeltaSigmaDef}).
Then, we can construct an unperturbed tetrad from Hartle's tetrad using the
following spin-boost transformation (also called class III):
\begin{subequations}
\begin{equation}
\vec l_{(0)} = N_l \vec l_{\rm H} \, , \quad \vec m_{(0)} = e^{i\Theta} 
\vec m_{\rm H}  \, , \quad \vec n_{(0)} = N_l^{-1} \vec n_{\rm H} \, ,
\end{equation}
where
\begin{equation}
N_l = \sqrt{\frac{\Sigma + 2Mr}{2\Sigma}} \, , \quad 
e^{i\Theta} = \frac{r + i a\cos\theta}{\sqrt \Sigma} \, .
\end{equation}
\end{subequations}

One can verify that the resulting orthonormal tetrad
\begin{subequations}
\begin{align}
\vec u^{\, (0)} & = \frac 1{\sqrt 2} (\vec l_{(0)} + \vec n_{(0)}) \, ,\\
\vec N^{\, (0)} & \equiv \vec e_{\hat r}^{\, (0)} 
= \frac 1{\sqrt 2} (\vec l_{(0)} - \vec n_{(0)}) \, ,\\
\vec e_{\hat \theta}^{\, (0)} & = \frac 1{\sqrt 2} (\vec m_{(0)} 
+ \vec m_{(0)}^*) \, ,\\
\vec e_{\hat \theta}^{\, (0)} & = \frac 1{i\sqrt 2} (\vec m_{(0)} 
- \vec m_{(0)}^*) \, ,
\end{align}
\end{subequations}
is exactly the ingoing-Kerr tetrad (\ref{eq:KSbasis}), when evaluated on the 
horizon, though away from the horizon it is not.

For the NP null tetrad to correspond, on the horizon, 
to the hypersurface-orthogonal 
$\{\vec u, \vec e_{\hat r}, \vec e_{\hat\theta}, \vec e_{\hat\phi}\}$
of Eqs.\ (\ref{eq:KSbasis}) and (\ref{eq:PerturbedEFTetrad})
via Eqs.\ (\ref{eq:NullAndOrtho}), 
we must choose the perturbative corrections to the tetrad to satisfy 
\begin{subequations}
\begin{align}
\vec l_{(1)} & = \frac 1{2\sqrt 2} [(h_{\hat 0\hat 0} \vec u_{(0)} 
- h_{\hat r\hat r} \vec e_{\hat r}^{\, (0)}) - 2 h_{\hat 0 \hat i} 
\vec e^{\, \hat i}_{(0)} - 2 h_{\hat r \hat A} \vec e^{\, \hat A}_{(0)}] \, ,\\
\vec n_{(1)} & = \frac 1{2\sqrt 2} [(h_{\hat 0\hat 0} \vec u_{(0)} + 
h_{\hat r\hat r} \vec e_{\hat r}^{\, (0)}) - 2h_{\hat 0 \hat i} 
\vec e^{\, \hat i}_{(0)} + 2 h_{\hat r \hat A} \vec e^{\, \hat A}_{(0)}] \, ,\\
\vec m_{(1)} & = -\frac 1{2\sqrt 2} [(h_{\hat \theta \hat \theta}
\vec e_{\hat \theta}^{\, (0)} + i h_{\hat \phi\hat \phi} 
\vec e_{\hat \phi}^{\, (0)}) + 2 h_{\hat \theta \hat \phi} 
\vec e_{\hat \phi}^{\, (0)}] \, .
\end{align}
\end{subequations}
Because $\vec e_{\hat r}$ is normal to surfaces of constant $r$ in slices of 
constant $\tilde t$ through perturbative order, we will need to choose our 
gauge so that the coordinate position of the horizon does not move from 
the constant value $r=r_+$.
Although Poisson \cite{Poisson2004} has shown that there are a wide class
of gauges that satisfy this property (horizon-locking gauges), for our 
calculation, we find it convenient to work in an ingoing radiation gauge based 
on the unperturbed tetrad vector $\vec l_{(0)}$,
\begin{equation}
h_{\mu\nu} l^\nu_{(0)} = 0 \, , \quad g^{\mu\nu} h_{\mu\nu}=0  \, .
\end{equation}
On the horizon, these gauge conditions imply that
\begin{subequations}
\begin{equation}
h_{\hat 0\hat 0} = h_{\hat r\hat r} = - h_{\hat 0\hat r} \, , 
\end{equation}
and that the null vector $\vec l$ undergoes a perturbative boost,
\begin{equation}
\vec l_{(1)} = \frac 12 h_{\hat 0\hat 0} \vec l_{(0)} \, .
\end{equation}
\end{subequations}
[To derive this, one should split 
$-2 h_{\hat 0 \hat i} \vec e^{\, \hat i}_{(0)}$ into a sum of two terms
$-2 (h_{\hat 0 \hat r} \vec e^{\, \hat r}_{(0)} + h_{\hat 0 \hat A} \vec 
e^{\, \hat A}_{(0)})$ and use the relation in Eq.\ (\ref{eq:NullAndOrtho}).]
In addition to keeping the horizon at a constant coordinate position $r=r_+$
(see \cite{Poisson2004}), using this gauge condition allows us to calculate
the perturbation to $\Psi_2$ in a much simpler way, as we describe in the next
subsection.

\subsection{Computing the horizon tendicity and vorticity from $\Psi_0$}
\label{App:HorizonFromPsi0}

Although the explicit expressions for spin coefficients in this tetrad are
somewhat lengthy (and, as a result, we do not give them here), through a direct
calculation we can verify that on the horizon
\begin{subequations}
\begin{equation}
\rho_{(0)} = \sigma_{(0)} = \kappa_{(0)} = 0 \, , \quad 
\epsilon_{(0)} \in \mathbb R \, .
\end{equation}
Moreover, because in this ingoing radiation gauge the perturbation to the 
vector $\vec l$ can be obtained by applying a boost (Class III) transformation 
to the tetrad, the perturbed value of $\kappa$ will also vanish,
\begin{equation}
\kappa_{(1)} = 0 \, .
\end{equation}
From Eq.\ (310a) of Ch.\ 1 of \cite{ChandrasekharBook} (which describes the
components of the Riemann tensor in the Newman-Penrose formalism), we see 
that the perturbation to the spin-coefficient $\rho$ satisfies an equation
\begin{equation}
{\bf D}_{(0)}\rho_{(1)} = 2\epsilon_{(0)} \rho_{(1)} \, ,
\nonumber
\end{equation}
where $\epsilon_{(0)} > 0$.
If $\rho_{(1)}$ is not zero, then the separated solution to this equation, 
$\rho_{(1)} = f(r,\theta)e^{-i(\omega \tilde t -m\tilde \phi)}$, 
implies the constraint that $2\epsilon_{(0)}+i N_l(\omega - m\omega_+) = 0$ 
[here $\omega_+=a/(2Mr_+)$ is the horizon angular velocity].
This condition is not satisfied for constant frequencies $\omega$, so 
the perturbation to the spin coefficient must vanish:
\begin{equation}
\rho_{(1)}=0 \, .
\end{equation}
\end{subequations}

From these conditions on the spin coefficients, and the fact that 
$\Psi_0^{(0)} = \Psi_1^{(0)} = 0$, we can write the Bianchi identities (see, 
e.g., Eqs.\ (321a) and (321b) of Ch.\ 1 of \cite{ChandrasekharBook}) as
\begin{subequations}
\begin{align}
({\bf D} -2\epsilon) \Psi_1 & = ({\bm \delta}^* + \pi - 4\alpha) \Psi_0 \, , \\
{\bf D} \Psi_2^{(1)} & = ({\bm \delta}^* + 2\pi - 2\alpha) \Psi_1 \, , 
\end{align}
\label{eq:DPsiEvoln}
\end{subequations}
where we have dropped the superscripts indicating perturbative orders on all 
differential operators and spin coeffients (because they are all background
quantities) and the Weyl scalars $\Psi_0$ and $\Psi_1$ (because they are
strictly perturbative quantities).
Note that we do not need the term of the form ${\bf D}_{(1)} \Psi_2^{(0)}$, 
because on the horizon the differential operator ${\bf D}_{(1)}$ contains only 
time and azimuthal-angle derivatives, but the background Weyl scalar 
$\Psi_2^{(0)}$ is only a function of $r$ and $\theta$.

By applying the differential operator $({\bf D} -2\epsilon)$ to the second Bianchi
identity and using the identity (valid on the horizon) that 
${\bm \delta}^* {\bf D} - {\bf D}{\bm \delta}^* = (\alpha + \beta^* - \pi) {\bf D}$, we find that
\begin{align}
({\bf D}-2\epsilon){\bf D}\Psi_2^{(1)} &= [{\bm \delta}^* 
+ 3(\pi-\alpha)-\beta^*)]({\bm \delta}^*+\pi-4\alpha)\Psi_0 \nonumber \\
& -({\bf D}-2\epsilon)(\lambda \Psi_0)\, ,
\end{align}
Using Geroch-Held-Penrose \cite{Geroch1973} notation, and the equation for
a component of the Riemann tensor (Eq.\ (310g) of  Ch.\ 1 of 
\cite{ChandrasekharBook}) restricted to the horizon 
\begin{equation}
{\bf D}\lambda - {\bm \delta}^*\pi = -2\epsilon\lambda 
+ \pi(\pi+\alpha-\beta^*) \, ,
\end{equation}
we find that
\begin{equation}
\mbox{\thorn\thorn}\Psi_2^{(1)} = (\eth'\eth' + 4\pi \eth' + 2\pi^2 
-\lambda \mbox{\thorn} )\Psi_0 \, .
\label{eq:Psi0DrivePsi2}
\end{equation}

Note that the $\Psi_0$ here is related to that which satisfies Teukolsky's
equation in the Kinnersley tetrad by
\begin{equation}
\Psi_0 = \frac{N_l^2 e^{2i\Theta} \Delta^2}{(r^2+a^2)^2} \Psi_0^{\rm K} 
\equiv A \Psi_0^{\rm K}\, .
\end{equation}
Starting from a modal solution for the Kinnersley $\Psi_0$ (denoted by 
$\Psi_{0\, lm\omega}^{\rm K}$), then we see that the corresponding perturbation
to $\Psi_2$ is given by
\begin{equation}
\label{eq:Psi2Mode}
\Psi_{2 \, lm\omega}^{(1)} = \frac{\eth'\eth' + 4\pi \eth' + 2\pi^2 +\lambda 
(iN_l\Omega + 4\epsilon)}{N_l\Omega(2i\epsilon - N_l\Omega)} 
(A\Psi_{0\, lm\omega}^{\rm K}) \, ,
\end{equation}
where $\Omega = \omega - m\omega_+$, and where we have used the fact that 
${\bf D} = N_l(\partial_{\tilde t} + \omega_+ \partial_{\tilde \phi})$ on the
horizon.

A Weyl scalar, $\Psi_0$ formed from the superposition of modes 
$\Psi_{0\, lm\omega}^{\rm K} \pm (-1)^m \Psi_{0\, l-m-\omega^*}^{\rm K}$, with
radial functions that obey 
${}_2R_{l-m-\omega^*} = \pm (-1)^m {}_2R_{lm\omega}^*$, transforms under parity
as $\Psi_0 \rightarrow \pm (-1)^l \Psi_0^*$.
The perturbation of $\Psi_2$ formed from superimposing Eq.\ 
(\ref{eq:Psi2Mode}) for the individual modes of $\Psi_0^{\rm K}$ above also 
transforms under parity as $\Psi_2 \rightarrow \pm (-1)^l \Psi_2^*$.
Using the relation $2\Psi_2 = \mathcal E_{NN} + i\mathcal B_{NN}$ and taking
the real and imaginary parts of $\Psi_2$, it becomes clear that 
$\mathcal E_{NN}$ and $\mathcal B_{NN}$ have definite parity.
Moreover, it is not difficult to see that $\mathcal E_{NN}$ of an 
electric-parity mode is equal to $i\mathcal B_{NN}$ of a magnetic-parity
mode of $i\Psi_{0\, lm\omega}^{\rm K}$, and $\mathcal E_{NN}$ of a 
magnetic-parity mode is $-i\mathcal B_{NN}$ of an electric-parity mode of 
$i\Psi_{0\, lm\omega}^{\rm K}$.

\emph{This demonstrates a perfect duality between electric-parity modes
and magnetic-parity modes, on the horizon of a Kerr black hole.}

\subsection{Relationship between $\Psi_2$ and the complex curvature}

As a final part of this appendix, we discuss how the relationship between the
complex curvature and $\Psi_2$,
\begin{equation}
\frac 14(\mathcal R + i\mathcal X) = -\Psi_2 + \mu\rho -\lambda \sigma \, ,
\label{eq:ComplexCurvature}
\end{equation}
simplifies for perturbations of Schwarzschild and Kerr black holes in the 
tetrad and gauge discussed in the sections of this appendix above.
First, we note that the spin coefficient $\lambda$ has as its unperturbed
value on the horizon
\begin{align}
\lambda_{(0)} =& -\frac{M r_+ \omega_+^2 \sin^2\theta e^{-2i\Theta}}
{N_l(r-ia\cos\theta)^3} \nonumber \\
& \times [4Mr_+ + (r_+-M)(r_+-ia\cos\theta)] \, ,
\end{align}
where we have made use of the fact that on the horizon $r_+^2+a^2=2Mr_+$.
For a Schwarzschild black hole, $\omega_+$ vanishes, and, therefore, the 
background values of all four spin coefficients $\mu$, $\rho$, $\lambda$, 
and $\sigma$ all vanish.
Through first-order in perturbation theory, therefore,
\begin{equation}
\mathcal R = -2 \mathcal E_{NN} \, , \quad \mathcal X = -2 \mathcal B_{NN} \, .
\end{equation}

[We briefly digress here to note that for Schwarzschild black hole, the spin
coefficient $\pi$ also vanishes, and Eq.\ (\ref{eq:Psi0DrivePsi2}) reduces to
\begin{equation}
\mbox{\thorn\thorn}\Psi_2^{(1)} = \eth'\eth' \Psi_0 \, .
\end{equation}
For a modal solution, Eq.\ (\ref{eq:Psi2Mode}) also simplifies to 
\begin{equation}
\Psi_{2 \, lm\omega}^{(1)} = \frac{4M \sqrt{2D}\alpha^4}
{\omega(i-4M\omega\sqrt 2)} {}_{2}R_{lm} Y_{lm} e^{-i\omega t} \, ,
\label{eq:Psi2SchwPsi0}
\end{equation}
where $\alpha^2 = 1-2M/r$, $D=(l+2)!/(l-2)!$, $\omega = \Omega$ (because 
$\omega_+ = 0$), and the radial function of $\psi_2$, ${}_{2}R_{lm}$, is 
evaluated at the horizon $r=2M$.
We have also used the fact that $\epsilon = 1/(8M)$ in this tetrad.
Because the spin coefficients vanish in Eq.\ (\ref{eq:ComplexCurvature}) for 
this perturbed Schwarzschild hole, the above expression is equivalent to minus 
one quarter of the complex curvature.]

For a Kerr black hole $\lambda_{(0)} \neq 0$, and we must compute the 
perturbation to $\sigma$.
It satisfies the differential equation
\begin{equation}
({\bf D}-2\epsilon) \sigma_{(1)} = \Psi_0 \, 
\end{equation}
[Eq.\ (310b) of Ch.\ 1 of \cite{ChandrasekharBook} specialized to our
tetrad and gauge]. 
For a modal solution of $\sigma_{(1)}$, we can solve this to find
\begin{equation}
\sigma_{(1)} = -\frac{\Psi_0}{iN_l\Omega + 2\epsilon} \, ,
\end{equation}
which implies that the perturbation to $\sigma$ does not vanish.
Thus, for a Kerr black hole, 
\begin{equation}
\frac 14(\mathcal R_{(1)} + i\mathcal X_{(1)}) = -\Psi_2^{(1)} -\lambda_{(0)} 
\sigma_{(1)} \, ,
\end{equation}
so the horizon tendicity and vorticity are no longer exactly equal the 
horizon's intrinsic and extrinsic scalar curvatures.

\section{Vortex and Tendex Lines of $(2,2)$ Perturbations of Schwarzschild and 
Kerr Black Holes with the Background Frame-Drag and Tidal Fields}
\label{sec:ExtraFigures}

\begin{figure*}
\includegraphics[width=0.6\columnwidth]{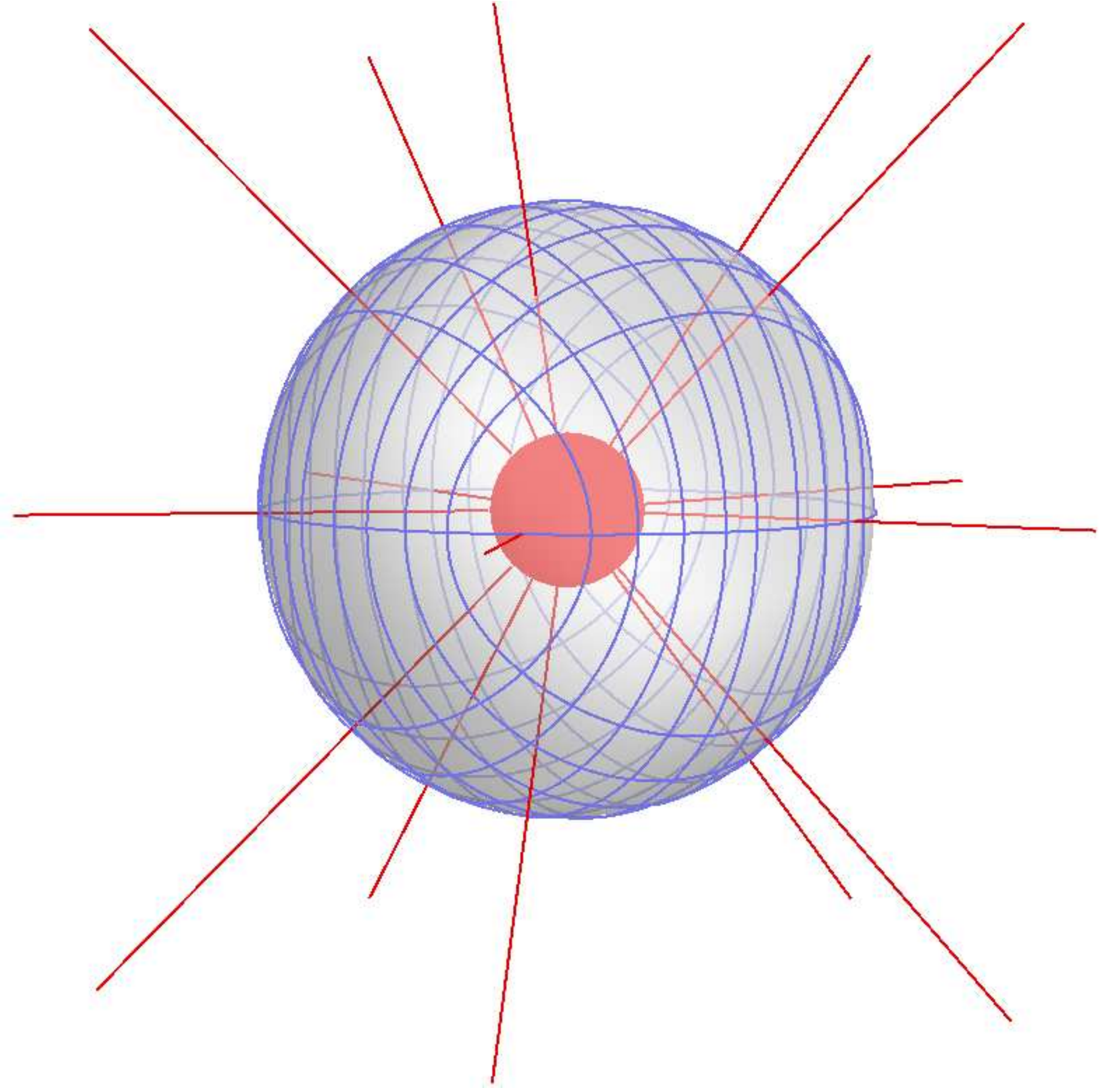}
\includegraphics[width=0.6\columnwidth]{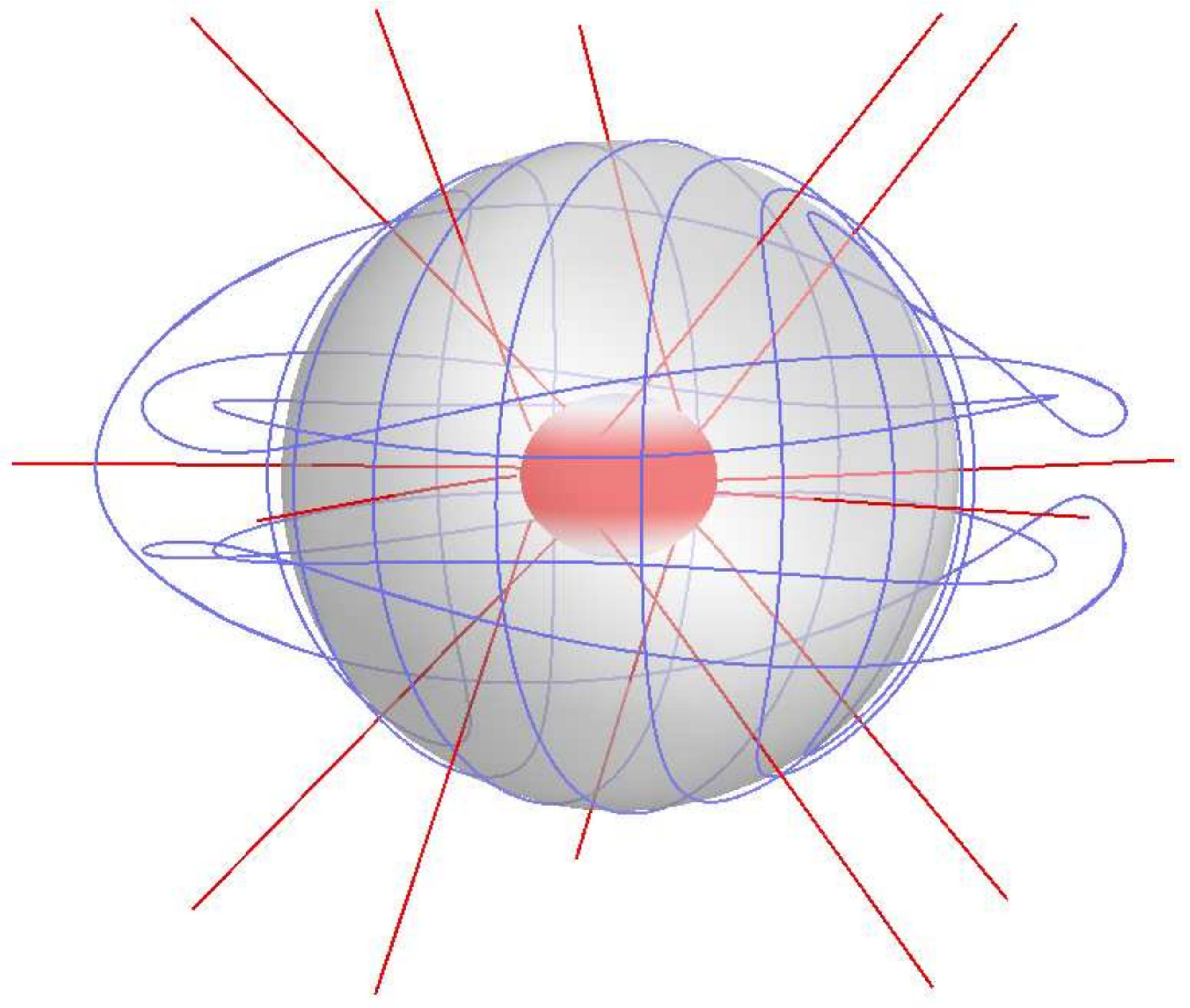}
\includegraphics[width=0.6\columnwidth]{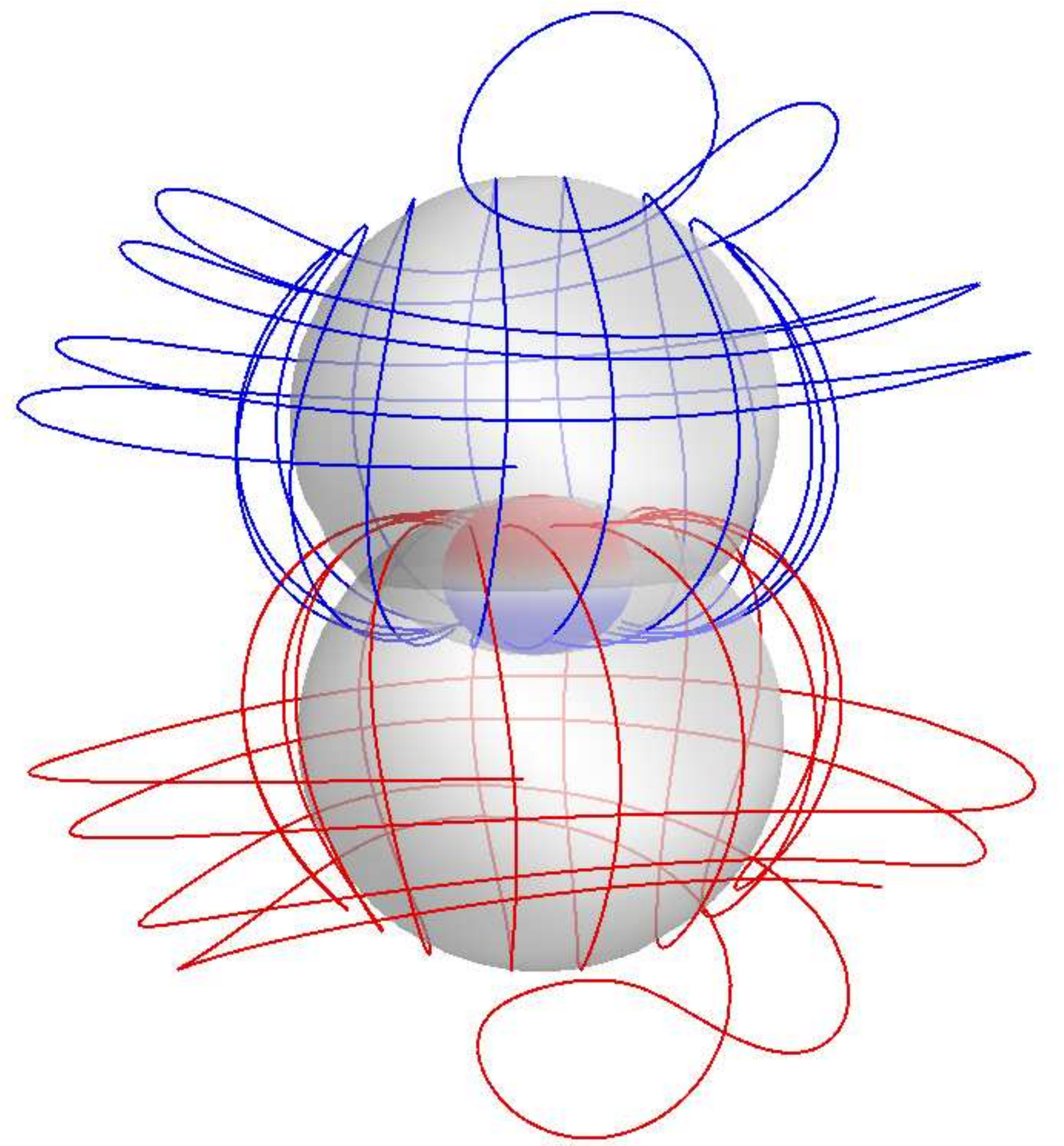}
\includegraphics[width=0.6\columnwidth]{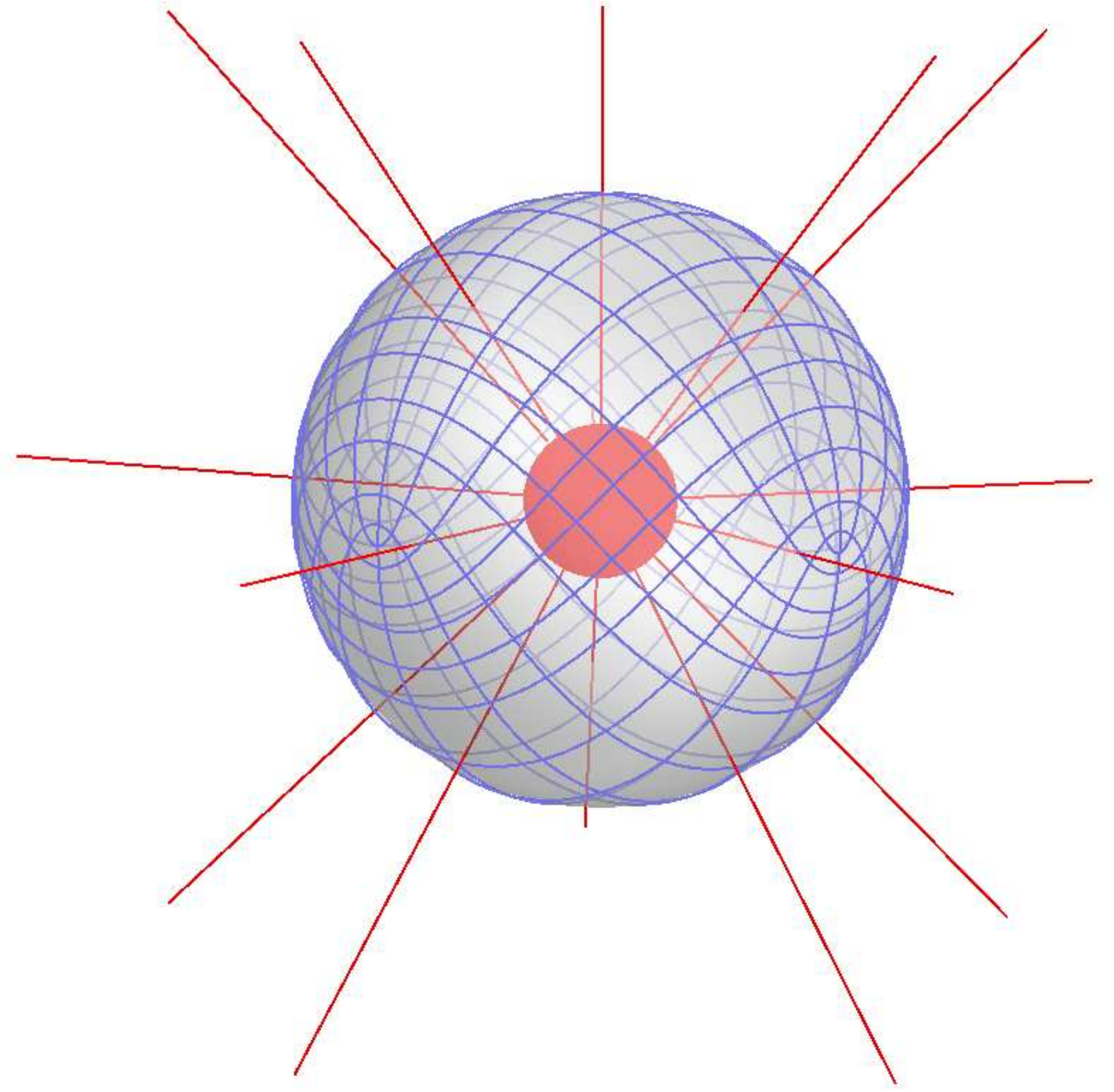}
\includegraphics[width=0.6\columnwidth]{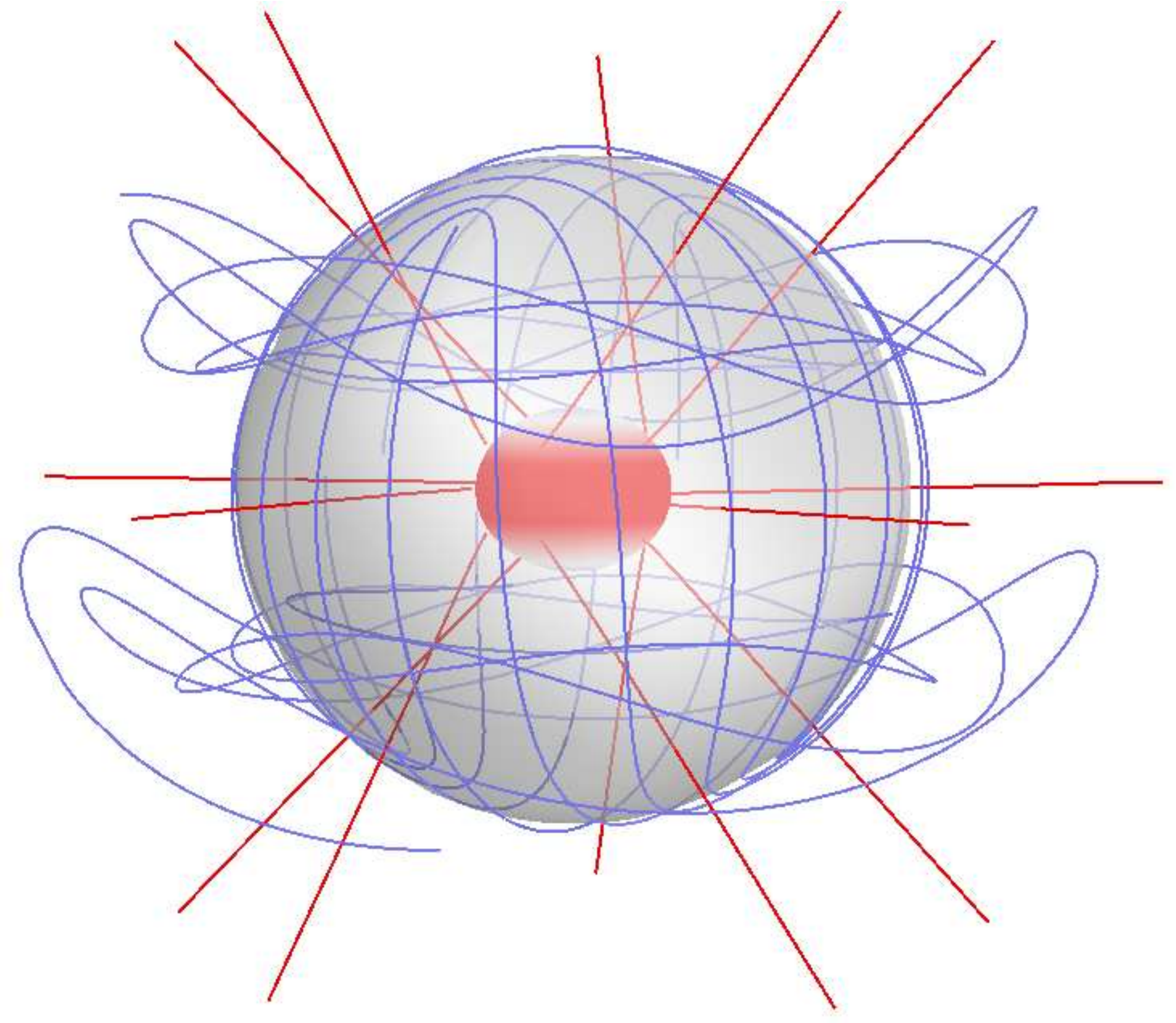}
\includegraphics[width=0.6\columnwidth]{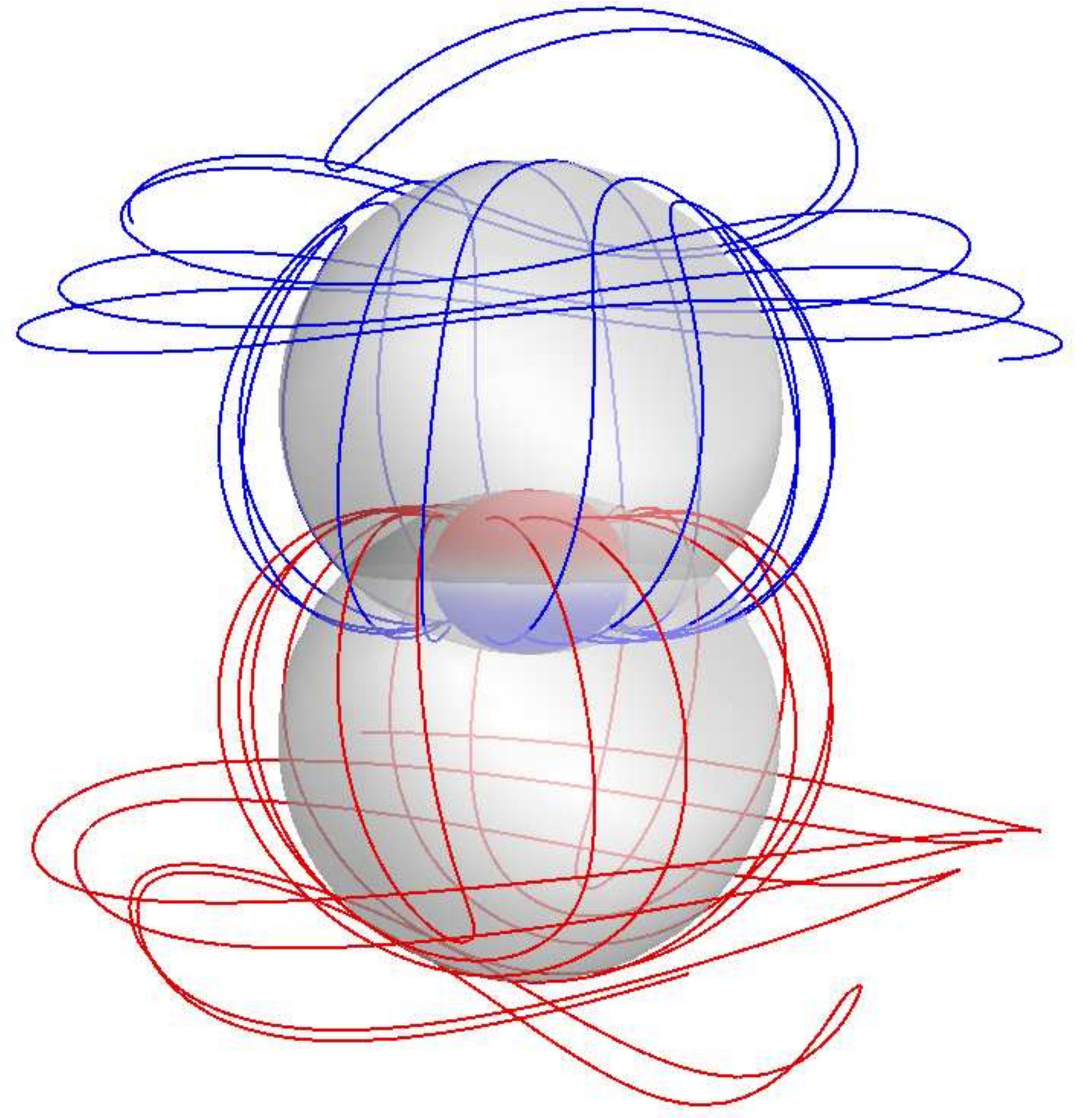}
\caption{(color online).
Tendex and vortex lines of Schwarzschild and Kerr black holes
(of spin $a/M=0.945$) perturbed by a $(2,2)$ mode of either electric or 
magnetic parity, without removing the background tidal or frame-drag fields.
The tendex lines and vortex lines are colored by the signs of their respective
tendicities and vorticities (blue [dark gray] for positive and red [light gray]
for negative).
The horizons are colored and shaded by their vorticities or tendicities,
and the transparent spheres have no physical significance, but they help to
add perspective to the figures.
The top panels are electric-parity perturbations and the bottom panels are 
magnetic-parity ones.
{\it Left column}: Tendex lines of Schwarzschild black holes.
{\it Middle column}: Tendex lines of Kerr black holes.
{\it Right column}: Vortex lines of Kerr black holes.}
\label{fig:KerrTendexVortexFull}
\end{figure*}

In this appendix, we show the tendex and vortex lines of Schwarzschild and
Kerr black holes when we plot a small $(2,2)$ perturbation of either electric
or magnetic parity on top of the background tidal or frame-drag fields in 
Fig.\ \ref{fig:KerrTendexVortexFull}.
Specifically, we plot the vortex and tendex lines of 
\begin{equation}
\boldsymbol{\mathcal E} = \boldsymbol{\mathcal E}^{(0)} 
+ \epsilon \boldsymbol{\mathcal E}^{(1)} \, , \quad
\boldsymbol{\mathcal B} = \boldsymbol{\mathcal B}^{(0)}
+ \epsilon \boldsymbol{\mathcal B}^{(1)} \, ,
\end{equation}
where $\boldsymbol{\mathcal E}^{(0)}$ and $\boldsymbol{\mathcal B}^{(0)}$
are the stationary, unperturbed background fields (visualized in Paper II),
$\boldsymbol{\mathcal E}^{(1)}$ and $\boldsymbol{\mathcal B}^{(1)}$ are the
perturbations (visualized by themselves in Fig.\ \ref{fig:TendexVortex3D}),
and $\epsilon$ is a constant that sets the scale of the perturbation.
To describe the strength of the perturbation, we will compare the perturbative
horizon tendicity or vorticity to the background tendicity (for Schwarzschild
holes) and the tendicity or vorticity (for Kerr black holes).
For the Schwarzschild black holes in Fig.\ \ref{fig:KerrTendexVortexFull},
we chose $\mathcal E_{NN}^{(1)}/\mathcal E_{NN}^{(0)} \approx 2\times10^{-4}$
for the electric-parity perturbations and 
$\mathcal B_{NN}^{(1)}/\mathcal E_{NN}^{(0)} \approx 2\times10^{-4}$ for the
magnetic-parity perturbations.
For electric-parity perturbations of Kerr holes, we chose
$\mathcal E_{NN}^{(1)}/\mathcal E_{NN}^{(0)} \approx 3.5\times10^{-3}$ and
$\mathcal B_{NN}^{(1)}/\mathcal B_{NN}^{(0)} \approx 3\times10^{-3}$, and
for the magnetic-parity perturbations the ratios we selected were
$\mathcal E_{NN}^{(1)}/\mathcal E_{NN}^{(0)} \approx 2.5\times10^{-3}$ and
$\mathcal B_{NN}^{(1)}/\mathcal B_{NN}^{(0)} \approx 5\times10^{-3}$.
We anticipate that these images may be useful for comparing with the results of
numerical-relativity simulations, in which there is more ambiguity about how to
separate a spacetime into a stationary background and dynamical perturbations,
and for which it may be more useful to visualize the full frame-drag and tidal
fields.

In the top panels of Fig.\ \ref{fig:KerrTendexVortexFull} are electric-parity
perturbations, and the bottom panels are magnetic-parity perturbations.
The left column of images are tendex lines of Schwarzschild black holes,
the center column are tendex lines of Kerr black holes of spin $a/M=0.945$,
and the right column are the corresponding vortex lines of the perturbed Kerr
black holes.
The lines are colored by the sign of their tendicity or vorticity (blue [dark
gray] for positive and red [light gray] for negative) and the horizons are 
colored by their tendicity or vorticity.
The transparent spheres are placed in the figures to help guide the eye, and
do not indicate any feature of the vortexes or tendexes.

In these figures, we must choose an amplitude for the perturbation (described
in the first paragraph above).
For the all the black holes, we make the perturbation sufficiently small 
that one cannot see the effect of the perturbation in either the horizon 
tendicity, or the red (light gray) radial tendex lines.
For the Kerr holes, we also require that the amplitude of the perturbation
is less than the difference of the tendicities of the two non-radial tendex 
lines at the equatorial plane and around the radius at which the angular
lines reach closest to the horizon.
With this choice, the angular tendex lines will retain some features of the
unperturbed lines before they become more distorted by the perturbation in
the regions near the poles.

First, we will describe the tendex lines of the Schwarzschild black holes.
An unperturbed Schwarzschild black hole is spherically symmetric, the 
tendicity on a sphere of constant radius is constant, and, therefore, any 
direction tangent to the sphere is a valid tendex line.
For a weakly perturbed Schwarzschild black hole, although the perturbation
may be small, the perturbation restricted to a sphere of constant $r$
completely determines the variation in the tendicity, and, furthermore, it
will determine the directions of the tendex lines.
This is analogous to degenerate perturbation theory in quantum mechanics, 
in which the eigenstates of the perturbing Hamiltonian restricted to the 
subspace spanned by the degenerate eigenstates are treated as the unperturbed
states within the degenerate subspace.
In directions that are not degenerate, however, its effects are negligible.

We can now use these facts about degeneracy to understand the tendex lines in 
the angular direction.
The tidal field in the strictly angular directions, Eq.\ 
(\ref{eq:AngularTidalFieldEven}) will determine the structure of the tendex 
lines on the sphere.
The angular dependence is determined by the transverse-traceless, 
electric-parity tensor harmonic (for the top-left panel), because the trace 
term in Eq.\ (\ref{eq:AngularTidalFieldEven}) is proportional to the identity 
and will not lift the degeneracy of the tendex lines.
We would expect, therefore, that the tendex lines in the angular direction 
would resemble those of transverse-traceless, $l=2$, $m=2$, gravitational 
waves generated by a time-dependent mass quadrupole.
These were shown in \cite{Zimmerman2011,Nichols:2011pu}, and the pattern of
the lines is nearly identical.
The tendicity along the lines is quite different from those of a gravitational 
wave, because for the perturbed Schwarzschild black hole, the tendicity is 
primarily determined by the constant unperturbed value on the sphere. 
Nevertheless, the tendex lines on the sphere show a striking similarity to 
those of gravitational waves at infinity.

For the magnetic-parity perturbation (the bottom-left panel), the tendex lines 
are determined by an $l=2$, $m=2$, magnetic-parity tensor harmonic; 
consequently, we would expect that the lines would resemble those of 
transverse-traceless gravitational waves at infinity, produced by a 
time-dependent, current-quadrupole source.
Those lines were shown in \cite{Zimmerman2011}, and they appear identical.
Once more, though, the value of the tendicity along the lines is set by the 
background Schwarzschild black hole for the lines in bottom-left panel of Fig.\ 
\ref{fig:KerrTendexVortexFull} (unlike the tendicity of the lines studied in 
\cite{Zimmerman2011}).

The degeneracy between the angular tendex and vortex lines can also be used 
to explain the tendex and vortex lines in the middle and right columns of 
Fig.\ \ref{fig:KerrTendexVortexFull}, respectively.
For both the tendex and vortex lines, when the lines are near the equatorial 
plane ($\theta=\pi/2$) they resemble the unperturbed lines, but as they head
toward the poles, they begin to become perturbed.
This happens because the perturbation is small compared to the difference in 
the eigenvalues near the equatorial plane, and the perturbations have little 
effect on the tendex or vortex lines.
Near the poles, however, the background vorticities and tendicities in the 
angular directions become degenerate (see the discussion at the end of Apps.~A and~B of Paper II), and the perturbation restricted to the degenerate 
subspace controls the lines' directions.
In the vicinity of the poles, the degenerate subspace is a plane parallel 
to the equatorial plane, and the perturbative tendex lines must form a
regular grid around these points.
When we combine this observation with the parity of the perturbation, we
see that the lines at the opposite poles must be either parallel or orthogonal.
Thus, these few simple constraints combine to explain the relatively simple
pattern of the vortex and tendex lines of the perturbation plus the background
frame-drag and tidal fields.

\bibliography{References/References}

\end{document}